%% file: Thesis.tex
\documentclass[oneside,12pt]{thesis}

\usepackage{lscape}
\usepackage{doublespace}
\usepackage{xspace}
\usepackage{amsthm}
\usepackage{amsmath}
\usepackage{amssymb}
\usepackage{epsfig}
\usepackage{pifont}
\usepackage{enumerate}
\usepackage{version}

\usepackage{float}

\renewcommand{\thesubsubsection}{\thesubsection.\alph{subsubsection}}

\theoremstyle{break}

\newtheorem{prop}{Proposition}[chapter]
\newtheorem{conj}{Conjecture}[chapter]

\def\lgth{[\,\mbox{length}\,]}

\def\eps{\epsilon}

\def\Om{\Omega}

\def\gam{\gamma}
\def\hsp5{\hspace{5mm}}

\def\poi{\theta}

\def\lb{\label}
\def\be{\begin{equation}}
\def\ee{\end{equation}}
\def\la{\langle}
\def\ra{\rangle}

\def\bigO{\mathcal{O}}
\def\bigt{\Theta}
\def\me{\mathbf{e}}
\def\udot{\dot{u}}
\def\parb{\pmb{\partial}}

\def\X{\mathbf{X}}
\def\Y{\mathbf{Y}}

\newcommand{\EEE}{E_1{}^1}
\newcommand{\Sp}{\Sigma_+}

\newcommand{\Udot}{\dot{U}}
\newcommand{\Sm}{\Sigma_-}
\newcommand{\Nc}{N_\times}
\newcommand{\Sc}{\Sigma_\times}
\newcommand{\St}{{\Sigma_2}}
\newcommand{\Stt}{{\Sigma_3}}
\newcommand{\Nm}{N_-}
\newcommand{\Np}{N_+}
\newcommand{\Oml}{\Omega_\Lambda}
\newcommand{\arctanh}{{\rm arctanh}}
\renewcommand{\varphi}{\phi}

\def\Smt{\tilde{\Sigma}_-}
\def\Sct{\tilde{\Sigma}_\times}
\def\Nmt{\tilde{N}_-}
\def\Nct{\tilde{N}_\times}
\def\ptl{\partial}
\def\Sph{{\Sigma_+^H}}
\def\Smh{{\Sigma_-^H}}

\def\Scb{\pmb{\Sc}}
\def\Nmb{\pmb{\Nm}}
\def\Npb{\pmb{\Np}}
\def\Stb{\pmb{\St}}

\def\Ep{\mathcal{E}_+}
\def\Em{\mathcal{E}_-}
\def\Ec{\mathcal{E}_\times}
\def\Et{{\mathcal{E}_2}}
\def\Ett{{\mathcal{E}_3}}

\def\Hp{\mathcal{H}_+}
\def\Hm{\mathcal{H}_-}
\def\Hc{\mathcal{H}_\times}
\def\Ht{{\mathcal{H}_2}}
\def\Htt{{\mathcal{H}_3}}

\def\fp{f_1}
\def\fpp{f_2}
\def\rty{\sqrt{3}}
\def\fz{f_0}

\def\foo{\Lambda \ell^2}

\newcommand{\dz}{\parb_0}
\newcommand{\di}{\parb_1}
\newcommand{\dt}{\partial_t}
\newcommand{\dx}{\partial_x}

\begin{document}
\bibliographystyle{ilnuerdo}

\begin{spacing}{1.0}

\newcommand{\etal}{\emph{et}\@ \emph{al}.\@\xspace}
\newcommand{\etc}{etc.\@\xspace}
\renewcommand{\vec}{\mathbf}
\newcommand{\dd}[3][\!]{\ensuremath{\frac{d^{#1}{#2}}{d {#3}^{#1}}}}
\newcommand{\didi}[3][\!]{\ensuremath{\frac{\partial^{#1}{#2}}{\partial
{#3}^{#1}}}}

\pagenumbering{roman}

\input{front}

\pagestyle{plain}

\thispagestyle{plain}
\tableofcontents
\thispagestyle{plain}
\listoftables
\thispagestyle{plain}
\listoffigures
\thispagestyle{plain}
\newpage

\pagenumbering{arabic}
\pagestyle{headings}

\input{intro}

\part{Dynamics of $G_2$ and SH cosmologies}\label{part:G2}

\input{off}

\input{G2_hierarchy}

\input{SH_dynamics}

\input{G2_dynamics}

\input{explicit_sol}

\input{past}

\part{Numerical exploration of $G_2$ cosmologies}\label{part:num}

\input{eqs}
\input{simulations}

\input{conclusion}



\appendix

\input{app_action}

\input{rotation}

\input{expansion}

\input{appG2}

\bibliography{cites}

\end{spacing}

\end{document}

%% file: front.tex
\thispagestyle{empty}
\begin{centering}
\vspace{1cm}

{\huge\sc The Dynamics of}

\vspace{0.5cm}

{\huge\sc Inhomogeneous Cosmologies}

\vspace{1cm}

by

\vspace{1cm}

{\Large Woei Chet Lim}

\vspace{1cm}

A thesis

presented to the University of Waterloo

in fulfilment of the 

thesis requirement for the degree of

Doctor of Philosophy

in

Applied Mathematics

\vspace{1cm}

Waterloo, Ontario, Canada, 2004

\vspace{1cm}

\Pisymbol{psy}{211} Woei Chet Lim 2004

\end{centering}
\pagebreak


\thispagestyle{plain}

%
%
%

\noindent
I hereby declare that I am the sole author of this thesis. This is a true 
copy of the thesis, including any required final revisions, as accepted by 
my examiners.

\

\noindent
I understand that my thesis may be electronically available to the public.

\pagebreak

\thispagestyle{plain}

\vspace*{3cm}

\begin{centering}

{\Large \bfseries   Abstract}

\end{centering}

\vspace{1cm}

In this thesis we investigate cosmological models more general than the
isotropic and homogeneous Friedmann-Lema\^{\i}tre models.
We focus on cosmologies with one spatial degree of freedom, whose
matter content consists of a perfect fluid and the cosmological constant.
We formulate the Einstein field equations as a system of quasilinear
first order partial differential equations, using scale-invariant
variables.

The primary goal is to study the dynamics in the two asymptotic regimes,
i.e. near the initial singularity and at late times.
We highlight the role of spatially homogeneous dynamics as the
background dynamics,
and analyze the inhomogeneous aspect of the dynamics.
We perform a variety of numerical simulations to support our analysis
and to explore new phenomena.

\pagebreak


\thispagestyle{plain}

\vspace*{3cm}

\begin{centering}

{\Large \bfseries   Acknowledgements}

\end{centering}

\vspace{1cm}

First of all, I want to thank my supervisor and mentor, John 
Wainwright, for his guidance over the course of my study (both Master's 
and 
doctorate). 
You are always enthusiastic and ready to help. 
I will do my best to become an exemplary scholar and teacher like you.

Special thanks go to my collaborators
Claes Uggla, Henk van Elst, Lars Andersson, Marsha Weaver and Ulf Nilsson,
with whom I have had many enlightening discussions on cosmology.
The timely $G_2$ project by Henk, Claes and John provided a bountiful 
field for this thesis.

I am grateful to Conrad Hewitt, Barry Collins and Ray McLenaghan 
for their helpful feedback on the thesis proposal.

I would like to acknowledge Alan Rendall, Achim Kempf and David Siegel for 
discussions on various analytical aspects of cosmology and differential 
equations.
On the numerical aspects, I want to thank Lars Andersson and Mattias 
Sandberg for their {\tt CLAWPACK} codes for the vacuum OT $G_2$ models, 
which helped me begin the captivating numerical project in this thesis.
Marsha's timely visit also helped speed up the project.
The brane cosmology project with Alan Coley and Yanjing He helped to
sharpen my skills in doing numerical simulations.
I would also like to acknowledge Matt Choptuik, David Garfinkle, Nathalie 
Lanson and Serge D'Alessio for discussions on numerical simulations. 

I have received all of my tertiary education at the University of 
Waterloo, and I would like to take this opportunity to
thank all the faculty members here who have 
taught me mathematics, physics and computer science.

Lastly, I dedicate this thesis to my parents in Malaysia, 
who have provided me with the best education they can afford.


%% file: intro.tex
	\chapter{Introduction}

\section{Background in cosmology}


The goal of relativistic cosmology is to describe the structure of
the observable universe on a sufficiently large scale, and to explain its
origins, using
Einstein's General Theory of Relativity.


The most fundamental cosmological observation is that light from
distant galaxies is redshifted (this was discovered by Hubble in the 
1920s). The
standard interpretation of the redshifts is that the universe is
expanding.
The second fundamental observation is that the universe is filled with
radiation (as discovered by Penzias \& Wilson in 1965), which has a
blackbody spectrum to a very high accuracy, corresponding to a present
temperature of about 3 K.
The standard interpretation
is that this radiation, called the cosmic microwave background (CMB), was
emitted when the universe was denser and hotter, and has travelled freely
to us ever since.
A third observation, that provides information about the physics of the
universe at early times, concerns the overall abundance of helium in the
universe (see Peebles 1993~\cite[pages 143--4]{book:Peebles} and Hoyle \&
Tayler 1964~\cite{art:HoyleTayler1964}). The constant abundance of 
helium in stars
and intergalactic matter ($\approx 25\%$ by mass) over time suggests that 
helium was formed during a very early epoch when the
universe was very hot.
These three observations form the physical basis for the so-called
hot big-bang
theory of the universe, which can be summarized in the following history
of the universe.

        The history begins at the so-called Planck time (see, for
example, Coles \& Lucchin 1995~\cite[page 111]{book:ColesLucchin1995}), 
the
time after
which classical general relativity may be assumed to be valid.
The corresponding Planck temperature is approximately $10^{32}$ K. At such
a high
temperature, photons have enough energy to produce matter and
anti-matter in collisions.
When the temperature fell to the order of $10^9$ K, collisions of photons
could no longer produce even the lightest known particles and
antiparticles
(electrons and positrons), and the radiation-dominated epoch began
(i.e. most of the energy was in the form of radiation).
Slight asymmetries between matter and anti-matter resulted in
all anti-matter being annihilated by matter, and the remaining matter
consisted of protons, neutrons, electrons, photons and neutrinos.
%
%
It was also cool enough at this temperature for most neutrons to be
rapidly tied to form helium nuclei in a process called nucleosynthesis
(see, for example, Weinberg 1977~\cite[page 110]{book:Weinberg1977}).
%
%
Radiation was ``coupled" to matter in the sense that photons were
frequently scattered by electrons and nuclei through Compton scattering,
and matter was kept ionized.
As the universe expanded and cooled, radiation lost energy faster than
matter did, with the result that eventually the universe became
matter-dominated.
When the temperature dropped to about 3000 K, photons no longer had enough
energy to constantly ionize matter.
In other words, the free electrons became bound to nuclei, with the result
that the photons could travel freely.
One says that the photons have decoupled from the matter.
These photons form the CMB radiation, which is detected at the present
time.
The CMB thus provides a snapshot of the universe at a temperature of
about 3000 K, but it is also affected by the geometry of spacetime as it
travels towards us.

Within this framework, the formation of galaxies occurs due to the growth
of density fluctuations under the influence of gravity.
The key point is that random fluctuations in the matter density do not
grow rapidly enough to form galaxies (see, for example, 
Peebles 1993~\cite[page 619]{book:Peebles}).
Hence one has to postulate the existence of a suitable spectrum of density
fluctuations at very early times.
The most popular, but not universally accepted, explanation of the origin
of density fluctuations is the theory of inflation in the early universe,
first proposed by Guth 1981~\cite{art:Guth1981}.
During the inflationary epoch, which is postulated to have occurred
shortly after the Planck time, the universe underwent exponential
expansion, and quantum fluctuations were hugely magnified to create
a spectrum of density fluctuations, which were sufficiently large to form
galaxies, but not so large as to produce too many black holes
(see also Peacock 1999~\cite[Chapter 11]{book:Peacock}).

        Over the past 20 years, as cosmological observations have become
more extensive and more accurate, an increasingly complicated picture of
the universe has emerged.
Redshift surveys (e.g. the Sloan Digital Sky
Survey~\cite{art:SDSS2001}, the Las Campanas
Redshift Survey~\cite{art:Linetal1996}) show
that  galaxies are not
distributed uniformly, but form clusters which in turn are grouped into
huge structures on a scale of 100 Mpc%
\footnote{1 parsec is about 3 lightyears;
the size of a typical galaxy is
1 Mpc and of a typical cluster is 10 Mpc.}%
%
,
e.g. the Great Attractor (probably a supercluster, with galaxy cluster  
Abell 3627 near its centre), which is attracting several other clusters,
including the Local Group.
These observations are being used to determine the power spectrum of the 
distribution of galaxies (see Strauss 2004~\cite{art:Strauss2004}, Tegmark
\etal 2004~\cite{art:Tegmarketal2004}) and to compare it with the CMB 
power spectrum (see Hu \& White 2004~\cite{art:HuWhite2004}).
Until recently,   
it was widely believed that the rate of expansion of the universe is
slowing down. This belief was challenged when the recent observations of
type Ia supernovae suggested an accelerating expansion (see, for example,
Riess \& Turner 2004~\cite{art:RiessTurner2004} and Turner \& Riess 
2002~\cite{art:TurnerRiess2002}). This observation, when 
combined with
 the observations of the power spectrum of the CMB temperature, restricts
the total density parameter%
\footnote{The density parameter is defined in 
Section~\ref{sec:hierarchy}.}
$\Omega$ to have a value close to 1 (see, for example,
Spergel \etal 2003~\cite{art:Spergeletal2003}).
However visible matter contributes a value of approximately $0.005$ to  
$\Omega$,
and non-luminous baryonic matter contributes a value of approximately
$0.035$ \cite{art:OstrikerSteinhardt2001}.
This dramatic conclusion suggests that the matter-energy content of the
universe is dominated by yet unidentified non-baryonic dark matter and/or
energy, indicating that our understanding of the physical universe is
still quite incomplete.

        The simplest mathematical models within the framework of classical
general relativity
 that describe an expanding
universe are the Friedmann-Lema\^{\i}tre (FL) models%
\footnote{These models are introduced in Section~\ref{sec:hierarchy}.}%
, which are spatially 
homogeneous and isotropic
(see Ellis 1989~\cite{inbook:Ellis1989} for a discussion of the historical
development).
An important question is this: Do cosmological observations support the
assumption of isotropy and homogeneity on a sufficiently large scale? In
view of the innate limitations in making cosmological observations (we can
observe only from one point in space and one instant in time on a
cosmological scale), and the difficulties in interpreting the
observations, the answer is not clear-cut.
It is thus desirable to investigate the dynamics of cosmological models 
more general than the FL models, and to determine
what restrictions observations impose on them.
This general area of investigation forms the basis of this thesis.

\section{Cosmological Hierarchy}\label{sec:hierarchy}

In this section, we define cosmological models and classify them according
to their symmetries.

\subsection*{Cosmological models}

A {\it cosmological model} $(\mathcal{M},\mathbf{g},\mathbf{u})$
is determined by a Lorentzian metric $\mathbf{g}$ defined on a manifold
$\mathcal{M}$, and a family of fundamental observers, whose congruence of
world-lines is represented by the unit timelike vector field $\mathbf{u}$,
which we often identify with the matter 4-velocity.
The dynamics of the model is governed by the Einstein field equations
(EFEs)
\be
        G_{ab} = T_{ab},
\footnote{We use geometrized units throughout, i.e. $c=1$ and $8\pi G=1$,
where $c$ is the speed of light in vacuum and $G$ is the graviational 
constant.}
\ee
with suitable matter content.

The matter content is assumed to be a perfect fluid with energy density
$\mu>0$, pressure $p$ and (unit timelike) fluid 4-vector $\mathbf{u}$:
\begin{equation}
        T_{ab}= \mu u_a u_b + p (g_{ab} + u_a u_b).
\end{equation}
We assume that the equation of state of the perfect fluid is of the form
$p=(\gamma-1)\mu$, where $1 \leq \gamma < 2$ is constant.
The cases $\gamma=1$ (dust) and $\gamma=\tfrac43$ (radiation) are of
primary physical interest.
The cosmological constant $\Lambda$ in the EFEs can be treated as a
perfect fluid with $\mu=\Lambda$ and $p=-\Lambda$.
A more formal introduction will be presented in the next chapter.

It is helpful to classify cosmological solutions of the EFEs using the
dimension of orbits of the symmetry group admitted by the metric. This
classification scheme forms a
hierarchy of cosmological models of increasing complexity, starting with
the FL models:
\begin{enumerate}[1)]
\item   Friedmann-Lema\^{\i}tre (FL) cosmologies ($G_6$)
\item   Spatially homogeneous (SH) cosmologies ($G_3$)
\item   $G_2$ cosmologies
\item   generic cosmologies ($G_0$)
\end{enumerate}

\noindent
An FL cosmology admits a local 6-parameter group $G_6$ of isometries
acting on spacelike 3-surfaces. i.e. it has 6 independent Killing vector
fields (KVFs) whose orbits form spacelike 3-surfaces.
An SH
cosmology is either a Bianchi cosmology, which admits a local 3-parameter
group $G_3$ of isometries acting simply transitively on spacelike
3-surfaces, or a Kantowski-Sachs cosmology, which admits a group $G_4$ of
isometries acting on spacelike 3-surfaces but does not admit a simply  
transitive $G_3$ subgroup.
A $G_2$
cosmology admits a local two-parameter Abelian group of isometries with
spacelike orbits, permitting one degree of freedom as regards
inhomogeneity.
From the viewpoint of analysis, the equations for $G_3$ cosmologies are 
ordinary differential equations
(ODEs), those
for $G_2$ cosmologies are 1+1 
partial differential equations
(PDEs), and those for $G_0$ cosmologies are 1+3 PDEs.

\subsection*{Friedmann-Lema\^{\i}tre cosmologies}

As mentioned in the introduction, FL models are spatially homogeneous and
isotropic. In
physical terms, ``spatially homogeneous" means ``the same at every point
at any given instant of time", while ``isotropic" means ``the same in 
every spatial direction". 
Mathematically, the requirements of spatial homogeneity and isotropy
imply that the line element admits the maximal group $G_6$ of isometries
acting on spacelike 3-surfaces. With an appropriate choice of
time coordinate, the
line element can be written as
\footnote{Latin indices run from 0 to 3; Greek indices from 1 to 3;
capital Latin indices run from 2 to 3.}
\begin{equation}
        ds^2 = -dt^2 +l(t)^2 \tilde{g}_{\alpha\beta}dx^\alpha dx^\beta,
\end{equation}
where $\tilde{g}_{\alpha\beta}dx^\alpha dx^\beta$ is the metric of a
(spacelike) 3-surface with constant curvature (see, for example,
Weinberg 1972~\cite[Section 13.5]{book:Weinberg1972}). The
function $l(t)$ is called the {\it length scale} of the model.
 
The two most important quantities describing an FL model are the Hubble
scalar $H$, where $H=\dot{l}/l$, and the total%
\footnote{In general, $\mu$ is a sum of terms, for example matter,
radiation and the energy density due to a cosmological constant.}
density parameter, defined by
\begin{equation}
        \Omega= \frac{\mu}{3H^2}.
\end{equation}
FL models are further separated into three types according to the
curvature of the spacelike 3-surfaces $t=const.$, which is
completely determined by $\Omega$:
\begin{xalignat*}{2}
        \text{Open FL:} &\text{ negative curvature,}& \Omega&<1
\\
        \text{Flat FL:} &\text{ zero curvature,}& \Omega&=1
\\
        \text{Closed FL:} &\text{ positive curvature,}& \Omega&>1
\end{xalignat*}

We want to be able to determine whether a given cosmological model is   
exactly FL or close to FL.
For this purpose it is necessary to introduce physical quantities called
the {\it kinematical quantities}
(see, for example,
Wainwright \& Ellis 1997 (WE)~\cite[page 18]{book:WainwrightEllis1997},
and
Ellis 1971~\cite[page 111]{art:Ellis1971}).
For a unit timelike vector $\mathbf{u}$, one can decompose
the covariant derivative $u_{a;b}$ of the corresponding 1-form into its
irreducible parts -- the
kinematical quantities
$\sigma_{ab}$, $\omega_{ab}$, $\Theta$ and $\dot{u}_a$. The scalar
$\Theta$ is the rate of expansion scalar and $\dot{u}_a$ is the
acceleration vector. The tensor $\sigma_{ab}$ is the rate of shear tensor
and $\omega_{ab}$ is the rate of vorticity tensor. If the vorticity is
zero, we
say that $\mathbf{u}$ is irrotational. Because we are working in a
cosmological context we shall usually replace $\Theta$ by the Hubble
scalar $H$ defined earlier, which is related to $\Theta$ according to
$H=\tfrac13\Theta$.
        The following theorem characterizes the FL models using the
kinematical quantities (see, for example, Krasi\'{n}ski 1997~\cite[page
11]{book:Krasinski1997}):

\begin{center}
\parbox{0.91\linewidth}{
{\bf Theorem 1} 
        
A solution of the EFEs with perfect fluid matter content is an
FL model if and only if
\begin{equation}
                \sigma_{ab}=0=\omega_{ab}=\dot{u}_a.
\end{equation}
}
\end{center}

The FL models can also be characterized using the
{\it Weyl conformal curvature tensor} $C_{abcd}$, which is interpreted as
describing the free gravitational field.
The Weyl tensor
 can be decomposed into its electric part $E_{ab}$ and magnetic part
$H_{ab}$ relative to $\mathbf{u}$. The following theorem characterizes
the FL models
using $E_{ab}$ and $H_{ab}$ (see, for example, Wainwright
1996~\cite[Section 4.1]{art:Wainwright1996}):

\begin{center}
\parbox{0.91\linewidth}{
{\bf Theorem 2}

A solution of the EFEs with perfect fluid matter content and equation of
state $p=p(\mu)$ is an FL model if and only if
\begin{equation}
                E_{ab}=0=H_{ab}.
        \label{c2}
\end{equation}
}
\end{center}

\subsection*{Spatially homogeneous cosmologies}
        
Spatially homogeneous cosmologies with an isometry group $G_3$ acting on 
spacelike 3-surfaces,
the so-called Bianchi cosmologies, are classified according to the 
structure of the $G_3$ group, as given in Table~\ref{tab:Bianchi}.
\footnote{The symmetric matrix $n_{\alpha\beta}$ in 
Table~\ref{tab:Bianchi} is defined in equation 
(\ref{comm2}).}

\begin{table}[h]
\begin{spacing}{1.1}
\caption{Classification of Bianchi cosmologies.}
                \label{tab:Bianchi}
\begin{center}
\begin{tabular*}{0.65\textwidth}%
        {@{\extracolsep{\fill}}ccccc}
\hline
\multicolumn{2}{c}{Group type} &
\multicolumn{3}{c}{Eigenvalues} \\
Class A & Class B & \multicolumn{3}{c}{of $n_{\alpha\beta}$}  \\
\hline
IX      &         & + & + & +  \\
VIII    &         & + & + &$-$ \\
VII$_0$ & VII$_h$ & + & + & 0  \\
VI$_0$  & VI$_h$  & + &$-$& 0  \\
II      & IV      & + & 0 & 0  \\
I       & V       & 0 & 0 & 0  \\
\hline
\end{tabular*}
\end{center}
\end{spacing}  
\end{table}

Bianchi cosmologies comprise two classes,
\begin{itemize}
\item[i)]	the \emph{non-tilted} models in which the 4-velocity 
		vector $\mathbf{u}$ is always orthogonal to the
		spacelike $G_3$ orbits, were first studied in detail by 
		Ellis \& MacCallum 1969~\cite{art:EllisMacCallum1969}.
\item[ii)]	the \emph{tilted} models in which $\mathbf{u}$ is never 
		orthogonal to the spacelike $G_3$ orbits,
		were first studied by 
		King \& Ellis 1973~\cite{art:KingEllis1973}.
\end{itemize}

The dynamics of non-tilted Bianchi cosmologies have been analyzed in 
detail (see WE, Hewitt \etal 2003~\cite{art:Hewittetal2003} and
Horwood \etal 2003~\cite{art:Horwoodetal2003}).
Much less is known about the dynamics of tilted Bianchi cosmologies and 
work is currently in progress.
\footnote{See Hervik 2004~\cite{art:Hervik2004} and Hervik \& Coley 
2004~\cite{com:HervikColey2004}.}
Indeed in this thesis we provide a formulation of the EFEs for analyzing 
the asymptotic and intermediate dynamics of the so-called $G_2$-compatible 
Bianchi cosmologies.

There exist $G_3$ subgroups within the $G_6$ isometry group of FL
cosmologies.  
Thus FL cosmologies appear as special cases of SH cosmologies -- open FL
as
Bianchi V and VII$_h$, flat FL as Bianchi I and VII$_0$, and closed FL as
Bianchi IX (see WE, page 37).
Notice that open and flat FL cosmologies appear as more than one type of
Bianchi cosmology. These are instances of \emph{multiple representation}
of special models in a more general class of models.

\subsection*{Spatially inhomogeneous cosmologies}

The simplest spatially inhomogeneous cosmologies are the $G_2$ 
cosmologies, which admit one spatial degree of freedom.
These models, which form the main focus of this thesis, will be classified 
in Chapter~\ref{chap:G2hie}. We do not consider generic spatially 
inhomogeneous cosmologies, i.e. the so-called $G_0$ cosmologies, in this 
thesis.
We refer to Uggla \etal 2003~\cite{art:Ugglaetal2003} and
Lim \etal 2004~\cite{art:Limetal2004} for a general framework for 
analyzing these models.

\section{Goals}\label{sec:goals}

        We now discuss three physically important goals of the study of
cosmological models more general than FL.
\begin{enumerate}[P1.]
 
\item   Describe the asymptotic dynamics at early times.

	Assuming that the universe is expanding from an initial 
singularity,
        the phrase ``dynamics at early times" refers to the dynamics in
the asymptotic regime associated with the initial singularity,
characterized by $l \rightarrow 0$, where $l$ is the overall length scale.
Classical general relativity breaks down at the singularity, and should 
be replaced by a theory of quantum gravity there. The time after which 
classical
general relativity is valid is called the Planck time. Thus in physical
terms the phrase ``dynamics at early times" refers to the evolution of
the universe in an epoch starting at the Planck time.
It is important to classify all possible asymptotic states at early times
that are permitted by the EFEs, in order to explain how the universe may
have evolved.

Two important conjectures about the asymptotic dynamics at early times
have been made by Belinskii, Khalatnikov \& Lifshitz, which we shall
refer
to as BKL I and BKL II.

\begin{center}
\parbox{0.81\linewidth}{
        {\bf BKL conjecture I}

        For a typical cosmological model the matter content is not
dynamically significant near the initial
singularity
(Lifshitz \& Khalatnikov 1963~\cite[page 200]{art:LK63},
BKL 1970~\cite[pages 532, 538]{art:BKL1970}).
}
\end{center}

        The implication of this conjecture is that one can use the vacuum
EFEs to approximate the evolution of a typical cosmological model near the
initial singularity. 
An example of atypical cosmologies is the set of models with an
isotropic singularity 
(Goode \& Wainwright 1985~\cite{art:GoodeWainwright1985},
Coley \etal 2004~\cite{art:Coleyetal2004}). 
Another example is the set of models in which
the matter content is a so-called stiff fluid ($p=\mu$)
(Andersson \& Rendall 2001~\cite{art:AnderssonRendall2001}).

\begin{center}
\parbox{0.81\linewidth}{
        {\bf BKL conjecture II}  

        Near the initial singularity, the spatial derivatives in the EFEs
for a typical cosmological model are not dynamically significant
(BKL 1982~\cite[page 656]{art:BKL1982}).
}
\end{center}

        The implication of this conjecture is that near the initial  
singularity the EFEs effectively reduce to ODEs, namely, the ODEs that
describe the evolution of SH cosmologies.
Hence it is plausible that SH models will play an important role in
analyzing spatially inhomogeneous models.

\item   Describe the asymptotic dynamics at late times.

        The phrase ``dynamics at late times" refers to the dynamics in the
asymptotic regime characterized by $l \rightarrow \infty$.
For example, the dynamics at late times in a radiation model would
describe the later stage of the radiation-dominated epoch before the
universe became matter-dominated.
Likewise, the dynamics at late times in an inflationary model would
describe the final stage of the inflationary epoch.

\enlargethispage{\baselineskip}

\item   Describe the dynamics of close-to-FL models.

As mentioned earlier,
it is usually assumed that deviations from exact isotropy and spatial
inhomogeneity can be described using linear perturbations of an FL model.%
\footnote{For a review of cosmological perturbations, see Mukhanov \etal 
1992~\cite{art:Mukhanovetal1992}. Other references that use a 
gauge-invariant approach are Bardeen 1980~\cite{art:Bardeen1980}, Bruni 
\etal 1992~\cite{art:Brunietal1992} and Goode 1989~\cite{art:Goode1989}.}
Linear perturbations have been used in studying, for example,
the evolution of the primordial density perturbations during the
radiation-dominated epoch, and the development of large-scale structure
in the distribution of galaxies.
        A key question is this: does linear perturbation theory correctly 
describe the dynamics of close-to-FL models?

\end{enumerate}

\section{Overview}

This thesis is divided into two parts.
The first part consists of Chapters~\ref{chap:off}--\ref{chap:past},
which provide a unified formulation of the EFEs,
the background and analytical results for the dynamics of $G_2$ 
and SH cosmologies.
The second part consists of Chapters~\ref{chap:G2}--\ref{chap:sim},
in which we present a numerical exploration of the dynamics of $G_2$ 
cosmologies.

We first establish a framework for the study of
inhomogeneous cosmologies in Chapter~\ref{chap:off}.
We use the orthonormal frame approach to formulate the EFEs,
and introduce the appropriate normalizations for $G_2$ and SH
models respectively, which lead to scale-invariant variables that are 
bounded at the initial 
singularity.

In Chapter~\ref{chap:G2hie} we present a hierarchy of subclasses 
within the $G_2$ cosmologies, and develop the links between $G_2$ and SH 
cosmologies.

A new, unified formulation for $G_2$-compatible SH cosmologies
is given in Chapter~\ref{chap:SH}.
We discuss the tilted SH cosmologies in detail, 
emphasizing the past asymptotic dynamics and giving conditions for the 
existence of Mixmaster dynamics.

Chapters~\ref{chap:G2_dyn}--\ref{chap:G2} deal with analytical aspects 
of $G_2$ cosmologies.
In Chapter~\ref{chap:G2_dyn} we complete the formulation for $G_2$ 
cosmologies by choosing appropriate temporal and spatial gauges.
We discuss the notions of the silent boundary and asymptotic silence, 
which play a key role in linking $G_2$ dynamics with SH dynamics.
Chapter~\ref{chap:explicit} presents two $G_2$ solutions 
-- one of which is new -- that develop large spatial gradients as they 
approach the initial singularity.
They serve as prototypes for such behaviour in typical $G_2$ solutions.
In Chapter~\ref{chap:past},
the results of Chapters~\ref{chap:SH} and~\ref{chap:explicit} are used to 
predict the past asymptotic dynamics of $G_2$ cosmologies.
In particular, we discuss the existence of Mixmaster dynamics and the 
validity of the BKL conjecture II.

Chapter~\ref{chap:G2} is devoted to various aspects of numerical 
simulations of $G_2$ cosmologies: choosing the initial condition, 
devising the numerical solvers, and comparing the errors of the solvers.

\enlargethispage{\baselineskip}

In Chapter~\ref{chap:sim} 
we present the results from a variety of 
numerical simulations to illustrate various facets of $G_2$ dynamics.
In contrast to the previous numerical work on vacuum models, we 
investigate non-vacuum models, in which the source of the gravitational 
field is a radiation fluid and a cosmological constant.
The main focus is on the past asymptotic regime, with a secondary focus 
on the future asymptotic dynamics with a positive cosmological constant 
and on the close-to-FL regime.

In the concluding Chapter~\ref{chap:conclusion} we summarize 
the new results in this thesis, and remark on future research.

As regards background material, we assume that the reader is familiar with 
the orthonormal frame formalism, although a brief review with references 
is given at the beginning of Chapter~\ref{chap:off}.
In addition we assume familiarity with some basic concepts in the theory 
of dynamical systems. We refer to WE (Chapter 4) for a summary and further 
references. For background material on the use of scale-invariant 
variables and dynamical systems methods in cosmology, we also refer to WE.

%% file: off.tex
	\chapter{The orthonormal frame formalism}\label{chap:off}

In this chapter we introduce the orthonormal frame formalism, which is 
based on a 1+3 decomposition of spacetime.
Our goal is to apply the formalism to give a 1+1+2 decomposition of 
spacetime, which provides the basis for our analysis of $G_2$ cosmologies.

The orthonormal frame formalism was first introduced in cosmology by
Ellis 1967~\cite{art:Ellis1967} and subsequently developed by
Ellis \& MacCallum 1969~\cite{art:EllisMacCallum1969} and
 MacCallum 1973~\cite{art:MacCallum1973}.
More recently, it was extended by
van Elst \& Uggla 1997~\cite{art:ElstUggla1997}.
Our standard references will be
MacCallum 1973~\cite{art:MacCallum1973}
 and
van Elst \& Uggla 1997~\cite{art:ElstUggla1997}.

\section{1+3 decomposition}\label{sec:1+3}

        In the orthonormal frame approach one does not use the metric
$\mathbf{g}$ directly (as done in the metric approach), but chooses at
each point of the spacetime manifold
$(\mathcal{M},\mathbf{g})$ a set of four linearly independent 1-forms
$\{\pmb{\omega}^a\}$ such that the line element can be locally expressed
as $ds^2 = \eta_{ab} \pmb{\omega}^a \pmb{\omega}^b$, where
$\eta_{ab}=\mathrm{diag}(-1,1,1,1)$. The corresponding vector fields
$\{\mathbf{e}_a\}$ are then mutually orthogonal and of unit length -- they
form an orthonormal basis, with $\mathbf{e}_0$ being timelike (and thus 
defining a timelike congruence).

The gravitational field is described by the commutation functions 
$\gamma^c_{\ ab}$ of the orthonormal frame, defined by
\begin{equation}
        [ \mathbf{e}_a, \mathbf{e}_b ] = \gamma^c_{\ ab} \mathbf{e}_c.
\end{equation}
The first step is to perform a 1+3 decomposition of the commutation 
functions as follows:%
\footnote{We follow the $\Omega_\alpha$ sign convention of 
van Elst \& Uggla 1997~\cite[page 2676]{art:ElstUggla1997}.}
\begin{align}
\label{comm1}
        [ \mathbf{e}_0, \mathbf{e}_\alpha ] &=
        \dot{u}_\alpha \mathbf{e}_0 - \left[ H {\delta_\alpha}^\beta
        +{\sigma_\alpha}^\beta - \epsilon_\alpha{}^{\beta\gamma}
        (\omega_\gamma-\Omega_\gamma)\right] \mathbf{e}_\beta\ ,
\\
\label{comm2}
        [ \mathbf{e}_\alpha, \mathbf{e}_\beta ] &=
        -2{\epsilon_{\alpha\beta}}^{\mu}\omega_\mu \mathbf{e}_0 + \left[
        \epsilon_{\alpha\beta\nu}n^{\mu\nu}  
        + a_\alpha {\delta_\beta}^\mu-a_\beta
        {\delta_\alpha}^\mu \right] \mathbf{e}_\mu\ .
\end{align}
The variables
in (\ref{comm1}) and (\ref{comm2})
 have physical or geometrical meanings, as follows.
The variable
$H$ is the Hubble scalar, $\sigma_{\alpha\beta}$ the rate of
 shear tensor,
$\dot{u}_\alpha$ the acceleration vector, and $\omega_\alpha$ the
rate of
vorticity vector of the timelike congruence defined by $\mathbf{e}_0$,
while $\Omega_\alpha$ is the angular velocity of the spatial frame
$\{ \mathbf{e}_\alpha \}$
with respect to a nonrotating frame ($\Omega_\alpha=0$).
The variables $n_{\alpha\beta}$ and $a_\alpha$ have no direct  
physical or geometrical meanings in general, 
but if the fundamental timelike vector field $\mathbf{e}_0$ is 
hypersurface-orthogonal, 
then they directly determine the geometry of the spacelike 3-surfaces 
orthogonal to $\mathbf{e}_0$ 
(see, for example WE, Section 1.6.3). 
We shall thus refer to $n_{\alpha\beta}$ and $a_\alpha$ as the {\it 
spatial curvature variables}.
Collectively, the variables above describe the gravitational field. We 
shall refer to them as the 
\emph{gravitational field variables}, and denote them 
by the state vector
\begin{equation}
	\X_{\rm grav} =
        (H, \sigma_{\alpha\beta}, \dot{u}_\alpha, \omega_\alpha,
\Omega_\alpha, n_{\alpha\beta}, a_\alpha),
\label{117}   
\end{equation}

The matter content of a cosmological model is described by the
stress-energy tensor $T_{ab}$,
which is decomposed into irreducible parts
 with respect to $\mathbf{e}_0$ in the following way
(let $\mathbf{e}_0=\mathbf{u}$ below):
\be
\label{pf_1}
        T_{ab} = \mu u_a u_b + 2 q_{(a}u_{b)} + p h_{ab} + \pi_{ab},
\ee
where
\[
        q_a u^a=0,\ \pi_{ab} u^b=0,\ \pi_a{}^a=0,\ \pi_{ab} = \pi_{ba},
\]
and $h_{ab} = g_{ab}+u_a u_b$ is the projection tensor which locally
projects into the 3-space orthogonal to $\mathbf{u}$.
Since we are using an orthonormal frame, we have $g_{ab} = \eta_{ab}$, 
$u^a
= (1,0,0,0)$, and $q_0 = 0 =\pi_{0a}$.
The variables $(\mu, p,q_\alpha,\pi_{\alpha\beta})$ have physical
meanings:
$\mu$ is the energy density, $p$ is the (isotropic) pressure, $q_\alpha$
is the energy flux density and $\pi_{\alpha\beta}$ is the anisotropic
pressure (see, for example, van Elst \& Uggla 1997~\cite[page
2677]{art:ElstUggla1997}).
We shall refer to these variables as the
\emph{matter variables}, and denote them
by the state vector
\be
\label{118}
	\X_{\rm matter} =
	(\mu, q_\alpha,p,\pi_{\alpha\beta}).
\ee

The dynamics of 
 the variables in (\ref{117}) and (\ref{118}) is 
described by the EFEs, the Jacobi identities (using $\me_a$) and the 
contracted Bianchi identities:
\begin{gather}
	G_{ab} + \Lambda g_{ab} = T_{ab}\ ,
\\
	\me_{[c} \gamma^d{}_{ab]} - \gamma^d{}_{e[c}\gamma^e{}_{ab]}=0\ ,
\\
	\text{and}\qquad \nabla_b T_a{}^b =0\ ,\qquad\qquad
\end{gather}
respectively.
The evolution of $p$ and $\pi_{\alpha\beta}$ has to be
specified by giving an equation of state for the matter content (e.g.
perfect fluid).
The variables $\udot_\alpha$ and $\Omega_\alpha$ correspond to the 
temporal and spatial gauge freedom respectively, and will be specified 
later.

Throughout this thesis we shall assume that
\emph{the fundamental congruence $\me_0$ has zero vorticity} 
($\omega_\alpha=0$), or equivalently that $\me_0$ is orthogonal to a 
family of spacelike 3-surfaces.
We now list the system of evolution and constraint equations 
(also see WE, page 33, but with the opposite sign
convention for $\Omega_\alpha$;
see van Elst \& Uggla 1997~\cite{art:ElstUggla1997} for the equations 
with vorticity).
The angle bracket used in (\ref{evo_sigma}) and (\ref{3S})
 denotes traceless symmetrization -- given
$V_{\alpha\beta}$, $V_{\la \alpha\beta \ra}$ is defined by
\be
        V_{\la \alpha\beta \ra} = V_{(\alpha\beta)} 
		- \tfrac{1}{3} V_\gamma{}^\gamma \delta_{\alpha\beta}.
\ee

\noindent
\begin{minipage}{\textwidth}
\subsection*{The general orthonormal frame equations ($\omega_\alpha=0$)
\footnote{The letter $C$ denotes a constraint, and the subscript denotes 
which one: Gauss, Codacci, vorticity and Jacobi.}
}

\noindent
{\it Einstein field equations:}
\begin{align}
\label{evo_H}
        \me_0 H &= -H^2
        -\tfrac{1}{3}\sigma_{\alpha\beta}\sigma^{\alpha\beta}
        +\tfrac{1}{3}(\me_\alpha
        +\udot_\alpha -2a_\alpha)\udot^\alpha-\tfrac{1}{6}(\mu+3p)
        +\tfrac{1}{3}\Lambda
\\
        \me_0 (\sigma_{\alpha\beta}) &=
        -3H \sigma_{\alpha\beta}
-2\epsilon^{\gamma\delta}{}_{(\alpha}\sigma_{\beta)\gamma}\Omega_\delta
        -{}^3\!S_{\alpha\beta}+\pi_{\alpha\beta}
        -\epsilon^{\gamma\delta}{}_{(\alpha}n_{\beta)\gamma}\udot_\delta
\notag\\
        &\qquad
        +(\me_{\la \alpha} + \udot_{\la \alpha} + a_{\la \alpha})
                \udot_{\beta \ra}
\label{evo_sigma}
\\
\label{c_g}
        0 &= (C_{\rm G}) = 6H^2 + {}^3\!R
        - \sigma_{\alpha\beta} \sigma^{\alpha\beta} -2\mu 
	-2\Lambda
\\
\label{c_c}
        0 &= (C_{\rm C})_\alpha =
        -2\me_\alpha H + \me_\beta \sigma_\alpha{}^\beta
        -3a_\beta \sigma_\alpha{}^\beta
        -\epsilon_\alpha{}^{\beta\gamma}
n_{\beta\delta}\sigma_\gamma{}^\delta
        + q_\alpha\ ,
\intertext{where ${}^3\!R$ and ${}^3\!S_{\alpha\beta}$ are the isotropic 
and anisotropic spatial curvature, and are given by}
        {}^3\!R &= 4\me_\alpha a^\alpha - 6 a_\alpha a^\alpha
	- n_{\alpha\beta} n^{\alpha\beta} 
	+ \tfrac{1}{2}(n_\gamma{}^\gamma)^2
\\
\label{3S}
        {}^3\!S_{\alpha\beta} &=
        \me_{\la \alpha}a_{\beta\ra}
        -(\me_\gamma -2a_\gamma)
                n_{\delta(\alpha} \epsilon_{\beta)}{}^{\gamma\delta}
	+ 2n_{\la\alpha}{}^\gamma n_{\beta\ra\gamma}    
	        - n_\gamma{}^\gamma n_{\la\alpha\beta\ra}\ .
%
\end{align} 
        
\noindent
{\it Jacobi identities:}
\begin{align}
\label{evo_n}
        \me_0 (n_{\alpha\beta}) &=
        -H n_{\alpha\beta} +2\sigma^\gamma{}_{(\alpha}n_{\beta)\gamma}
        -2\epsilon^{\gamma\delta}{}_{(\alpha}n_{\beta)\gamma}\Omega_\delta
\notag\\
        &\qquad
        -(\me_\gamma+\udot_\gamma)
         (\epsilon^{\gamma\delta}{}_{(\alpha}\sigma_{\beta)\delta}
          -\delta^\gamma{}_{(\alpha}\Omega_{\beta)}
          +\delta_{\alpha\beta}\Omega^\gamma )
\\
        \me_0 (a_\alpha)
        &=  (- H {\delta_\alpha}^\beta -{\sigma_\alpha}^\beta -
        \epsilon_\alpha{}^{\beta\gamma} \Omega_\gamma) a_\beta
\notag\\
        &\qquad
        -\tfrac{1}{2}(\me_\beta+\udot_\beta)
        (2 H {\delta_\alpha}^\beta -{\sigma_\alpha}^\beta -
        \epsilon_\alpha{}^{\beta\gamma} \Omega_\gamma)
\\
        0 &= (C_{\omega})^\alpha =
        [\epsilon^{\alpha\beta\gamma}(\me_\beta-a_\beta)
                -n^{\alpha\gamma}]\udot_\gamma
\\
        0 &= (C_{\rm J})_\alpha = \me_\beta(n_\alpha{}^\beta 
        +\epsilon_\alpha{}^{\beta\gamma}a_\gamma)
        -2a_\beta n_\alpha{}^\beta\ .
\end{align}
        
\noindent
{\it Contracted Bianchi identities:}
\begin{align}
        \me_0 (\mu) &= -3H(\mu+p) -\sigma_{\alpha\beta}\pi^{\alpha\beta}
                        -(\me_\alpha+2\udot_\alpha-2a_\alpha)q^\alpha
\\
        \me_0 (q_\alpha) &=
	(-4H \delta_\alpha{}^\beta - \sigma_\alpha{}^\beta 
	- \epsilon_\alpha{}^{\beta\gamma} \Omega_\gamma) q_\beta
        -(\mu+p)\udot_\alpha
        -\me_\alpha p
\notag\\
        &\qquad
        -(\me_\beta+\udot_\beta-3a_\beta)\pi_\alpha{}^\beta
        +\epsilon_\alpha{}^{\beta\gamma}n_{\beta\delta}\pi_\gamma{}^\delta
        \ .
\label{evo_q}
\end{align}

\begin{figure}[H]
\begin{center}
\setlength{\unitlength}{1mm}
\begin{picture}(120,0)(0,0)
\put(60,90){\oval(150,175)[t]}
\put(60,90){\oval(150,170)[b]}
\end{picture}
\end{center}
\end{figure}
\end{minipage}

\newpage

\noindent
{\it Perfect fluid}
\vspace{3mm}

This thesis focuses on cosmological models whose matter content includes a 
perfect fluid, which we now introduce.
A \emph{perfect fluid} has a 4-velocity $\tilde{\mathbf{u}}$ which is not
necessarily aligned with the 4-velocity $\mathbf{e}_0=\mathbf{u}$ of our
fundamental observers. The stress-energy tensor $T_{ab}$ is
\be
        T_{ab} = \tilde{\mu} \tilde{u}_a \tilde{u}_b + \tilde{p}
        (g_{ab}+\tilde{u}_a \tilde{u}_b),\quad \tilde{u}_a \tilde{u}^a=-1.
\label{pf_11}
\ee
A \emph{perfect fluid with a linear barotropic equation of state}
specifies $\tilde{p}$:
\be
\label{eos}
        \tilde{p} = \tilde{p}(\tilde{\mu}) = (\gamma-1)\tilde{\mu},
\ee
where $\gamma$ is a constant parameter.
The range
\be
\lb{gam}
1 \leq \gam < 2
\ee
is of particular physical interest, since it ensures that the
perfect fluid satisfies the dominant and strong energy conditions
and the causality requirement that the speed of sound should be
less than that of light. The values $\gam = 1$ and $\gam =
\tfrac{4}{3}$ correspond to pressure-free matter (``dust'') and
 radiation, respectively.
Since $\tilde{\mathbf{u}}$ and $\mathbf{u}$ are not necessarily aligned,
we write
\be
\label{pf_4}
        \tilde{u}^a = \Gamma(u^a + v^a),
\ee
where
\[	u_\alpha=0,\quad
        v_0=0,\quad \Gamma = (1-v^2)^{-\frac{1}{2}},\quad 
	v^2=v_\alpha v^\alpha.
\]
It follows from (\ref{pf_1}), (\ref{pf_11})--(\ref{pf_4}) that 
$(\mu, p,q_\alpha,\pi_{\alpha\beta})$ 
are given by
\begin{gather}
        \mu = \frac{G_+}{1-v^2} \tilde{\mu},
\\
        p = \frac{(\gamma-1)(1-v^2)+\tfrac{1}{3}\gamma v^2}{G_+}\mu,\quad
        q_\alpha = \frac{\gamma \mu}{G_+} v_\alpha,\quad
        \pi_{\alpha\beta} 
	= \frac{\gamma \mu}{G_+} v_{\la \alpha} v_{\beta \ra},  
\label{perfect_fluid}
\end{gather}
where $G_+ = 1 + (\gamma-1)v^2$.
The basic variables that we use are $\mu$ and $v^\alpha$.
The vector
$v^\alpha$ is called the \emph{tilt} of the fluid, and has three degrees 
of freedom.

\newpage

\noindent
{\it Weyl tensor}
\vspace{3mm}

It is useful to keep track of the Weyl curvature tensor $C_{abcd}$, 
decomposed with respect to $\mathbf{u}=\me_0$ according to (see WE, page 
19):
\[
	E_{ac} = C_{abcd} u^b u^d,
	\quad
	H_{ac} = \tfrac{1}{2} \eta_{ab}{}^{st} C_{stcd} u^b u^d,
\]
and are given in terms of the gravitational and matter variables as 
follows (See WE, page 35):
\begin{align}
\label{Weyl_E}
	E_{\alpha\beta} 
	&= H \sigma_{\alpha\beta}
	-(\sigma_\alpha{}^\gamma \sigma_{\gamma\beta} 
		-\tfrac{1}{3} \sigma_{\gamma\delta} \sigma^{\gamma\delta} 
			\delta_{\alpha\beta})
	+ {}^3\!S_{\alpha\beta}
	- \tfrac{1}{2} \pi_{\alpha\beta}
\\
\label{Weyl_H}
	H_{\alpha\beta}
	&= (\me_\gamma-a_\gamma) 
		\sigma_{\delta(\alpha} \epsilon_{\beta)}{}^{\gamma\delta}
	- 3 \sigma^\gamma{}_{(\alpha} n_{\beta)\gamma}
	+ n_{\gamma\delta} \sigma^{\gamma\delta} \delta_{\alpha\beta}
	+ \tfrac{1}{2} n_\gamma{}^\gamma \sigma_{\alpha\beta}\ .
\end{align}

\vspace{3mm}
\noindent
{\it Local coordinates}
\vspace{3mm}

For quantitative analyses, we need to 
express the differential operators $\me_0$ and $\me_\alpha$ as partial 
differential operators in terms of $t$ and $x^i$.
\footnote{We reserve the indices $i,j,k$, running from 1 to 3, for the 
local coordinate components with respect to $x^i$.}
The 1+3 decomposition for $\{\me_0,\me_\alpha\}$ is as follows:
\be
\label{ptl1}
        \me_0=N^{-1} \partial_t\ ,\quad
        \me_\alpha = M_\alpha \ptl_t + e_\alpha{}^i \partial_i\ .
\ee
The assumption that $\me_0$ is hypersurface-orthogonal allows us to set
\be
	M_\alpha =0.
\ee
It follows that the 3-surfaces%
\footnote{For brevity we shall refer to these 3-surfaces as 
``slices".}
 to which $\me_0$ is orthogonal are 
given by
$t=const.$.
See 
van Elst \& Uggla 1997~\cite[Section 2.7]{art:ElstUggla1997}
for details.
The variables
\be
\label{frame_coeff}
        (N,\ e_\alpha{}^i)
\ee
are called the \emph{frame coefficients}.
When the commutators (\ref{comm1}) and (\ref{comm2}) act on $t$ and $x^i$, 
they provide the equations that relate the frame coefficients to the 
variables in (\ref{117}).

\vspace{3mm}
\noindent
{\it Commutators acting on $t$ and $x^i$:}
\begin{align}
\label{evo_eai}
        \me_0 (e_\alpha{}^i)
        &= (- H {\delta_\alpha}^\beta -{\sigma_\alpha}^\beta -
        \epsilon_\alpha{}^{\beta\gamma} \Omega_\gamma) e_\beta{}^i
\\
\label{c_udot}
        0 &= (C_{\udot})_\alpha = N^{-1} \me_\alpha(N) -\udot_\alpha
\\
        0 &= (C_{\rm com})^i{}_{\alpha\beta} =
        2(\me_{[\alpha}-a_{[\alpha})e_{\beta]}{}^i
        -\epsilon_{\alpha\beta\delta}n^{\gamma\delta}e_\gamma{}^i\ .
\end{align}
The frame coefficients are directly related to the metric
$\mathbf{g}$ through
\be
        g^{00} = -N^{-2},\quad
        g^{0i} = 0,\quad
        g^{ij} = \delta^{\alpha\beta} e_\alpha{}^i e_\beta{}^j.
\ee
This ties the variables in (\ref{117}) to the metric.

All these equations give an autonomous system of first order PDEs 
involving
temporal and spatial derivatives, and a set of constraint equations which
involve only spatial derivatives, for the variables (\ref{117}), 
(\ref{118}) and (\ref{frame_coeff}).
The system is under-determined, however, since there are no evolution
equations for $N$, $\dot{u}_\alpha$, $\Omega_\alpha$, $p$ and 
$\pi_{\alpha\beta}$.
A perfect fluid matter source 
with linear barotropic equation of state
gives (\ref{perfect_fluid}),
while temporal and spatial gauges determine ($N$,$\dot{u}_\alpha$) and 
$\Omega_\alpha$ respectively.

The general orthonormal frame equations (\ref{evo_H})--(\ref{evo_q}) 
describe the dynamics of $G_0$ cosmologies.
They are also ideal for describing the dynamics of SH cosmologies of 
Bianchi types VIII and IX, for which
the $t =const.$ slices are chosen to coincide with the $G_3$ group 
orbits (and the orthonormal frame is chosen to be \emph{group-invariant}), 
so that
\be
\label{SH}
	\me_\alpha(\X_{\rm grav})=0,\quad 
	\me_\alpha(\X_{\rm matter})=0,\quad
	\udot_\alpha=0.
\ee

\section{1+1+2 decomposition for $G_2$ cosmologies}\label{sec:1+1+2}

By a $G_2$ cosmology, we mean a cosmological model which admits an Abelian 
group $G_2$ of isometries whose orbits are spacelike 2-surfaces.
We derive the evolution and constraint equations for $G_2$ cosmologies
 by specializing the 
general orthonormal frame equations in the previous section.
We shall preserve the vorticity-free $\me_0$
 while taking full advantage of the $G_2$
structure to simplify the equations, namely to make the variables depend 
on $t$ and $x$ only.


Let $\pmb{\xi}_1$, $\pmb{\xi}_2$ be two independent  
Killing vector fields (KVFs) of the Abelian group of a $G_2$ 
cosmology.
First, we choose a \emph{group-invariant} orthonormal frame,
i.e.
\be
\label{group_inv}
	[ \mathbf{e}_a, \pmb{\xi}_1 ]=\mathbf{0}
        =[ \mathbf{e}_a, \pmb{\xi}_2 ].
\ee
Secondly, we also adapt the orthonormal frame $\{ \mathbf{e}_a \}$ so that
$\mathbf{e}_2$ and $\mathbf{e}_3$ are tangent to the $G_2$ orbits.
The existence of such a frame is established in the proof of Theorem 3.1 
in Wainwright 1979~\cite{art:Wainwright1979} (see Hewitt 
1989~\cite[page 120]{thesis:Hewitt1989} for a more general theorem).
For a group-invariant orbit-aligned frame it follows 
(Wainwright 1979~\cite{art:Wainwright1979})
that
\begin{gather}
\label{foo_1}
	\mathbf{e}_A(\gamma^a{}_{bc})=0,
\\
        \gamma^0{}_{0A} = \gamma^1{}_{0A}=0,\quad
        \gamma^0{}_{1A} = \gamma^1{}_{1A}=0,\quad
        \gamma^0{}_{23} = \gamma^1{}_{23}=0.
\label{foo_2}
\end{gather}
Recall that Latin indices run from 0 to 3; Greek indices 
from 1 to 3; capital Latin indices run from 2 to 3.
Writing (\ref{foo_2}) in terms of the irreducible 
variables in (\ref{117})  gives
\begin{equation}
        \dot{u}_A=0,\quad
	\Omega_A = - \epsilon_{AB} \sigma^{1B},\quad
        \omega_\alpha=0,\quad  
        a_A=0,\quad
        n_{1\alpha}=0.
\label{eq:giab}
\end{equation}
Note that $\Omega_A$ is now expressed in terms of $\sigma_{1A}$, and 
$\epsilon_{AB}$ is the anti-symmetric permutation tensor 
($\epsilon_{23}=-\epsilon_{32}=1$).

For convenience we introduce the trace-free quantity
\be
	\tilde{n}_{AB} = n_{AB} - \tfrac{1}{2} n_C{}^C \delta_{AB},
\ee
and its dual
\be
\lb{intro_dual}
	{}^*\tilde{n}_{AB} = \epsilon_{AC}\tilde{n}_B{}^C,
\ee
and similarly for $\sigma_{AB}$, ${}^3\!S_{AB}$ and $\pi_{AB}$.
The resulting system involves the following variables:
\begin{gather}
\label{117_2}
	\X_{\rm grav}=(H,\sigma_{11},\sigma_{1A},\tilde{\sigma}_{AB},
	n_C{}^C,\tilde{n}_{AB},a_1,\udot_1,\Omega_1),
\\
\label{118_2}
	\X_{\rm matter}=
	(\mu,q_1,q_A,p,\pi_{11},\pi_{1A},\tilde{\pi}_{AB}).
\end{gather}

\enlargethispage{\baselineskip}

By performing a 1+2 spatial decomposition of
the general orthonormal frame equations in the previous section, we obtain
the following equations for $G_2$ cosmologies.%
\footnote{These equations generalize the system of equations in WE, pages 
48--49, which are valid for OT $G_2$ cosmologies.}
These equations arise in a fairly obvious way from the general orthonormal 
frame equations (\ref{evo_H})--(\ref{evo_q}), with $\Omega_A$ 
replaced using (\ref{eq:giab}).

\noindent
\begin{minipage}{\textwidth}
\subsection*{The 1+1+2 equations for $G_2$ cosmologies}

\noindent
{\it Einstein field equations:}
\begin{align}
\label{AB_H}
        \me_0 H &= -H^2
        -\tfrac{1}{3}\sigma_{\alpha\beta}\sigma^{\alpha\beta}
        +\tfrac{1}{3}(\me_1
        +\udot_1 -2a_1)\udot_1-\tfrac{1}{6}(\mu+3p)
        +\tfrac{1}{3}\Lambda
\\
        \me_0 (\sigma_{11}) &=
        -3H \sigma_{11} -2 \sigma_{1A} \sigma^{1A}
        +\tfrac{2}{3}(\me_1+\udot_1+a_1)\udot_1
        -{}^3\!S_{11}+\pi_{11}
\\
\label{evo_sigma_1A}
        \me_0 (\sigma_{1A}) &=
        -3(H-\tfrac{1}{2}\sigma_{11}) \sigma_{1A} 
	- (\tilde{\sigma}_{AB}
           - \Omega_1 \epsilon_{AB})\sigma^{1B}
	+\pi_{1A}
\\
        \me_0 (\tilde{\sigma}_{AB}) &=
        -3H \tilde{\sigma}_{AB} + 2 \sigma_{1A} \sigma_{1B}
        - \sigma_{1C} \sigma^{1C} \delta_{AB}
        -2\Omega_1 {}^*\tilde{\sigma}_{AB}
        - \udot_1 {}^*\tilde{n}_{AB}
\notag\\
	&\qquad
        -{}^3\!\tilde{S}_{AB}+\tilde{\pi}_{AB}
\label{evo_sigma_AB}
\\
        0 &= (C_{\rm G}) = 6H^2 + {}^3\!R
        - \sigma_{\alpha\beta} \sigma^{\alpha\beta} -2\mu 
	-2\Lambda
\\
\label{AB_c_c}
        0 &= (C_{\rm C})_1 =
        -2\me_1( H - \tfrac{1}{2} \sigma_{11})
        -3 a_1 \sigma_{11}
        - {}^* \tilde{\sigma}_{AB} \tilde{n}^{AB} + q_1
\\
\label{c_c_A}
        0 &= (C_{\rm C})_A =
        (\me_1 -3 a_1) \sigma_{1A}
        - ({}^* \tilde{n}_{AB}
           + \tfrac{1}{2} n_C{}^C \epsilon_{AB}) \sigma^{1B} + q_A\ ,
\intertext{where}
        \sigma_{\alpha\beta}\sigma^{\alpha\beta} &=
        \tfrac{3}{2} (\sigma_{11})^2 + 2 \sigma_{1A}\sigma^{1A}
        + \tilde{\sigma}_{AB} \tilde{\sigma}^{AB}
\\
        {}^3\!R &= 4\me_1 a_1 - 6 a_1^2
                - \tilde{n}_{AB}\tilde{n}^{AB}
\\
        {}^3\!S_{11} &=
        \tfrac{2}{3} \me_1 a_1 -\tfrac{2}{3} \tilde{n}_{AB}\tilde{n}^{AB}
	,\qquad
        {}^3\!S_{1A} = 0
\\
        {}^3\!\tilde{S}_{AB} &=
        (\me_1 -2a_1){}^*\tilde{n}_{AB} + n_C{}^C \tilde{n}_{AB}\ .
\end{align}
  
\noindent
{\it Jacobi identities:}
\begin{align}
\label{evo_n_AA}
        \me_0 (n_A{}^A) &=
        (-H - \sigma_{11}) n_A{}^A
        + 2 \tilde{\sigma}^{AB} \tilde{n}_{AB}
        - 2 (\me_1 + \udot_1) \Omega_1
\\
\label{evo_n_AB}
        \me_0 (\tilde{n}_{AB}) &=
        (-H - \sigma_{11}) \tilde{n}_{AB}
        + n_C{}^C \tilde{\sigma}_{AB}
        - 2\Omega_1{}^*\tilde{n}_{AB}
        +(\me_1 + \udot_1) {}^*\tilde{\sigma}_{AB}
\\
\label{evo_a_1}
        \me_0 (a_1)
        &=  (-H - \sigma_{11}) a_1
        -(\me_1+\udot_1)(H-\tfrac{1}{2}\sigma_{11})\ .
\end{align}

\noindent
{\it Contracted Bianchi identities:}
\begin{align}
\label{evo_mu_AB}
        \me_0 (\mu) &= -3H(\mu+p)
        -\tfrac{3}{2}\sigma_{11} \pi_{11}
        -2\sigma_{1A} \pi^{1A}
        -\tilde{\sigma}_{AB} \tilde{\pi}^{AB}
\notag\\
	&\qquad
        -(\me_1+2\udot_1-2a_1)q_1
\\
        \me_0 (q_1) &=
        (-4H - \sigma_{11}) q_1 -2\sigma_{1A} q^A
        -\me_1 p -(\mu+p)\udot_1
\notag\\
        &\qquad
        -(\me_1+\udot_1-3a_1)\pi_{11}
        -{}^*\tilde{n}_{AB}\tilde{\pi}^{AB}
\label{evo_q1}
\\
        \me_0 (q_A) &=
        (-4H + \tfrac{1}{2} \sigma_{11}) q_A
        - (\tilde{\sigma}_{AB}
           + \Omega_1 \epsilon_{AB}) q^B
\notag\\
\label{evo_qA}
        &\qquad
        -(\me_1+\udot_1-3a_1)\pi_{1A}
        + ({}^*\tilde{n}_{AB}
		+\tfrac{1}{2}n_C{}^C \epsilon_{AB}) \pi^{1B}.
\end{align}   

\begin{figure}[H]
\begin{center}
\setlength{\unitlength}{1mm}
\begin{picture}(120,0)(0,0)
\put(60,90){\oval(150,230)[t]}
\put(60,90){\oval(150,170)[b]}
\end{picture}
\end{center}
\end{figure}
\end{minipage}

\newpage

\noindent
{\it Perfect fluid}
\vspace{3mm}

The 1+2 spatial decomposition of (\ref{perfect_fluid}) gives
\begin{gather}
        p = \frac{(\gamma-1)(1-v^2)+\tfrac{1}{3}\gamma v^2}{G_+}\mu,\quad
        q_1 = \frac{\gamma \mu}{G_+} v_1,\quad
        q_A = \frac{\gamma \mu}{G_+} v_A,
\notag\\
        \pi_{11} = \frac{\gamma \mu}{G_+} (v_1^2 -\tfrac{1}{3}v^2),\quad
	\pi_{1A} = \frac{\gamma \mu}{G_+} v_1 v_A,
\notag\\
	\tilde{\pi}_{AB} = \frac{\gamma \mu}{G_+} 
			(v_A v_B - \tfrac{1}{2} v_C v^C \delta_{AB})\ ,
\label{pf}
\end{gather}
where 
\be
	G_+ = 1 + (\gamma-1)v^2,\quad 
	v^2 = v_\alpha v^\alpha = v_1^2 + v_A v^A.
\ee

\vspace{3mm}
\noindent
{\it Weyl tensor}
\vspace{3mm}

The 1+2 spatial decomposition of (\ref{Weyl_E})--(\ref{Weyl_H}) gives
\begin{align}
\label{Weyl_E11}
	E_{11} &= H \sigma_{11} 
	- \tfrac{1}{2}\sigma_{11}{}^2 -\tfrac{1}{3} \sigma^{1A}\sigma_{1A}
	+ \tfrac{1}{3} \tilde{\sigma}_{AB} \tilde{\sigma}^{AB} 
	+ {}^3\!S_{11} -\tfrac{1}{2}\pi_{11}
\\
	E_{1A} &= H \sigma_{1A}
	-(\tfrac{1}{2}\sigma_{11}\sigma_{1A} 
		+ \sigma^{1B} \tilde{\sigma}_{AB})
	- \tfrac{1}{2}\pi_{1A}
\\
	\tilde{E}_{AB}
	&= H \tilde{\sigma}_{AB}
	- ( \sigma_{1A}\sigma_{1B} 
		-\tfrac{1}{2}\sigma_{1C}\sigma^{1C}\delta_{AB}
	-\sigma_{11}\tilde{\sigma}_{AB})
        + {}^3\!\tilde{S}_{AB} -\tfrac{1}{2}\tilde{\pi}_{AB}
\\
	H_{11} &= \tilde{n}_{AB} \tilde{\sigma}^{AB}
\\
	H_{1A} &= -\tfrac{1}{2}(\parb_1 - a_1)\sigma_{1B} \epsilon_A{}^B
	-\tfrac{3}{2}\sigma^{1B}\tilde{n}_{AB} 
	-\tfrac{1}{4}n_C{}^C\sigma_{1A} 
\\
	\tilde{H}_{AB} &=
	-(\parb_1 - a_1)\, {}^*\tilde{\sigma}_{AB} 
	- n_C{}^C \tilde{\sigma}_{AB}
	+ \tfrac{3}{2}\sigma_{11}\tilde{n}_{AB}\ .
\label{Weyl_HAB}
\end{align}

\vspace{3mm}
\noindent
{\it Local coordinates}
\vspace{3mm}

We choose the $x$-coordinate to be constant on the $G_2$ 
orbits.
The 1+1+2 coordinate decomposition of the frame coefficients 
(\ref{frame_coeff})
 is as follows:
\be
	\me_0 = N^{-1} \ptl_t,\quad
	\me_1 = e_1{}^1 \ptl_x + e_1{}^I \ptl_{x^I},\quad
	\me_A = e_A{}^1 \ptl_x + e_A{}^I \ptl_{x^I}\ .
\footnote{We reserve the indices $I,J,K$, running from 2 to 3, for the 
local coordinate components with respect to $y$ and $z$.}
\label{1+2_coord}
\ee
That $x$ is constant on the $G_2$ orbits means $\me_A(x)=0$, which implies
\be
	e_A{}^1=0,
\label{eAI_0}
\ee
It follows from (\ref{foo_1}), (\ref{1+2_coord}) and (\ref{eAI_0}) that
\be
\label{yz_0}
	\ptl_{x^I}(\X_{\rm grav})=0,\quad
	\ptl_{x^I}(\X_{\rm matter})=0,
\ee
which implies that
in the evolution and constraint equations
 the operator $\me_1$ is effectively
\be
        \me_1 = e_1{}^1 \partial_x\ .
\ee
In conjunction with (\ref{eq:giab}),
the commutators (\ref{comm1}) and (\ref{comm2}) acting on $t$, $x$ and 
$x^I$ give the following
equations for the remaining frame coefficients
\be
\label{119_2}
	(N,e_1{}^1,e_1{}^I,e_A{}^I).
\ee

\vspace{3mm}
\noindent
{\it Commutators acting on $t$, $x$ and $x^I$:}
\begin{align}
\label{c_udot1}
        0 &= (C_{\udot})_1 = N^{-1} \me_1(N) -\udot_1
\\
\label{evo_e11}
        \me_0 (e_1{}^1)
        &= (- H - \sigma_{11}) e_1{}^1
\\
	\me_0 (e_1{}^I)
	&= (-H - \sigma_{11}) e_1{}^I -2\sigma^{1A} e_A{}^I
\\
	\me_0 (e_A{}^I)
	&= -H e_A{}^I - \sigma_A{}^B e_B{}^I 
		- \Omega_1 \epsilon_A{}^B e_B{}^I
\\
\label{c_com1A}
        0 &= (C_{\rm com})^I{}_{1A} =
	(\me_1 -a_1) e_A{}^I 
	- ({}^*\tilde{n}_A{}^B
	   +\tfrac{1}{2}n_C{}^C \epsilon_A{}^B) e_B{}^I.
\end{align}
Note that $e_1{}^I$ and $e_A{}^I$ \emph{decouple} from the 
system (\ref{AB_H})--(\ref{evo_qA}).

Equations (\ref{AB_H})--(\ref{evo_qA}), (\ref{c_udot1}) and 
(\ref{evo_e11})
give a system of first order PDEs in $t$ and $x$ for the variables in 
(\ref{117})--(\ref{118}) and the pair $(N,e_1{}^1)$.
The system is under-determined, however, since there are no evolution
equations for $N$, $\dot{u}_1$, $\Omega_1$, $p$ and
$\pi_{\alpha\beta}$.
A perfect fluid matter source 
with linear barotropic equation of state
gives (\ref{pf}),
while temporal and spatial gauges determine ($N$,$\dot{u}_1$) and
$\Omega_1$ respectively.
These equations are used to analyze the dynamics of $G_2$ cosmologies and 
of SH cosmologies whose $G_3$ group has an Abelian $G_2$ subgroup.

\section{Scale-invariant variables}\label{sec:invariant}

The variables in (\ref{117}), (\ref{118}) and (\ref{frame_coeff}) are 
scale-dependent and dimensional, and are unsuitable for describing the 
asymptotic behaviour of cosmological models near the initial singularity, 
since they typically diverge.
It is thus essential to introduce scale-invariant (dimensionless)
variables, which one hopes will be bounded as the initial singularity is 
approached.
As a motivating example, consider an FL model, in which the matter density 
$\mu \rightarrow \infty$ at the initial singularity. 
One wants to introduce a corresponding scale-invariant, dimensionless
density parameter $\Omega=\mu/(3H^2)$ such 
that $\Omega \rightarrow 1$.%
\footnote{Although it is sufficient to require that $\Omega$ tends to a 
non-zero constant, it is customary to normalize $\mu$ such that
$\Omega \rightarrow 1$.}
Another example is the Kasner models, in which the shear 
$\sigma_{\alpha\beta}$ satisfies
$\sigma^2 \equiv \tfrac{1}{2}\sigma_{\alpha\beta} \sigma^{\alpha\beta}
 \rightarrow \infty$ at the initial 
singularity. 
One wants to introduce a corresponding shear parameter 
$\Sigma_{\alpha\beta}= \sigma_{\alpha\beta}/H$ such that
$\Sigma^2 \equiv \sigma^2/(3H^2) \rightarrow 1$.
These examples motivate the use of the Hubble scalar $H$ as the 
normalizing factor.

\subsection*{1+3 Hubble-normalized variables}

We define the Hubble-normalized gravitational and matter variables 
respectively as follows:
\begin{gather}
\label{Hubble_1}
	(\Sigma_{\alpha\beta}, \Udot_\alpha, R_\alpha, N_{\alpha\beta}, 
	A_\alpha)
	= (\sigma_{\alpha\beta}, \udot_\alpha, \Omega_\alpha, 
	n_{\alpha\beta}, a_\alpha)/H
\\
\label{Hubble_2}
	(\Omega, Q_\alpha, P, \Pi_{\alpha\beta},\Omega_\Lambda)
	= (\mu, q_\alpha, p, \pi_{\alpha\beta},\Lambda) / (3H^2)\ .
\end{gather}
The differential operators
(\ref{ptl1}) and the frame coefficients (\ref{frame_coeff})
 are also redefined to be scale-invariant:
\begin{gather}
\label{Hubble_3}
        \parb_0 = \frac{1}{H} \me_0,\quad 
	\parb_\alpha = \frac{1}{H} \me_\alpha,
\\
	\mathcal{N} = N H,\quad
	E_\alpha{}^i = \frac{e_\alpha{}^i}{H},
\label{Hubble_3b}
\end{gather}
since the frame variables have the same dimension as $H$.
In order to transform the evolution and constraint equations, 
it is necessary to introduce the 
\emph{deceleration parameter} $q$ and the 
\emph{Hubble gradient} $r_\alpha$ according to
\begin{align}
\label{def_q}
        q+1 &= - \frac{\parb_0 H}{H}
\\
\label{def_r}
        r_\alpha &= - \frac{\parb_\alpha H}{H}\ .
\end{align}   
In order to transform the equations, it is convenient to use the following 
identities.
For the gravitational field variables $\X_{\rm grav}$ in (\ref{117}), 
which have the same dimension as $H$, we use the identities
\begin{align}
	\frac{1}{H^2} \me_0 \X_{\rm grav}
	&=
	[ \parb_0 - (q+1) ] \left( \frac{\X_{\rm grav}}{H} \right) 
\\
        \frac{1}{H^2} \me_\alpha \X_{\rm grav}
	&=
	[ \parb_\alpha - r_\alpha ] \left( \frac{\X_{\rm grav}}{H} \right),
\end{align}
which follow from (\ref{Hubble_1}), (\ref{Hubble_3})--(\ref{def_r}).
Similarly, for the matter variables $\X_{\rm matter}$ in (\ref{118}), 
which have the dimension $[H^2]$, we use the identities
\begin{align}
        \frac{1}{3H^3} \me_0 \X_{\rm matter}
	&=
        [ \parb_0 - 2(q+1) ] \left( \frac{\X_{\rm matter}}{3H^2} \right)
\\
        \frac{1}{3H^3} \me_\alpha \X_{\rm matter}
	&=
        [ \parb_\alpha - 2r_\alpha ] \left( \frac{\X_{\rm matter}}{3H^2} 
		\right).
\end{align}
Then equation (\ref{evo_H}) for $\parb_0 H$ decouples, and through 
(\ref{def_q}) gives the following expression for $q$:
\be
\label{q_exp}
	q = \tfrac{1}{3} \Sigma_{\alpha\beta} \Sigma^{\alpha\beta}
	+ \tfrac{1}{2}(\Omega+3P) - \Oml
	-\tfrac{1}{3}(\parb_\alpha 
	-r_\alpha+\Udot_\alpha-2A_\alpha)\Udot^\alpha,
\ee
i.e. $q$ is expressed in terms of the Hubble-normalized variables 
(\ref{Hubble_1}) and (\ref{Hubble_2}).

Equations (\ref{perfect_fluid}) in Hubble-normalized form are
\be
\label{G0_perfect_fluid}
        P = \frac{(\gamma-1)(1-v^2)+\tfrac{1}{3}\gamma v^2}{G_+}\Omega,\quad
        Q_\alpha = \frac{\gamma \Omega}{G_+} v_\alpha,\quad
        \Pi_{\alpha\beta}
        = \frac{\gamma \Omega}{G_+} v_{\la \alpha} v_{\beta \ra}\ .
\ee

The spatial curvature variables ${}^3\!R$ and ${}^3\!S_{\alpha\beta}$
are normalized according to
\be
	\Om_k = - \frac{{}^3\!R}{6H^2}\ ,\quad
	\mathcal{S}_{\alpha\beta} = \frac{{}^3\!S_{\alpha\beta}}{3H^2}\ .
\ee
The Weyl curvature variables $E_{\alpha\beta}$ and $H_{\alpha\beta}$
are normalized according to
\be
\label{Weyl_Hubble}
	(\mathcal{E}_{\alpha\beta} , \mathcal{H}_{\alpha\beta})
	= (E_{\alpha\beta},H_{\alpha\beta})/(3H^2).
\ee

The commutators (\ref{comm1})--(\ref{comm2}) in Hubble-normalized form are
\begin{align}
\label{Comm1}
        [ \parb_0, \parb_\alpha ] &=
        (\dot{U}_\alpha-r_\alpha) \parb_0 
	- \left( -q {\delta_\alpha}^\beta
        +{\Sigma_\alpha}^\beta + \epsilon_\alpha{}^{\beta\gamma}
        R_\gamma \right) \parb_\beta\ ,
\\
\label{Comm2}
        [ \parb_\alpha, \parb_\beta ] &=
        \left[ \epsilon_{\alpha\beta\nu}N^{\mu\nu}
        + (A_\alpha+r_\alpha) {\delta_\beta}^\mu-(A_\beta+r_\beta)
        {\delta_\alpha}^\mu \right] \parb_\mu\ .
\end{align}

The 1+3 Hubble-normalized variables and equations are well-suited for 
analyzing $G_0$ cosmologies.
They are also well-suited for SH cosmologies of types VIII and IX, in 
which (\ref{SH}) implies
\be
\label{Hubble_last}
	\parb_\alpha(\X_{\rm grav}/H)=0,\quad
	\parb_\alpha(\X_{\rm matter}/(3H^2) )=0,\quad 
	\Udot_\alpha=0,\quad r_\alpha=0.
\ee
It then follows from (\ref{c_udot}) and 
(\ref{Hubble_3})--(\ref{Hubble_3b}) that 
we can set
\be
\label{SH_N_1}
	\mathcal{N}=1,\quad\text{so that}\quad \parb_0 = \ptl_t\ .
\ee

\subsection*{1+1+2 Hubble-normalized variables for
                $G_2$-compatible SH cosmologies}

For $G_2$-compatible SH cosmologies, we define the Hubble-normalized 1+1+2 
decomposed gravitational and matter variables as follows:
\begin{gather}
\label{1+1+2_1}
        (\Sigma_{11},\Sigma_{1A},\tilde{\Sigma}_{AB},R,
                N_C{}^C,\tilde{N}_{AB},A)
        =(\sigma_{11},\sigma_{1A},\tilde{\sigma}_{AB},\Omega_1,
        n_C{}^C,\tilde{n}_{AB},a_1)/H
\\
        (\Omega,Q_1,Q_A,P,\Pi_{11},\Pi_{1A},\tilde{\Pi}_{AB},
                \Omega_\Lambda) =
        (\mu,q_1,q_A,p,\pi_{11},\pi_{1A},\tilde{\pi}_{AB},\Lambda) / (3H^2).
\label{1+1+2_2}
\end{gather}
Equivalently, one can view this as the 1+1+2 decomposed version of 
the 1+3 Hubble-normalized variables in (\ref{Hubble_1})--(\ref{Hubble_2}).
Equations (\ref{Hubble_3})--(\ref{Hubble_last}) are similarly decomposed.

\subsection*{1+1+2 $\beta$-normalized variables for 
		$G_2$ cosmologies}

For $G_2$ cosmologies,
the area expansion of the $G_2$ orbits is described by
\be
        \Theta_{AB} \equiv H \delta_{AB} + \sigma_{AB}.
\ee
The average expansion rate of the $G_2$ orbits is described by
\be
\label{H_beta_off}
        \beta \equiv \tfrac{1}{2} \Theta_C{}^C = H - \tfrac{1}{2} \sigma_{11}.
\ee
It turns out that there are some advantages in using $\beta$ 
instead of $H$ as the normalizing factor
(see van Elst \etal 2002~\cite{art:vEUW2002}).
We define the $\beta$-normalized variables as follows:
\begin{gather}
	(\Sigma_{11},\Sigma_{1A},\tilde{\Sigma}_{AB},\Udot,R,
		N_C{}^C,\tilde{N}_{AB},A)
        =(\sigma_{11},\sigma_{1A},\tilde{\sigma}_{AB},\udot_1,\Omega_1,
	n_C{}^C,\tilde{n}_{AB},a_1)/\beta
\\
        (\Omega,Q_1,Q_A,P,\Pi_{11},\Pi_{1A},\tilde{\Pi}_{AB},
		\Omega_\Lambda) =
	(\mu,q_1,q_A,p,\pi_{11},\pi_{1A},\tilde{\pi}_{AB},\Lambda) / (3\beta^2).
\end{gather}
The differential operators are likewise normalized as follows:
\begin{gather}
        \parb_0 = \frac{1}{\beta} \me_0,\quad
        \parb_1 = \frac{1}{\beta} \me_1, 
\\
        \mathcal{N} = N \beta,\quad
        \EEE = \frac{e_1{}^1}{\beta},
\label{N_beta}
\end{gather}
We define the analogous
deceleration parameter $q$, and the $\beta$ gradient $r$ according to
\be
\label{q_r_beta}
        q+1 = - \frac{\parb_0 \beta}{\beta},\quad
        r   = - \frac{\parb_1 \beta}{\beta}\ .
\ee
The equations are similarly transformed using the following identities:
\begin{align}
        \frac{1}{\beta^2} \me_0 \X_{\rm grav} 
	&=
        [ \parb_0 - (q+1) ] \left( \frac{\X_{\rm grav}}{\beta} \right)
\\
        \frac{1}{\beta^2} \me_1 \X_{\rm grav}
	&=
        [ \parb_1 - r ] \left( \frac{\X_{\rm grav}}{\beta} 
	\right)
\\
        \frac{1}{3\beta^3} \me_0 \X_{\rm matter}
	&=
        [ \parb_0 - 2(q+1) ] \left( \frac{\X_{\rm matter}}{3\beta^2} 
	\right)
\\
        \frac{1}{3\beta^3} \me_1 \X_{\rm matter}
	&=
        [ \parb_1 - 2r ] \left( \frac{\X_{\rm 
	matter}}{3\beta^2}\right).
\end{align}
Consider the evolution equation for $\beta$, in which $\me_1 a_1$ is 
replaced using $(C_{\rm G})$:
\begin{multline}
	\me_0 \beta 
	= -\tfrac{3}{2}\beta^2 
	+\tfrac{1}{2}\sigma_{1A}\sigma^{1A}
	-\tfrac{1}{4}(\tilde{\sigma}_{AB} \tilde{\sigma}^{AB}
			+ \tilde{n}_{AB} \tilde{n}^{AB})
	+\tfrac{1}{2}a_1^2 - a_1 \udot_1
\\
	-\tfrac{1}{2}(p+\pi_{11}) + \tfrac{1}{2}\Lambda.
\label{beta_evo}
\end{multline}
This defines the corresponding area deceleration parameter $q$:
\begin{multline}
\label{beta_q}
	q = \tfrac{1}{2} - \tfrac{1}{2}\Sigma_{1A}\Sigma^{1A}
        +\tfrac{1}{4}(\tilde{\Sigma}_{AB} \tilde{\Sigma}^{AB}
                        + \tilde{N}_{AB} \tilde{N}^{AB})
        -\tfrac{1}{2}A^2 + A \Udot
\\
        +\tfrac{3}{2}(P+\Pi_{11}) - \tfrac{3}{2}\Oml.
\end{multline}
The evolution equation for $\beta$ decouples from the main system.
The $(C_{\rm C})_1$ constraint (\ref{AB_c_c}) can be solved for $r$:
\be
\label{beta_r}
	r = \tfrac{3}{2} (  A \Sigma_{11}
		+ {}^* \tilde{\Sigma}_{AB} \tilde{N}^{AB} - Q_1 ).
\ee

The spatial curvature variables ${}^3\!R$,
${}^3\!S_{11}$
 and ${}^3\!S_{AB}$
are normalized according to
\be
        \Om_k = - \frac{{}^3\!R}{6\beta^2}\ ,\quad
        \mathcal{S}_{11} = 
		\frac{{}^3\!S_{11}}{3\beta^2}\ ,\quad
        \mathcal{S}_{AB} =
                \frac{{}^3\!S_{AB}}{3\beta^2}\ .
\ee
The Weyl curvature variables 
are normalized according to
\be
\label{Weyl_beta}
        (\mathcal{E}_{11},
	\mathcal{E}_{1A},
	\tilde{\mathcal{E}}_{AB},
	\mathcal{H}_{11}, 
        \mathcal{H}_{1A},
        \tilde{\mathcal{H}}_{AB})
	=
	(E_{11},E_{1A},\tilde{E}_{AB},
	H_{11},H_{1A},\tilde{H}_{AB})/(3\beta^2).
\ee



The commutator $[\parb_0,\parb_1]$ is given by
\be
\label{dzdi}
	[\parb_0,\parb_1]
	= (\Udot-r)\parb_0 + (q-\tfrac{3}{2}\Sigma_{11}) \parb_1 
	-2\Sigma_1{}^A \parb_A\ .
\ee

%% file: G2_hierarchy.tex
	\chapter{The $G_2$ and SH hierarchies}\label{chap:G2hie}

In this chapter we give a classification of $G_2$ cosmologies, and of the 
SH cosmologies that arise as limiting cases, using the action of the $G_2$ 
group of isometries, and the tilt degrees of freedom of the perfect fluid.
Our goal is to provide a framework for describing the remarkably diverse 
range of dynamical behaviour that is possible within the class of $G_2$ 
cosmologies, only a small part of which has appeared in the literature to 
date.

\section{The $G_2$ hierarchy}\label{sec:G2tree}

It is useful to classify $G_2$ cosmologies according to the action of the 
$G_2$ group, as determined by the properties of the KVFs associated with 
the $G_2$ group.
There are four restrictions relating to the KVFs,
which determine four special classes of $G_2$ cosmologies.
If none of these restrictions is satisfied we shall refer to the $G_2$ 
cosmology as \emph{generic}.
We now give the above-mentioned restrictions,
and then characterize the subclasses using the gravitational field 
variables (\ref{117_2}).

\

\noindent
(1){\it Hypersurface-orthogonal}

This case is specified by requiring that
the $G_2$ admits a hypersurface-orthogonal KVF $\pmb{\xi}$,
\be
	\xi_{[a;b} \xi_{c]} =0.
\ee
The following proposition provides a convenient way to check for the 
existence of a hypersurface-orthogonal KVF.

\begin{prop}\lb{prop3.1}
Consider a $G_2$ cosmology presented in a group-invariant orbit-aligned 
frame
with $\mathbf{e}_2$ and $\mathbf{e}_3$ tangent to the $G_2$ orbits.
Then the $G_2$ admits a hypersurface-orthogonal KVF if and only if
the gravitational field variables (\ref{117_2}) satisfy
\be
\label{HO_grav}
        \tilde{\sigma}_{AB} \tilde{n}^{AB}=0,\quad
        \tilde{n}_{AB} \sigma^{1A} \sigma^{1B} =0,\quad
        {}^*\tilde{\sigma}_{AB} \sigma^{1A} \sigma^{1B} =0.\
\footnote{{\rm If $\tilde{n}_{AB}$ and $\sigma^{1A}$ are non-zero, then 
any two of the conditions in (\ref{HO_grav}) imply the third.
Furthermore, for SH cosmologies, we also have $n_C{}^C=0$.}}
\ee
In addition, the matter variables (\ref{118_2}) satisfy
\be
        \epsilon^{AB} q_A \sigma_{1B} =0,\quad
        \epsilon^{AB} \pi_{1A} \sigma_{1B} =0,\quad
        {}^* \tilde{\pi}_{AB} \sigma^{1A} \sigma^{1B} =0.\
\footnote{{\rm It also follows from (\ref{HO_grav}) and 
(\ref{HOKVF_matter}) that
\[
	{}^* \tilde{\pi}_{AB} \tilde{\sigma}^{AB}=0,\quad
	\tilde{\pi}_{AB} \tilde{n}^{AB}=0.
\]}}
\label{HOKVF_matter}
\ee
\end{prop}

\begin{proof}
 See Appendix~\ref{app:HO}.
\end{proof}

\

\noindent
(2){\it Orthogonally transitive}

This case is specified by requiring that
the $G_2$ acts orthogonally transitively, 
\be
	\xi_{[a;b}\xi_c \eta_{d]} =0,\quad
	\eta_{[a;b}\eta_c \xi_{d]} =0,
\ee
where $\pmb{\xi}$ and $\pmb{\eta}$ are two linearly independent KVFs
(Carter 1973~\cite[page 160]{inbook:Carter1973}).
In other words, the 2-spaces
orthogonal to the orbits of the $G_2$ are surface-forming.
This is equivalent to (see WE, page 46)
\be
	\sigma_{1A} =0.
\label{OT_grav}
\ee
In this class, 
equations (\ref{evo_sigma_1A}) and (\ref{c_c_A}) imply that
the matter variables (\ref{118}) are restricted as follows:
\be
	q_A=0,\quad \pi_{1A}=0.
\label{OT_matter}
\ee

\noindent
(3){\it Diagonal}

This case is specified by requiring that
the $G_2$ admits two hypersurface-orthogonal KVFs. 
This class is the intersection of the first two classes, and is thus
characterized by
\be
\label{diag_grav}
        \sigma_{1A} =0,\quad \tilde{\sigma}_{AB} \tilde{n}^{AB}=0,
\ee
as follows from (\ref{HO_grav}) and (\ref{OT_grav}).
In this class the matter variables (\ref{118}) are restricted as follows:
\be
        q_A=0,\quad \pi_{1A}=0,\quad
	{}^*\tilde{\pi}_{AB} \tilde{\sigma}^{AB} =0.
\ee
The name
``diagonal" reflects the fact that the line element can be written in
diagonal form in this class.

\

\noindent
(4)
{\it Plane symmetric or locally rotationally symmetric (LRS)}

In the diagonal $G_2$ class, if there exists in addition a one-parameter 
isotropy group, then the model is
 locally rotationally symmetric (WE, page 43). 
This class is characterized by
\be
\label{LRS_grav}
        \sigma_{1A} =0 = \tilde{\sigma}_{AB} = \tilde{n}_{AB}\ ,
\ee
since otherwise there would exist a preferred direction in the 23-space.
In this class the matter variables (\ref{118}) are restricted as follows:
\be
        q_A=0,\quad \pi_{1A}=0,\quad
        \tilde{\pi}_{AB} =0.
\ee

\begin{figure}
\begin{center}
    \epsfig{file=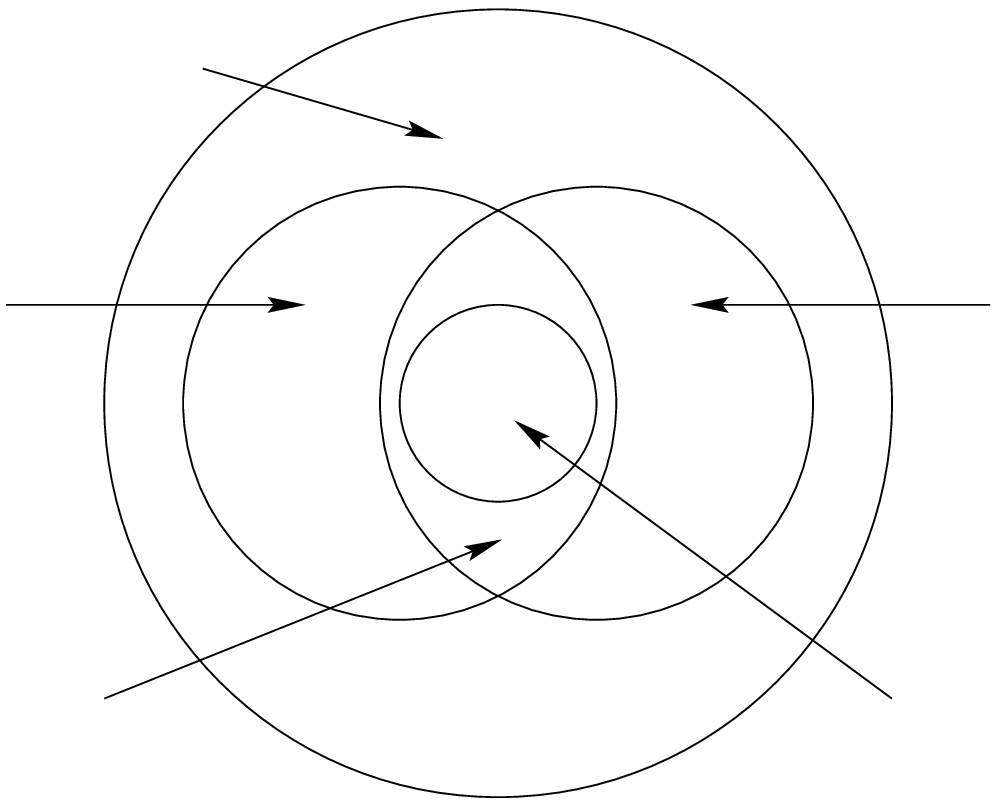,width=8cm}
\setlength{\unitlength}{1mm}
\begin{picture}(120,0)(0,0)
\put(10,10){\makebox(0,0)[l]{diagonal}}
\put(90,10){\makebox(0,0)[l]{LRS}}     
\put(0,45){\makebox(0,0)[l]{HO KVF}}     
\put(100,45){\makebox(0,0)[l]{OT}}     
\put(20,65){\makebox(0,0)[l]{generic}}     

\end{picture}
\end{center}
\caption{Classification of $G_2$ cosmologies.}\label{fig:G2_action}
\end{figure}

\

Figure~\ref{fig:G2_action} summarizes the classification of $G_2$ 
cosmologies 
using the action of the $G_2$ group and the existence of an LRS group
(see WE, page 44).
On account of (\ref{HO_grav}) and (\ref{OT_grav}) we can say that a $G_2$ 
cosmology is generic if and only if
\be
	\sigma_{1A} \sigma^{1A} \neq 0,\quad
        (\tilde{\sigma}_{AB} \tilde{n}^{AB})^2
        + (\tilde{n}_{AB} \sigma^{1A} \sigma^{1B})^2
        + ({}^*\tilde{\sigma}_{AB} \sigma^{1A} \sigma^{1B})^2 \neq 0.
\ee

\newpage

Assuming a perfect fluid matter content,
we now show that $G_2$ cosmologies can be further classified using 
the 
\emph{tilt degrees of freedom} of the perfect fluid,
as described by the tilt vector $v_\alpha$
(see Section~\ref{sec:1+3}).
From (\ref{evo_sigma_1A}), (\ref{evo_qA}), and using (\ref{c_c_A}) and 
(\ref{pf}), one finds that the evolution equation for 
$\epsilon^{AB} q_A \sigma_{1B}$ is homogeneous:
\be
	\me_0 (\epsilon^{AB} q_A \sigma_{1B}) = 
	(-7H + 2 \sigma_{11} - \me_1 v_1 - v_1 \udot_1 + 6v_1 a_1 - v_1 
	\me_1) (\epsilon^{AB} q_A \sigma_{1B}).
\label{312}
\ee
Recalling from (\ref{pf}) that $q_A = \gamma G_+^{-1} \mu v_A$,
equation (\ref{312}) implies that
$\epsilon^{AB} v_A \sigma_{1B}=0$ defines an invariant set.
Likewise, the evolution equation for $q_A$ is homogeneous, on account of 
(\ref{pf}), implying that $v_A=0$ forms an invariant set.
These considerations lead to the hierarchy presented in
Table~\ref{tab:pf_hierarchy}.%
\footnote{By inspection of equations (\ref{evo_q1})--(\ref{evo_qA}) and 
using (\ref{pf}), one finds that cases where $v_1=0$, $v_A\neq0$ are not 
invariant.}

\newpage

\begin{table}[h]
\begin{spacing}{1.1}
\caption{Perfect fluid hierarchy in the presence of a $G_2$ group.}
                \label{tab:pf_hierarchy}
\begin{center}
\begin{tabular*}{\textwidth}%
        {@{\extracolsep{\fill}}llc}
\hline
Restriction	& Description	& tilt degrees of freedom
\\
\hline
$\epsilon^{AB} v_A \sigma_{1B}\neq0$ & 
$v_A$ not parallel to $\sigma_{1A}$ & 3
\\
$\epsilon^{AB} v_A \sigma_{1B} = 0,\ v_A \neq0$ &
$v_A$ parallel to $\sigma_{1A}$ & 2
\\
$v_A =0,\ v_1\neq 0$ &
$v_\alpha$ orthogonal to $G_2$ orbits & 1
\\
$v_\alpha =0$ & non-tilted fluid & 0
\\
\hline
\end{tabular*}
\end{center}
\end{spacing}
\end{table}

\enlargethispage{2\baselineskip}

\begin{figure}
\begin{center}
\setlength{\unitlength}{1mm}
\begin{picture}(120,140)(0,0)
\put(30,70){\framebox(60,70)[c]{}}
\put(60,135){\makebox(0,0)[t]{generic $G_2$ action}}
\put(35,115){\framebox(50,10)[c]{with $\epsilon^{AB} v_A 
		\sigma_{1B}\neq0$}}
\put(35,95){\framebox(50,10)[c]{with $\epsilon^{AB} v_A \sigma_{1B}=0$}}
\put(35,75){\framebox(50,10)[c]{with $v_A=0$}}

\put(-10,15){\framebox(60,50)[c]{}}
\put(20,60){\makebox(0,0)[t]{one HO KVF}}
\put(-5,40){\framebox(50,10)[c]{with $\epsilon^{AB} v_A
                \sigma_{1B}=0$}}
\put(-5,20){\framebox(50,10)[c]{with $v_A=0$}}

\put(80,20){\framebox(40,10)[100,20]{OT $G_2$ ($v_A=0$)}}
\put(35,0){\framebox(50,10)[50,0]{diagonal $G_2$ ($v_A=0$)}}
\put(60,75){\line(4,-1){40}}
\put(60,95){\line(-4,-1){40}}
\put(60,10){\line(4,1){40}}
\put(60,10){\line(-4,1){40}}
\put(60,115){\line(0,-1){10}}
\put(60,95){\line(0,-1){10}}
\put(20,85){\line(0,-1){20}}
\put(20,40){\line(0,-1){10}} 
\put(100,65){\line(0,-1){35}}
\end{picture}
\end{center}
\caption{Action of the $G_2$ group and the tilt degrees
of freedom.}\label{fig:G2_action_tilt}
\end{figure}
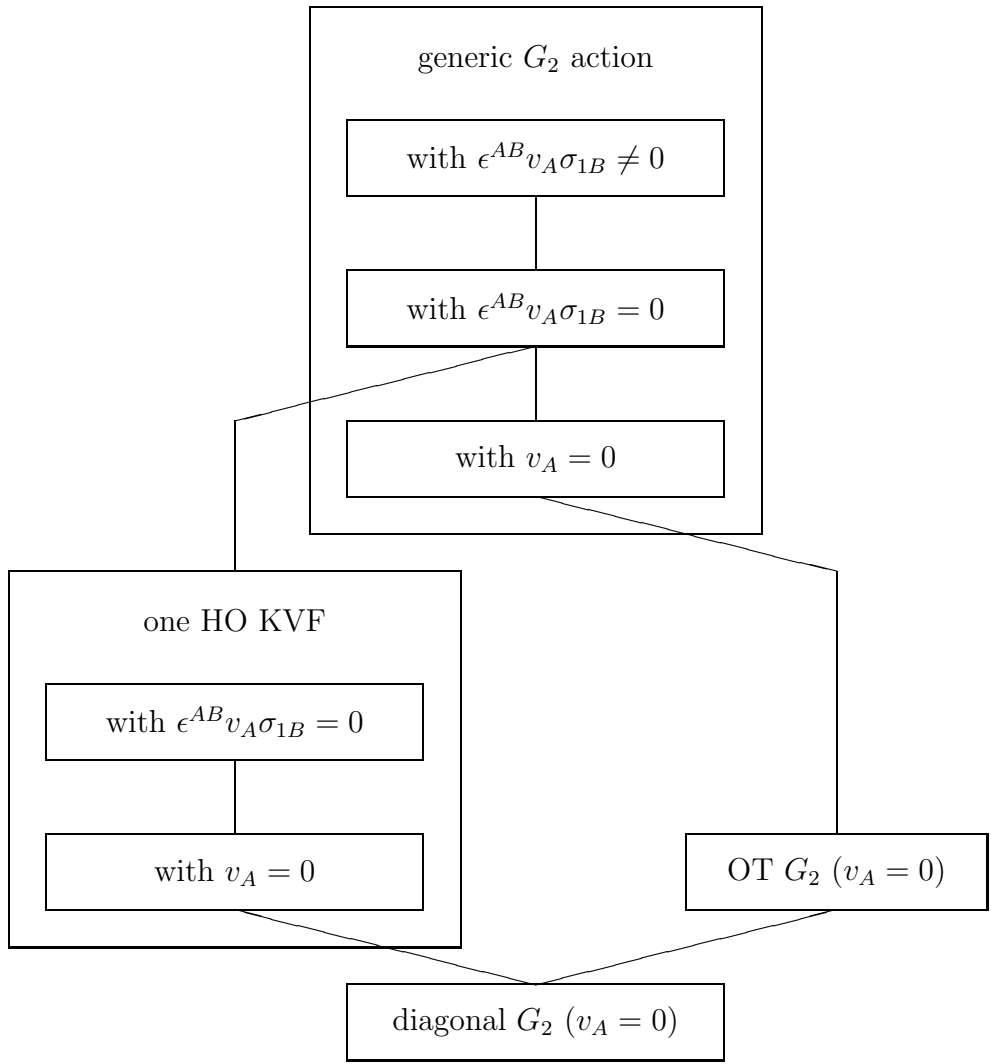

Figure~\ref{fig:G2_action_tilt} summarizes the classification of $G_2$ 
cosmologies using the action of the $G_2$ group and the tilt degrees
of freedom. 
Note that the matter restriction (\ref{HOKVF_matter}) of
the class of hypersurface-orthogonal $G_2$ implies
$\epsilon^{AB} v_A \sigma_{1B}=0$,
and restriction (\ref{OT_matter}) of the class of OT $G_2$ implies
$v_A=0$.
The class of hypersurface-orthogonal $G_2$ also contains $v_A=0$ as an 
invariant subset.

\section{$G_2$-compatible SH cosmologies}\label{sec:SHtree}

Perfect fluid SH cosmologies of Bianchi type I--VII admit an Abelian $G_2$ 
subgroup
(this can be inferred from Table 11.2 in 
MacCallum 1979~\cite{inbook:MacCallum1979}).
The properties of $\det n_{AB}$ and $a_1$ for $G_2$--compatible SH
cosmologies are summarized in Table~\ref{tab:G2_Bianchi}.
These cosmologies can be further classified according to the $G_2$ action
(see Figure~\ref{fig:G2_action})
and the tilt degrees of freedom,
as in Table~\ref{tab:pf_hierarchy}.%
\footnote{The parameter $h$ in Table~\ref{tab:pf_hierarchy} will be 
introduced in equation (\ref{intro_h}).}

\begin{table}[h]
\begin{spacing}{1.1}
\caption{$G_2$--compatible SH cosmologies.}
                \label{tab:G2_Bianchi}
\begin{center}
\begin{tabular*}{0.65\textwidth}%
        {@{\extracolsep{\fill}}ccc}
\hline
	& $a_1=0$ & $a_1\neq0$
\\
\hline
det $n_{AB} >0$ & VII$_0$ & VII$_h$ \\
det $n_{AB} <0$ & VI$_0$  & VI$_h$  \\
det $n_{AB} =0$ & II      & IV      \\
    $n_{AB} =0$ & I       & V       \\
\hline
\end{tabular*}  
\end{center}
\end{spacing}
\end{table}

\subsection*{SH cosmologies with generic $G_2$}

\begin{figure}[h]
\begin{center}
    \epsfig{file=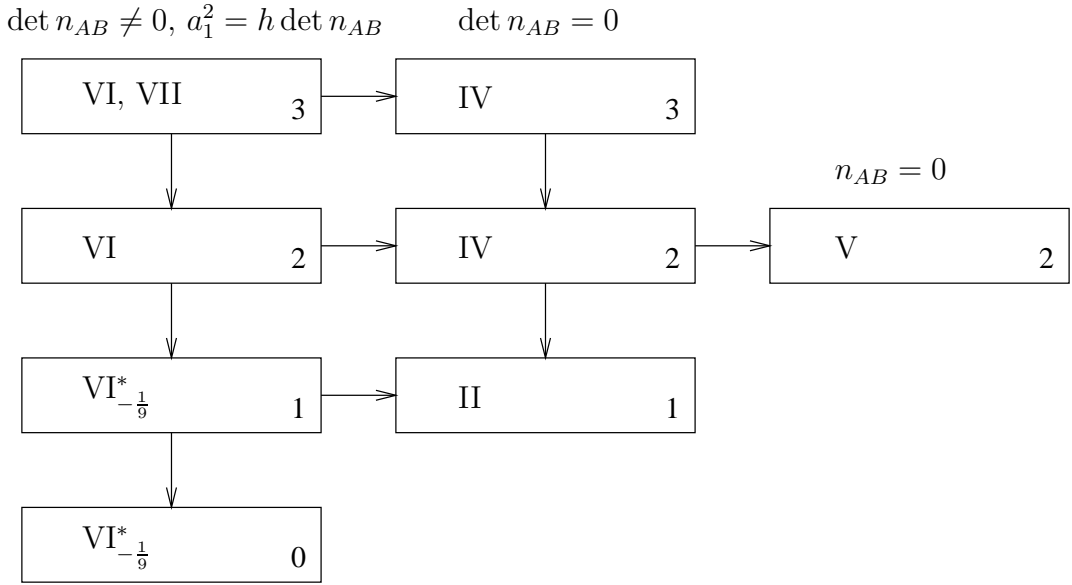,width=14cm}
\setlength{\unitlength}{1mm}  
\begin{picture}(120,0)(0,0)

\put(-10,80){\makebox(0,0)[l]{$\det n_{AB} \neq 0$, $a_1^2 = h \det 
n_{AB}$}}
\put(0,70){\makebox(0,0)[l]{VI, VII}}
\put(0,50){\makebox(0,0)[l]{VI}}   
  
\put(0,30){\makebox(0,0)[l]{VI$^*_{-\frac{1}{9}}$}}
\put(0,10){\makebox(0,0)[l]{VI$^*_{-\frac{1}{9}}$}}

\put(100,60){\makebox(0,0)[l]{$n_{AB}=0$}}  
\put(100,50){\makebox(0,0)[l]{V}}

\put(50,80){\makebox(0,0)[l]{$\det n_{AB}=0$ }}
\put(50,70){\makebox(0,0)[l]{IV}}
\put(50,50){\makebox(0,0)[l]{IV}}
\put(50,30){\makebox(0,0)[l]{II}}

\end{picture}
\end{center}
\caption[SH cosmologies with generic $G_2$ action.]
{SH cosmologies with generic $G_2$ action. The numbers 0,1,2,3 
indicate the tilt degrees of freedom.
}\label{fig:gen_G2}
\end{figure}

Consider the SH cosmologies with a generic $G_2$ subgroup.
We now show that the number of tilt degrees of freedom imposes strong 
restrictions on the possible Bianchi type, and that conversely, for a 
given Bianchi type, the tilt degrees of freedom are restricted.
The results are summarized in Figure~\ref{fig:gen_G2}.
The number on the right of each box in Figure~\ref{fig:gen_G2}
 indicates the tilt degrees of freedom, as defined in 
Table~\ref{tab:pf_hierarchy}.

We now justify the results in Figure~\ref{fig:gen_G2}.
Bianchi V cosmologies with generic $G_2$ action necessarily have 
$\sigma_{1A}$ parallel to $q_A$, as follows from (\ref{c_c_A}).
Note that Bianchi VII cosmologies with generic $G_2$ action necessarily 
have the most general tilt, as shown in the following proposition.

\begin{prop}
Consider a $G_2$-compatible SH cosmology presented in a group-invariant 
orbit-aligned frame with $\mathbf{e}_2$ and $\mathbf{e}_3$ tangent to the 
$G_2$ orbits, and $\me_1$ tangent to the $G_3$ orbits.
If
\[	\text{det } n_{AB} >0
	\quad\text{and}\quad \sigma_{1A}\sigma^{1A}\neq0,
\]
then
\[
	\epsilon^{AB} v_A \sigma_{1B} \neq 0.
\]
\end{prop}

\begin{proof} Suppose $\epsilon^{AB} v_A \sigma_{1B} =0$. Contract
$(C_{\rm C})_A$ with $\epsilon^{AB}\sigma_{1B}$ to obtain
\be
	n_C{}^C\sigma_{1A}\sigma^{1A} 
	= 2 \tilde{n}_{AB} \sigma^{1A} \sigma^{1B}.
\label{VII_proof_1}
\ee
Assuming $\sigma_{1A}\sigma^{1A}\neq0$,
represent $\sigma_{1A}$ using polar coordinates as follows:
\[
	\sigma_{1A} = \sigma_\perp K_{1A},
\]
where $K_{12}=\cos\phi$, $K_{13} =\sin\phi$.
Then it follows from (\ref{VII_proof_1}) that
\be
	n_C{}^C = 2 \tilde{n}_{AB} K^{1A} K^{1B}.
\label{polar_K}
\ee
The determinant of $n_{AB}$, given by
\be
\label{det_n}
	\det(n_{AB}) = \tfrac{1}{4} (n_C{}^C)^2 - \tfrac{1}{2} 
			\tilde{n}_{AB} \tilde{n}^{AB},
\ee
is then expanded and simplified using
(\ref{polar_K}) and
 trigonometric identities to yield
\be
\label{VII_proof_det}
	\det(n_{AB}) = 
	- ({}^*\tilde{n}_{AB} K^{1A} K^{1B})^2 \leq 0.
\ee
where ${}^*\tilde{n}_{AB}$ is defined by (\ref{intro_dual}).
\end{proof}

\subsubsection*{The $h$ parameter for Bianchi VI$_h$ and VII$_h$ 
cosmologies}

We now define the $h$ parameter for Bianchi VI$_h$ and VII$_h$
cosmologies.
Note that the evolution equation for $\det(n_{AB})$ is given by
\be
	\me_0 (\det n_{AB}) = 2(-H-\sigma_{11}) \det n_{AB},
\ee
as follows from (\ref{det_n}) and (\ref{evo_n_AA})--(\ref{evo_n_AB}).
Comparing with the evolution equation (\ref{evo_a_1}) for $a_1$:
\be
	\me_0 (a_1) = (-H-\sigma_{11}) a_1,
\ee
it follows that if $\det(n_{AB})\neq 0$, there exists a constant $h$ such 
that
\be
\lb{intro_h}
	a_1^2 =  h \det(n_{AB}).
\ee
Bianchi VII$_h$ cosmologies are characterized by
\[
	\det(n_{AB})> 0,\quad a_1 \neq 0,
\]
requiring that $h$ satisfies $h>0$,
while Bianchi VI$_h$ cosmologies are characterized by
\[ 
        \det(n_{AB})< 0,\quad a_1 \neq 0,
\]
requiring that $h$ satisfies $h<0$.

\subsubsection*{The exceptional Bianchi VI$^*_{-\frac{1}{9}}$ 
cosmologies}

We now explain the origin of the exceptional Bianchi VI$^*_{-\frac{1}{9}}$ 
cosmologies.
For Bianchi VI$_h$ models that also satisfy $\epsilon^{AB} v_A
\sigma_{1B} = 0$, we can use (\ref{VII_proof_det}) to
 write
\be
	a_1 = \pm \sqrt{ -h}\ {}^*\tilde{n}_{AB} K^{1A} K^{1B}.
\ee
Notice that from (\ref{c_c_A}), $(C_{\rm C})_A \sigma^{1A}$ yields
\be
\label{cc_A_s1A}
	0 = -3[ a_1 + \tfrac{1}{3}{}^*\tilde{n}_{AB} K^{1A} K^{1B}] + 
	\frac{q_AK^{1A}}{\sigma_\perp}
\ee
for $\sigma_\perp \neq 0$.
Thus, choosing $a_1 + \tfrac{1}{3}{}^*\tilde{n}_{AB} K^{1A} K^{1B}=0$, 
which results in $h = -\tfrac{1}{9}$, also forces $q_A K^{1A}=0$. In 
conjunction with $\epsilon^{AB} v_A \sigma_{1B} = 0$ and $\sigma_\perp \neq 
0$, this yields
\be
	v_A=0.
\ee
This class of models is referred to as the class of exceptional Bianchi 
VI$^*_{-\frac{1}{9}}$ cosmologies.
In other words, the exceptional Bianchi
VI$^*_{-\frac{1}{9}}$ cosmologies
are characterized by a non-OT $G_2$ action ($\sigma_\perp \neq 0$) and a 
fluid congruence that is orthogonal to the $G_2$ orbits ($v_A=0$).
Note that choosing $a_1 - \tfrac{1}{3}{}^*\tilde{n}_{AB} K^{1A} K^{1B}=0$
also results in $h = -\tfrac{1}{9}$, but does not permit $q_A K^{1A}=0$. 
This class consists of the ordinary Bianchi VI$_{-\frac{1}{9}}$
cosmologies.

\subsubsection*{Bianchi IV and II cosmologies}

Note that it follows from (\ref{VII_proof_det}) and (\ref{cc_A_s1A}) that 
\be
	\det n_{AB} =0,\quad v_A=0
\ee
implies
\be
	a_1 =0,
\ee
i.e. there is no Bianchi IV with generic $G_2$ and one tilt degree of 
freedom.

Lastly, we show that
\be
	\det n_{AB}=0,\quad a_1=0 \quad \Rightarrow \quad v_A=0,
\ee
i.e. Bianchi II cosmologies with generic $G_2$ have one tilt degree of 
freedom only.
To see this, choose the spatial gauge (\ref{e2_align_S3}). Then $\det 
n_{AB}=0$ and (\ref{c_c_A}) imply that $\Nc=0$ and $v_A=0$.
But $v_A=0$ is rotation-independent, so it is not an artifact of the 
spatial gauge choice.

\

Bianchi cosmologies with a generic $G_2$ subgroup are interesting because 
all of them except Bianchi V exhibit the Mixmaster dynamics, as we shall 
show in Section~\ref{sec:IX}.

\subsection*{Bianchi cosmologies with non-generic $G_2$}

\begin{figure}[H]
\begin{center}
    \epsfig{file=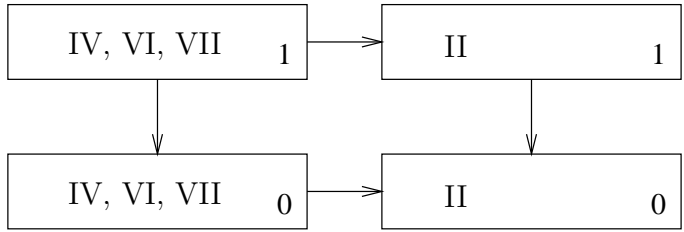,width=14cm}
\setlength{\unitlength}{1mm}
\begin{picture}(120,0)(0,0)
\put(0,30){\makebox(0,0)[l]{IV, VI, VII}}

\put(0,10){\makebox(0,0)[l]{IV, VI, VII}}

\put(50,30){\makebox(0,0)[l]{II}}
\put(50,10){\makebox(0,0)[l]{II}}

\end{picture}
\end{center}
\caption[SH cosmologies with OT $G_2$ action.]
{SH cosmologies with OT $G_2$ action.
The numbers 0,1 indicate the tilt degrees of freedom. 
}\label{fig:OT_G2}
\end{figure}
\begin{figure}[H]
\begin{center}
    \epsfig{file=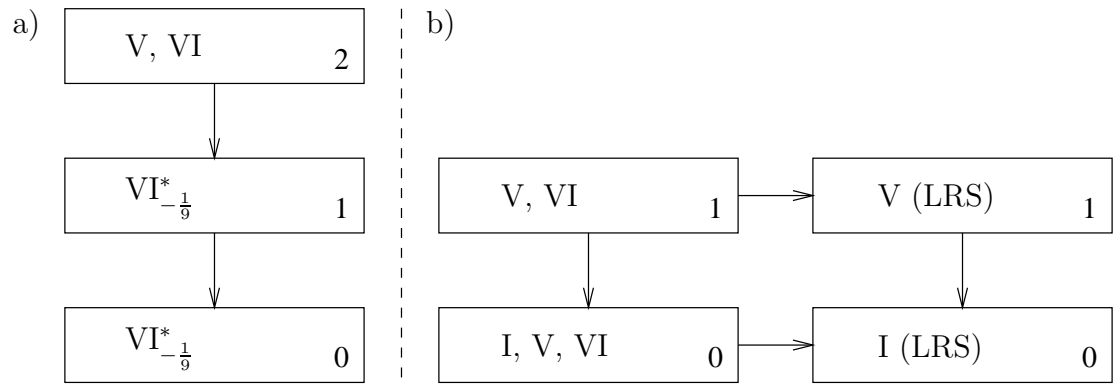,width=14cm}
\setlength{\unitlength}{1mm}
\begin{picture}(120,0)(0,0)
\put(-15,53){\makebox(0,0)[l]{a)}}
\put(40,53){\makebox(0,0)[l]{b)}}   

\put(0,50){\makebox(0,0)[l]{V, VI}}

\put(0,30){\makebox(0,0)[l]{VI$^*_{-\frac{1}{9}}$}}
\put(0,10){\makebox(0,0)[l]{VI$^*_{-\frac{1}{9}}$}}

\put(50,30){\makebox(0,0)[l]{V, VI}}
\put(100,30){\makebox(0,0)[l]{V (LRS)}}

\put(50,10){\makebox(0,0)[l]{I, V, VI}}
\put(100,10){\makebox(0,0)[l]{I (LRS)}}

\end{picture}
\end{center}
\caption[a) SH cosmologies whose $G_2$ admits one HO KVF.
        b) SH cosmologies with diagonal $G_2$ action.]
{a) SH cosmologies whose $G_2$ admits one HO KVF.
	b) SH cosmologies with diagonal $G_2$ action.
The numbers 0,1,2 indicate the tilt degrees of freedom.
}\label{fig:HO_diag}
\end{figure}

For completeness, Bianchi cosmologies with special $G_2$ action are 
classified in Figures~\ref{fig:OT_G2}--\ref{fig:HO_diag}.

%% file: SH_dynamics.tex
	\chapter{SH dynamics}\label{chap:SH}

Spatially homogeneous cosmologies are relevant in the study of 
$G_2$ cosmologies because their dynamics serves as the background 
dynamics of the inhomogeneous cosmologies
in regimes in which the spatial derivatives are small compared with the 
time derivatives.
In this chapter, 
we discuss certain aspects of the asymptotic dynamics of SH cosmologies, 
in particular the occurrence of Mixmaster dynamics.
We introduce a unified system of evolution equations for the $G_2$ 
compatible SH cosmologies (models of Bianchi types I--VII), which enables 
us to predict in which classes of models Mixmaster dynamics will occur.
We also mention some of the unsolved problems concerning the dynamics of 
SH cosmologies,
which can be addressed using these evolution equations.

\section{The standard Mixmaster dynamics}\label{sec:IX}

The standard Mixmaster dynamics arises in vacuum SH cosmologies of Bianchi 
types VIII and IX.
The justification for considering vacuum cosmologies is that for typical 
SH cosmologies with a non-tilted perfect fluid, the matter density is 
dynamically insignificant in the sense that
\be
	\lim_{t \rightarrow -\infty} \Omega =0,
\ee
where $t$ is an appropriately defined time.
(See
Ringstr\"{o}m 2001~\cite{art:Ringstrom2001}
for a proof of this result
 for the class of non-tilted Bianchi type 
IX cosmologies.)

\subsection*{The evolution equations}

The evolution equations for vacuum Bianchi VIII and IX cosmologies
can be derived from the 
system (\ref{evo_H})--(\ref{evo_q}) by choosing a $G_3$-invariant 
orthonormal frame
and introducing Hubble-normalized variables as in 
Section~\ref{sec:invariant}.

Recall from Table~\ref{tab:Bianchi} that $A_\alpha=0$ for models of 
Bianchi types VIII and IX.
We use a
 time-dependent rotation of the spatial frame vectors $\{ \me_\alpha \}$
to diagonalize $N_{\alpha\beta}$:
\be
        N_{\alpha\beta} = \text{diag}(N_{11}, N_{22}, N_{33}).
\ee
The $(C_{\rm C})_1$ constraint (\ref{c_c}) implies that the shear is 
also diagonal.
It then follows from (\ref{evo_n}) that 
the spatial frame vectors are
Fermi-propagated, i.e. $R_\alpha=0$.
As in WE, Section 6.1.1, we introduce the notation
\be
	\Sp = \tfrac{1}{2}(\Sigma_{22}+\Sigma_{33}),\quad
	\Sm = \tfrac{1}{2\sqrt{3}}(\Sigma_{22}-\Sigma_{33}).
\ee
The state vector (\ref{Hubble_1}) reduces to
\be
        \X = (\Sp,\Sm,N_{11},N_{22},N_{33}).
\ee

The evolution equations
for  $\X$ are obtained by taking the appropriate linear combinations of
 (\ref{evo_sigma}) and 
by specializing
(\ref{evo_n}):
\begin{align}
\label{IX_1}
	\Sp' &= (q-2)\Sp - \tfrac{1}{6}\left[
	(N_{22}-N_{33})^2 - N_{11}(2N_{11}-N_{22}-N_{33})\right]
\\
	\Sm' &= (q-2)\Sm - \tfrac{1}{2\sqrt{3}}(N_{33}-N_{22})
		(N_{11}-N_{22}-N_{33})
\\
\label{IX_3}
	N_{11}' &= (q-4\Sp)N_{11}
\\
	N_{22}' &= (q+2\Sp+2\sqrt{3}\Sm)N_{22}
\\	
	N_{33}' &= (q+2\Sp-2\sqrt{3}\Sm)N_{33}\ ,
\label{IX_5}
\end{align}
where $'$ denotes the time derivative, and
\be
	q = 2 \Sigma^2,\quad \Sigma^2 = \Sp^2 + \Sm^2\ .
\label{IX_q}
\ee
The only remaining constraint is the Gauss constraint (\ref{c_g}), which 
reads
\begin{align}
	 0 &= (\mathcal{C}_{\rm G}) = 1 - \Omega_k - \Sigma^2  ,
\label{IX_C_G}
\intertext{where}
	\Omega_k &= \tfrac{1}{12}\left[ N_{11}^2 +N_{22}^2 + N_{33}^2
                -2 (N_{11}N_{22} + N_{22}N_{33} + N_{33}N_{11}) \right]\ .
\end{align}

\subsection*{The Kasner circle}

The Mixmaster dynamics is generated by the stability properties of the 
Kasner solutions. These vacuum solutions appear as a circle of 
equilibrium points of the evolution equations
(\ref{IX_1})--(\ref{IX_5}), given by
\be
	N_{\alpha\alpha}=0,\quad \Sp^2+\Sm^2=1.
\label{IX_Kasner}
\ee
It is customary to refer to this circle as the \emph{Kasner circle}, 
denoted $\mathcal{K}$.
It follows from (\ref{IX_q}) and (\ref{IX_Kasner}) that the Kasner 
solutions satisfy
\be
	q = 2.
\label{Kasner_q}
\ee

We note that the metric for the Kasner solutions has the simple form:
\be
        \ell_0^{-2} ds^2 = -dT^2 + T^{2p_1} dx^2 + T^{2p_2} dy^2 + 
				T^{2p_3} dz^2,
\ee
where the constants $p_\alpha$ satisfy
\be
        p_1+p_2+p_3 = 1 = p_1^2 + p_2^2 + p_3^2.
\ee
The link between the $p_\alpha$ and the Hubble-normalized shear is
(see WE, equation (6.17))
\be
	p_1 = \tfrac{1}{3}(1-2\Sp),\quad
        p_2 = \tfrac{1}{3}(1+\Sp+\sqrt{3}\Sm),\quad
        p_3 = \tfrac{1}{3}(1+\Sp-\sqrt{3}\Sm).
\ee

\begin{figure}[h]
\begin{center}
    \epsfig{file=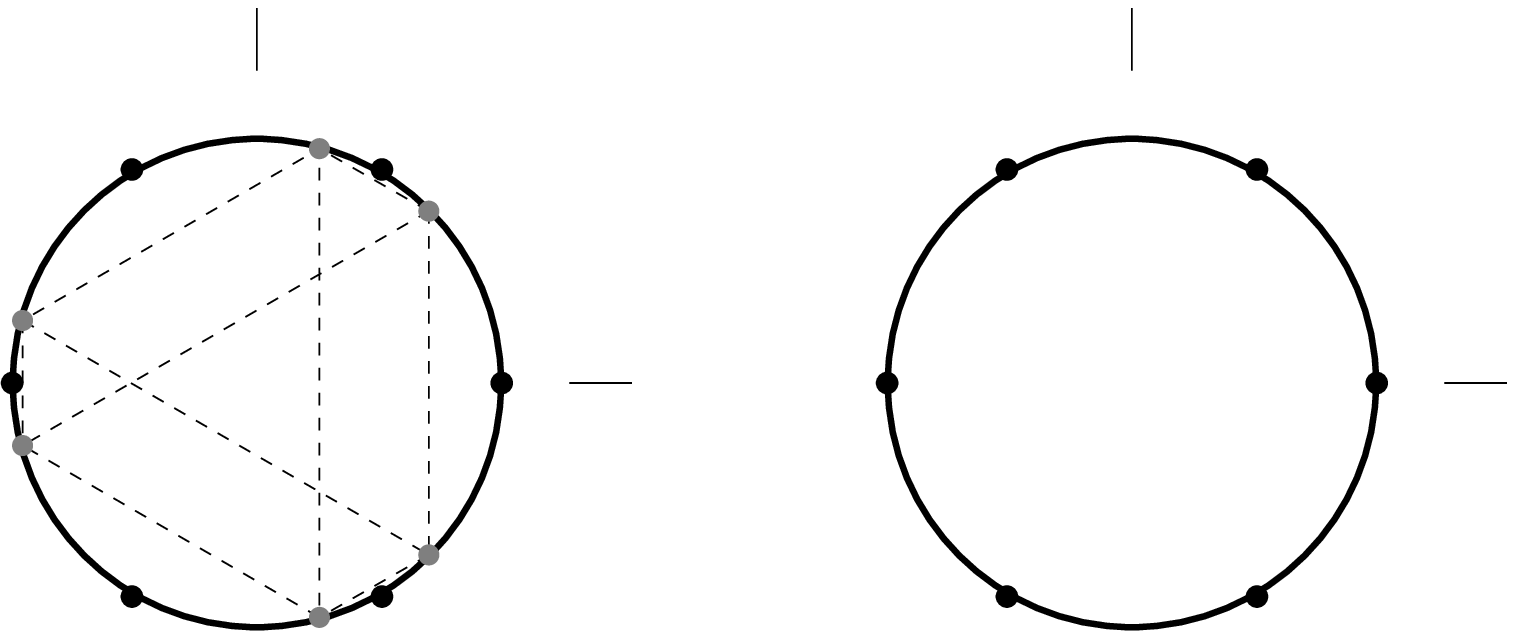,width=4.9in}
\setlength{\unitlength}{1mm}
\begin{picture}(120,0)(0,0)   
\put(19,68){\makebox(0,0)[t]{$\Sm$}} 
\put(54,33){\makebox(0,0)[t]{$\Sp$}}

\put(-5,33){\makebox(0,0)[t]{$T_1$}}
\put(30,55){\makebox(0,0)[t]{$T_2$}}
\put(30,11){\makebox(0,0)[t]{$T_3$}}

\put(43,32){\makebox(0,0)[t]{$Q_1$}}
\put(8,55){\makebox(0,0)[t]{$Q_3$}}
\put(8,11){\makebox(0,0)[t]{$Q_2$}}

\put(-8,28){\makebox(0,0)[t]{$(-1,0)$}}
\put(30,62){\makebox(0,0)[t]{$(\tfrac{1}{2},\tfrac{\sqrt{3}}{2})$}} 
\put(30,6){\makebox(0,0)[t]{$(\tfrac{1}{2},-\tfrac{\sqrt{3}}{2})$}} 

\put(44,27){\makebox(0,0)[t]{$(1,0)$}}
\put(5,62){\makebox(0,0)[t]{$(-\tfrac{1}{2},\tfrac{\sqrt{3}}{2})$}}    
\put(5,6){\makebox(0,0)[t]{$(-\tfrac{1}{2},-\tfrac{\sqrt{3}}{2})$}}



\put(91,68){\makebox(0,0)[t]{$\Sm$}}
\put(126,33){\makebox(0,0)[t]{$\Sp$}}

\put(67,33){\makebox(0,0)[t]{$T_1$}}
\put(102,55){\makebox(0,0)[t]{$T_2$}}
\put(102,11){\makebox(0,0)[t]{$T_3$}} 

\put(113,45){\makebox(0,0)[t]{$N_{11}$}}
\put(113,22){\makebox(0,0)[t]{$N_{11}$}}

\put(70,22){\makebox(0,0)[t]{$N_{22}$}}
\put(91,9){\makebox(0,0)[t]{$N_{22}$}}

\put(70,45){\makebox(0,0)[t]{$N_{33}$}}
\put(91,56){\makebox(0,0)[t]{$N_{33}$}}

\put(-6,70){\makebox(0,0)[t]{a)}}
\put(70,70){\makebox(0,0)[t]{b)}}

\end{picture}
\end{center}
\caption{a) The Kasner circle and its six-fold symmetry. b) The triggers 
	 $N_{11}$, $N_{22}$ and $N_{33}$ and their unstable Kasner 
	arcs.}\label{fig:Kasner_circle}
\end{figure}

A typical Kasner solution is in fact represented six times on the Kasner 
circle.
Figure~\ref{fig:Kasner_circle}a illustrates this symmetry of the Kasner   
circle.
The grey dots connected by dashed lines are six different representations
of a typical Kasner solution.
The exceptions are the so-called 
Taub Kasner solutions 
($p_\alpha = (1,0,0)$ and cycle),
labelled $T_\alpha$ on the Kasner circle in 
Figure~\ref{fig:Kasner_circle}a,
and the LRS Kasner solutions
($p_\alpha = (-\tfrac{1}{3},\tfrac{2}{3},\tfrac{2}{3})$ and cycle), 
labelled $Q_\alpha$.
See WE, Sections 6.2.2 and 9.1.1 for details.%

\subsection*{Kasner instability and Mixmaster dynamics}

In order to proceed with the analysis we restrict our consideration to
solutions that are close to a Kasner solution at some time,
i.e. to solutions whose orbit in the Hubble-normalized state space enters 
an arbitrarily small neighbourhood of the Kasner circle $\mathcal{K}$.
Ringstr\"{o}m 2001~\cite{art:Ringstrom2001}
proved that all Bianchi IX solutions except a set of measure zero
have this property.

\begin{figure}[h]
\begin{center}
    \epsfig{file=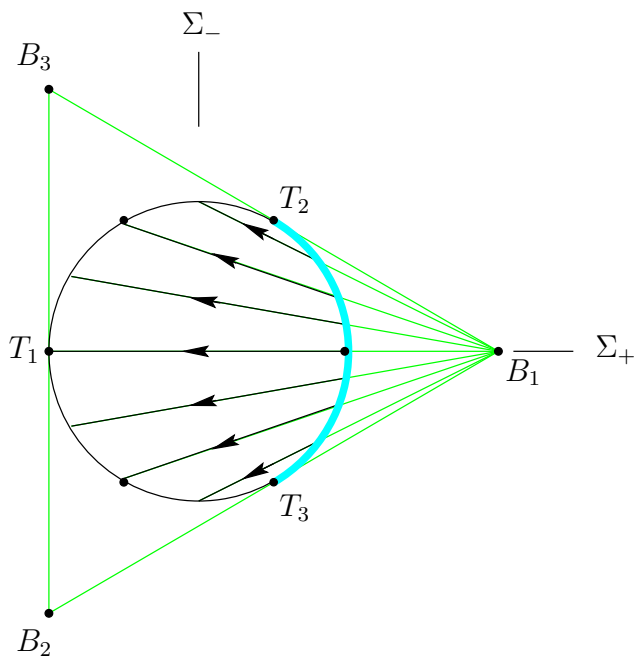,width=7.1cm}
\setlength{\unitlength}{1mm}
\begin{picture}(120,0)(0,0)
\put(43,84){\makebox(0,0)[l]{$\Sm$}}
\put(98,41){\makebox(0,0)[l]{$\Sp$}}

\put(86,38){\makebox(0,0)[l]{$B_1$}}
\put(21, 2){\makebox(0,0)[l]{$B_2$}}
\put(21,80){\makebox(0,0)[l]{$B_3$}}

\put(20,41){\makebox(0,0)[l]{$T_1$}}
\put(56,61){\makebox(0,0)[l]{$T_2$}}
\put(56,20){\makebox(0,0)[l]{$T_3$}}
   
\end{picture}
\end{center}
\caption[The curvature transition set and the arc of instability (into the
        past) of the trigger $N_{11}$.]%
	{The curvature transition set and the arc of instability (into the 
	past) of the trigger $N_{11}$. The transition sets for $N_{22}$ 
	and $N_{33}$ are obtained by symmetry,
	using the points $B_2$ and $B_3$ respectively.}\label{fig:curv}
\end{figure}

Consider linearizing (\ref{IX_3})--(\ref{IX_5}) on the Kasner circle 
(\ref{IX_Kasner}).
Using (\ref{Kasner_q}), we obtain
\begin{align}
        N_{11}' &= 2(1-2\Sp)N_{11}
\\
        N_{22}' &= 2(1+\Sp+\sqrt{3}\Sm)N_{22}
\\
        N_{33}' &= 2(1+\Sp-\sqrt{3}\Sm)N_{33}\ ,
\end{align}
which imply that the
$N_{\alpha\alpha}$ are unstable into the past on the following arcs:
\[
	\text{$N_{11}$ on arc($T_2 T_3$), $N_{22}$ on arc($T_3 T_1$),
	$N_{33}$ on arc($T_1 T_2$).}
\]
These arcs are non-overlapping and cover all the Kasner equilibrium points 
except the Taub Kasner points (see Figure~\ref{fig:Kasner_circle}b).
As a result, each of these Kasner equilibrium points is unstable into the 
past.
In the SH state space,
the unstable manifold of a Kasner equilibrium point is an orbit 
lying on one of the ellipsoids
\be
	1 = \tfrac{1}{12}N_{\alpha\alpha}^2 + \Sp^2 + \Sm^2\ ,\quad
	\alpha = \text{1, 2 or 3,}
\label{IX_ellipsoid}
\ee
as follows from the Gauss constraint (\ref{IX_C_G}).
In fact, the orbits represent the Taub vacuum Bianchi 
II solutions, and appear as straight lines 
when projected onto the $(\Sp,\Sm)$ plane, as illustrated in 
Figure~\ref{fig:curv}.
We shall refer to these orbits as the
\emph{curvature transition sets}, denoted $\mathcal{T}_N$.
Each transition is a result of the growth (into the past) of one of 
$N_{\alpha\alpha}$. 
Therefore we call $N_{\alpha\alpha}$ the ``triggers".
Because $N_{\alpha\alpha}$ represent the spatial curvature, we call them
the ``curvature triggers". 

\enlargethispage{-\baselineskip}

The evolution of a Bianchi VIII or IX cosmology
 is represented by an orbit in the Hubble-normalized state space.
As $t \rightarrow -\infty$, a typical orbit enters a neighbourhood of
 the Kasner circle, but cannot be asymptotic to a Kasner equilibrium point 
in general. Instead it shadows the unstable manifolds of 
the Kasner equilibrium points, going from one Kasner equilibrium point to 
another.
In this way, the three families of
curvature
 transition sets sustain an endless sequence 
of transitions, called \emph{Mixmaster dynamics}, 
in recognition of the work of Misner 1969~\cite{art:Misner1969b}.%
\footnote{Misner discovered this behaviour by using the Hamiltonian 
formulation of the EFEs.}
For a more detailed discussion, see
Ma \& Wainwright~\cite{inbook:MaWainwright1992}
and Section 6.4 of WE.
The essential point is that
Mixmaster dynamics will occur if for each typical Kasner equilibrium point 
$P$
there is a trigger variable that renders it unstable into the past, and
with the resulting unstable manifold joining $P$ to another 
Kasner equilibrium point.

\

To summarize, the occurrence of Mixmaster dynamics and the identification 
of the past attractor is established by proving three results
(see Ringstr\"{o}m 2001~\cite{art:Ringstrom2001}):
\begin{alignat}{2}
	&\text{i)} &&\quad
	\lim_{t \rightarrow -\infty} \Omega =0,\quad
	\lim_{t \rightarrow -\infty} \Oml =0.
\\
	&\text{ii)} &&\quad
	\text{A typical orbit enters an arbitrarily small neighbourhood}
\notag\\
	&&&\quad
	\text{of the Kasner circle, and}
\notag\\
	&\text{iii)} &&\quad
	\lim_{t \rightarrow -\infty} 
	( N_{11}N_{22}, N_{22}N_{33}, N_{33}N_{11} )=\mathbf{0}.
\end{alignat}
Mixmaster dynamics is initiated by Property ii), which we shall refer to 
as the \emph{Kasner Attractivity Conjecture}.
\footnote{This is stated formally later as Conjecture~\ref{KAC4.1}.}
It appears to play a fundamental role in the dynamics of SH cosmologies 
and more generally, in inhomogeneous cosmologies, although it has not been 
proved in more general situations.
Property iii) states that the curvature transitions cannot occur 
simultaneously.

\

For Bianchi VIII and IX cosmologies,
 the triggers are the eigenvalues of $N_{\alpha\beta}$.
If one of the triggers is zero, e.g. $N_{11}=0$, then there is a stable 
arc on $\mathcal{K}$ which is the past attractor.
For $G_2$-compatible SH cosmologies (Bianchi I--VII), at least one of the
eigenvalues of $N_{\alpha\beta}$ is zero (see Table~\ref{tab:Bianchi}).
Can Mixmaster dynamics occur in these cosmologies?
At first sight, it does not seem so.
But it has been shown that in fact it
can occur, although in general it requires the presence of a 
perfect fluid with non-zero tilt,
as we shall show in Section~\ref{sec:general_mixmaster}.
But first we need to derive the Hubble-normalized evolution equations for 
$G_2$-compatible SH cosmologies, which we do in Section~\ref{sec:SH_var}.

\section{The $G_2$-compatible SH evolution equations}\label{sec:SH_var}

The 1+1+2 Hubble-normalized variables (\ref{1+1+2_1})--(\ref{1+1+2_2})
were introduced for the purpose of
 analyzing $G_2$-compatible SH cosmologies.
It is necessary to write out the Hubble-normalized
evolution equations in component 
form (i.e. the 1+1+1+1 form!) rather than in the 1+1+2 form,
so that we can impose a suitable spatial gauge condition.
For convenience we introduce the following notation:
\be
        N_C{}^C = 2 N_+,\quad
        \tilde{N}_{AB} = \sqrt{3}
                \left(
                \begin{matrix}
                        N_- & N_\times
                \\
                        N_\times & -N_-   
                \end{matrix}
                \right),
\label{23_subspace}
\ee
and
\be
        \Sigma_{12} = \sqrt{3}\Sigma_3,\quad
        \Sigma_{13} = \sqrt{3}\Sigma_2.
\ee
The Hubble-normalized equations in component form 
are obtained from
the 1+1+2 system (\ref{AB_H})--(\ref{evo_qA})
by expressing them in terms of the Hubble-normalized variables 
(\ref{1+1+2_1})--(\ref{1+1+2_2}), using 
(\ref{Hubble_3})--(\ref{Hubble_3b}), 
and are given below:

\noindent
\begin{minipage}{\textwidth}
\vspace{3mm}
\subsection*{$G_2$-compatible SH system (Hubble-normalized)}

\noindent
{\it Evolution equations for the gravitational field:}
\begin{align}
\label{SH_Sp}
	\Sp' &= (q-2)\Sp + 3(\St^2+\Stt^2)-2(\Nm^2+\Nc^2)
	-\frac{3}{2} \frac{\gamma\Omega}{G_+}(v_1^2-\tfrac{1}{3}v^2)
\\
	\Sm' &= (q-2)\Sm + 2A\Nc -2R\Sc -2\Np\Nm + \sqrt{3}(\Stt^2-\St^2)
\notag\\
	&\qquad
	+ \frac{\sqrt{3}}{2} \frac{\gamma\Omega}{G_+} (v_2^2-v_3^2)
\label{SH_Sm}
\\
	\Sc' &= (q-2)\Sc -2A\Nm + 2R\Sm -2\Np\Nc + 2\sqrt{3}\Stt\St
\notag\\
        &\qquad
		+\sqrt{3} \frac{\gamma\Omega}{G_+} v_2 v_3
\label{SH_Sc}
\\
\label{SH_Stt}
	\Stt' &= (q-2-3\Sp-\sqrt{3}\Sm)\Stt -(R+\sqrt{3}\Sc)\St
		+ \sqrt{3} \frac{\gamma\Omega}{G_+} v_1 v_2
\\
\label{SH_St}
        \St' &= (q-2-3\Sp+\sqrt{3}\Sm)\St +(R-\sqrt{3}\Sc)\Stt
                + \sqrt{3} \frac{\gamma\Omega}{G_+} v_1 v_3
\\
\label{SH_Np} 
	\Np' &= (q+2\Sp)\Np + 6\Sm\Nm + 6\Sc\Nc
\\
	\Nm' &= (q+2\Sp)\Nm + 2\Np\Sm - 2R\Nc
\\
        \Nc' &= (q+2\Sp)\Nc + 2\Np\Sc + 2R\Nm
\\
\label{SH_A}
	A' &= (q+2\Sp)A\ ,
\end{align}
where
\begin{align}
\label{SH_q}
        q &= 2\Sigma^2
        + \tfrac{1}{2} G_+^{-1} [(3\gamma-2)(1-v^2)+2\gamma v^2]\Omega
        -\Oml
\\
        \Sigma^2 &= \Sp^2 + \Sm^2 + \Sc^2 + \St^2 + \Stt^2
\\
        G_\pm &= 1 \pm (\gamma-1)v^2,\quad v^2 = v_1^2 + v_2^2 
+v_3^2\ .
\end{align}
{\it Constraint equations:}
\begin{align}
\label{SH_C_G}
	0 &= (\mathcal{C}_{\rm G}) = 1 - \Nm^2 - \Nc^2 - A^2 - \Sigma^2
		- \Omega - \Oml
\\
	0 &= (\mathcal{C}_{\rm C})_1 =
	2A\Sp + 2(\Sm\Nc-\Sc\Nm) + \frac{\gamma\Omega}{G_+} v_1
\\
\label{SH_CC2}
        0 &= (\mathcal{C}_{\rm C})_2 =
	- (3A + \sqrt{3}\Nc) \Stt 
	-  (\Np-\sqrt{3}\Nm) \St
	+ \sqrt{3} \frac{\gamma\Omega}{G_+} v_2
\\
        0 &= (\mathcal{C}_{\rm C})_3 =
        - (3A - \sqrt{3}\Nc) \St
        + (\Np+\sqrt{3}\Nm) \Stt
        + \sqrt{3}\frac{\gamma\Omega}{G_+} v_3\ .
\end{align}

\begin{figure}[H]
\begin{center}
\setlength{\unitlength}{1mm}
\begin{picture}(120,0)(0,0)
\put(60,90){\oval(150,230)[t]}
\put(60,90){\oval(150,170)[b]}
\end{picture}
\end{center}
\end{figure}
\vspace{1cm}
\end{minipage}

\vspace{5mm}
\noindent
{\it Evolution equations for the matter:} 
\begin{align}
	\Omega' &= G_+^{-1} [ 2G_+q-(3\gamma-2)-(2-\gamma)v^2
		- \gamma \Sigma_{\alpha\beta} v^\alpha v^\beta
		+2 \gamma A v_1 ] \Omega
\\
	v_1' &= (M+2\Sp)v_1 -2\sqrt{3}(\Stt v_2 + \St v_3)
	+ \sqrt{3} \Nc (v_3^2-v_2^2) + 2\sqrt{3}\Nm v_2 v_3 - A v^2
\label{v_1}
\\
	v_2' &= (M-\Sp-\sqrt{3}\Sm + \sqrt{3}\Nc v_1) v_2
	- [ R+\sqrt{3}\Sc - (\Np-\sqrt{3}\Nm)v_1]v_3
\label{v_2}
\\
        v_3' &= (M-\Sp+\sqrt{3}\Sm - \sqrt{3}\Nc v_1) v_3
        + [ R-\sqrt{3}\Sc - (\Np+\sqrt{3}\Nm)v_1]v_2
\label{v_3}
\\
	\Oml' &= 2(q+1) \Oml\ ,
\label{SH_Oml}
\end{align}
where
\begin{align}
	M &= G_-^{-1} \Big[ (3\gamma-4)(1-v^2) 
		+ (2-\gamma)\Sigma_{\alpha\beta}v^\alpha v^\beta
\notag\\
	&\qquad
		+ [G_+ - 2(\gamma-1)]Av_1 \Big]
\\
	\Sigma_{\alpha\beta}v^\alpha v^\beta &=
	\Sp (v_2^2+v_3^2-2v_1^2) 
	+ \sqrt{3}\Sm(v_2^2-v_3^2) + 2\sqrt{3} \Sc v_2 v_3
\notag\\        
        &\qquad 
	+ 2 \sqrt{3} (\Stt v_2 + \St v_3) v_1\ .
\end{align}

\begin{figure}[H]
\begin{center}
\setlength{\unitlength}{1mm}
\begin{picture}(120,0)(0,0)
\put(60,50){\oval(150,125)[t]}
\put(60,50){\oval(150,90)[b]}
\end{picture}
\end{center}
\end{figure}

\vspace{-1cm}

In order to derive the $v_\alpha'$ equations%
\footnote{See van Elst \& Uggla 1997~\cite{art:ElstUggla1997}, page 2682.}
 from the evolution equations 
(\ref{evo_q1})--(\ref{evo_qA}) for $q_\alpha$, 
we first obtain $\Omega'$ and $Q_\alpha'$ by
converting (\ref{evo_mu_AB})--(\ref{evo_qA}) to Hubble-normalized form.
On differentiating $
Q_\alpha = \gamma G_+^{-1}\Omega v_\alpha$,
we obtain
\be
         Q_\alpha'
        = \frac{\gamma v_\alpha}{G_+}  \Omega'
        + \frac{\gamma \Omega}{G_+}  v_\alpha'
        - \frac{\gamma \Omega}{G_+{}^2}v_\alpha ( G_+ )'\ .
\label{expand}
\ee
Contracting with $v_\alpha$ and simplifying gives
\be
	v^\alpha Q_\alpha' =
	\frac{\gamma v^2}{G_+} \Omega'
	+ \frac{\gamma G_- \Omega}{G_+{}^2}( v^2 )'\ .
\ee
The resulting equation for $v^2$ is
\be
	(v^2)' = \tfrac{2}{G_-}(1-v^2)\left[
		(3\gamma-4)v^2 - \Sigma_{\alpha\beta} v^\alpha v^\beta
		-2(\gamma-1)v^2 A v_1  \right]\ ,
\label{vsq}
\ee
which we use to evaluate the $\parb_0 G_+$ term in (\ref{expand})
to obtain $v_\alpha'$.

It follows from (\ref{Weyl_E11})--(\ref{Weyl_HAB}) and (\ref{Weyl_Hubble})  
that
the Hubble-normalized Weyl curvature variables are given by
\begin{align}
        \Ep &= \tfrac{1}{3}(1+\Sp)\Sp + \tfrac{1}{6}(\St^2+\Stt^2)
                - \tfrac{1}{3}(\Sm^2+\Sc^2)
\notag\\
        &\quad
                + \tfrac{2}{3}(\Nc^2+\Nm^2)
                + \frac{1}{4} \frac{\gamma \Omega}{G_+}
                                        (v_1^2-\tfrac{1}{3}v^2)
\label{SH_Ep}
\\
        \Em &= \tfrac{1}{3}(1-2\Sp)\Sm
                - \tfrac{1}{2\sqrt{3}}(\Stt^2-\St^2)
\notag\\
        &\quad
                -\tfrac{2}{3}A\Nc + \tfrac{2}{3}\Np\Nm
                + \frac{1}{2\sqrt{3}}\frac{\gamma \Omega}{G_+}(v_2^2-v_3^2)
\label{SH_Em}
\\
        \Ec &= \tfrac{1}{3}(1-2\Sp)\Sc
                - \tfrac{1}{\sqrt{3}}\Stt\St
\notag\\
        &\quad
                +\tfrac{2}{3}A\Nm + \tfrac{2}{3}\Np\Nc
                + \frac{1}{\sqrt{3}}\frac{\gamma \Omega}{G_+} v_2 v_3
\label{SH_Ec}
\\ 
        \Ett &= \tfrac{1}{3}(1+\Sp-\sqrt{3}\Sm)\Stt - \tfrac{1}{\sqrt{3}}\Sc\St
        - \frac{1}{2} \frac{\gamma \Omega}{G_+} v_1 v_2
\\
        \Et &= \tfrac{1}{3}(1+\Sp+\sqrt{3}\Sm)\St - \tfrac{1}{\sqrt{3}}\Sc\Stt
        - \frac{1}{2} \frac{\gamma \Omega}{G_+} v_1 v_3
\\
        \Hp &= -\Nm\Sm-\Nc\Sc
\\
        \Hm &= \tfrac{1}{3}A\Sc-\tfrac{2}{3}\Np\Sm-\Sp\Nm
\label{SH_Hm}
\\
        \Hc &= -\tfrac{1}{3}A\Sm -\tfrac{2}{3}\Np\Sc-\Sp\Nc
\label{SH_Hc}
\\
        \Htt &= -\tfrac{1}{3}(A+\sqrt{3}\Nc)\St
                - \tfrac{1}{\sqrt{3}}\Nm\Stt
                + \frac{1}{2\sqrt{3}}\frac{\gamma \Omega}{G_+}v_3
\\
        \Ht &= \tfrac{1}{3}(A-\sqrt{3}\Nc)\Stt
                + \tfrac{1}{\sqrt{3}}\Nm\St
                - \frac{1}{2\sqrt{3}}\frac{\gamma \Omega}{G_+}v_2\ .
\label{SH_Ht}
\end{align}
Note that the $(\mathcal{C}_{\rm C})_2$ and $(\mathcal{C}_{\rm C})_3$
constraints have been used to simplify $\Ht$ and $\Htt$ respectively.

Recall from Section~\ref{sec:SHtree} that
for 
SH cosmologies of
Bianchi types VI and VII,
the spatial curvature variables are related via
\be
	A^2 = h \det N_{AB},\quad
	\det N_{AB} = \Np^2 -3(\Nm^2+\Nc^2)\neq 0.
\label{VI_VII_h}
\ee
Thus the Bianchi VI and VII state space consists of 
``layers" 
of Bianchi VI$_h$ and VII$_h$ state spaces.

For SH cosmologies with generic $G_2$ action, 
a simple way to specify the spatial gauge is to use a time-dependent 
rotation of the spatial frame vectors $\{\me_A\}$
to make $\me_2$ parallel to a KVF. It then follows from Lemma 1 in 
Appendix~\ref{app:HO} that
\be
        \Np = \sqrt{3}\Nm,\quad
         R = -\sqrt{3} \Sc\ .  
\label{KSG}
\ee
We shall refer to this gauge as the \emph{Killing spatial gauge}.
This gauge, however, has the disadvantage that it cannot be used for 
Bianchi VII 
cosmologies, because it entails $\det N_{AB}\leq0$, as follows from 
(\ref{VI_VII_h}).

A second possibility is
 to rotate the frame $\{\me_A\}$ so that $\Stt=0$. 
The evolution equation 
(\ref{SH_Stt}) and the constraint (\ref{SH_CC2}) then give
\be
\label{SH_spatial_gauge}
        \Sigma_{3}=0,\quad R = -\sqrt{3} \Sc + (\Np -\sqrt{3}\Nm)v_1\ .
\ee
We shall refer to this gauge as the \emph{shear spatial gauge}.
For the invariant set $\{ v_2=0,\St \neq 0\}$, the $(\mathcal{C}_{\rm 
C})_2$ constraint (\ref{SH_CC2}) implies $\Np = \sqrt{3} \Nm$, and 
(\ref{SH_spatial_gauge}) simplifies to
\be
\label{e2_align_S3}
        \Sigma_{3}=0,\quad
        \Np = \sqrt{3}\Nm,\quad
	 R = -\sqrt{3} \Sc\ .
\ee
On the other hand, for the invariant set $\St=0$ corresponding 
to SH cosmologies with OT 
$G_2$ action, the $(\mathcal{C}_{\rm C})_2$ constraint (\ref{SH_CC2}) 
implies $v_2=0$, leaving (\ref{SH_spatial_gauge}) the same.

A third alternative is the \emph{tilt spatial gauge}, where
one rotates the frame $\{\me_A\}$ so that $v_2=0$.
The $v_2'$ equation (\ref{v_2}) then gives
\be
\label{tilt_spatial_gauge}
        v_2=0,\quad R = -\sqrt{3} \Sc + (\Np -\sqrt{3}\Nm)v_1\ .
\ee
Observe that
the tilt spatial gauge and the shear spatial gauge have the same 
expression for $R$.

We shall use the shear spatial gauge 
(\ref{SH_spatial_gauge})
rather than the tilt spatial gauge
because it has the advantage of displaying the tilt degrees of freedom 
explicitly.
With this choice, the Hubble-normalized variables are
\begin{gather} 
        \X=
	(\Sigma_+,\Sigma_-,\Sigma_\times,\Sigma_2,
        N_+,N_-,N_\times,A,
        \Omega,v_1,v_2,v_3,\Omega_\Lambda),
\label{SH_state_vector}
\end{gather}
together with the parameter $\gamma$. 
There are 13 variables and 4 constraints, making the state space 
9-dimensional.
\footnote{The state space of Bianchi VI$_h$ or VII$_h$ cosmologies for 
each fixed $h$ is 8-dimensional, taking into account the extra 
constraint (\ref{VI_VII_h}).}

\subsubsection*{Constraints}

The stability of the constraints is of great concern when doing numerical
simulations.
From Uggla \etal 2003~\cite[equations (A2)--(A3)]{art:Ugglaetal2003} 
we obtain the evolution equations for the constraints:
\begin{align}
	(\mathcal{C}_{\rm G})' 
	&= 2q(\mathcal{C}_{\rm G})-\tfrac{2}{3}A(\mathcal{C}_{\rm C})_1
\\
	(\mathcal{C}_{\rm C})_1'
	&= (2q-2+2\Sp)(\mathcal{C}_{\rm C})_1
	-2\sqrt{3}\Stt(\mathcal{C}_{\rm C})_2
	-2\sqrt{3}\St (\mathcal{C}_{\rm C})_3
\\
	(\mathcal{C}_{\rm C})_2'
	&= (2q-2-\Sp-\sqrt{3}\Sm)(\mathcal{C}_{\rm C})_2
	-(R+\sqrt{3}\Sc)(\mathcal{C}_{\rm C})_3
\\
        (\mathcal{C}_{\rm C})_3'
        &= (2q-2-\Sp+\sqrt{3}\Sm)(\mathcal{C}_{\rm C})_3
        +(R-\sqrt{3}\Sc)(\mathcal{C}_{\rm C})_2\ .
\end{align}
We see that the constraints are stable into the past near the Kasner 
circle (as the coefficients of the first terms are positive).

\section{Generalized Mixmaster dynamics}\label{sec:general_mixmaster}

We now predict under what circumstances Mixmaster dynamics can occur in 
$G_2$-compatible SH cosmologies. 
We use the gauge choice (\ref{SH_spatial_gauge}), but we shall show that 
the main result, Proposition~\ref{prop1}, is independent of this choice.

We begin by describing the Kasner equilibrium points.
Within the state space of $G_2$-compatible SH cosmologies, it is 
straightforward to verify that the Kasner circle $\mathcal{K}$, now given 
by
\begin{gather}
	(\Sc,\St,\Np,\Nm,\Nc,A)=\mathbf{0},
\label{VII_trig}
\\
        \Sp^2+\Sm^2=1,
\\
        \Omega=0=\Oml,\quad
	v_\alpha=0,
\end{gather}
is a set of equilibrium points, and satisfies $q=2$.
In fact, by assuming
\be
	\Omega=0=\Oml,\quad (\Np,\Nm,\Nc,A)=\mathbf{0},
\label{Kasner_1}
\ee
one quickly arrives at $\Sigma^2=1$ and $q=2$ through
(\ref{SH_C_G}) and (\ref{SH_q}).
Then (\ref{SH_Sp})--(\ref{SH_St}) imply
\be
	\St=\Sc=0.
\label{Kasner_2}
\ee
Evolution equations
(\ref{v_1})--(\ref{v_3}) and (\ref{vsq}) with
(\ref{Kasner_1}) and (\ref{Kasner_2}) give
\begin{align}
\label{SH_v_1_linear}
        v_1' &= (M+2\Sp)v_1
\\
        v_2' &= (M-\Sp-\sqrt{3}\Sm)v_2
\\
        v_3' &= (M-\Sp+\sqrt{3}\Sm)v_3
\label{SH_v_3_linear}
\\ 
        (v^2)' &= \tfrac{2}{G_-}(1-v^2)[(3\gamma-4)v^2
                        -\Sigma_{\alpha\beta}v^\alpha v^\beta]\ ,
\end{align}
where
\begin{align}
        M &= G_-^{-1} [ (3\gamma-4)(1-v^2)
                +(2-\gamma) \Sigma_{\alpha\beta}v^\alpha v^\beta ]
\\
        \Sigma_{\alpha\beta}v^\alpha v^\beta &=
        \Sp(v_2^2+v_3^2-2v_1^2) + \sqrt{3}\Sm(v_2^2-v_3^2)\ .
\end{align}   
Equilibrium points with $v_\alpha=0$ give the Kasner circle $\mathcal{K}$.
Equilibrium points with one non-zero tilt variable and $v^2=1$ give
\be
        v_\alpha = (\pm1,0,0),\ (0,\pm1,0),\ (0,0,\pm1).
\ee
Hence we have 6 more Kasner circles $\mathcal{K}_{\pm\alpha}$ of
equilibrium points, given by
\begin{gather}
        (\Sc,\St,\Np,\Nm,\Nc,A)=\mathbf{0},
\label{VII_trig_alpha}
\\
        \Sp^2+\Sm^2=1,
\\
        \Omega=0=\Oml,\quad
        v_\alpha= (\pm1,0,0),\ (0,\pm1,0),\ (0,0,\pm1).
\end{gather}
The $\mathcal{K}_{\pm\alpha}$ equilibrium points are unphysical due to the
extreme tilt $v^2=1$, but we shall see that they
nevertheless play a significant role
dynamically.
Equilibrium points with one non-zero tilt variable and $v^2<1$ gives
\be
        \begin{cases}
        \Sp = -\tfrac{1}{2}(3\gamma-4) & \text{if $v_1 \neq 0$}
        \\
        \Sp + \sqrt{3}\Sm = 3\gamma-4  & \text{if $v_2 \neq 0$}
        \\
        \Sp - \sqrt{3}\Sm = 3\gamma-4  & \text{if $v_3 \neq 0$.}
        \end{cases}
\ee
There are equilibrium points with two non-zero tilt variables, but they
occur at $Q_\alpha$ and $T_\alpha$ and not on the entire Kasner circle.
There are no equilibrium points with $v_1 v_2 v_3 \neq 0$, since
(\ref{SH_v_1_linear})--(\ref{SH_v_3_linear}) require that
\[
        M + 2\Sp = 0 = M-\Sp-\sqrt{3}\Sm = M-\Sp+\sqrt{3}\Sm\ ,
\]
which
implies $\Sp=\Sm=0$, contradicting $\Sp^2+\Sm^2=1$.
The Kasner solutions are in fact characterized by (\ref{Kasner_1}), and 
thus have other (non-equilibrium) representations, which we shall 
encounter very soon in this chapter.

In order to proceed with the analysis we first restrict our consideration 
to solutions that satisfy
\be
        \lim_{t \rightarrow -\infty} \Omega =0,\quad
        \lim_{t \rightarrow -\infty} \Oml =0,
\ee
based on the BKL conjecture I (see Section~\ref{sec:goals}).
Second, based on the Kasner Attractivity Conjecture below,
we restrict our considerations to solutions whose orbit enter an 
arbitrarily small 
neighbourhood of one of the Kasner circles.
Numerical simulations suggest that this property is satisfied, and hence 
we formalize it as a conjecture:

\begin{conj}[Kasner Attractivity Conjecture]\lb{KAC4.1}
The orbits of all $G_2$-compatible SH cosmologies, except for a set of 
measure zero, enter an arbitrarily small
 neighbourhood of one of the Kasner circles.
\end{conj}

Consider linearizing the evolution equations
(\ref{SH_Sc}), (\ref{SH_St})--(\ref{SH_A})
for the spatial curvature and off-diagonal shear variables 
(\ref{VII_trig})
 on the above Kasner circles:
\footnote{Recall from (\ref{23_subspace}) that $N_{22}=\Np+\sqrt{3}\Nm$, 
$N_{33} = \Np-\sqrt{3}\Nm$.}
\begin{align}
\label{lin1}
        \Sc' &= -2\sqrt{3}\Sm \Sc
\\
        \St' &= (-3\Sp+\sqrt{3}\Sm) \St
\label{lin2}
\\
        N_{22}' &=
        2(1+\Sp+\sqrt{3}\Sm)N_{22}
\label{lin3}
\\
        N_{33}' &=
        2(1+\Sp-\sqrt{3}\Sm)N_{33}
\label{lin4}
\\
	\Nc' &= 2(1+\Sp)\Nc
\\
	A' &= 2(1+\Sp)A\ ,
\end{align}
where $(\Sp,\Sm)=(\cos \phi,\sin\phi)$, and $\phi$ is a constant.
It follows that the first four variables are unstable on some arc of the 
Kasner 
circles, i.e. will act as triggers, while $\Nc$ and $A$ are stable on the 
Kasner circles, into the past. It is plausible that the stable variables 
tend to zero:
\be
\label{Nc_A_limit}
	\lim_{t \rightarrow -\infty} (\Nc,A) =\mathbf{0},
\ee
and this conclusion is strongly supported by numerical simulations.

\begin{figure}
\begin{center}
    \epsfig{file=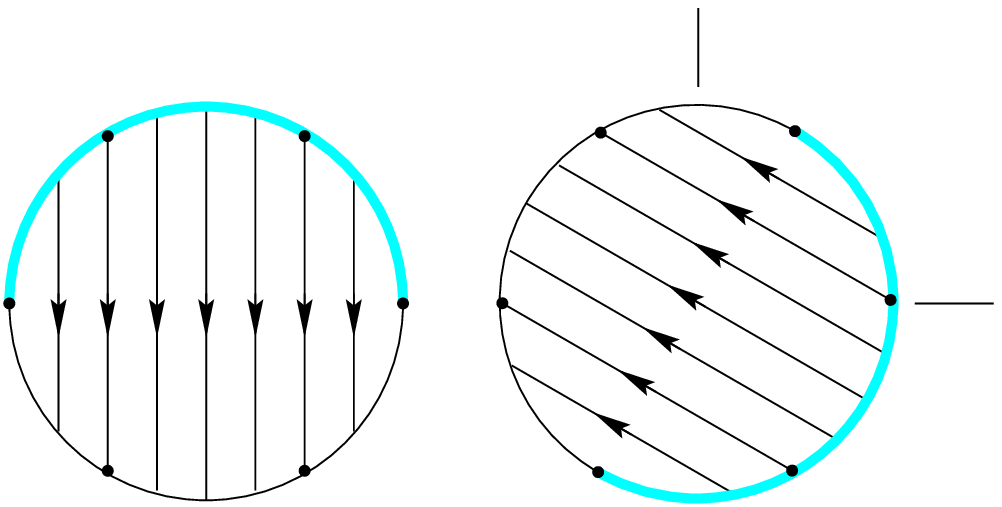,width=10.1cm}
\setlength{\unitlength}{1mm}
\begin{picture}(120,0)(0,0)
\put(78,60){\makebox(0,0)[l]{$\Sm$}}
\put(111,26){\makebox(0,0)[l]{$\Sp$}}

\put(20,50){\makebox(0,0)[l]{$\Sc$ unstable}}
\put(100,10){\makebox(0,0)[l]{$\St$ unstable}}

\end{picture}
\end{center}
\caption[The frame transition sets corresponding to the triggers
$\Sc$ and $\St$, respectively.]%
	{The frame transition sets corresponding to the triggers 
$\Sc$ and $\St$, respectively.
The arrows indicate evolution into the past.}\label{fig:frame}
\end{figure}

We recognize that $N_{22}$ and $N_{33}$ are two of the curvature triggers 
introduced in Section~\ref{sec:IX}.
In the present situation, however, the curvature transition sets 
$\mathcal{T}_{N_{22}}$
and
$\mathcal{T}_{N_{33}}$
also describe transitions on the Kasner circles 
$\mathcal{K}_{\pm\alpha}$, in addition to $\mathcal{K}$.
The new triggers here are $\Sc$ and $\St$: the off-diagonal shear 
components.
The $\Sc$ transition orbits lie on the sphere
$\Sp^2+\Sm^2+\Sc^2=1$.
Evaluating (\ref{SH_Sp}) on the sphere gives $\Sp'=0$, which implies
\be
	\Sp = const.
\ee
on each $\Sc$ transition orbit.
Similarly, the $\St$ transition orbits lie on the sphere 
$\Sp^2+\Sm^2+\St^2=1$.
Evaluating (\ref{SH_Sp})--(\ref{SH_Sm}) on the sphere gives
$(\Sp+\sqrt{3}\Sm)'=0$, which implies
\be
	\Sp+\sqrt{3}\Sm = const.
\ee
on each $\St$ transition orbit.
The $\Sc$ and $\St$ transition sets are shown in
Figure~\ref{fig:frame}.
By comparing Figure~\ref{fig:frame} with
the dashed lines in
Figure~\ref{fig:Kasner_circle}a,
we see that these transition sets connect different representations of
the same Kasner solutions.
These transitions are in fact an artifact of a rotating spatial frame
(non-zero $\Omega_\alpha$), and the transition sets actually represent the
Kasner solutions.
Therefore
we shall call the off-diagonal shear components the \emph{frame triggers},
with corresponding \emph{frame transition sets}, denoted 
$\mathcal{T}_{\Sc}$ and $\mathcal{T}_{\St}$.
\footnote{This artifact is beneficial.
If one chooses a Fermi-propagated frame, i.e. $\Omega_\alpha=0$, then
 the Kasner equilibrium points form a hypersphere
\[
        \Sp^2+\Sm^2+\Sc^2+\St^2+\Stt^2=1
\]
instead of a circle $\Sp^2+\Sm^2=1$.}

\begin{figure}
\begin{center}
    \epsfig{file=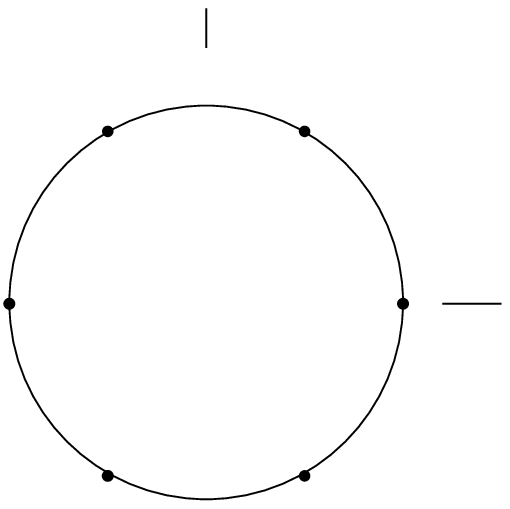,width=5.1cm}
\setlength{\unitlength}{1mm}
\begin{picture}(120,0)(0,0)
\put(55,63){\makebox(0,0)[t]{$\Sm$}}
\put(90,31){\makebox(0,0)[t]{$\Sp$}}


        
\put(38,29){\makebox(0,0)[l]{$T_1$}}
\put(64,43){\makebox(0,0)[l]{$T_2$}}
\put(64,17){\makebox(0,0)[l]{$T_3$}}

\put(69,29){\makebox(0,0)[l]{$Q_1$}}
\put(44,43){\makebox(0,0)[l]{$Q_3$}}
\put(44,17){\makebox(0,0)[l]{$Q_2$}}

\put(50,52){\makebox(0,0)[l]{$\Sc,N_{33}$}}
\put(22,40){\makebox(0,0)[l]{$\Sc,N_{33}$}}
\put(30,20){\makebox(0,0)[l]{$N_{22}$}}
\put(50,6){\makebox(0,0)[l]{$\St,N_{22}$}}
\put(76,20){\makebox(0,0)[l]{$\St$}}
\put(76,40){\makebox(0,0)[l]{$\St,\Sc$}}

\end{picture}
\end{center} 
\caption{The unstable Kasner arcs and their triggers.}\label{fig:arcs2}
\end{figure}

The 
triggers that lead to the instability of the various
 Kasner arcs are shown in Figure~\ref{fig:arcs2}, which is obtained using 
Figures~\ref{fig:curv}--\ref{fig:frame}.
The essential point is that

\begin{center}
\parbox{0.81\linewidth}{
	Mixmaster dynamics occurs if and only if each Kasner arc has at 
	least one trigger,
	and so
	if an arc has no triggers, then it will form part of the past 
	attractor, and Mixmaster dynamics will not occur.
}
\end{center}

Table~\ref{tab:SH_variables} lists the variables for the subclasses with 
generic $G_2$ action and the additional restrictions.

\begin{table}[H]
{\small
\begin{spacing}{1.1}

\caption[The Hubble-normalized variables for the $G_2$-compatible SH
cosmologies with $\Lambda=0$ and generic $G_2$ action
(see Figure~\ref{fig:gen_G2}), relative to the shear spatial gauge.]
	{The Hubble-normalized variables for the $G_2$-compatible SH 
cosmologies with $\Lambda=0$ and generic $G_2$ action
(see Figure~\ref{fig:gen_G2}), relative to the shear spatial gauge.
Trigger variables appear in bold.
Recall that $N_{22}=\Np+\sqrt{3}\Nm$, $N_{33}=\Np-\sqrt{3}\Nm$.}
                \label{tab:SH_variables}
\begin{center}
\begin{tabular}{lll}
\hline
Subclass & State vector $\X$ & Additional restriction
\\
\hline
\\
\multicolumn{2}{l}{With 3 tilt degrees of freedom:} &
\\
VI, VII & $(\Sigma_+,\Sigma_-,
	\Scb,\Stb,
        \Npb,\Nmb,N_\times,A,
        \Omega,v_1,v_2,v_3)$
	& none
\\
IV	& $(\Sigma_+,\Sm,
	\Scb,\Stb,     
        \Npb,\Nmb,
        N_\times,A,
        \Omega,v_1,v_2,v_3)$
	& $\Np^2-3(\Nm^2+\Nc^2)=0$
\\
\hline
\\
\multicolumn{2}{l}{With 2 tilt degrees of freedom: $v_2=0$} 
	& $\Np=\sqrt{3}\Nm$ 
\\
VI	& $(\Sigma_+,\Sigma_-,
	\Scb,\Stb,\Nmb,
	N_\times,A,
        \Omega,v_1,v_3)$
	& $\Nc=-\sqrt{3}A$ if $h=-\frac{1}{9}$
\\
IV	& $(\Sigma_+,\Sigma_-,
	\Scb,\Stb,\Nmb,
        A,
        \Omega,v_1,v_3)$
	& $\Nc=0$
\\
V	& $(\Sigma_+,\Sigma_-,
	\Scb,\Stb,
        A, 
        \Omega,v_1,v_3)$
	& $\Np=\Nm=\Nc=0$
\\
\hline
\\
\multicolumn{2}{l}{With 1 tilt degree of freedom: $v_2 = v_3=0$}
	& $\Np=\sqrt{3}\Nm$, $\Nc=\sqrt{3}A$
\\
VI$^*_{-\frac{1}{9}}$ & $(\Sigma_+,\Sigma_-,
        \Scb,\Stb,\Nmb,
        A,    
        \Omega,v_1)$
	& none
\\
II	& $(\Sigma_+,\Sigma_-,
        \Scb,\Stb,\Nmb,
        \Omega,v_1)$
	& $\Nc=A=0$
\\
\hline
\\
\multicolumn{2}{l}{Non-tilted: $v_\alpha =0$}
	& $\Np=\sqrt{3}\Nm$, $\Nc=\sqrt{3}A$
\\
VI$^*_{-\frac{1}{9}}$ & $(\Sigma_+,\Sigma_-,
        \Scb,\Stb,\Nmb,
        A,
        \Omega)$
	& none
\\
\hline
\end{tabular}
\end{center}
\end{spacing}
}
\end{table}

Figure~\ref{fig:arcs2} and Table~\ref{tab:SH_variables}
provide the basis for the following proposition.

\begin{prop}\lb{prop1}

Consider the class of perfect fluid $G_2$-compatible SH cosmologies which 
satisfy the Kasner Attractivity Conjecture.
\begin{itemize}
\item[i)]	
	If the $G_2$ action is generic, then Mixmaster dynamics 
	occur if and only if the anisotropic spatial curvature 
	$\mathcal{S}_{\alpha\beta}$ is non-zero.
\item[ii)]
	If the $G_2$ action is non-generic, then Mixmaster dynamics does 
	not occur.
\end{itemize}

\end{prop}

\begin{proof}
From Figure~\ref{fig:arcs2} and Table~\ref{tab:SH_variables},
we see that for models of all Bianchi types with a generic $G_2$ action, 
each arc on the Kasner circles has at least one trigger, unless 
$\Np=\Nm=\Nc=0$ (Bianchi type V, where $\mathcal{S}_{\alpha\beta}=0$).
It follows that Mixmaster dynamics
occur if and only if $\mathcal{S}_{\alpha\beta}\neq0$.

On the other hand, if the $G_2$ action is non-generic,
the state vector $\X$ in (\ref{SH_state_vector}) is restricted as follows.
For SH cosmologies with OT $G_2$ action, from (\ref{OT_grav}) we have 
\be
\label{SH_OT}
	\St=0.
\ee
For SH cosmologies with one HO KVF, from (\ref{HO_grav}) we have 
\be
\label{SH_HOKVF}
	\Np=\Nm=0=\Sc.
\ee
For SH cosmologies with diagonal $G_2$ action, we have 
\be
\label{SH_diag}
	\St=0,\quad \Np=\Nm=0=\Sc.
\ee
Figure~\ref{fig:arcs2} shows that in each case there is an arc with no 
triggers, implying that Mixmaster dynamics does not occur.

\end{proof}

\noindent
{\bf Comment:}
To show that this result is independent of using a particular gauge, 
namely the shear spatial gauge (\ref{SH_spatial_gauge}), we 
note that the infinite sequences of transitions that arise lead to 
infinite sequences of changes in physically significant scalars, 
for example, the
 scalars formed from the Weyl curvature tensor, as we describe later.
Since these changes are independent of the choice of gauge, the occurrence 
of Mixmaster dynamics is independent of the choice of gauge.
However, the choice of spatial gauge does affect which triggers occur and 
hence the details of the dynamics.%
\footnote{The real, physical dynamics, and artifacts of the rotating 
spatial frame both affect the triggers and other gravitational field 
variables. To separate these effects, we observe rotation-invariant 
scalars formed these variables (see (\ref{scalars})--(\ref{scalars3}) 
below).}
For example, in the Killing spatial gauge (\ref{KSG}), the
triggers are $N_{22}$, $\Sc$, $\St$ and $\Stt$.
It should be kept in mind that the frame transitions are merely an 
artifact of the rotating spatial frame, and do not lead to a change of 
Kasner state.%
\footnote{As an extreme example,
an artificial type of Mixmaster dynamics consisting of an
endless sequence of frame transitions occurs for models of Bianchi type V 
if the
Killing spatial gauge (\ref{KSG}) is used,
since the only triggers are $\Sc$, $\St$ and $\Stt$.
The rotational freedom can be used to set
$\Stt=0$ to eliminate this artificial Mixmaster dynamics
(the arc $(T_1 Q_2)$ becomes stable).}

\

In addition to (\ref{Nc_A_limit}), there appear to be other limits that 
help to characterize the asymptotic dynamics.
From the linearized equations (\ref{lin1})--(\ref{lin4}) we obtain
\begin{align}
	(N_{22} N_{33})' &= 4(1+\Sp) (N_{22} N_{33})
\\
	(N_{22} \Sc)' &= 2 (1+\Sp) (N_{22} \Sc)
\\
	(N_{33} \St)' &= (2-\Sp-\sqrt{3}\Sm) (N_{33} \St)\ .
\end{align}
It follows that these products are stable on the Kasner circles into the 
past, suggesting that they tend to zero as $t \rightarrow -\infty$:
\be
	\lim_{t \rightarrow -\infty} 
	( N_{22} N_{33}, N_{22} \Sc, N_{33} \St ) = \mathbf{0}\ .
\ee
This claim is strongly supported by numerical simulations.
These limits can be interpreted as stating that the two triggers in each 
product cannot be active simultaneously.

\subsubsection*{Multiple transitions}

Returning to Figure~\ref{fig:arcs2},
notice that there is more than one trigger on some Kasner arcs, which
means that two triggers can become active simultaneously, 
leading to more complicated transitions between Kasner states.
These so-called \emph{multiple transitions}
were first studied in detail by
Hewitt \etal 2003~\cite{art:Hewittetal2003}
in the context of non-tilted exceptional
Bianchi VI$^*_{-1/9}$ cosmologies,
The orbit of a multiple
transition is not a straight line when projected on the $(\Sp,\Sm)$ plane.
See Figures 3 and 4 in \cite{art:Hewittetal2003}.
Two of the three
 possible multiple transitions involve the pairs $(N_{22}, \St)$ and 
$(N_{33}, \Sc)$, which describe the Taub vacuum Bianchi II solutions 
relative to a non-Fermi-propagated frame.
The third one, involving the pair $(\St, \Sc)$, describes the Kasner 
solutions relative to a non-Fermi-propagated frame.
The possibility of multiple transitions 
is a
significant difference from the ``standard" Mixmaster 
dynamics (of Bianchi VIII and IX).
On the basis of detailed numerical simulations,
Hewitt \etal conjecture that the probability of multiple transitions
occurring tends to zero as $t \rightarrow -\infty$.
The numerical experiments suggest that multiple transitions keep occurring 
as $t \rightarrow -\infty$, but that their strength, as measured by
$|N_{22} \St|$, $|N_{33} \Sc|$ and $|\St \Sc |$, keeps decreasing, i.e. 
that
\be
	\lim_{t \rightarrow -\infty} 
	( N_{22} \St, N_{33} \Sc, \St \Sc ) =\mathbf{0}\ .
\ee
We caution that this result is not very conclusive, and
we shall revisit multiple transitions in the context of
$G_2$ dynamics in Chapters~\ref{chap:past} and \ref{chap:sim}.

\subsubsection*{Tilt}

\begin{figure}[h]
\begin{center}
    \epsfig{file=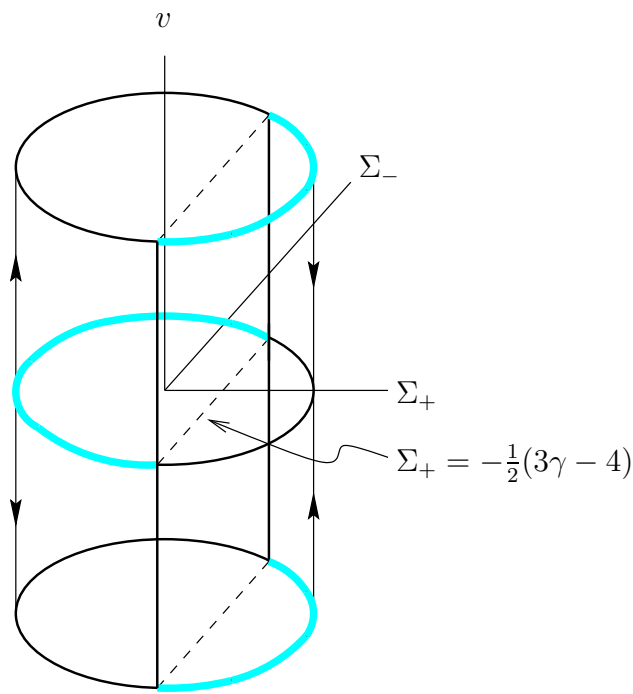,width=5.1cm}
\setlength{\unitlength}{1mm}
\begin{picture}(120,0)(0,0)
\put(54,95){\makebox(0,0)[l]{$v$}}
\put(81,75){\makebox(0,0)[l]{$\Sm$}}
\put(86,45){\makebox(0,0)[l]{$\Sp$}}
\put(86,36){\makebox(0,0)[l]{$\Sp= -\tfrac{1}{2}(3\gamma-4)$}}
        
\end{picture}   
\end{center}
\caption{The tilt transition for $v_1$.}\label{fig:tilt}
\end{figure}
\begin{figure}
\begin{center}
    \epsfig{file=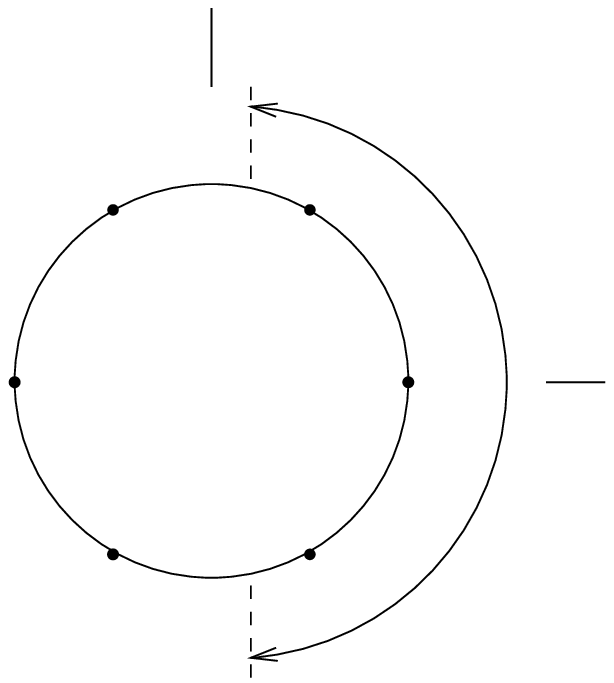,width=7.0cm}
\setlength{\unitlength}{1mm}
\begin{picture}(120,0)(0,0)
\put(55,78){\makebox(0,0)[t]{$\Sm$}}
\put(100,37){\makebox(0,0)[t]{$\Sp$}}
        
\put(30,37){\makebox(0,0)[t]{$T_1$}}
\put(66,60){\makebox(0,0)[t]{$T_2$}}
\put(66,16){\makebox(0,0)[t]{$T_3$}}
   
\put(80,37){\makebox(0,0)[t]{$Q_1$}}
\put(45,60){\makebox(0,0)[t]{$Q_3$}}
\put(45,16){\makebox(0,0)[t]{$Q_2$}}

\put(55,53){\makebox(0,0)[t]{$\Sc,N_{33}$}} 
\put(45,46){\makebox(0,0)[t]{$\Sc,N_{33}$}}
\put(41,31){\makebox(0,0)[t]{$N_{22}$}}
\put(55,21){\makebox(0,0)[t]{$\St,N_{22}$}}
\put(70,31){\makebox(0,0)[t]{$\St$}}
\put(67,46){\makebox(0,0)[t]{$\St,\Sc$}}

\put(85,21){\makebox(0,0)[l]{$\Sp > -\tfrac{1}{2}(3\gamma-4)$}}
\put(85,16){\makebox(0,0)[l]{$v_1$ is stable for $\mathcal{K}$,}} 
\put(85,11){\makebox(0,0)[l]{unstable for $\mathcal{K}_{\pm1}$}}
\end{picture}
\end{center}
\caption{The unstable arcs on $\mathcal{K}$ and $\mathcal{K}_{\pm1}$
	 and their triggers.}\label{fig:arcs}
\end{figure}

What happens to the tilt variables $v_\alpha$?
To determine the behaviour of $v_\alpha$, linearize 
(\ref{v_1})--(\ref{v_3}) on $\mathcal{K}$,
 and also
(\ref{v_2})--(\ref{v_3}) and (\ref{vsq}) on $\mathcal{K}_{\pm1}$:
\begin{alignat}{2}
\label{v_1_linear}
	&\text{on $\mathcal{K}$:}\quad&        
	v_1' &= (3\gamma-4 + 2\Sp) v_1
\\
	&&
        v_2' &= (3\gamma-4 -\Sp-\sqrt{3}\Sm) v_2
\\
	&&
        v_3' &= (3\gamma-4 -\Sp+\sqrt{3}\Sm) v_3
\label{v_3_linear}
\\
\notag\\
	&\text{on $\mathcal{K}_{\pm1}$:}\quad&
        v_2' &= (-3\Sp-\sqrt{3}\Sm) v_2
\\
	&&
        v_3' &= (-3\Sp+\sqrt{3}\Sm) v_3
\\
\label{omvsq_linear}
	&&
        (1-v^2)' &= -2 \frac{3\gamma-4 + 2\Sp}{2-\gamma}(1-v^2)\ .
\end{alignat}
For the purpose of this thesis, we shall focus on
models with only one tilt degree of freedom, i.e. $v_1\neq0$, $v_2=v_3=0$.
Equations (\ref{v_1_linear}) and (\ref{omvsq_linear}) give the linearized 
solutions
\begin{align}
\label{v_1_growth}
        v_1 &= (v_1)_0 e^{(3\gamma-4 + 2\Sp)t}
\\
        \sqrt{1-v^2} &= (\sqrt{1-v^2})_0
	 e^{ -\frac{(3\gamma-4 + 2\Sp)}{(2-\gamma)} t}\ .
\label{w_growth}
\end{align}  
Hence, the tilt is unstable into the past on the following arcs:
\be
        \mathcal{K}\ \text{arc}(\Sp < -\tfrac{1}{2}(3\gamma-4)),\quad
        \mathcal{K}_{\pm1}\ \text{arc}(\Sp > -\tfrac{1}{2}(3\gamma-4)).
\ee
This creates transition orbits between $\mathcal{K}$ and 
$\mathcal{K}_{\pm1}$, as illustrated in Figure~\ref{fig:tilt}.   
The unstable arcs are highlighted, and
the arrows indicate evolution into the past.
For convenience we augment Figure~\ref{fig:arcs2} by adding the unstable 
arcs for $v_1$ to produce Figure~\ref{fig:arcs}.
There are analogous transition orbits between $\mathcal{K}$ and 
$\mathcal{K}_{\pm2}$, and between $\mathcal{K}$ and
$\mathcal{K}_{\pm3}$.
There are also transition orbits between the extreme tilt Kasner circles.
\footnote{We refer the reader to the work by
Uggla, van Elst, Wainwright \& Ellis 2003~\cite{art:Ugglaetal2003}
in the context of $G_0$ cosmologies.}

\noindent
{\bf Comments:}
The nature of the tilt transition means 
\emph{we cannot make
non-Mixmaster dynamics into Mixmaster by adding only tilt transitions.}
However adding tilt degrees of freedom to a non-tilted model also adds 
off-diagonal shear variables and it is these variables that lead to the 
occurrence of Mixmaster dynamics.
From Table~\ref{tab:SH_variables}, we see that
 the tilted Bianchi II cosmologies 
form the prototype for Mixmaster dynamics with tilt.
We refer to the
analysis of tilted Bianchi II cosmologies by
Hewitt \etal 2001~\cite{art:Hewittetal2001}. 
\footnote{The spatial gauge used in
this paper is a permutation of the one used in this thesis.
The indices they use are ``one lower" than ours, e.g.
they have
\be
        N_{11} \neq 0,\quad v_3 \neq 0,\quad \Sigma_{31} =0,
\ee
while we have
\be
        N_{22} \neq 0,\quad v_1 \neq 0,\quad \Sigma_{12} =0.
\ee
They define $\Sigma_\pm$ relative to their 23-subspace, as we do in
(\ref{23_subspace}).
}
The class of tilted exceptional Bianchi~VI$^*_{-1/9}$ cosmologies has not 
been analyzed, but we can now predict, without additional work, that
its Mixmaster dynamics is that of the tilted Bianchi II cosmologies.

\subsubsection*{Physical aspects of Mixmaster dynamics}

The physical aspects of Mixmaster dynamics are described by 
rotation-invariant scalars, such as the shear scalar $\Sigma$ and the 
electric and magnetic Weyl scalars $\mathcal{E}$ and $\mathcal{H}$:
\begin{align}
\label{scalars}
	\Sigma^2 &= \Sp^2+\Sm^2+\Sc^2+\St^2+\Stt^2
\\
        \mathcal{E}^2 &= \Ep^2+\Em^2+\Ec^2+\Ett^2+\Et^2
\\
        \mathcal{H}^2 &= \Hp^2+\Hm^2+\Hc^2+\Htt^2+\Ht^2.
\label{scalars3}
\end{align}
Furthermore, the scalar $\mathcal{E}^2-\mathcal{H}^2$ is boost-invariant.
The shear scalar $\Sigma$ is close to 1 except during curvature 
transitions. The other invariant of $\Sigma_{\alpha\beta}$ is the 
determinant $\det \Sigma_{\alpha\beta}$, which is also equal to
$\tfrac{1}{3}\Sigma_{\alpha\beta} \Sigma^{\beta\gamma} 
\Sigma_{\gamma}{}^{\alpha}$. $\det \Sigma_{\alpha\beta}$ is close to 
constant
except during curvature transitions.
The electric Weyl scalar $\mathcal{E}$ is equal to
$\sqrt{2-\det \Sigma_{\alpha\beta}}$ when $\Sigma^2=1$. It is close to 
constant
except during curvature transitions.
The magnetic Weyl scalar $\mathcal{H}$ is close to zero except during 
curvature transitions.
We refer to WE, Section 11.4 for the graphs of these scalars under 
the standard Mixmaster dynamics.
There are other rotation-invariant scalars, such as $\Om_k$, 
$\mathcal{S}_{\alpha\beta} \mathcal{S}^{\alpha\beta}$ and $\det 
\mathcal{S}_{\alpha\beta}$.
The principal feature of Mixmaster dynamics is that the Hubble-normalized 
Weyl scalars are bounded but do not have a limit in the past asymptotic 
regime.

Another physical aspect is the fluid vorticity.
Although the timelike congruence $\me_0$ is vorticity-free, 
the fluid congruence is not necessarily so.
From Appendix C of Hewitt \etal 2001~\cite{art:Hewittetal2001}, 
the fluid vorticity for $G_2$-compatible SH cosmologies is given by
\begin{align}
        W_1 &= \frac{1}{2B} \frac{1}{1-v^2} N^{CD} v_C v_D v_1
\\
        W_A &= \frac{1}{2B} \left[ (N_A{}^C+ \eps_A{}^C A) v_C
                + \frac{1}{1-v^2} N^{CD} v_C v_D v_A \right]\ ,
\end{align} 
where
\[
        B = \frac{1-\tfrac{1}{3}(v^2 + \Sigma_{\alpha\beta} v^\alpha 
v^\beta
        + 2A v_1)}{G_-\ \sqrt{1-v^2}}\ .
\]
Observe that
the fluid vorticity is zero if $v_2=v_3=0$.

For the case $v_2=0$, $v_3\neq0$, the spatial gauge is given by
(\ref{e2_align_S3}), 
which implies $N^{CD} v_C v_D =0$.
It follows that the only non-zero vorticity component is
\be
        W_2 = \frac{1}{2B} (\sqrt{3}\Nc+A)v_3\ .
\ee
The limit (\ref{Nc_A_limit}) then implies that $W_2$ tends to
zero.

For the case with three tilt degrees of freedom,   
since the limit of $N^{CD} v_C v_D$ does not exist,
the limit of the fluid vorticity does
not exist as $t \rightarrow -\infty$,
i.e. Mixmaster dynamics is reflected in the fluid vorticity.

Thus we have the following proposition.

\begin{prop}
Consider $G_2$-compatible SH cosmologies with generic $G_2$ action.
\begin{itemize}	
\item	If there is one tilt degree of freedom (i.e. $v_A=0$), then the 
	fluid vorticity is zero.
\item	If there are two tilt degrees of freedom, then the fluid vorticity
	tends to zero as $t \rightarrow -\infty$.
\item	If there are three tilt degrees of freedom, then the
	limit of the fluid vorticity does
	not exist as $t \rightarrow -\infty$.
\end{itemize}

\end{prop}

\begin{figure}
\begin{center}
    \epsfig{file=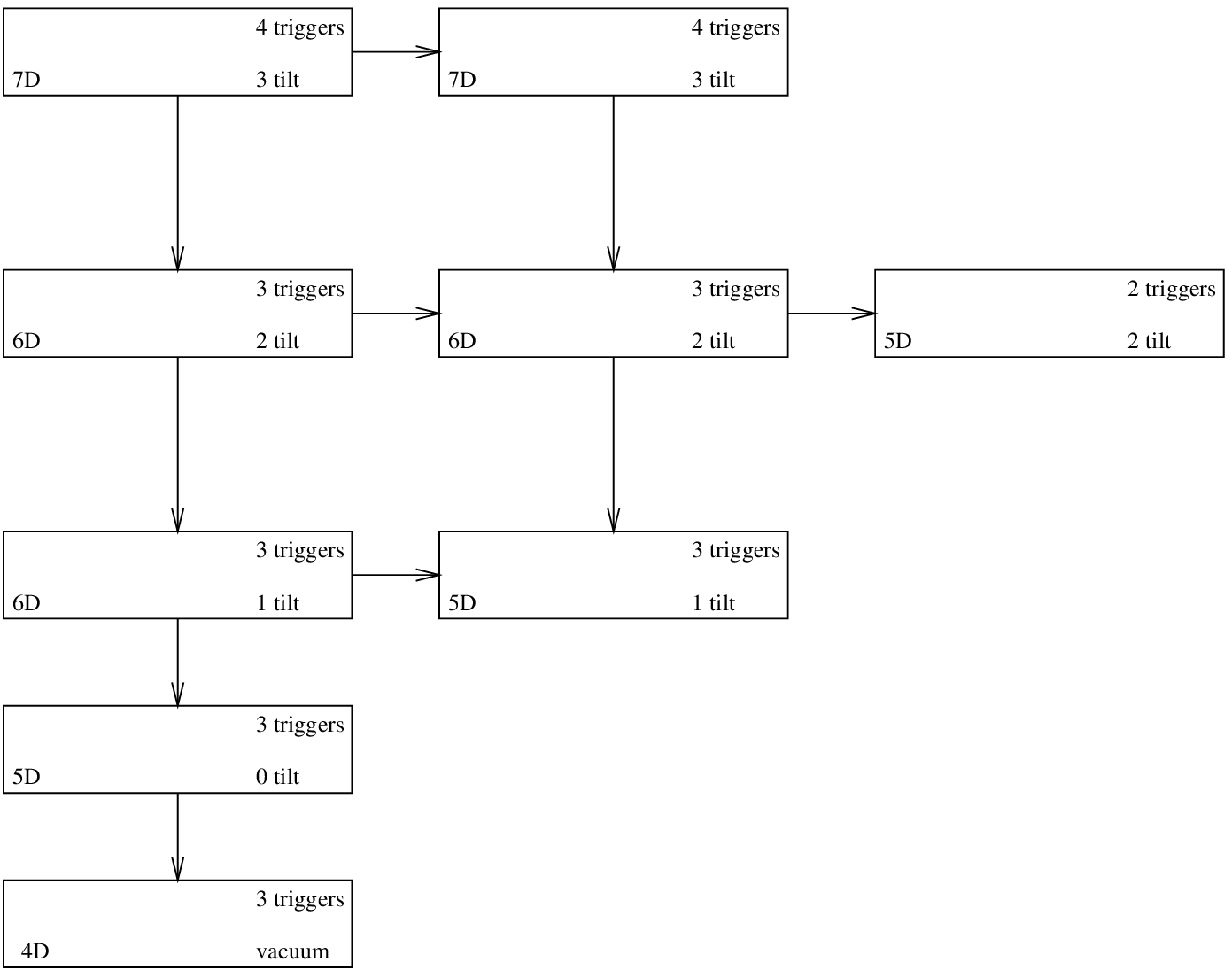,width=14cm}
\setlength{\unitlength}{1mm}
\begin{picture}(120,0)(0,0)

\put(-10,120){\makebox(0,0)[l]{$\det n_{AB} \neq 0$, $a_1^2 = h \det n_{AB}$}}
\put(0,110){\makebox(0,0)[l]{VI, VII}}  
\put(0,80){\makebox(0,0)[l]{VI}}

\put(0,50){\makebox(0,0)[l]{VI$^*_{-\frac{1}{9}}$}}
\put(0,30){\makebox(0,0)[l]{VI$^*_{-\frac{1}{9}}$}}
\put(0,10){\makebox(0,0)[l]{VI$^*_{-\frac{1}{9}}$}}

\put(100,90){\makebox(0,0)[l]{$n_{AB}=0$}}
\put(100,80){\makebox(0,0)[l]{V}}

\put(50,120){\makebox(0,0)[l]{$\det n_{AB}=0$ }}
\put(50,110){\makebox(0,0)[l]{IV}}
\put(50,80){\makebox(0,0)[l]{IV}}
\put(50,50){\makebox(0,0)[l]{II}}

\put(90,40){\makebox(0,0)[l]{$W=0$}}
\put(90,65){\makebox(0,0)[l]{$W\not\equiv0$, 
	$\lim\limits_{t \rightarrow -\infty} W =0$}}
\put(90,110){\makebox(0,0)[l]{$\lim\limits_{t \rightarrow -\infty} W$ 
	does not exist}}

\put(36,110){\oval(98,30)[t]}
\put(36,110){\oval(98,20)[b]}

\put(61,80){\oval(148,30)[t]}
\put(61,80){\oval(148,20)[b]}

\put(36,30){\oval(98,60)[t]}
\put(36,30){\oval(98,20)[b]}

\end{picture}
\end{center}
\caption[SH cosmologies with generic $G_2$ action (more detailed).]
	{SH cosmologies with generic $G_2$ action, indicating the Bianchi 
group type, the dimension of the Hubble-normalized state space, the number 
of triggers and the number of tilt degrees of freedom.
Here $W$ is the Hubble-normalized fluid vorticity scalar, defined by
$W^2 = W_1^2 + W_A W^A$.}\label{fig:gen_G2_trig}
\end{figure}

We revisit Figure~\ref{fig:gen_G2} and include the total number of 
curvature and frame triggers, the dimension of the state space, and the 
asymptotic behaviour of the fluid vorticity (see 
Figure~\ref{fig:gen_G2_trig}).
Note that for Bianchi VI and VII cosmologies, the dimension listed is for
the state space with a fixed parameter $h$.
For $\Lambda>0$, increase the dimension by one.

\

Before giving a summary we note that
there is some uncertainty as regards the behaviour of the tilt
as $t \rightarrow -\infty$.
The analysis of the triggers and of the tilt transitions suggests that 
along a typical infinite sequence of transitions, the tilt scalar $v$ will 
not approach a limit:
\be
	\lim_{t \rightarrow -\infty} v \quad\text{does not exist.}
\ee
However, while performing numerical simulations of Bianchi II models with
$\gamma =\tfrac{4}{3}$, Bridson 1997~\cite{cal:Bridson1997}
discovered that,
despite the existence of tilt transitions from $\mathcal{K}_{\pm1}$
to $\mathcal{K}$
that transform $v$ from $\pm1$ to $0$,
$v$ tends to get stuck at $\pm1$.
He argued that, due to the arrangement of the transition orbits,
the state vector $\X$ spends more time close to
the
part of $\mathcal{K}_{\pm1}$ where $1-v^2$ is
stable and hence $1-v^2$ is forced even closer to zero, so that when it
has an opportunity
 to activate,
i.e. when $\X$ is close to the unstable arc on
$\mathcal{K}_{\pm1}$ for $1-v^2$,
it needs more time to do so and loses out to the
curvature and frame triggers.
This tendency will, however, depend significantly on the value of 
$\gamma$.
If $\gamma$ is close to 2,
then $v$ should have a strong tendency to stay close to zero, for two   
reasons: $(1-v^2)$ is unstable on most of $\mathcal{K}_{\pm1}$, and
also has
a high growth rate into the past due to the $(2-\gamma)$ factor in the
denominator (see equation (\ref{w_growth})).
Hence
one expects that
somewhere between $\gamma = \tfrac{4}{3}$ and $\gamma=2$ there
is a critical value of $\gamma$ where one tendency replaces
the other.

\subsubsection*{Summary}

To recapitulate, the \emph{generalized Mixmaster dynamics}
that we have discussed in this section
 can be described by 
the following limits:
\begin{alignat}{2}
\label{Mix_1}
	&\text{BKL conjecture I:}
	&&
	\lim_{t \rightarrow -\infty} \Omega=0,\quad
	\lim_{t \rightarrow -\infty} \Oml =0,
\\
\label{Mix_2}
	&\text{Stable variables:}
	&&
	\lim_{t \rightarrow -\infty} (A,\Nc)=\mathbf{0},
\\
\label{Mix_3}
	&\text{Non-overlapping triggers:}\
	&&
	\lim_{t \rightarrow -\infty} 
		(N_{22}N_{33}, N_{22}\Sc, N_{33}\St)=\mathbf{0},
\\
\label{Mix__4}
        &\text{Trigger variables:}
        &&
        \lim_{t \rightarrow -\infty} (N_{22},N_{33},\St,\Sc) \quad
        \text{do not exist.}
\end{alignat}

Unlike the case of standard Mixmaster dynamics discussed in 
Section~\ref{sec:IX}, to date no proofs
 have been given concerning generalized Mixmaster 
dynamics (e.g. the limits (\ref{Mix_1})--(\ref{Mix_3}) which serve to 
restrict the past attractor).
The non-tilted Bianchi~VI$^*_{-1/9}$ solutions and the
tilted Bianchi II solutions
are prototypes for the new features of generalized Mixmaster dynamics. 
Understanding these two special cases, and having our formulation of the
evolution equations for the whole hierarchy,
gives qualitative information about the dynamics of the whole class and
provides equations for doing detailed analyses, and perhaps for providing 
proofs of some of the above limits.

\section{Non-Mixmaster dynamics}

In this section, we discuss the past asymptotic dynamics of SH subclasses 
with non-generic $G_2$ action.
The conditions for OT $G_2$ action, HO KVF and diagonal $G_2$ action are 
given by (\ref{SH_OT})--(\ref{SH_diag}) respectively.
Figures~\ref{fig:OT_G2}--\ref{fig:HO_diag} summarized the hierarchy for 
these SH subclasses.
Tables~\ref{tab:OT}--\ref{tab:diag} list the variables 
and the additional restrictions.

{\small
\begin{spacing}{1.1}
\begin{table}[h]
\caption[The Hubble-normalized variables for the $G_2$-compatible SH
cosmologies with $\Lambda=0$ and OT $G_2$ action (see
Figure~\ref{fig:OT_G2}).]
	{The Hubble-normalized variables for the $G_2$-compatible SH
cosmologies with $\Lambda=0$ and OT $G_2$ action (see 
Figure~\ref{fig:OT_G2}).
Triggers appear in bold.}
                \label{tab:OT}
\begin{center}
\begin{tabular}{lll}
\hline
Subclass & State vector $\X$ & Additional restriction
\\
\hline
\\
\multicolumn{2}{l}{With 1 tilt degree of freedom: $v_2 = v_3=0$}
        & $\St=0$
\\
VII	& $(\Sigma_+,\Sigma_-,
	\Scb,\Npb,\Nmb,
	\Nc,A,\Omega,v_1)$
	& none
\\
VI      & $(\Sigma_+,\Sigma_-,
	\Scb,\Nmb,
	\Nc,A,\Omega,v_1)$
        & $\Np = \sqrt{3}\Nm$
\\
IV	& $(\Sigma_+,\Sigma_-,
        \Scb,\Nmb,
	A,\Omega,v_1)$
	& $\Np=\sqrt{3}\Nm,\ \Nc=0$
\\
II	& $(\Sigma_+,\Sigma_-,
        \Scb,\Nmb,
	\Omega,v_1)$
	& $\Np=\sqrt{3}\Nm,\ \Nc=0=A$
\\
\hline
\\
\multicolumn{2}{l}{Non-tilted: $v_\alpha=0$}
        & $\St=0$
\\
VII     & $(\Sigma_+,\Sigma_-,
        \Scb,\Npb,\Nmb,
	\Nc,A,\Omega)$
        & none
\\
VI      & $(\Sigma_+,\Sigma_-,
        \Scb,\Nmb,
	\Nc,A,\Omega)$
        & $\Np = \sqrt{3}\Nm$
\\
IV      & $(\Sigma_+,\Sigma_-,
        \Scb,\Nmb,
	A,\Omega)$
        & $\Np=\sqrt{3}\Nm,\ \Nc=0$
\\
II      & $(\Sigma_+,\Sigma_-,
        \Nmb,
	\Omega)$
        & $\Np=\sqrt{3}\Nm,\ \Nc=0=A=\Sc$
\\
\hline
\end{tabular}
\end{center}
\end{table}
\end{spacing}
}

{\small
\begin{spacing}{1.1}
\begin{table}[h]
\caption[The Hubble-normalized variables for the $G_2$-compatible SH
cosmologies with $\Lambda=0$ and one HO KVF (see
Figure~\ref{fig:HO_diag}).]
	{The Hubble-normalized variables for the $G_2$-compatible SH
cosmologies with $\Lambda=0$ and one HO KVF (see 
Figure~\ref{fig:HO_diag}).
$\St$ is the only trigger.}
                \label{tab:HO}
\begin{center}
\begin{tabular}{lll}   
\hline
Subclass & State vector $\X$ & Additional restriction
\\
\hline
\\
\multicolumn{2}{l}{With 2 tilt degrees of freedom: $v_2=0$}
        & $\Np=\sqrt{3}\Nm=0=\Sc$
\\
VI      & $(\Sigma_+,\Sigma_-,
	\Stb,
	N_\times,A,\Omega,v_1,v_3)$
        & $\Nc=-\sqrt{3}A$ if $h=-\frac{1}{9}$
\\
V       & $(\Sigma_+,\Sigma_-,\Stb,A,\Omega,v_1,v_3)$
        & $\Nc=0$
\\
\hline
\\
\multicolumn{2}{l}{With 1 tilt degree of freedom: $v_2 = v_3=0$}
        & $\Np=\sqrt{3}\Nm=0=\Sc$, $\Nc=\sqrt{3}A$
\\
VI$^*_{-\frac{1}{9}}$ & 
	$(\Sigma_+,\Sigma_-,\Stb,A,\Omega,v_1)$
        & none
\\
\hline
\\
\multicolumn{2}{l}{Non-tilted: $v_\alpha=0$}
        & $\Np=\sqrt{3}\Nm=0=\Sc$, $\Nc=\sqrt{3}A$,
\\
	& & $\Sp=-\sqrt{3}\Sm$
\\
VI$^*_{-\frac{1}{9}}$ & 
        $(\Sigma_-,\Stb,A,\Omega)$
        & none
\\
\hline
\end{tabular}
\end{center}
\end{table}
\end{spacing}
}

{\small
\begin{spacing}{1.1}
\begin{table}[h]
\caption{The Hubble-normalized variables for the $G_2$-compatible SH
cosmologies with $\Lambda=0$ and diagonal $G_2$ action (see 
Figure~\ref{fig:HO_diag}).}
                \label{tab:diag}
\begin{center}
\begin{tabular}{lll}
\hline
Subclass & State vector $\X$ & Additional restriction
\\
\hline
\\
\multicolumn{2}{l}{With 1 tilt degree of freedom: $v_2 = v_3=0$}
        & $\St=0$, $\Np=\sqrt{3}\Nm=0=\Sc$
\\
VI      & $(\Sigma_+,\Sigma_-,\Nc,A,\Omega,v_1)$
        & none
\\
V       & $(\Sigma_+,\Sigma_-,A,\Omega,v_1)$
        & $\Nc=0$
\\
V (LRS) & $(\Sigma_+,A,\Omega,v_1)$
        & $\Nc=0=\Sm$
\\
\hline
\\
\multicolumn{2}{l}{Non-tilted: $v_\alpha=0$}
        & $\St=0$, $\Np=\sqrt{3}\Nm=0=\Sc$
\\
VI      & $(\Sigma_+,\Sigma_-,\Nc,A,\Omega)$
        & none               
\\
V       & $(\Sigma_-,A,\Omega)$
        & $\Nc=0=\Sp$                  
\\
I	& $(\Sigma_+,\Sigma_-,\Omega)$
	& $\Nc=0=A$
\\
I (LRS) & $(\Sigma_+,\Omega)$
        & $\Nc=0=A=\Sm$
\\
\hline
\end{tabular}
\end{center}
\end{table}
\end{spacing}
}

For each entry in Tables~\ref{tab:OT}--\ref{tab:diag}, we combine the 
unstable arcs of the curvature, frame and tilt
transition sets in Figure~\ref{fig:arcs}
and look for the remaining stable arcs on the Kasner 
circles $\mathcal{K}$ and $\mathcal{K}_{\pm1}$.
In each case, sequences of transition sets 
will terminate at some stable arc.
For each class the past attractor is an arc or union of arcs on the Kasner 
circles.
The result is that each orbit, except for a set of measure zero, is past 
asymptotic to a Kasner equilibrium point.

For SH cosmologies with OT $G_2$ action (\ref{SH_OT}), the $\St$ trigger 
is zero,
making the Kasner arc $(T_3 Q_1)$ on $\mathcal{K}$ in 
Figure~\ref{fig:arcs}  stable.
The corresponding arc on 
$\mathcal{K}_{\pm1}$ is unstable due to the tilt.
Thus the conjectured past attractor is given by
\be
        \mathcal{A}^- = \mathcal{K}\ \text{arc}(T_3 Q_1).
\label{SH_OT_at}
\ee
The past attractor for the non-tilted case (i.e. $v_1=0$) is also given by 
(\ref{SH_OT_at}), 
except for Bianchi II, whose past attractor is a bigger set
\be
	\mathcal{A}^- = \mathcal{K}\ \text{arc}(T_1 Q_3 T_2 Q_1 T_3).
\ee

For SH cosmologies with one HO KVF (\ref{SH_HOKVF}), there is only one 
trigger, $\St$.
With one tilt degree of freedom,
from Figure~\ref{fig:arcs}, we see that the conjectured past 
attractor is given by
\begin{multline}
        \mathcal{A}^- =
        \left[
        \mathcal{K}\ {\rm arc}(\Sp >-\tfrac{1}{2}(3\gamma-4) )\
        \cap\
        \mathcal{K}\ {\rm arc}(Q_2 T_1 Q_3 T_2)\
        \right]
\\
        \cup\
        \left[
        \mathcal{K}_{\pm1}\ {\rm arc}(\Sp <-\tfrac{1}{2}( 3\gamma-4) )\
        \cap\
        \mathcal{K}_{\pm1}\ {\rm arc}(Q_2 T_1 Q_3 T_2)\ \right]\ .
\label{SH_HOKVF_tilt_1}
\end{multline}
For non-tilted models, the past attractor is the point
$(\Sp,\Sm)=(- \tfrac{\sqrt{3}}{2}, \tfrac{1}{2})$ on $\mathcal{K}$.
Finally, we note that the past attractor for the case
with two tilt degrees of freedom is also given by (\ref{SH_HOKVF_tilt_1}).
\footnote{It turns out that the set for stable arcs on 
$\mathcal{K}_{\pm3}$, given by
\[
	\mathcal{K}_{\pm3}\ {\rm arc}(\Sp - \sqrt{3}\Sm > 3\gamma-4 )\
        \cap\
        \mathcal{K}_{\pm3}\ {\rm arc}(Q_3 T_2 Q_1 T_3)\
        \cap\
        \mathcal{K}_{\pm3}\ {\rm arc}(Q_2 T_1 Q_3 T_2)\ ,
\]
is empty, and the arcs for which 
$v_3=0$ is stable on $\mathcal{K}_{\pm1}$ is identical to 
$\mathcal{K}_{\pm1}\ 
{\rm arc}(Q_2 T_1 Q_3 T_2)$ (compare (\ref{v_3_linear}) with 
(\ref{lin2})). Finally, the stable arcs on $\mathcal{K}$, given by
\[
	\mathcal{K}\ {\rm arc}(\Sp >-\tfrac{1}{2}(3\gamma-4) )\
        \cap\
	\mathcal{K}\ {\rm arc}(\Sp - \sqrt{3}\Sm < 3\gamma-4)\
	\cap\
        \mathcal{K}\ {\rm arc}(Q_2 T_1 Q_3 T_2)\ ,
\]
simplifies to
\[
	\mathcal{K}\ {\rm arc}(\Sp >-\tfrac{1}{2}(3\gamma-4) )\
        \cap\
        \mathcal{K}\ {\rm arc}(Q_2 T_1 Q_3 T_2)\ .
\]
}

For SH cosmologies with diagonal $G_2$ action (\ref{SH_diag}), 
there are no triggers. From Figure~\ref{fig:tilt}, we see that
for tilted models ($v_1 \neq 0$)
 the 
conjectured past attractor is given by
\be
        \mathcal{A}^- = \mathcal{K}\
        {\rm arc}(\Sp >-\tfrac{1}{2}(3\gamma-4) )\
           \cup\ \mathcal{K}_{\pm1}\
        {\rm arc}(\Sp <-\tfrac{1}{2}( 3\gamma-4) ).
\label{SH_diag_at}
\ee
With zero tilt, the past attractor is given by
\be
	\mathcal{A}^- = \mathcal{K}\ ,
\ee
except for Bianchi V, whose past attractor is a smaller set
\be
	\mathcal{A}^- = \{ \Sp=0,\Sm=\pm1 \}.
\ee

\section{Future asymptotics}\label{sec:SH_future}

In this section we discuss the future asymptotic 
dynamics of $G_2$-compatible SH cosmologies.

\subsection*{The role of $\Lambda$}

The presence of a cosmological constant $\Lambda>0$ simplifies the future 
asymptotic dynamics of SH cosmologies dramatically, 
as shown by Wald 1983~\cite{art:Wald1983}.
He proved, 
for all SH cosmologies excluding Bianchi type IX,
subject to certain restrictions on the matter content, that
\be
	\lim_{t \rightarrow \infty} H = \sqrt{\tfrac{\Lambda}{3}},\quad
	\lim_{t \rightarrow \infty} \sigma_{\alpha\beta}=0,
\ee
where $H$ is the Hubble scalar and $\sigma_{\alpha\beta}$ is the  shear 
tensor.
In other words, the cosmological models are asymptotic to the de Sitter 
solution.

Starobinski\v{\i} 1983~\cite{art:Starobinskii1983} 
discussed the asymptotic dynamics of cosmologies without symmetry 
assumptions and provided a conjecture for behaviour of
the tilt variables, an issue that Wald had not addressed.
He called the positive cosmological constant ``the best isotropizer".

Our goal is to describe the effect of 
including a positive cosmological constant
$\Lambda$ from the perspective of 
the Hubble-normalized state space.
First, including $\Lambda$ increases the dimension of the 
Hubble-normalized state space by one, by adjoining the Hubble-normalized 
variable $\Oml$.
Second, the Gauss constraint (\ref{SH_C_G}) implies
\be
	0 < \Oml \leq 1.
\label{Oml_range}
\ee
The condition
\be
	\Oml =1
\label{Oml_1}
\ee
with (\ref{SH_C_G}) and (\ref{SH_q}) implies that
\be
	\Om=0,\quad 
	\Sigma^2=0,\quad
	\Nm=0=\Nc=A,\quad
	q=-1,
\label{Oml_2}
\ee
and the evolution equation (\ref{SH_Oml}) for $\Oml$
is identically satisfied.
Equations
(\ref{Oml_1})--(\ref{Oml_2}) thus define an invariant set.
The evolution equation (\ref{SH_Np}) evaluated on the invariant set
(\ref{Oml_1})--(\ref{Oml_2}) gives
\be
        \Np' = - \Np\ ,
\ee
which implies that equilibrium points must satisfy
\be
        \Np = 0.
\label{Oml_4}
\ee
Evolution equations (\ref{v_1})--(\ref{v_3}) and (\ref{vsq}) evaluated on 
this invariant set give
\begin{align}
	v_\alpha' &= \frac{(3\gamma-4)}{G_-}(1-v^2)v_\alpha
\\
	(v^2)' &= \frac{2(3\gamma-4)}{G_-}(1-v^2)v^2\ ,
\end{align}
which imply that the equilibrium points are
\be
	\begin{cases}
	v_\alpha=0 \quad \text{or}\quad v^2=1 & 
			\text{if $\gamma \neq \tfrac{4}{3}$}
	\\
	v_\alpha=const.,\quad 0 \leq v^2 \leq 1 &
			\text{if $\gamma = \tfrac{4}{3}$.}
	\end{cases}
\label{Oml_3}
\ee
We shall refer to the set
(\ref{Oml_1})--(\ref{Oml_2}), (\ref{Oml_4}) and 
(\ref{Oml_3}) as the \emph{de Sitter invariant set}.

The evolution equation (\ref{SH_Oml}) for $\Oml$, (\ref{SH_q}) and 
(\ref{Oml_range}) imply that
\be
	\Oml' = (q+1)\Oml >0\ ,
\ee
which implies that $\Oml$ is monotone increasing, and
\be
	\lim_{t \rightarrow \infty} \Oml=1,\quad
	\lim_{t \rightarrow -\infty} \Oml=0.
\ee
The presence of $\Oml$ thus has no effect on the past asymptotic dynamics, 
but destabilizes all equilibrium points on the set $\Oml=0$ into the 
future.
That $\lim\limits_{t \rightarrow \infty} \Oml=1$ implies
\be
	\lim_{t \rightarrow \infty} \Omega =0,\quad
	\lim_{t \rightarrow \infty} \Sigma^2=0,\quad
	\lim_{t \rightarrow \infty} (\Np,\Nm,\Nc,A)=\mathbf{0}.
\ee
As a result, the Weyl curvature variables also tend to zero:
\be
        \lim_{t \rightarrow \infty} \mathcal{E}_{\alpha\beta}=0,\quad
        \lim_{t \rightarrow \infty} \mathcal{H}_{\alpha\beta}=0.
\ee
Since there are no equilibrium points with $0 < \Oml < 1$, as $t 
\rightarrow \infty$ 
the solutions must approach the de Sitter invariant set.
The tilt variables have the following limits:
\be
	\begin{cases}
	\lim\limits_{t \rightarrow \infty} v_\alpha=0 
		&\text{if $\gamma<\tfrac{4}{3}$}
	\\
	\lim\limits_{t \rightarrow \infty} v_\alpha=const.
		 &\text{if $\gamma=\tfrac{4}{3}$}
	\\
	\lim\limits_{t \rightarrow \infty} v_\alpha=const.,\quad
        \lim\limits_{t \rightarrow \infty} v^2 =1
                 &\text{if $\tfrac{4}{3}<\gamma<2$.}
	\end{cases}
\ee
See Lim \etal 2004~\cite{art:Limetal2004}, Appendix B for a proof.

\subsection*{The case $\Lambda=0$}

The future dynamics in the case $\Lambda=0$ is much more complicated than 
in the case $\Lambda>0$, and depends significantly on the 
Bianchi type and on the equation of state parameter $\gamma$.
The future dynamics for the non-tilted case has been completely analyzed,
\footnote{See WE,
Wainwright, Hancock \& Uggla 1999~\cite{art:Wainwrightetal1999},
Nilsson, Hancock \& Wainwright 2000~\cite{art:Nilssonetal2000},
Horwood \etal 2003~\cite{art:Horwoodetal2003}
and Hewitt \etal 2003~\cite{art:Hewittetal2003}.}
but much remains to be done as regards the tilted case.
Based on the known cases -- the Bianchi II and VI$_0$ subclasses 
(see \cite{art:Hewittetal2001} and \cite{art:Hervik2004} respectively),
non-tilted equilibrium points in the interior ($\Omega>0$) of the state 
space, which act as the future attractor for non-tilted models,
 are destabilized by the tilt as $\gamma$ increases.
The resulting bifurcations lead to the creation of tilted equilibrium 
points that act as the future attractor.

Table~\ref{tab:ref} gives an overview of the work on
tilted $G_2$-compatible SH cosmologies (all with $\Lambda=0$)
using Hubble-normalized variables.
\footnote{See WE for references to work on tilted SH cosmologies using 
other approaches.}
Hervik and Coley 2004~\cite{com:HervikColey2004}
 are currently analyzing the future dynamics of tilted Bianchi
VI$_h$ cosmologies, and a preliminary result suggests that a 
complicated, periodic orbit attractor exists for a certain range of 
$\gamma$.

\begin{table}
\begin{spacing}{1.1}
\caption{Previous work on $G_2$-compatible SH cosmologies with tilt and
$\Lambda=0$.}
                \label{tab:ref}
\begin{center}
\begin{tabular}{lll}
\hline
Bianchi type & $G_2$ action & paper
\\
\hline
\\
Bianchi V & generic $G_2$   & Harnett 1996~\cite{cal:Harnett1996}
\\
& HO KVF         & Harnett 1996~\cite{thesis:Harnett1996}
\\
& tilted diagonal & Hewitt \& Wainwright
                        1992~\cite{art:HewittWainwright1992}
\\
& LRS tilted diagonal   & Collins \& Ellis
                                1979~\cite{art:CollinsEllis1979}
\\
\hline
\\
Bianchi II & generic $G_2$   & Hewitt \etal 2001~\cite{art:Hewittetal2001}
\\
\hline
\\
Bianchi VI$_0$ & generic $G_2$   & Hervik 2004~\cite{art:Hervik2004}
\\
\hline
\end{tabular}
\end{center}
\end{spacing}
\end{table}



The class of SH cosmologies of Bianchi type VII is of particular interest 
because it contains the flat and open FL cosmologies as isotropic special 
cases (flat FL as a special case of VII$_0$ and open FL as a special case 
of VII$_h$).

The Bianchi VII$_0$ state space is unbounded: $\Np$ is not bounded by 
the Gauss constraint (\ref{SH_C_G}).
This is not an artifact of a bad choice of spatial frame,
since the Weyl curvature scalars
become unbounded as $\Np \rightarrow \infty$
($\Np$ appears in (\ref{SH_Em})--(\ref{SH_Ec}) and 
(\ref{SH_Hm})--(\ref{SH_Hc})).
In other words, the Hubble-normalized Weyl curvature scalars are unbounded 
on the Bianchi VII$_0$ state space.
The detailed analysis of the non-tilted Bianchi VII$_0$ cosmologies shows 
that 
if $\gamma \leq \tfrac{4}{3}$,
they isotropize as regards the shear (i.e. $\Sigma_{\pm} \rightarrow0$ 
as $r \rightarrow \infty$) even though $\Np \rightarrow \infty$,
but that the Weyl curvature diverges if $\gamma >1$
\cite{art:Wainwrightetal1999} 
\cite{art:Nilssonetal2000}.
Preliminary investigation 
of the tilted models
\cite{cal:wclim2000}
indicates that no tilted equilibrium points exist for Bianchi VII$_0$ 
cosmologies. We expect that if $\gamma \leq \tfrac{4}{3}$, they isotropize 
as regards the shear.
This matter requires further investigation.

%% file: G2_dynamics.tex
	\chapter{The state space for $G_2$ cosmologies}\label{chap:G2_dyn}


In this chapter, we use the 1+1+2 orthonormal frame formalism as developed 
in Section~\ref{sec:1+1+2} to formulate
the evolution equations for $G_2$ cosmologies.
We also introduce
the notion of \emph{asymptotic silence}
and
 the \emph{silent boundary},
which provides the link between $G_2$ dynamics and SH dynamics.

\section{The evolution equations}\label{sec:G2}

The starting point for deriving the evolution
 equations for $G_2$ cosmologies is
the 1+1+2 form of the orthonormal frame equations.
As discussed
in Section~\ref{sec:invariant} we normalize the variables for $G_2$ 
cosmologies using the area expansion rate $\beta$, as motivated 
by van Elst \etal 2002~\cite{art:vEUW2002}.
$\beta$-normalization is advantageous only when used in conjunction with 
the so-called \emph{separable area gauge},
introduced by van Elst \etal 2002~\cite{art:vEUW2002} in the context of OT
$G_2$ cosmologies.

The separable area gauge condition is given by
\be
\label{separable_area}
        \Udot = r.
\ee
By using
 the $(C_{\udot})_1$ constraint  (\ref{c_udot1})
 we find that $\mathcal{N}$, introduced in (\ref{N_beta}),
 is independent of $x$, which implies that we 
may
 reparametrize $t$ to set 
\be
\label{separable_area_2}
	\mathcal{N}=1.
\ee

A special case of the separable area gauge is the
\emph{timelike area gauge}, for which
\be
\label{timelike_area_G_2}
        \Udot = r, \quad \mathcal{N}=1,\quad A=0.
\ee
This gauge is equivalent to the one that
 has been widely used in the study of vacuum OT $G_2$ cosmologies using 
the metric approach
(see, for example, Rendall \& Weaver 2001~\cite{art:RendallWeaver2001}).
Although the timelike area gauge is simpler than the separable area gauge 
in a number of respects (see Sections~\ref{sec:IC},
\ref{sec:hyperbolic} and \ref{sec:acc}),
we shall resist setting $A=0$, since the separable area gauge enables 
one to include all $G_2$-compatible SH cosmologies as special cases, as 
will be explained in Section~\ref{sec:G2_invariant_subset}.

As in Section~\ref{sec:SH_var}, 
for convenience we introduce the following notation:
\be
        N_C{}^C = 2 N_+,\quad
        \tilde{N}_{AB} = \sqrt{3}  
                \left(
                \begin{matrix}
                        N_- & N_\times
                \\
                        N_\times & -N_-
                \end{matrix}
                \right),
\ee
and
\be
        \Sigma_{12} = \sqrt{3}\Sigma_3,\quad
        \Sigma_{13} = \sqrt{3}\Sigma_2.
\ee
The 1+1+1+1 $\beta$-normalized evolution equations and constraints then 
follow from the system (\ref{AB_H})--(\ref{evo_qA}),
 and are given below.
In the process, 
we replace $H$ by $\beta$ using the relation
\be
	H = (1-\Sigma_+)\beta
\ee
(see (\ref{H_beta_off})),
and normalize the variables using 
(\ref{1+1+2_1})--(\ref{1+1+2_2}).
The variables
$q$ and $r$ are introduced and are given by 
(\ref{beta_q}) and (\ref{beta_r}) respectively.
We remind the reader that
\be
	\parb_1 = \EEE \ptl_x\ .
\ee

\noindent
\begin{minipage}{\textwidth}
{\small
\subsection*{$G_2$ system ($\beta$-normalized)}

\noindent
{\it Evolution equations for the gravitational field:}\vspace{-2mm}
\begin{align}
\label{be_sys}
	\ptl_t \EEE &= (q+3\Sp) \EEE
\\
	\ptl_t A &= (q+3\Sp)A 
\\
	\ptl_t \Sp &= -(q+3\Sp)(1-\Sp) + 2(\Sp+\Sm^2+\Sc^2+\St^2+\Stt^2)
\notag\\
	&\quad
	+ \tfrac{1}{2} G_+^{-1} [ (3\gamma-2)+(2-\gamma)v^2 ] \Omega 
	-\Oml 
	- \tfrac{1}{3}(\parb_1-2A)r
\\
	\ptl_t \Np &= (q+3\Sp)\Np + 6\Sm\Nm+6\Sc\Nc 
	- \parb_1 R 
\\
\label{G2_Sm}
	\ptl_t \Sm + \parb_1 \Nc 
	&= (q+3\Sp-2)\Sm + 2A\Nc -2R\Sc -2\Np\Nm
\notag\\
        &\quad
	+ \sqrt{3}(\Stt^2-\St^2)
        + \frac{\sqrt{3}}{2} \frac{\gamma\Omega}{G_+} (v_2^2-v_3^2)
\\
        \ptl_t \Nc + \parb_1 \Sm 
	&= (q+3\Sp)\Nc + 2\Np\Sc + 2R\Nm
\\
        \ptl_t \Sc - \parb_1 \Nm 
	&= (q+3\Sp-2)\Sc -2A\Nm + 2R\Sm -2\Np\Nc 
\notag\\
        &\quad
	+ 2\sqrt{3}\Stt\St
        +\sqrt{3} \frac{\gamma\Omega}{G_+} v_2 v_3
\label{G2_Sc}
\\
\label{G2_Nm}
        \ptl_t \Nm - \parb_1 \Sc 
	&= (q+3\Sp)\Nm + 2\Np\Sm - 2R\Nc
\\
\label{G2_Stt}
        \ptl_t \Stt 
	&= (q-2-\sqrt{3}\Sm)\Stt -(R+\sqrt{3}\Sc)\St
                + \sqrt{3} \frac{\gamma\Omega}{G_+} v_1 v_2
\\
\label{G2_St}
        \ptl_t \St 
	&= (q-2+\sqrt{3}\Sm)\St +(R-\sqrt{3}\Sc)\Stt
                + \sqrt{3} \frac{\gamma\Omega}{G_+} v_1 v_3\ ,
\end{align}
where\vspace{-2mm}
\begin{align}
\label{q}
        q &= \tfrac{1}{2} \Big[
                1 + 3(-\St^2-\Stt^2 + \Sm^2+\Sc^2 + \Nm^2+\Nc^2)
                -A^2 +2Ar
\notag\\
        &\quad
                + 3 G_+^{-1}[(\gamma-1)(1-v^2)+\gamma v_1^2]\Omega -3\Oml
                \Big]
\\
\label{r}
        r &= -3A\Sp -3(\Nc\Sm -\Nm\Sc)
                -\frac{3}{2} \frac{\gamma \Omega}{G_+} v_1
\\
        G_\pm &= 1 \pm (\gamma-1)v^2,\quad v^2 = v_1^2 + v_2^2 +v_3^2\ .
\end{align}
{\it Constraint equations:}\vspace{-2mm}
\begin{align}
\label{C_G}
        0 &= (\mathcal{C}_{\rm G}) = 
	1 
	- 2\Sp - \Sm^2 - \Sc^2 - \St^2 - \Stt^2
        - \Omega - \Oml \hspace{5cm}
\notag\\
        &\qquad\qquad\qquad
	- \Nm^2 - \Nc^2 + \tfrac{2}{3} (\parb_1 -r)A - A^2
\\
        0 &=\! (\mathcal{C}_{\rm C})_2 =
	(\parb_1 -r)\Stt
	- (3A + \sqrt{3}\Nc) \Stt
        - (\Np-\sqrt{3}\Nm) \St
        + \!\sqrt{3} \frac{\gamma\Omega}{G_+} v_2
\label{CC_2}
\\
        0 &=\! (\mathcal{C}_{\rm C})_3 =
        (\parb_1 -r)\St
        - (3A - \sqrt{3}\Nc) \St
        + (\Np+\sqrt{3}\Nm) \Stt
        + \!\sqrt{3} \frac{\gamma\Omega}{G_+} v_3
\label{CC_3}
\\
\label{C_beta}
	0 &= (\mathcal{C}_\beta) =
	(\parb_1 - 2r)\Oml\ .
\end{align}

\begin{figure}[H]
\begin{center}
\setlength{\unitlength}{1mm}
\begin{picture}(120,0)(0,0)
\put(60,90){\oval(150,240)[t]}
\put(60,90){\oval(150,170)[b]}
\end{picture}
\end{center}
\end{figure}
}
\end{minipage}

\noindent
\begin{minipage}{\textwidth} 
        
\noindent
{\it Evolution equations for the matter:}
\begin{align}
	\ptl_t \Omega 
	&= -\frac{\gamma v_1}{G_+} \parb_1 \Omega 
	+ \frac{\gamma \Omega}{G_+} \Big[
	2\frac{G_+}{\gamma}(q+1)
	-3(1-\Sp)(1+\tfrac{1}{3}v^2) 
\notag\\
        &\quad
	- \Sigma_{\alpha\beta} v^\alpha v^\beta
	-\parb_1 v_1 + v_1 \parb_1 \ln G_+
	+2Av_1 \Big]
\label{G2_Omega}
\\
	\ptl_t v_1 &= -v_1 \parb_1 v_1 + \parb_1 \ln G_+
		-\frac{\gamma-1}{\gamma}(1-v^2)(\parb_1\ln\Omega-2r) -r
\notag\\
        &\quad
	+(M+2\Sp)v_1 -2\sqrt{3}(\Stt v_2 + \St v_3)
\notag\\
        &\quad
        + \sqrt{3} \Nc (v_3^2-v_2^2) + 2\sqrt{3}\Nm v_2 v_3 - A v^2
\label{G2_v_1}
\\
	\ptl_t v_2 &= - v_1 \parb_1 v_2
	+ (M-\Sp-\sqrt{3}\Sm + \sqrt{3}\Nc v_1) v_2
\notag\\
        &\quad
        - [ R+\sqrt{3}\Sc - (\Np-\sqrt{3}\Nm)v_1]v_3
\\
	\ptl_t v_3 &= - v_1 \parb_1 v_3
	+ (M-\Sp+\sqrt{3}\Sm - \sqrt{3}\Nc v_1) v_3
\notag\\
        &\quad
        + [ R-\sqrt{3}\Sc - (\Np+\sqrt{3}\Nm)v_1]v_2
\\
        \ptl_t \Oml &= 2(q+1) \Oml\ ,
\label{ee_sys}
\end{align}
where
\begin{align}
	M &= G_-^{-1} \Big[
		(\gamma-1)(1-v^2)\parb_1 v_1 
		- (2-\gamma) v_1 \parb_1 \ln G_+
\notag\\
        &\quad\qquad
		+ \frac{\gamma-1}{\gamma}(2-\gamma)(1-v^2)v_1
			(\parb_1 \ln \Omega -2r) + G_- r v_1
\notag\\
        &\quad\qquad
		+ (3\gamma-4)(1-v^2)(1-\Sp)
                + (2-\gamma)\Sigma_{\alpha\beta}v^\alpha v^\beta
\notag\\
        &\quad\qquad
                + [G_+ - 2(\gamma-1)]Av_1 
	\Big]\ ,
\\
        \Sigma_{\alpha\beta}v^\alpha v^\beta &=
        \Sp (v_2^2+v_3^2-2v_1^2)
        + \sqrt{3}\Sm(v_2^2-v_3^2) + 2\sqrt{3} \Sc v_2 v_3
\notag\\
        &\quad
        + 2 \sqrt{3} (\Stt v_2 + \St v_3) v_1.
\end{align}

\begin{figure}[H]
\begin{center}
\setlength{\unitlength}{1mm}
\begin{picture}(120,0)(0,0)
\put(60,90){\oval(150,138)[t]}
\put(60,90){\oval(150,170)[b]}
\end{picture}
\end{center}
\end{figure}
\end{minipage}

Note that we have used
 the $(C_{\rm C})_1$ constraint (\ref{AB_c_c}) to 
solve for $r$, giving 
(\ref{r}).
The $(\mathcal{C}_\beta)$
constraint (\ref{C_beta}) is an immediate 
consequence of the definitions (\ref{q_r_beta}) and $\Oml= 
\Lambda/(3\beta^2)$.
Equation (\ref{be_sys}) is the $\beta$-normalized 
form of equation (\ref{evo_e11}).
The evolution equations for $v_\alpha$ are derived in the same manner as 
explained in Section~\ref{sec:SH_var} for SH cosmologies.
They were first derived by Weaver 2002~\cite{cal:Weaver2002} 
for generic $G_2$ cosmologies.

Three useful auxiliary equations are
\begin{align}
        q + 3\Sp &= 2(1-A^2) + \di A -3(\St^2+\Stt^2)
        -\tfrac{3}{2}(2-\gamma)\frac{(1-v^2)}{G_+}\Omega -3\Omega_\Lambda
\label{qts}   
\\
	\ptl_t(v^2) &= -v_1\parb_1 (v^2) +\frac{2}{G_-}(1-v^2)\Big[
	v_1 \parb_1 \ln G_+ +(\gamma-1)v^2 \parb_1 v_1
\notag\\
        &\qquad
	- \frac{\gamma-1}{\gamma}(1-v^2)v_1(\parb_1\ln\Omega-2r)
	-G_- r v_1 
\notag\\
        &\qquad
	+(3\gamma-4)v^2(1-\Sp)
	-\Sigma_{\alpha\beta}v^\alpha v^\beta
	-2(\gamma-1)v^2 A v_1
	\Big]
\label{G2_vsq}
\\
        \ptl_t r &= (q+3\Sp) r +\parb_1 q\ .
\label{dtr}  
\end{align}
Equation (\ref{qts}) is derived from (\ref{q}) and (\ref{C_G}),
while (\ref{dtr}) is derived from the 
commutator $[\dz,\ \di ]$ in (\ref{dzdi}) acting on
$\beta$. Equation (\ref{qts}) becomes algebraic and convenient to use in 
the timelike area gauge (\ref{timelike_area_G_2}). 
Equation (\ref{dtr}) is not used for numerical simulations, but is useful 
for analysis.

It follows from (\ref{Weyl_E11})--(\ref{Weyl_HAB}) and (\ref{Weyl_beta})
that the Weyl curvature variables are given by
\begin{align}
\label{Ep}
	\Ep &= \tfrac{1}{3}\Sp + \tfrac{1}{6}(\St^2+\Stt^2)
		- \tfrac{1}{3}(\Sm^2+\Sc^2)
\notag\\
	&\quad
		- \tfrac{1}{9}(\parb_1-r)A + \tfrac{2}{3}(\Nc^2+\Nm^2)
		+ \frac{1}{4} \frac{\gamma \Omega}{G_+}
					(v_1^2-\tfrac{1}{3}v^2)
\\
	\Em &= \tfrac{1}{3}(1-3\Sp)\Sm 
		- \tfrac{1}{2\sqrt{3}}(\Stt^2-\St^2)
\notag\\
        &\quad
		+\tfrac{1}{3}(\parb_1-r-2A)\Nc + \tfrac{2}{3}\Np\Nm
		+ \frac{1}{2\sqrt{3}}\frac{\gamma \Omega}{G_+}(v_2^2-v_3^2)
\\
	\Ec &= \tfrac{1}{3}(1-3\Sp)\Sc
                - \tfrac{1}{\sqrt{3}}\Stt\St
\notag\\
        &\quad
                -\tfrac{1}{3}(\parb_1-r-2A)\Nm + \tfrac{2}{3}\Np\Nc
                + \frac{1}{\sqrt{3}}\frac{\gamma \Omega}{G_+} v_2 v_3
\\
	\Ett &= \tfrac{1}{3}(1-\sqrt{3}\Sm)\Stt - \tfrac{1}{\sqrt{3}}\Sc\St 
	- \frac{1}{2} \frac{\gamma \Omega}{G_+} v_1 v_2
\\        
	\Et &= \tfrac{1}{3}(1+\sqrt{3}\Sm)\St - \tfrac{1}{\sqrt{3}}\Sc\Stt
        - \frac{1}{2} \frac{\gamma \Omega}{G_+} v_1 v_3
\end{align}
\begin{align}
	\Hp &= -\Nm\Sm-\Nc\Sc
\\
	\Hm &= -\tfrac{1}{3}(\parb_1-r-A)\Sc-\tfrac{2}{3}\Np\Sm-\Sp\Nm
\\
	\Hc &= \tfrac{1}{3}(\parb_1-r-A)\Sm -\tfrac{2}{3}\Np\Sc-\Sp\Nc
\\
	\Htt &= -\tfrac{1}{3}(A+\sqrt{3}\Nc)\St 
		- \tfrac{1}{\sqrt{3}}\Nm\Stt 
		+ \frac{1}{2\sqrt{3}}\frac{\gamma \Omega}{G_+}v_3
\\
	\Ht &= \tfrac{1}{3}(A-\sqrt{3}\Nc)\Stt
                + \tfrac{1}{\sqrt{3}}\Nm\St
                - \frac{1}{2\sqrt{3}}\frac{\gamma \Omega}{G_+}v_2\ .
\label{Ht}
\end{align}
Note that the $(\mathcal{C}_{\rm C})_2$ and $(\mathcal{C}_{\rm C})_3$
constraints have been used to simplify $\Ht$ and $\Htt$ respectively.

At this stage the $\beta$-normalized state vector is
\begin{gather}
        (\EEE,\Sigma_+,\Sigma_-,\Sigma_\times,\Sigma_2,\Sigma_3,
        N_+,N_-,N_\times,A,R,
        \Omega,v_1,v_2,v_3,\Omega_\Lambda).
\end{gather}
The variable $R$ is to be determined by choosing a spatial gauge,
and the final form of the state vector depends on this choice, as we now 
describe.

\subsubsection*{The spatial gauge}

We use a $(t,x)$-dependent rotation
(see Appendix~\ref{app_rotation})
 to rotate the spatial frame vectors $\{\me_A\}$ so that $\Stt=0$.
The evolution equation
(\ref{G2_Stt}) and the constraint (\ref{CC_2}) then give
\be
\label{G2_spatial_gauge}
        \Sigma_{3}=0,\quad R = -\sqrt{3} \Sc + (\Np -\sqrt{3}\Nm)v_1\ .
\ee
This is the \emph{shear spatial gauge} (\ref{SH_spatial_gauge}) introduced 
for $G_2$-compatible SH cosmologies.
Alternatively, we can use the Killing spatial gauge
(\ref{KSG}):
\be
        \Np = \sqrt{3}\Nm,\quad
         R = -\sqrt{3} \Sc\ .
\ee
Both gauges specialize to (\ref{e2_align_S3}) for
generic $G_2$ cosmologies with fewer than three tilt degrees of freedom.
The motivation for using the shear spatial gauge rather than the Killing
spatial gauge is the compatibility of the former with all $G_2$-compatible
SH cosmologies.

Using the shear spatial gauge (\ref{G2_spatial_gauge}),
the final $\beta$-normalized state vector for $G_2$ cosmologies is
\be
\label{state_vector}
	\X =(\EEE,\Sigma_+,\Sigma_-,\Sigma_\times,\Sigma_2,
        N_+,N_-,N_\times,A,
        \Omega,v_1,v_2,v_3,\Omega_\Lambda),  
\ee
together with the parameter $\gamma$.

Separating $\EEE$ from the rest of the variables as follows,
\begin{align}
\label{X_Y}
        \X &= \EEE \oplus \Y\ ,
\intertext{where}
        \Y &= (\Sigma_+,\Sigma_-,\Sigma_\times,\Sigma_2,
        N_+,N_-,N_\times,A,
        \Omega,v_1,v_2,v_3,\Omega_\Lambda),
\label{Y}
\end{align}
we can write the system of evolution equations and constraints
(\ref{be_sys})--(\ref{ee_sys})
in a concise symbolic form.

\noindent
{\it Evolution equations and constraints:}
\begin{gather}
\label{G2_EEE}
	\dt \EEE = B(\Y) \EEE
\\
\label{G2_1}
        \dt \Y + M(\Y) \EEE \partial_x \Y = g(\Y)
\\
        \mathcal{C}(\Y, \EEE \ptl_x \Y)=0\ .
\label{G2_2}        
\end{gather}
In addition we write the expression (\ref{r}) for
the spatial gradient $r$ of the normalization factor $\beta$
symbolically as
\be
\label{G2_3}
	r = F(\Y).
\ee

\subsubsection*{Features of $\beta$-normalization and separable area 
gauge}

The principal advantage of using $\beta$-normalized variables and the 
associated separable area gauge is that
it leads to evolution equations that are first order in the spatial 
derivatives as well as the time derivatives, i.e. a first order 
quasi-linear system of PDEs, which is desirable for numerical simulations.
In contrast, if we were to use Hubble-normalized variables and the 
associated separable volume gauge (see Uggla \etal 
2003~\cite{art:Ugglaetal2003}), then the evolution equations would contain 
second spatial derivative terms. The reason for this difference is that
(\ref{r}) gives an algebraic expression for $r$, whereas when using 
Hubble-normalized variables and the
associated separable volume gauge, the expression for $r$ contains first 
order spatial derivatives, which lead to second spatial derivatives since 
the equation for $\ptl_t \Sp$ contains the spatial derivative $\parb_1 r$.
A minor variation of the formulation (\ref{X_Y})--(\ref{G2_3}) is possible 
due to the fact that (\ref{C_G}) is linear in $\Sp$, allowing one to 
replace $\Sp$ by
\begin{multline}
        \Sp = \Big[ 1-A^2+\tfrac{2}{3}\di A
        + A \frac{\gamma v}{G_+} \Omega
        + 2A(\Nc\Sm-\Nm\Sc)
\\
         -\Sm^2-\Sc^2-\St^2-\Nc^2-\Nm^2-\Omega -\Oml
                \Big]/ [2(1-A^2)] 
\end{multline}
(substitute (\ref{r}) in (\ref{C_G}) and solve for $\Sp$).
\footnote{We restrict to models with $A^2 < 1$, i.e. models with a 
timelike gradient vector of the $G_2$ area density. See van Elst \etal 
2002~\cite{art:vEUW2002}, equation (62).}
Coupled with the fact that $\parb_1 \Sp$ does not appear in 
the evolution equations, this 
again leads to a first order quasi-linear system of PDEs.
We shall make use of this possibility when doing numerical simulations.

\section{Invariant subsets of the $\beta$-normalized state space}
	\label{sec:G2_invariant_subset}

In this section, we explain the important invariant sets and equilibrium 
points within the $G_2$ state space, and the link with the SH state space.

\subsubsection*{The $G_2$ hierarchy}

We now revisit the hierarchy 
of the action of the $G_2$ group and the tilt degrees of freedom,
as shown in Figure~\ref{fig:G2_action_tilt},
and give the defining conditions in terms of the state variables 
(\ref{Y}).

For OT $G_2$ cosmologies, (\ref{OT_grav}) and (\ref{G2_spatial_gauge}) 
imply
\be
\label{OT_G2}
	\St=0.
\ee
For $G_2$ cosmologies with one HO KVF,
(\ref{HO_grav}) and (\ref{G2_spatial_gauge})
imply
\be
\label{HO_G2}
	\Np=\Nm=0=\Sc\ .
\ee
For diagonal $G_2$ cosmologies, (\ref{OT_G2}) and (\ref{HO_G2}) imply
\be
\label{diag_G2}
	\St=0,\quad \Np=\Nm=0=\Sc\ .
\ee

The hierarchy of invariant sets is summarized in
Table~\ref{tab:G2_sets}, which lists the variables and the additional 
restrictions.
We have omitted $\Oml$ in Table~\ref{tab:G2_sets} for simplicity and ease 
of comparison with Tables~\ref{tab:SH_variables}--\ref{tab:diag} 
for $G_2$-compatible SH cosmologies.

\begin{table}
{\small
\begin{spacing}{1.1}

\caption{The $\beta$-normalized variables for $G_2$ cosmologies.}
	\label{tab:G2_sets}
\begin{center}  
\begin{tabular}{lll}
\hline
\multicolumn{2}{l}{Invariant set $\qquad$ State vector $\Y$} & Additional 
restriction
\\ 
\hline
\\
\multicolumn{2}{l}{Generic $G_2$ with 3 tilt degrees of freedom:}
        & none
\\
$\qquad\qquad$
&
	$(\Sigma_+,\Sigma_-,
        \Sc,\St,
        \Np,\Nm,N_\times,A,
        \Omega,v_1,v_2,v_3)$
\\
\multicolumn{2}{l}{Generic $G_2$ with 2 tilt degrees of freedom: $v_2=0$}
        & $\Np=\sqrt{3}\Nm$
\\
&
	$(\Sigma_+,\Sigma_-,
        \Sc,\St,\Nm,
        N_\times,A,
        \Omega,v_1,v_3)$
\\
\multicolumn{2}{l}{Generic $G_2$ with 1 tilt degree of freedom: $v_2 = v_3=0$}
        & $\Np=\sqrt{3}\Nm$
\\
&
        $(\Sigma_+,\Sigma_-,
        \Sc,\St,\Nm, 
        N_\times,A,
        \Omega,v_1)$
\\
\hline
\\
\multicolumn{2}{l}{OT $G_2$ with 1 tilt degree of freedom: $v_2 = v_3=0$}
        & $\Np=\sqrt{3}\Nm$, $\St=0$
\\
&
	$(\Sigma_+,\Sigma_-,
        \Sc,\Np,\Nm,
        \Nc,A,\Omega,v_1)$
\\
\hline
\\
\multicolumn{2}{l}{One HO KVF with 2 tilt degrees of freedom: $v_2=0$}
        & $(\Np,\Nm,\Sc)=\mathbf{0}$
\\
&
	$(\Sigma_+,\Sigma_-,
        \St,
        N_\times,A,\Omega,v_1,v_3)$
\\
\multicolumn{2}{l}{One HO KVF with 1 tilt degree of freedom: $v_2=v_3=0$}
        & $(\Np,\Nm,\Sc)=\mathbf{0}$
\\
&  
        $(\Sigma_+,\Sigma_-,
        \St,
        N_\times,A,\Omega,v_1)$
\\
\hline
\\
\multicolumn{2}{l}{Diagonal $G_2$ with 1 tilt degree of freedom: $v_2 = v_3=0$}
        & $(\Np,\Nm,\Sc,\St)=\mathbf{0}$

\\
&
	$(\Sigma_+,\Sigma_-,\Nc,A,\Omega,v_1)$
\\
\hline
\end{tabular}
\end{center}
\end{spacing} 
}
\end{table}

\subsubsection*{SH dynamics again}

To specialize to $G_2$-compatible SH models (see 
Figures~\ref{fig:gen_G2}--\ref{fig:HO_diag}), one sets
\be
\label{G2_2_SH}
        \ptl_x \Y=0,\quad r=0.
\ee
Then the separable area gauge (\ref{separable_area}) coincides with the 
$G_3$-adapted gauge of SH cosmologies.
The system (\ref{G2_1})--(\ref{G2_3}) simplifies to
\begin{gather}
\label{G2_SH_beta_1}
	\ptl_t \Y = g(\Y)
\\
	\mathcal{C}(\Y,0)=0,\quad F(\Y)=0,
\label{G2_SH_beta_2}
\end{gather}
and (\ref{G2_EEE}) decouples.
The system (\ref{G2_SH_beta_1})--(\ref{G2_SH_beta_2}), given in detail by
(\ref{be_sys})--(\ref{ee_sys}) subject to (\ref{G2_2_SH}),
 describes the 
dynamics of $G_2$-compatible SH cosmologies, but in $\beta$-normalized 
variables,
and is fully equivalent to the Hubble-normalized system, given by 
(\ref{SH_Sp})--(\ref{SH_Oml}).

\subsubsection*{The silent boundary}

The evolution equations (\ref{be_sys}) and (\ref{dtr}),
\be
	\ptl_t \EEE = (q+3\Sp) \EEE,\quad
	\ptl_t r = (q+3\Sp) r + \EEE \ptl_x q
\ee
imply that the conditions
\be
\label{silent_definition}   
        \EEE=0,\quad r=0
\ee
define an invariant subset of the evolution equations, which we shall call 
the \emph{silent boundary}.
\footnote{van Elst \etal 2002~\cite{art:vEUW2002} define the silent
boundary to be $\EEE=0$, with no restrictions on $r$.
Using this definition,
Andersson \etal 2004~\cite{art:Anderssonetal2004} relate
the vacuum $G_2$ dynamics to the dynamics of \emph{spatially
self-similar}
cosmologies (see sets (2) and (3) on page S39 of
\cite{art:Anderssonetal2004}).
The difference here does not affect the past asymptotic dynamics, as
spatially
self-similar cosmologies and SH cosmologies have the same past asymptotic
dynamics.
}

It follows by setting $\EEE=0$ and $r=0$ in (\ref{G2_1})--(\ref{G2_3}), 
that 
the SH evolution equations
(\ref{G2_SH_beta_1})--(\ref{G2_SH_beta_2}) describe the dynamics in 
the silent boundary, but now with
\be
	\ptl_x \Y \quad \text{unrestricted (subject to 
				(\ref{G2_SH_beta_2})).}
\ee
In Section~\ref{sec:past_as} we shall present evidence that the dynamics 
in 
the silent boundary approximates the dynamics of $G_2$ cosmologies in the 
asymptotic regimes
($t \rightarrow -\infty$ or, $\Lambda>0$ and $t \rightarrow \infty$)
 in the sense that $\EEE$ and $r$ tend to zero 
asymptotically.
An important aspect of this approximation is that
equilibrium points for the SH evolution equations are also
equilibrium points for the $G_2$  evolution equations on the silent 
boundary.

\begin{figure}[h]
\begin{center}
    \epsfig{file=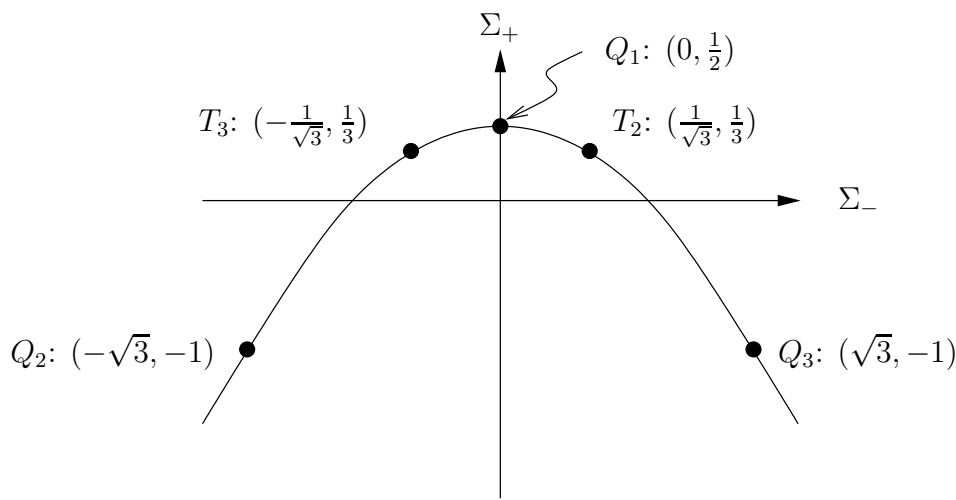,width=8cm}
\setlength{\unitlength}{1mm}
\begin{picture}(120,0)(0,0)
\put(105,45){\makebox(0,0)[l]{$\Sm$}}
\put(60,70){\makebox(0,0)[t]{$\Sp$}}

\put(97,25){\makebox(0,0)[l]{$Q_3$: $(\sqrt{3},-1)$}} 
\put(-5,25){\makebox(0,0)[l]{$Q_2$: $(-\sqrt{3},-1)$}}

\put(75,55){\makebox(0,0)[l]{$T_2$: $(\tfrac{1}{\sqrt{3}},\tfrac{1}{3})$}}
\put(20,55){\makebox(0,0)[l]{$T_3$: 
$(-\tfrac{1}{\sqrt{3}},\tfrac{1}{3})$}}

\put(74,65){\makebox(0,0)[l]{$Q_1$: $(0,\tfrac{1}{2})$}}

\end{picture}
\end{center}
\caption{The Kasner parabola in the $\beta$-normalized state space.}
	\label{fig:Kasner_parabola}
\end{figure}

\subsubsection*{The Kasner equilibrium points}

On the silent boundary in the $\beta$-normalized $G_2$ state space,
the Kasner solutions are 
described by a parabola of equilibrium points (see 
Figure~\ref{fig:Kasner_parabola}), 
given by
\begin{gather}
	\Omega=0=\Oml\ ,\quad
	(\Np,\Nm,\Nc,A,\Sc,\St)=0\ ,
\end{gather}
and
\be
\label{parabola}
	0 = 1-2\Sp -\Sm^2\ .
\ee
The latter equation follows from (\ref{C_G}).
From (\ref{qts}), the deceleration parameter $q$ is given by
\be
	q = -3\Sp\ .
\ee
The parabola (\ref{parabola}) is equivalent to the circle
\be
        (\Sph)^2+(\Smh)^2=1
\ee
in the Hubble-normalized state space, as follows from 
(\ref{beta_to_Hubble}) below.
The Taub Kasner point $T_1$ (where $\beta=0$) is not represented on the 
parabola, because 
the $\beta$-normalized variables are undefined there.
The corresponding transitions and Mixmaster dynamics can be described in 
the $\beta$-normalized state space, but their representation in 
the Hubble-normalized state space is simpler geometrically.

\subsubsection*{The de Sitter equilibrium points}

On the silent boundary in the $\beta$-normalized $G_2$ state space, the de 
Sitter equilibrium points are given by
\be
\label{de_Sitter_1}
	\Oml=1,\quad
	\Om=0,\quad
        \Sigma^2=0,\quad
        \Np=\Nm=0=\Nc=A
\ee
\be
\label{de_Sitter_2}
        \begin{cases}
        v_\alpha=0 \quad \text{or}\quad v^2=1 &
                        \text{if $\gamma \neq \tfrac{4}{3}$}
        \\
        v_\alpha=v_\alpha(x),\quad 0 \leq v^2 \leq 1 &
                        \text{if $\gamma = \tfrac{4}{3}$,}
        \end{cases}
\ee
which are the same as (\ref{Oml_1})--(\ref{Oml_2}), (\ref{Oml_4}) and 
(\ref{Oml_3}), except
$v_\alpha$ are functions of $x$ rather than constants.
From (\ref{q}), the deceleration parameter is given by
\be
	q = -1.
\ee
The $\beta$-normalized and the Hubble-normalized variables coincide here 
because $\Sp=0$ (see equation (\ref{H_beta}) below).

\subsubsection*{Visualization of SH and $G_2$ cosmologies}

In the SH state space, an SH cosmology at a fixed time is represented by a 
point, and its evolution in time by an orbit.
A $G_2$ cosmology at a fixed time can be represented by an arc of 
points in the SH state space, parametrized by the spatial variable $x$.
The evolution of a $G_2$ cosmology along one timeline is represented by an 
orbit in the SH state space, and
the full evolution of a $G_2$ cosmology 
is represented by a 
moving arc of points.
We shall use this representation in Chapters~\ref{chap:explicit}
and \ref{chap:sim} to illustrate $G_2$ solutions in the SH state space.

As discussed in Section~\ref{sec:G2},
$\beta$-normalized variables
have the major advantage of leading to evolution equations that are first
order in the spatial derivatives,
 and will thus be
used for analysis and numerical simulations.
On the other hand,
Hubble-normalized variables are
directly related to the standard observational parameters, and give a
simpler representation of the Kasner solutions.
We shall thus use them
 for displaying the state space and most of the plotted figures in 
Chapter~\ref{chap:sim}.

To avoid
 possible confusion between the notation of the Hubble-normalized
variables in Section~\ref{sec:SH_var} and of the $\beta$-normalized
variables here, we shall use a superscript $H$ for Hubble-normalized
variables for the rest of Part~\ref{part:G2} and Part~\ref{part:num}
 of this thesis.
Note that from (\ref{H_beta_off}),
\be
\label{H_beta}
	\frac{H}{\beta} = 1-\Sp = \frac{1}{1+\Sph}\ .
\ee
Using this relation,
$\beta$-normalized variables can be converted into Hubble-normalized 
variables as follows:
\be
	\Smh = \frac{\Sm}{1-\Sp},\quad
	\Omega^H = \frac{\Omega}{(1-\Sp)^2},\ \text{\etc}
\label{beta_to_Hubble}
\ee
In addition,
the gradients of $H$ and $\beta$, defined by
(\ref{def_q})--(\ref{def_r}) and (\ref{q_r_beta}),
 are related according to
\begin{align}
	q^H + 1 &= \frac{q+1}{1-\Sp} + \frac{\ptl_t \Sp}{(1-\Sp)^2}
\\
	r^H &= \frac{r}{1-\Sp} + \frac{\EEE \ptl_x \Sp}{(1-\Sp)^2}\ .
\end{align}
It is important to note, however, that
converting $\beta$-normalized variables to Hubble-normalized variables
 does not 
change the temporal or spatial gauge (we are still using the separable 
area gauge and the shear spatial gauge).

When the separable area gauge
(\ref{separable_area})--(\ref{separable_area_2})
 and the $G_3$-adapted gauge coincide,
\footnote{This happens for SH cosmologies, and in the silent boundary of 
$G_2$ cosmologies (see below).}
the time variable $t$ for the separable 
area gauge is related to the time variable $t^H$ for the $G_3$-adapted 
gauge according to
\be
\label{t_H_t}
	t^H = (1-\Sp) t\ .
\ee

\section{Asymptotic silence}\label{sec:past_as}

There is considerable analytical and numerical evidence that the silent 
boundary plays a significant role in determining the asymptotic dynamics 
of $G_2$ cosmologies in the following sense:
$\EEE$ and $r$ tend to zero along almost all timelines as $t \rightarrow 
-\infty$, and if there is a positive cosmological constant, as $t 
\rightarrow \infty$.
The significance of this behaviour is that the asymptotic dynamics of 
$G_2$ 
cosmologies will then be approximated locally, i.e. along individual 
timelines, 
by the dynamics in the silent boundary, i.e by SH dynamics.

\subsubsection*{Past asymptotic silence}

We now define past asymptotic silence as
\be
\label{beta_silent}
	\lim_{t \rightarrow -\infty} \EEE =0,\quad
	\lim_{t \rightarrow -\infty} r =0.\
\footnote{We note that, away from the Taub Kasner point $T_1$ (where 
$\beta=0$), Hubble-normalized and $\beta$-normalized variables only differ 
by a finite factor. Equation (\ref{beta_silent}) is thus equivalent to
\[
       \lim_{t \rightarrow -\infty} (\EEE)^H =0,\quad
        \lim_{t \rightarrow -\infty} r^H =0,
\]
since the limits are zero.}
\ee
On the basis of numerical simulations and physical considerations about 
the existence of particle horizons (see below),
 we conjecture that generic $G_2$ cosmologies are asymptotically silent 
into the past:

\begin{conj}[Past asymptotic silence]
A typical $G_2$ cosmology is asymptotically silent into the past
along almost all timelines.
\end{conj}

Some analytical evidence for asymptotic silence is provided by the 
evolution equations.
In the special case of vacuum OT $G_2$ cosmologies with $A=0$, the 
evolution equation (\ref{be_sys}) for $\EEE$ simplifies to
\be
        \ptl_t \EEE = 2 \EEE\ ,
\ee
which gives
\be
        \EEE = (\EEE)_0 e^{2t}\ \rightarrow 0 \quad
        \text{as $t \rightarrow -\infty$,}
\ee
where $(\EEE)_0$ is a positive function of $x$.
We note that for a timeline whose orbit is past asymptotic to the 
Taub Kasner point $T_1$, $t$ tends to a finite limit $t_s$ (see
Section~\ref{sec:lrs_dust}).
So $\EEE$ tends to zero along orbits that
are not past asymptotic to $T_1$.
The evolution equation (\ref{dtr}) for $r$ also simplifies to
\be
        \ptl_t r = 2 r + \EEE \ptl_x q\ .
\ee
On the invariant set $\EEE=0$, this equation reduces to
\be
	\ptl_t r = 2r,
\ee
and hence along orbits on the invariant set $\EEE=0$, $r \rightarrow 0$ as 
$t \rightarrow -\infty$.

\subsubsection*{Future asymptotic silence}

Similarly, we define future asymptotic silence as
\be
        \lim_{t \rightarrow \infty} \EEE =0,\quad
        \lim_{t \rightarrow \infty} r =0.
\ee
On the basis of numerical simulations,
we expect that 
typical
$G_2$ cosmologies with $\Lambda>0$ are future asymptotic to 
the de Sitter solution.
Evaluating (\ref{be_sys}) and (\ref{dtr}) on the de Sitter invariant set
(\ref{de_Sitter_1})--(\ref{de_Sitter_2}), 
we obtain
\be
	\ptl_t \EEE = - \EEE\ ,\quad
	\ptl_t r = - r\ ,
\ee
which imply that
\be
	\lim_{t \rightarrow \infty} \EEE = 0,\quad
	\lim_{t \rightarrow \infty} r =0.
\ee
Thus we expect that $G_2$ cosmologies that are asymptotic to the de Sitter
invariant set are asymptotically silent into the future.

\subsubsection*{Physical interpretation of asymptotic silence}

The coordinate speed of light in the $x$-direction is given by
\be
\label{slope}
	\frac{dx}{dt} = 
	\pm \mathcal{N} \EEE\ ,\
\footnote{The easiest way to see this is from the metric
\[
	ds^2 = - N^2 dt^2 +(e_1{}^1)^{-2} dx^2 + \cdots\ .
\]
For null geodesics in the $x$-direction, set $ds=0=dy=dz$, and we obtain
\[
	\frac{dx}{dt} = \pm N e_1{}^1 = \pm \mathcal{N} \EEE\ .
\]
}
\ee
and it depends on the choice of temporal gauge and the 
parametrization of the $t$- and $x$-coordinates.
In using the separable area gauge
(\ref{separable_area})--(\ref{separable_area_2}), 
we have set $\mathcal{N}=1$.
We also note that the slope of the light cone in the $x$-direction is the 
inverse of the coordinate speed of light:
\be
	\frac{dt}{dx} =
        \pm \frac{1}{\mathcal{N} \EEE}\ .
\ee
The physical speed of light has been set to unity by using geometrized
units.

If there exists a particle horizon in the $x$-direction, 
then its coordinate distance (from the observer at $x=0$) at $t_0$ is 
given by
\be
\label{particle_horizon}
	x_H(t_0) = \int_{-\infty}^{t_0} \EEE \mathcal{N} dt.
\footnote{More precisely, $x_H(t_0)$ is given by
$\int_{-\infty}^{t_0} \EEE(t,x_{\rm light}(t))\,
        \mathcal{N}(t,x_{\rm light}(t))\, dt$,
where $x_{\rm light}(t)$ is the path of the light that reaches the
observer at $t_0$.
See Lim \etal 2004~\cite{art:Limetal2004} for the equations of the null
geodesics.}
\ee
The physical distance of the horizon at $t_0$ from the observer is given 
by
\be
	d_H(t_0) = \int_0^{x_H(t_0)} \frac{1}{\beta(t_0, x) \EEE(t_0,x) } 
	\,dx\ ,
\label{d_H}
\ee
where $x_H(t_0)$ is given by (\ref{particle_horizon}).

With $\mathcal{N}=1$, if a particle horizon exists, then $x_H(t_0)$ is 
finite and (assuming a bounded derivative) $\EEE$ must 
tend to zero as $t \rightarrow -\infty$. If we also have $r$ tending to 
zero, then we have past asymptotic silence.
Thus, 
\emph{past asymptotic silence is interpreted as the existence of a 
particle horizon.}

Similarly, if there exists an event horizon in the $x$-direction, then its
coordinate and physical distances are given by
\begin{align}
	x_{EH}(t_0) &= \int^{\infty}_{t_0} \EEE \mathcal{N} dt
\\
	d_{EH}(t_0) &= \int_0^{x_{EH}(t_0)} \frac{1}{\beta(t_0,x) 
	\EEE(t_0,x)} \,dx\ .
\end{align}
With $\mathcal{N}=1$, if an event horizon exists, then $x_{EH}(t_0)$ is
finite and (assuming a bounded derivative) $\EEE$ must
tend to zero as $t \rightarrow \infty$. If we also have $r$ tending to
zero, then we have future asymptotic silence.
Thus,
\emph{future asymptotic silence is interpreted as the existence of an 
event horizon.}

\subsubsection*{The significance of the spatial average of $1/\EEE$}

At a fixed time $t$,
consider a spatial inhomogeneity with a coordinate wavelength of 
$\Delta x=2\pi$.
Its physical wavelength is given by
\be
	L(t) = \int_0^{2\pi} \frac{1}{\beta(t, x) \EEE(t,x)}\, dx\ .
\label{L_t}
\ee
Suppose that a particle horizon exists.
We can make a rough approximation of the ratio of
this physical wavelength to the distance to the particle horizon.
By assuming that $\beta\EEE$ does not vary dramatically with $x$,
it follows from (\ref{d_H}) and (\ref{L_t}) that
\be
	\frac{L}{d_H} \approx \frac{2\pi}{x_H}\ .
\label{L_d_H}
\ee
Under a change of $x$-coordinate
\be
	\hat{x} = f(x)\ ,
\label{hat_x}
\ee
$\EEE$ transforms as follows:
\be
	\hat{E}_1{}^1 = f'(x) \EEE \ .
\ee
Note that the spatial average of $\tfrac{1}{\EEE}$ 
(on $x \in [0 ,2\pi]$ for example), given by
\be
	\left\langle \frac{1}{\EEE} \right\rangle 
	= \frac{\int_0^{2\pi} \frac{1}{\EEE} dx}{\int_0^{2\pi} \ dx}\ ,
\ee
is preserved under (\ref{hat_x}) provided that
\be
\label{f_condition} 
        f(0)=0,\quad
        f(2\pi)=2\pi,\quad
        f'(x) >0 \quad \text{on $x\in[0,2\pi]$.}
\ee
This average has physical significance, as we now explain.

We choose $f(x)$ such that
\be
	\frac{1}{\hat{E}_1{}^1} = const. 
	= \left\langle \frac{1}{\EEE} \right\rangle \ .
\ee
Assuming that $\EEE$ has an exponential growth rate
from $t=-\infty$ up to time $t$ (e.g. in the past asymptotic regime),
(\ref{particle_horizon}) reduces to
$x_H \approx \EEE$, and (\ref{L_d_H}) reduces to
\be
        \frac{L}{d_H} \approx 2\pi\, \left\langle \frac{1}{\EEE} 
		\right\rangle\ .
\ee
The spatial average of $\tfrac{1}{\EEE}$
is thus a measure of
the ratio of the scale of
spatial inhomogeneities and the distance to the particle 
horizon.
The bigger $\EEE$ is, the shorter the 
wavelength is relative to the horizon distance.

This understanding will be helpful in
interpreting the numerical simulations in Chapter~\ref{chap:sim}, in which
we prescribe periodic boundary conditions
by identifying $x=2\pi$ with $x=0$.
\footnote{In doing so, we do not regard the universe as being
spatially finite, but 
rather as an infinite universe with repetitive patterns. In this way, we 
consider the particle horizon still exists even when $x_H>2\pi$.}
In this situation
the maximum coordinate wavelength is $\Delta x = 2\pi$.

%% file: explicit_sol.tex
        \chapter{Explicit $G_2$ cosmologies}\label{chap:explicit}

In this chapter we give two explicit $G_2$ solutions
that can
develop large spatial gradients on approach to the initial singularity.
In the first solution, the asymptotic form of the (Hubble-normalized) 
state vector can have jump discontinuities or ``steps", while in the 
second 
solution the asymptotic form can have ``spike" discontinuities.
The explicit solutions allow us to 
analyze the development of large spatial gradients in detail.

\section{An explicit solution with ``steps"}\label{sec:lrs_dust}

In this section, we discuss an explicit LRS $G_2$ dust solution, which
can develop step-like structures as it approaches the initial singularity.

\subsubsection*{The framework}

We consider the class of LRS $G_2$ cosmologies which are defined by
the condition (\ref{LRS_grav}):
\be
	\St=0=\Stt=\Sm=\Sc = \Nm=\Nc,\quad v_2=0=v_3,
\label{LRS_1}
\ee
and we drop the index on $v_1$.
We use the timelike area gauge (\ref{timelike_area_G_2}):
\be
        \Udot = r, \quad \mathcal{N}=1,\quad A=0.
\ee
Recall that the spatial gauge is given by (\ref{SH_spatial_gauge}), which 
now 
simplifies to
\be
	\Np=0,\quad R=0.
\ee
With these restrictions the evolution equations 
(\ref{be_sys})--(\ref{ee_sys})
simplify dramatically.
Firstly the Gauss constraint (\ref{C_G}) defines $\Sp$ according to
\be
        \Sp = \tfrac{1}{2}(1-\Omega-\Oml),
\label{Sp_LRS}
\ee
and the remaining constraints are satisfied identically.
Secondly, the expressions (\ref{q}) and (\ref{r}) for $q$ and $r$ become
\begin{align}
\label{q_LRS}
        q       &= \tfrac{1}{2}\Big[ 1
        + 3 G_+^{-1}[(\gamma-1)(1-v^2)+\gamma v^2]\Omega -3\Oml
                \Big]
\\
        r       &= -\frac{3}{2}\frac{\gamma \Omega}{G_+} v,\quad
		G_\pm = 1\pm(\gamma-1)v^2.
\label{r_LRS}
\end{align}
The state vector (\ref{state_vector}) reduces to
\be
        \X =(\EEE,\Omega,v,\Oml).
\ee
Using (\ref{LRS_1})--(\ref{r_LRS}) the evolution equations (\ref{be_sys}),
(\ref{G2_Omega}), (\ref{G2_v_1}) and (\ref{ee_sys}) for these variables 
reduce to the following:
\begin{align}
\label{EEE_LRS}
	\dt \EEE &= (q+3\Sp) \EEE
\\
        \dt \Omega &+ \frac{\gamma v}{G_+} \di \Omega
                        + \frac{\gamma G_-}{{G_+}^2} \Omega \di v
\notag\\
                &=\frac{2\gamma}{G_+}\Omega\left[
                        \frac{G_+}{\gamma}(q+1)-\tfrac12(1-3\Sp)(1+v^2)-1
                                 \right]
\\
        \dt v &+ \frac{(\gamma-1)}{\gamma G_-}(1-v^2)^2 \di \ln \Omega
                 - [(3\gamma-4)-(\gamma-1)(4-\gamma)v^2]\frac{v}{G_+G_-}\di v
\notag\\
                &= -\frac{(1-v^2)}{\gamma G_-} \Big[
                        (2-\gamma)G_+ r
                        - \gamma v [3\gamma-4+3(2-\gamma)\Sp] \Big]
\label{v_LRS}
\\
        \dt \Omega_\Lambda &= 2(q+1)\Omega_\Lambda\ .
\label{Oml_LRS_0}
\end{align}

\subsubsection*{An explicit solution}

Consider the ansatz $v=0$, and with $\parb_1 \Omega\neq0$.
Equation (\ref{r_LRS}) implies $r=0$,
and then (\ref{v_LRS}) implies $\gamma=1$
(otherwise $\Omega=\Omega(t)$). 
Equation (\ref{q_LRS}) simplifies to
\be
	q = \tfrac{1}{2}(1-3\Oml).
\ee
Note that $r=0$ 
also
implies $\beta=\beta(t)$ and $\Oml=\Oml(t)$, 
in accordance 
with equations (\ref{q_r_beta}) and (\ref{C_beta}).
Equations (\ref{EEE_LRS})--(\ref{Oml_LRS}) reduce to
 the following evolution equations:
\begin{align}
        \dt \EEE &= (2-\tfrac{3}{2}\Omega-3\Oml) \EEE
\\
	\dt \Omega &= \tfrac{3}{2}(1-\Omega-3\Oml)\Omega
\label{Om_LRS}
\\
	\dt \Oml &= 3(1-\Oml)\Oml\ .
\label{Oml_LRS}
\end{align}

These equations can be solved explicitly, yielding
\begin{align}
\label{solution_1}
        \Oml &= \frac{\foo}{\foo + e^{-3t}}
\\
\label{solution_2}  
        \Omega &= \frac{e^{-3t}}{(\foo+e^{-3t})
                        \left[1+f(x)\sqrt{\foo + e^{-3t}}\,\right]}
\\
        \EEE &= \frac{e^{-t}}{\sqrt{\foo + e^{-3t}}
                        \left[1+f(x)\sqrt{\foo + e^{-3t}}\,\right]}
\label{solution_3}
\end{align}
where $\ell$ is a constant with dimension $\lgth$, and
$f(x)$ is an essential arbitrary function, which is required to 
satisfy
\be
	1 + f(x)\sqrt{\foo} > 0.
\ee
This condition ensures that $\Omega>0$ for $t$ sufficiently large.
If $f(x)>0$, then $t$ assumes all real values while if $f(x)<0$, then $t$ 
satisfies
\be
	t_s(x) < t<\infty,
\ee
where
\be
	t_s(x) = -\tfrac{1}{3}\ln(f(x)^{-2} - \foo)\ .
\ee
An arbitrary multiplicative function of $x$ in (\ref{solution_3}) has been 
absorbed by means of $x$-reparametrization.
The shear is given by
\be
	\Sp = \frac{\tfrac{1}{2}f(x)e^{-3t}}{\sqrt{\foo + e^{-3t}}
			\left[1+f(x)\sqrt{\foo + e^{-3t}}\,\right]}\ .
\label{solution_4}
\ee


If $f(x)=const.$, the solution is an LRS Bianchi I dust solution 
with $\Lambda>0$.
In addition, if $f(x)\equiv0$, the solution is a flat FL solution with 
$\Omega+\Oml=1$.

\subsubsection*{Structure of the singularity}

For simplicity we shall analyze the structure of the singularity using the 
explicit solution (\ref{solution_1})--(\ref{solution_3}).
To show that an initial singularity exists, we show that the matter
density $\mu\rightarrow \infty$. 
Using the definition of $\Omega$ and $\Oml$ and equations 
(\ref{solution_1})--(\ref{solution_2}), it follows that
\be
        \mu \ell^2 = 
        \frac{e^{-3t}}{1+f(x) \sqrt{\foo + e^{-3t}} }
        \rightarrow
        \begin{cases}
        \infty & \text{as $t \rightarrow t_s(x)^+$, if $f(x)<0$}
        \\
        \infty & \text{as $t \rightarrow -\infty$, if $f(x)\geq0$.}
        \end{cases}
\ee 

In order to determine the asymptotic state of the singularity,
it is helpful to calculate the Hubble-normalized shear.
Equations (\ref{solution_4}) and (\ref{beta_to_Hubble}) lead to
\be
	\Sph = \frac{e^{-3t} f(x) }{
	2\sqrt{\foo + e^{-3t}} +f(x)(2\foo + e^{-3t}) } \ .
\ee
It follows that
\be
	\Sph \rightarrow \begin{cases}
	\text{$-1$ as $t \rightarrow t_s(x)^+$} 
	& \text{for $f(x)<0$ (Taub Kasner)}
\\
        \text{0 as $t \rightarrow -\infty$} 
	& \text{for $f(x)=0$ (flat FL)}
\\
        \text{1 as $t \rightarrow -\infty$} 
	& \text{for $f(x)>0$ (LRS Kasner)}.
	\end{cases}
\ee
Crucially, if $f(x)$ has both signs, the limit of $\Sph$ at the 
singularity has 
a jump discontinuity, i.e. a \emph{``step"}.
Thus the solution can develop step-like structures as it approaches the 
initial singularity.

\begin{figure}
\begin{center}
    \epsfig{file=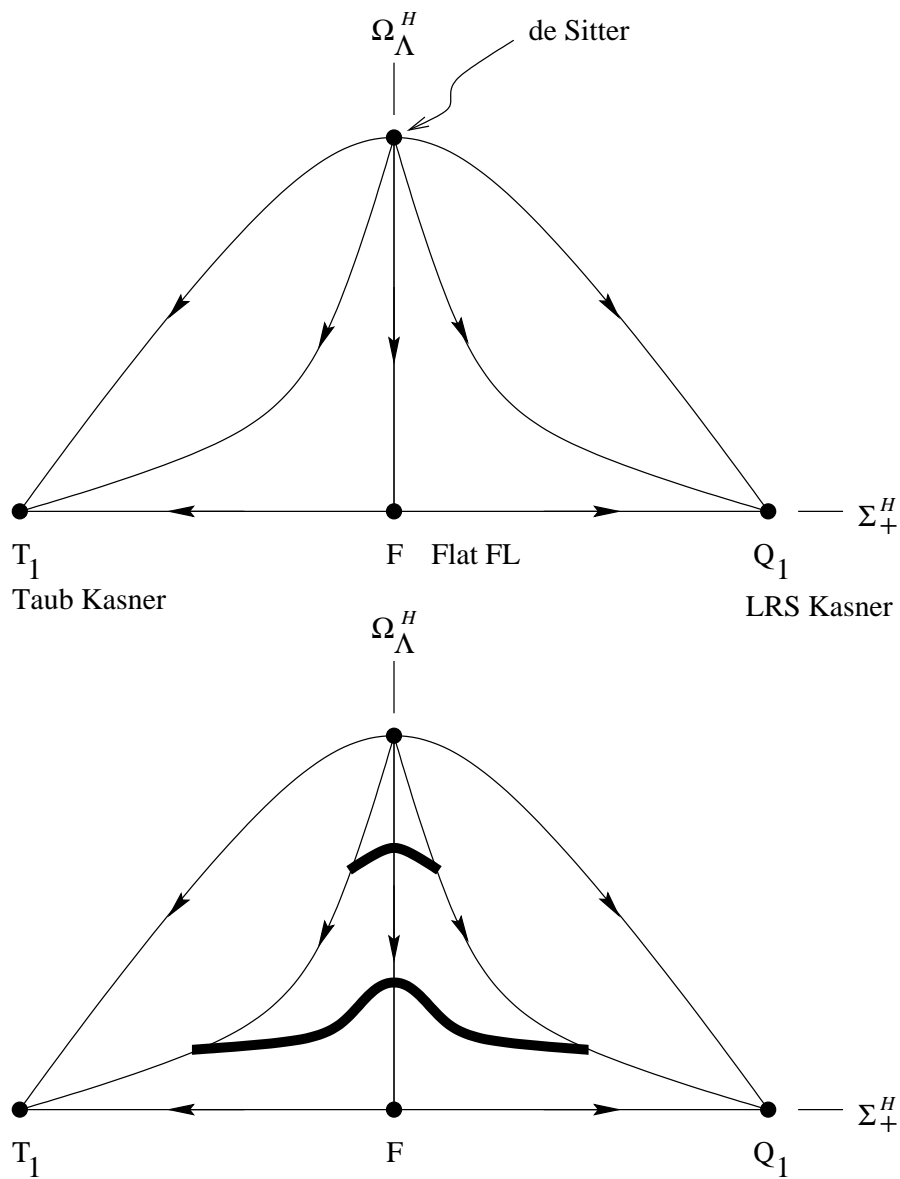,width=11.9cm}
\end{center}
\caption[The LRS Bianchi I state space ($\Lambda>0$), and an LRS
dust solution as an ``inhomogenized" solution at two different times.]
	{The LRS Bianchi I state space ($\Lambda>0$), and an LRS
dust solution as an ``inhomogenized" solution at two different times.
Arrows indicate evolution into the past.}\label{fig:stretch_2}
\end{figure}

One can gain insight into the creation of ``steps" by thinking of the LRS 
$G_2$ dust solution as an ``inhomogenized" LRS Bianchi I dust solution.
The Hubble-normalized state space for the LRS Bianchi I solutions is 
shown%
\footnote{The state space follows from a routine analysis of equations 
(\ref{Om_LRS})--(\ref{Oml_LRS}), and then making use of (\ref{Sp_LRS}).}
in Figure~\ref{fig:stretch_2}.
In terms of evolution into the past,
the flat FL equilibrium point is a saddle point, and the de
Sitter equilibrium point is a source.
The line going from the de Sitter equilibrium point to the flat FL
equilibrium point is the stable manifold of the flat FL equilibrium
point.
At a given time,
an LRS $G_2$ dust solution is represented by a segment of 
points in the LRS Bianchi I state space.
If $f(x)>0$ for all $x$, then $\Sph>0$ and hence
 the segment lies to the right of the 
stable manifold of the FL equilibrium point, and follows the arrows to the 
LRS Kasner equilibrium point as $t\rightarrow -\infty$.
Likewise,
if $f(x)<0$ for all $x$, then the segment lies to the left of the
stable manifold of the FL equilibrium point, and follows the arrows to the
Taub Kasner equilibrium point as $t\rightarrow t_s(x)$.
Crucially, if $f(x)$ has both signs, then the segment straddles the
stable manifold of the FL equilibrium point. As $t$ tends to $t_s(x)$ or 
$-\infty$, the segment is stretched out more and more.
The part of the segment straddling the stable manifold of the flat FL 
equilibrium point is being stretched the most, giving a step-like 
structure in the spatial profile of $\Sph$.

This solution is a special case of the Szekeres solution 
(see Krasinski 1997~\cite[page 25]{book:Krasinski1997}, the case
$k=\beta'=0$).
It is first given explicitly by Barrow \& Stein-Schabes 
1984~\cite[case $k=\beta'=0$]{art:BarrowStein-Schabes1984}.
They however did not discuss the past asymptotic dynamics.
A special case of this solution, with $\Lambda=0$, is first given 
explicitly by 
Bonnor \& Tomimura 1976~\cite[Model PI with 
$\mu\neq0$, $\beta=0$]{art:BonnorTomimura1976}.
They discussed the past asymptotic dynamics, but overlooked the 
possibility that models can be past asymptotic to the Taub Kasner point.

\section{An explicit solution with ``spikes"}\label{sec:WMspike}

In this section, we obtain a new explicit vacuum OT $G_2$ solution that 
develops 
``spikes" as $t \rightarrow -\infty$.
Spikes are narrow spatial structures with steep slopes appearing on an 
otherwise mild profile of a variable of a cosmological solution 
(see Berger \& Moncrief 1993~\cite{art:BergerMoncrief1993}).
To generate this spiky solution, we apply a transformation introduced by
Rendall \& Weaver 2001~\cite{art:RendallWeaver2001} to the
Wainwright-Marshman solution
(see Wainwright \& Marshman 1979~\cite{art:WainwrightMarshman1979},
Case I with $m=-\tfrac{3}{16}$).

\subsubsection*{The framework}

We consider the class of vacuum OT $G_2$ cosmologies 
which are defined by the conditions
\be
\label{wm_1}
	\Omega=0,\quad 
	v_\alpha=0,\quad
	\Oml=0,\quad
	\Sigma_2=0
\ee
(see (\ref{OT_G2}) for the OT condition).
We use the timelike area gauge (\ref{timelike_area_G_2}):
\be
	\Udot=r,\quad \mathcal{N}=1,\quad	A=0.
\label{timelike_area}
\ee
Recall that the spatial gauge is (\ref{SH_spatial_gauge}):
\be
	\Np = \sqrt{3} \Nm,\quad R = -\sqrt{3} \Sc.
\label{spatial_gauge}
\ee
With these restrictions the evolution equations 
(\ref{be_sys})--(\ref{ee_sys}) simplify dramatically.
Firstly the Gauss constraint (\ref{C_G}) defines $\Sp$ according to
\be
	\Sp = \tfrac{1}{2}(1-\Sm^2-\Sc^2-\Nm^2-\Nc^2),
\ee	
and the remaining constraints are satisfied identically.
Secondly, the expressions (\ref{q}) and (\ref{r}) for $q$ and $r$ become
\begin{align}
        q       &= 2 - 3 \Sp
\\
        r       &= -3(\Nc\Sm-\Nm\Sc)\ .
\label{wm_6}
\end{align}
The state vector (\ref{state_vector}) reduces to
\be
        \X =(\EEE,\Sigma_-,\Sigma_\times,N_-,N_\times).
\ee
Using (\ref{wm_1})--(\ref{wm_6}) the evolution equations (\ref{be_sys}), 
(\ref{G2_Sm})--(\ref{G2_Nm}) for these variables reduce to the
following equations:
\begin{align}
        \dt \EEE                &= 2 \EEE
\label{eq:EEE}
\\
        \dt \Sm + \EEE \dx \Nc  &= 2 \sqrt{3} (\Sc^2-\Nm^2)
\label{WM2}
\\      
        \dt \Nc + \EEE \dx \Sm  &= 2 \Nc
\label{WM3}
\\
        \dt \Sc - \EEE \dx \Nm  &= -2 \sqrt{3} (\Nm\Nc+ \Sm\Sc)
\label{WM4}
\\
        \dt \Nm - \EEE \dx \Sc  &= 2 \Nm + 2\sqrt{3}(\Sm\Nm+\Nc\Sc)\ .
\label{WM5}
\end{align} 
Equation (\ref{eq:EEE}) can be solved explicitly to give
\be
	\EEE = e^{g(x)} e^{2t},
\ee
and we reparametrize $x$ to set $g(x)=0$ for simplicity.
Since $\EEE \rightarrow 0$ as $ t\rightarrow -\infty$, the solutions are 
asymptotically silent (see Section~\ref{sec:G2_invariant_subset}).
The metric in this temporal and spatial gauge can be written in this form
(see, e.g. van Elst \etal 2002~\cite[Appendix A.3]{art:vEUW2002}):
\begin{equation}
         ds^2 = \beta^{-2} [ - dt^2 + (\EEE)^{-2} dx^2]
        + \ell_0^2 e^{2t}  \left[ e^{P(t,x)} (dy+Q(t,x)dz)^2
                        + e^{-P(t,x)} dz^2 \right]\ ,
\label{PQ_metric}
\end{equation}
where
the variables $(\Sm,\Nc,\Sc,\Nm)$ are defined in terms of $P$ and $Q$ 
(see van Elst \etal
2002~\cite[equations (180) and (181)]{art:vEUW2002}, which are derived 
from the constraint (\ref{c_com1A})):
\be
\label{PQ_rel}
	(\Sm,\Nc) = \tfrac{1}{2\sqrt{3}}(\dt,-\EEE\dx)P,\quad
	(\Sc,\Nm) = \tfrac{1}{2\sqrt{3}}e^P(\dt,\EEE\dx)Q.
\ee

\subsubsection*{The transformation method}

Rendall \& Weaver 2001~\cite{art:RendallWeaver2001}
introduce a two-step solution-generating method.

The first step of the method is
to perform the following transformation
on the metric functions $P$ and $Q$:
\footnote{We have changed the sign of (\ref{Qhat}) from equation (4) of 
\cite{art:RendallWeaver2001}.}
\begin{align}
        e^{\hat{P}} &= e^{-P} [(Q e^P)^2+1]
\label{Phat}
\\
        \hat{Q} &= -\frac{Q e^{2P}}{(Q e^P)^2+1}\ .
\label{Qhat}
\end{align}

The second step is
to perform the so-called
Gowdy-to-Ernst transformation.
Rendall \& Weaver 2001~\cite{art:RendallWeaver2001})
define this transformation  in terms of its effect on $P$ and $Q$:
\begin{align}
\label{G_E_1}
        \hat{P} &= -P-2t
\\
        \hat{Q}_t &= e^{2(P+2t)} Q_x
\\
        \hat{Q}_x &= e^{2P} Q_t\ ,
\label{G_E_3}
\end{align}
which generates a new solution,%
\footnote{This is the same transformation as equation (7) of 
Rendall \& Weaver 2001~\cite{art:RendallWeaver2001}, but looks slightly 
different because
 the time 
variable $\tau$
in \cite{art:RendallWeaver2001}
 is related to our time variable $t$ via $\tau = -2t$.}
although in general $\hat{Q}$ cannot be obtained explicitly.

We now describe the transformations in the framework of the orthonormal 
frame formalism.
First, we show that the transformation (\ref{Phat})--(\ref{Qhat}) is a 
rotation of the frame
vectors $\{ \mathbf{e}_2,\ \mathbf{e}_3\}$, and describe it in terms of 
$\beta$-normalized variables.

Consider a rotation of the frame vectors 
$\{ \mathbf{e}_2,\ \mathbf{e}_3\}$ (see Appendix~\ref{app_rotation}),
\begin{equation}
\left(\begin{array}{c}
      \hat{\mathbf{e}}_{2} \\
      \hat{\mathbf{e}}_{3}
      \end{array}\right)
= \left(\begin{array}{cc}
        \cos\phi & \sin\phi \\
       -\sin\phi & \cos\phi
        \end{array}\right)
\left(\begin{array}{c}
       \mathbf{e}_{2} \\
       \mathbf{e}_{3}
       \end{array}\right) \ .
\label{rotate}
\end{equation}
To preserve the spatial gauge condition (\ref{spatial_gauge}), 
it follows from (\ref{R_rot})--(\ref{N_rot}) that
the angle $\phi$ must 
satisfy
\begin{align}
\label{preserve_1}
	\dt \phi &=
	\sqrt{3}(- \sin 2\phi\, \Sm + \cos 2\phi\, \Sc) 
	 -\sqrt{3}\Sc
\\
	\EEE \dx \phi &=
	\sqrt{3}(\cos 2\phi\, \Nm + \sin 2\phi\, \Nc) 
	- \sqrt{3} \Nm\ .
\label{preserve_2}
\end{align}
It turns out that these equations are 
satisfied if the rotation angle $\phi$ is related to $P$ and $Q$ according 
to
\begin{equation}
        \cos 2\phi = \frac{(Qe^P)^2-1}{(Qe^P)^2+1}\, ,\quad
        \sin 2\phi = \frac{2 Qe^P}{(Qe^P)^2+1}\ ,
\label{eq:phi} 
\end{equation}
as follows from differentiating (\ref{eq:phi}) and using (\ref{PQ_rel}).
The transformation (\ref{Sigma_rot})--(\ref{N_rot}), with $\phi$ given by
(\ref{eq:phi}), is in fact
equivalent to performing the transformation 
(\ref{Phat})--(\ref{Qhat}).
This statement can be verified by 
differentiating (\ref{Phat})--(\ref{Qhat}) and expressing them in terms of 
$(\Sm,\Nc,\Sc,\Nm)$ using (\ref{PQ_rel}), then comparing with 
(\ref{Sigma_rot})--(\ref{N_rot}) to obtain (\ref{eq:phi}).

The Gowdy-to-Ernst transformation (\ref{G_E_1})--(\ref{G_E_3}) is
 much simpler in terms of $(\Sm,\Nc,\Sc,\Nm)$, as follows
from (\ref{PQ_rel}):
\begin{align}
\label{GE_1}
        \hat{\Sigma}_- &= -\Sm-\tfrac1{\sqrt{3}}
\\
        \hat{N}_\times &= -\Nc
\\
        \hat{\Sigma}_\times &= \Nm
\\
        \hat{N}_- &= \Sc\ .
\label{GE_4}
\end{align}
It is easily verified that 
this transformation preserves the form of the
evolution equations (\ref{WM2})--(\ref{WM5}).

Composing the transformations (\ref{Sigma_rot})--(\ref{N_rot}) with 
(\ref{eq:phi}) and 
(\ref{GE_1})--(\ref{GE_4}) leads to
\begin{align}
        \Smt &= - (\cos 2\phi\, \Sm + \sin 2\phi\, \Sc) -
                \tfrac1{\sqrt{3}}
\label{eq:Smt}
\\
        \Nct &=\quad \sin 2\phi\, \Nm - \cos 2\phi\, \Nc
\\
        \Sct &= \quad \cos 2\phi\, \Nm + \sin 2\phi\, \Nc
\\
        \Nmt &= -\sin 2\phi\, \Sm + \cos 2\phi\, \Sc
\label{eq:Nmt}
\end{align}
where $\phi$ is given by (\ref{eq:phi}).

In summary, 
\emph{given a solution $(\Sm,\Nc,\Sc,\Nm)$ of 
(\ref{WM2})--(\ref{WM5}),
the metric functions $P$ and $Q$ are determined by (\ref{PQ_rel}).
Then (\ref{eq:Smt})--(\ref{eq:Nmt}), with $\phi$ determined by $P$ and $Q$ 
according to (\ref{eq:phi}), determines a new solution.
}

\subsubsection*{The Wainwright-Marshman solution}

Consider the ansatz
\be
	\Sc + \Nm=0\quad (\Nm \not\equiv 0).
\ee
It follows from (\ref{WM4})--(\ref{WM5}) that 
$\Sm=\Nc-\tfrac{1}{2\sqrt{3}}$. 
Then (\ref{WM2})--(\ref{WM3}) imply $\Nc=0$.
Solving (\ref{WM4}) then gives $\Sc= \tfrac1{\sqrt{3}} e^t f'(e^{2t}-2x)$, 
where
$f: \mathbb{R} \rightarrow \mathbb{R}$ is freely specifiable,
and $e^{2t}-2x$ is the argument of the function $f'$.
To recapitulate, we have
\be
        \Sm = -\tfrac1{2\sqrt{3}},\quad
        \Nc = 0,\quad
        \Sc = -\Nm = \tfrac1{\sqrt{3}} e^t f'(e^{2t}-2x).
\label{WM_sol}
\ee
We can now obtain $P$ and $Q$ via (\ref{PQ_rel}):
\be
        P = -t\, ,\quad Q=f( e^{2t}-2x).
\ee
This is the Wainwright-Marshman (WM) solution (see Wainwright \& Marshman 
1979~\cite{art:WainwrightMarshman1979},
Case I with $m=-\tfrac{3}{16}$).
\footnote{The Gowdy-to-Ernst transformation leaves the WM solution 
essentially invariant -- it changes only the sign of $f$.}

\subsubsection*{The transformed Wainwright-Marshman solution}

We apply the transformation (\ref{eq:Smt})--(\ref{eq:Nmt}) to the WM
solution and obtain a new explicit solution
of the evolution equations (\ref{eq:EEE})--(\ref{WM5}):
\begin{align}
        \Smt &= - (-\tfrac1{2\sqrt{3}}\cos 2\phi + 
			\tfrac1{\sqrt{3}}\sin 2\phi\,
              		e^t f') - \tfrac1{\sqrt{3}}
\label{transformed_WM_1}
\\
        \Nct &= - \tfrac1{\sqrt{3}}\sin 2\phi\,  e^t f'
\\
        \Sct &= - \tfrac1{\sqrt{3}}\cos 2\phi\,  e^t f'
\\
        \Nmt &= \tfrac1{2\sqrt{3}}\sin 2\phi + \tfrac1{\sqrt{3}}\cos 2\phi\,
                e^t f'
\label{transformed_WM_4}
\end{align}
where
\be
        \cos 2\phi = \frac{(fe^{-t})^2-1}{(fe^{-t})^2+1}\, ,\quad
        \sin 2\phi = \frac{2 fe^{-t}}{(fe^{-t})^2+1}\, ,\quad
	f = f(e^{2t}-2x).
\label{WM_cs}
\ee

\subsubsection*{Structure of the singularity}

\begin{figure}
\begin{center}
    \epsfig{file=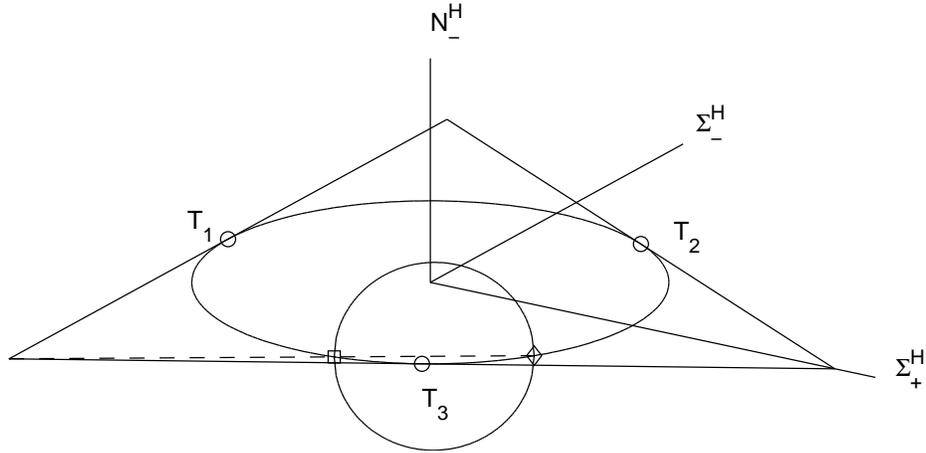,width=.9\textwidth}
\end{center}
\caption[The two $\Nm^H$ transition orbits shadowed by the transformed WM
        solution.]
	{The two $\Nm^H$ transition orbits shadowed by the transformed WM 
	solution.
	$(\Sph,\Smh)=(\tfrac{1}{7},-\tfrac{4\sqrt{3}}{7})$
	is marked by a square, and
	$(\Sph,\Smh)=(\tfrac{11}{13},-\tfrac{4\sqrt{3}}{13})$
	is marked by a diamond.}
        \label{fig:WM_3d}
\end{figure}

What solution(s) does the transformed WM solution tend to as $t\rightarrow 
-\infty$?
From (\ref{WM_cs}), we see that for a fixed $x$,
\be
        \lim_{t \rightarrow -\infty} \cos 2\phi
        = \begin{cases}
                -1 & \text{if $f(-2x) =0$}
                \\
                1 & \text{if  $f(-2x) \neq 0$,}
        \end{cases}
\ee
and $\sin2\phi \rightarrow 0$ only pointwise because
\be
        \sin2\phi = \pm 1\quad\text{whenever} \quad 
	f(e^{2t}-2x)e^{-t} = \pm1.
\ee
Then from (\ref{transformed_WM_1})--(\ref{WM_cs}), we obtain (assuming 
that $f(e^{2t}-2x)$ is smooth):
\be
	\lim_{t\rightarrow-\infty} (\Smt,\Nct,\Sct,\Nmt) = 
	\begin{cases}
	(-\tfrac{\sqrt{3}}{2},0,0,0) & \text{for $f(-2x)=0$}
	\\
	(-\tfrac{1}{2\sqrt{3}},0,0,0) & \text{for $f(-2x)\neq0$.}
	\end{cases}
\ee
In terms of Hubble-normalized variables, we obtain
\be
\label{attractor_WM}
	\lim_{t\rightarrow-\infty} (\tilde{\Sigma}_+^H,\tilde{\Sigma}_-^H) 
	= \begin{cases}
	(\tfrac{1}{7},-\tfrac{4\sqrt{3}}{7}) & \text{for $f(-2x)=0$}
	\\
	(\tfrac{11}{13},-\tfrac{4\sqrt{3}}{13}) & \text{for $f(-2x)\neq0$.}
	\end{cases}
\ee
i.e. the transformed WM solution tends pointwise to two distinct Kasner 
solutions if $f$ has both signs.
These two Kasner solutions are shown in the SH state space in 
Figure~\ref{fig:WM_3d}, and are in fact connected by two $\Nm^H$ 
transition orbits. 
Recall that it follows from (\ref{IX_ellipsoid}) that the two 
$\Nm^H$ transition orbits lie on the sphere
\be
\label{Nm_sphere}
	1 = (\Nm^H)^2 + (\Sp^H)^2 + (\Sm^H)^2,
\ee
going from
$(\tilde{\Sigma}_+^H,\tilde{\Sigma}_-^H) = 
(\tfrac{1}{7},-\tfrac{4\sqrt{3}}{7})$
to
$(\tilde{\Sigma}_+^H,\tilde{\Sigma}_-^H) =
(\tfrac{11}{13},-\tfrac{4\sqrt{3}}{13})$
as $t\rightarrow-\infty$.
We can think of the transformed WM solution as represented by a segment of 
points in the SH state space in Figure~\ref{fig:WM_3d}.
If $f$ has both signs, then this segment gets stretched along the two 
$\Nm^H$ transition orbits as $t\rightarrow -\infty$, thus creating spiky 
structures in the variables.

\newpage

\begin{figure}
\begin{center}
    \epsfig{file=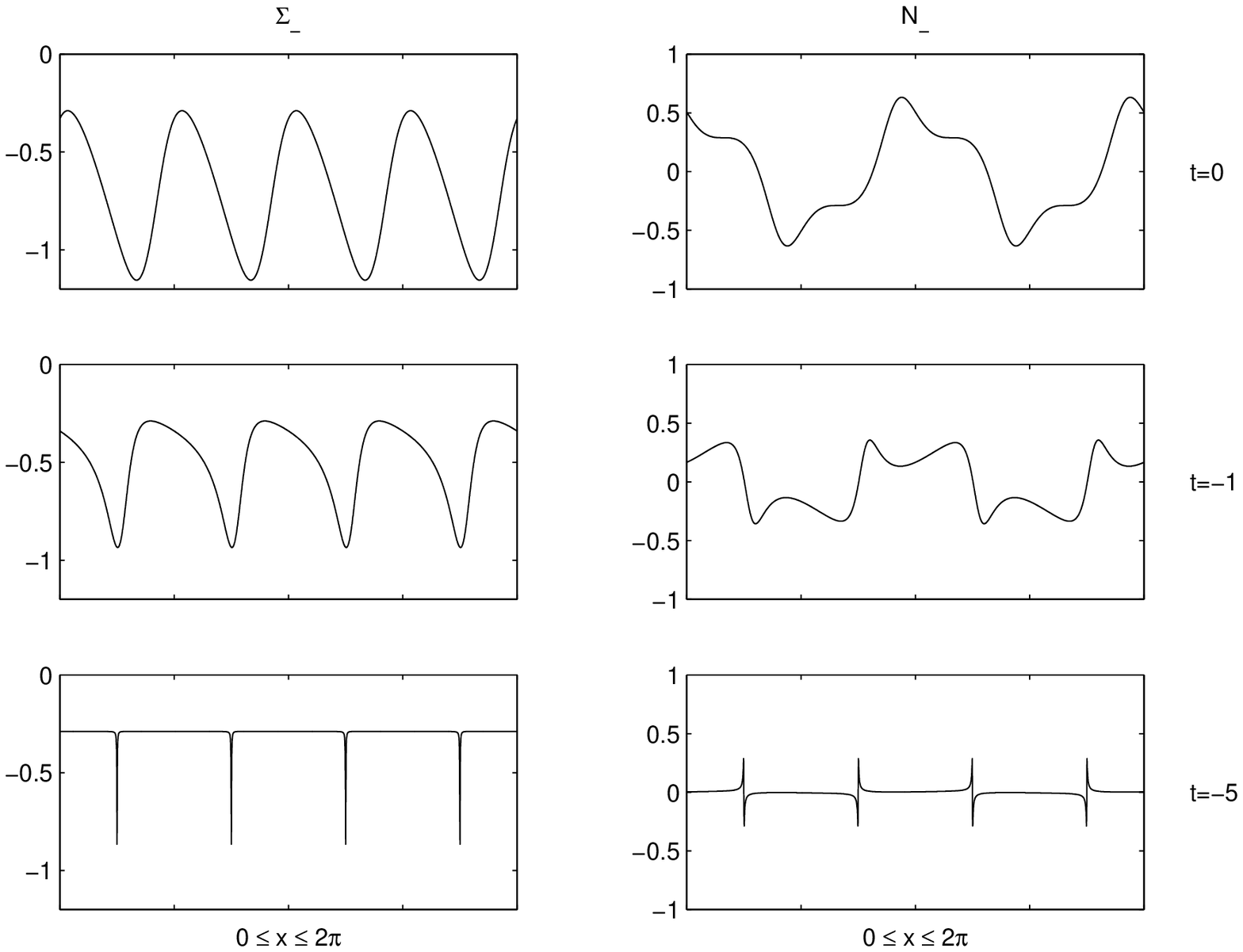,width=\textwidth}
\end{center}
\caption{Snapshots of the variables $\Smt$ and $\Nmt$
 for $f(\bullet)=\cos(\bullet)$.}
        \label{fig:WM_spike}
\end{figure}

To see this clearly,
consider the example $f(e^{2t}-2x)=\cos(e^{2t}-2x)$.
Figure~\ref{fig:WM_spike} shows the snapshots of $\Smt$ 
and $\Nmt$ at $t=0,-1,-5$.
The variable $\Smt$ tends to $-\tfrac{\sqrt{3}}{2}$ for 
$x=\tfrac{\pi}{4},\tfrac{3\pi}{4},\tfrac{5\pi}{4},\tfrac{7\pi}{4}$, 
and $-\tfrac{1}{2\sqrt{3}}$ otherwise.
The variable
$\Nmt$ tends to zero pointwise, but not uniformly. The height of the 
spikes in $\Nmt$ is $\tfrac{1}{2\sqrt{3}}$.
We note that spikes also form in $\Nct$ and $\Sct$, but these variables 
still tend to zero uniformly.

\

In Section~\ref{sec:acc} we shall use both the WM solution and the 
transformed WM solution to test the accuracy of three numerical schemes.
The transformed WM solution also provides a simple way to analyze the 
nature of spikes. We shall return to this in Section~\ref{sec:spike}.

\section{Spiky and step-like structures}

\begin{figure}[h]
\begin{center}
    \epsfig{file=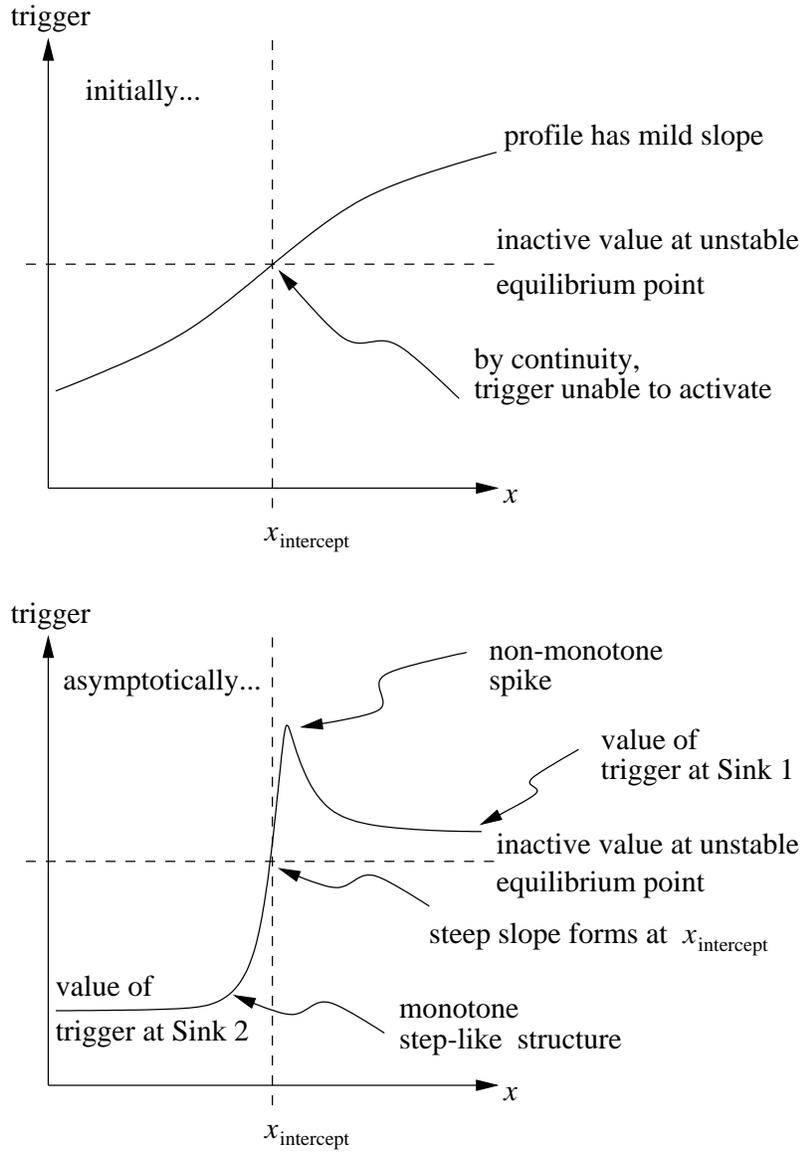,width=10.9cm}
\end{center}
\caption{Formation of spiky and step-like structures.}
        \label{fig:steep}
\end{figure}

Spikes and step-like structures are governed by the same mechanism.
They have one 
thing in common -- they form because
the solution shadows two
disjoint branches of the unstable manifold (of an unstable 
equilibrium point).
The meaning of a ``trigger", introduced in 
Section~\ref{sec:IX}, can 
be extended to include the variable whose axis is parallel to the 
directions of the branches above.
In the language of triggers, consider a trigger
of an unstable equilibrium point.
It assumes a certain value (we call it an ``inactive value") at the 
unstable 
equilibrium point.
Now consider a $G_2$ solution in which the
trigger of an unstable equilibrium point straddles its inactive value,
e.g. the LRS dust solution in which $\Sph$ straddles its inactive value 
of zero, and the transformed WM solution in which $\Nmt$ straddles its 
inactive value of zero.
Then the trigger activates on both sides of the inactive value, but by 
continuity there exists some point $x_{\rm intercept}$ in the middle where 
it cannot activate.
Over time, a steep slope forms at $x_{\rm intercept}$.
How the profile of the trigger extends further is determined by the
unstable manifolds of the saddle point, whose orbits may lead directly to 
the sinks or indirectly through a sequence of saddle points.
If these orbits are such that the trigger develops a monotone profile, 
then we see a step-like structure. 
Otherwise the profile resembles a non-monotone, spiky structure.
Figure~\ref{fig:steep} illustrates the initial smooth profile 
and the asymptotic formation of a 
steep slope at $x_{\rm intercept}$, 
and either a step-like structure or a generic spiky structure. 

We end this chapter with the following conjecture on what causes the 
formation of spiky or step-like structures.

\begin{conj}
Consider a class of $G_2$ cosmologies and their invariant 
subset of SH cosmologies.
In an asymptotically silent regime, the $G_2$ cosmologies 
locally approach the attractor of their SH subset.
Step-like or spiky structures will form if a $G_2$ solution, represented 
by a segment of points in the SH state space, straddles the stable 
manifold of a saddle point.
\end{conj}

%% file: past.tex
	\chapter{Dynamics in the past asymptotic regime}\label{chap:past}

In this chapter, we make predictions about the dynamics
 in the past 
asymptotic regime of $G_2$ cosmologies, 
based on a heuristic analysis of the equations.
Recall that we wrote the system (\ref{be_sys})--(\ref{ee_sys})
 symbolically as (\ref{G2_EEE})--(\ref{G2_3}):
\begin{gather}
        \dt \EEE = B(\Y) \EEE
\\
        \dt \Y + M(\Y) \EEE \partial_x \Y = g(\Y)
\label{symbolic_eq1}
\\
        \mathcal{C}(\Y, \EEE \ptl_x \Y)=0\ ,\quad r = F(\Y),
\end{gather}
where $\Y$ is the state vector in (\ref{state_vector}).
In analyzing the dynamics, we shall consider three types of 
behaviour:
\begin{itemize}
\item	the background dynamics,
\item	the effects of saddle points, and
\item	the effects of spatial derivative terms.
\end{itemize}
The background dynamics is governed by the SH system
(\ref{G2_SH_beta_1})--(\ref{G2_SH_beta_2}):
\begin{gather}
        \ptl_t \Y = g(\Y)
\\
        \mathcal{C}(\Y,0)=0,\quad F(\Y)=0,
\end{gather}
which we regard as governing the dynamics in the silent boundary 
(\ref{silent_definition}):
\be
        \EEE=0,\quad r=0.
\ee
We shall show that the saddle points of the SH system can combine with 
inhomogeneity to initiate some interesting effects in this system.
Furthermore,
under the right conditions, the term $M(\Y) \EEE \partial_x \Y$ 
becomes active and the system~(\ref{symbolic_eq1}) exhibits some 
fascinating behaviour.
We explain each type of dynamical behaviour in the sections below,
and present numerical simulations that illustrate the behaviour in
Chapter~\ref{chap:sim}.

\section{Background dynamics: role of algebraic terms}

In Section~\ref{sec:past_as}, we conjectured that a typical $G_2$ 
cosmology is asymptotically silent into the past, suggesting that SH 
dynamics plays a major role in the past asymptotic dynamics of $G_2$ 
cosmologies.
The separable area gauge for $G_2$ cosmologies also coincides with the 
$G_2$ adapted gauge for SH cosmologies as $t \rightarrow -\infty$, since 
asymptotic silence entails $r \rightarrow 0$.

\subsection*{Mixmaster dynamics}

With asymptotic silence, it is plausible that $G_2$ cosmologies, like SH 
cosmologies, also satisfy the BKL conjecture I and the Kasner Attractivity 
Conjecture, which will now refer to tilted Kasner circles 
$\mathcal{K}_{\pm\alpha}$ as well as the standard Kasner circle 
$\mathcal{K}$.

\vspace{3mm}
\noindent
{\bf BKL conjecture I}

\noindent
{\it
Along almost all local orbits of a typical $G_2$ cosmology,
\be
\label{BKL_I_past}
	\lim_{t \rightarrow -\infty} \Omega=0,\quad
	\lim_{t \rightarrow -\infty} \Oml=0.
\ee
}

\noindent
{\bf Kasner Attractivity Conjecture}

\noindent
{\it
Almost all local orbits of a typical $G_2$ cosmology enter
a neighbourhood of one of the Kasner circles.
}
\vspace{3mm}

We emphasize that at this time there has been no progress in proving these 
conjectures for $G_2$ cosmologies.
They are, however, strongly supported by our numerical simulations.
We believe that they are fundamentally important as regards the dynamics 
in the past asymptotic regime.

\

Once the $G_2$ orbits enter a neighbourhood of one of the 
Kasner circles, Mixmaster dynamics takes over along each individual orbit, 
in accordance with SH dynamics.
Proposition~\ref{prop1} concerning the occurrence of Mixmaster dynamics in 
$G_2$-compatible SH cosmologies now leads directly to the following 
proposition. 

\begin{prop}\lb{prop1_G_2}
Consider $G_2$ cosmologies that satisfy the Kasner Attractivity 
Conjecture.
\begin{itemize}
\item[i)]
	If the $G_2$ is generic, Mixmaster dynamics occurs
	along typical timelines.%
	\footnote{{\rm We note that in contrast to the case of SH 
	cosmologies, 
	$N_{\alpha\beta}=0$ is not an invariant set in $G_2$ cosmologies,
	due to the presence of the terms $\parb_1 \Sm$ and $\parb_1 \Sc$ 
	in the $\ptl_t \Nc$ and $\ptl_t \Nm$ equations.}}
\item[ii)]
	If the $G_2$ is non-generic, Mixmaster 
	dynamics does not occur.
\end{itemize}

\end{prop}

For generic $G_2$ cosmologies, the background dynamics is that of SH 
cosmologies of Bianchi type VI and VII.

\

To recapitulate,
the limits (\ref{Mix_2})--(\ref{Mix__4}) for $G_2$-compatible SH 
cosmologies should also hold locally (i.e. along typical timelines) for 
$G_2$ cosmologies:
\begin{alignat}{2}
\label{G2_Mix__1}
        &\text{Stable variables:}
        &&   
        \lim_{t \rightarrow -\infty} (A,\Nc)=0,
\\
\label{G2_Mix_2}
        &\text{Non-overlapping triggers:}\
        &&
        \lim_{t \rightarrow -\infty}
                (N_{22}N_{33}, N_{22}\Sc, N_{33}\St)=0,
\\
        &\text{Trigger variables:}
        &&
        \lim_{t \rightarrow -\infty} (\Nm,\St,\Sc) \quad
        \text{do not exist.}
\label{G2_Mix__3}
\end{alignat}

Again we emphasize that no progress has been made in proving Proposition 
\ref{prop1_G_2} and the limits (\ref{G2_Mix__1})--(\ref{G2_Mix__3}).
The limits  (\ref{BKL_I_past}), (\ref{G2_Mix__1}) and (\ref{G2_Mix_2}), 
which assert that almost all local $G_2$ orbits shadow the unstable 
manifolds of the Kasner equilibrium points, provide the foundation of 
Mixmaster dynamics in $G_2$ cosmologies.

\subsection*{Non-Mixmaster dynamics}

For more special $G_2$ actions, there are too few triggers to 
sustain Mixmaster dynamics as some of the arcs on the Kasner circle 
become sinks (into the past) in the smaller state space.

For OT $G_2$ cosmologies, 
the background dynamics is that of OT SH cosmologies of type VI and VII.
Thus we expect the local past attractor to be given by (\ref{SH_OT_at}):
\be
\label{G2_OT_at}
	\mathcal{A}^- = \mathcal{K}\ \text{arc}(T_3 Q_1).
\ee
By the phrase ``local past attractor" we mean that almost all local orbits 
of a typical OT $G_2$ cosmology are past asymptotic to the arc 
(\ref{G2_OT_at}).
As an example, the transformed WM solution tends to two distinct Kasner 
solutions, as given by (\ref{attractor_WM}). The local orbit of a
spike point tends to the Kasner solution
	$(\Sph,\Smh)=(\tfrac{1}{7},-\tfrac{4\sqrt{3}}{7})$,
which does not lie on $\mathcal{K}\ \text{arc}(T_3 Q_1)$, but other local 
orbits tend to the Kasner solution
	$(\Sph,\Smh)=(\tfrac{11}{13},-\tfrac{4\sqrt{3}}{13})$,
which lies on $\mathcal{K}\ \text{arc}(T_3 Q_1)$.

For $G_2$ cosmologies with one HO KVF,
the background dynamics is that of SH cosmologies of type VI with one HO 
KVF.
We expect the local past attractor to be given by (\ref{SH_HOKVF_tilt_1}):
\begin{multline}
	\mathcal{A}^- = 
	\left[
	\mathcal{K}\ {\rm arc}(\Sph >-\tfrac{1}{2}(3\gamma-4) )\
	\cap\ 
	\mathcal{K}\ {\rm arc}(Q_2 T_1 Q_3 T_2)\
	\right]
\\
        \cup\
	\left[ 
	\mathcal{K}_{\pm1}\ {\rm arc}(\Sph <-\tfrac{1}{2}( 3\gamma-4) )\
        \cap\
        \mathcal{K}_{\pm1}\ {\rm arc}(Q_2 T_1 Q_3 T_2)\ \right]\ .
\end{multline}

For diagonal $G_2$ cosmologies, 
the background dynamics is that of diagonal SH cosmologies of type VI.
Thus we expect the local past attractor to be
given by (\ref{SH_diag_at}):
\be
\label{G2_diag_at}
        \mathcal{A}^- = \mathcal{K}\
        {\rm arc}(\Sph >-\tfrac{1}{2}(3\gamma-4) )\
           \cup\ \mathcal{K}_{\pm1}\
        {\rm arc}(\Sph <-\tfrac{1}{2}( 3\gamma-4) ).
\ee
Isenberg \& Moncrief 1990~\cite{art:IsenbergMoncrief1990} proved that 
$\mathcal{K}$ contains
 the past attractor for vacuum diagonal $G_2$ cosmologies.
Anguige 2000~\cite{art:Anguige2000} gave the asymptotic expansions for
a generic class of perfect fluid diagonal $G_2$ cosmologies that are past 
asymptotic to $\mathcal{K}\ {\rm arc}(\Sph >-\tfrac{1}{2}(3\gamma-4))$.

\section{Effects of spatial inhomogeneity I: role of saddles}
	\label{sec:spike}

Recall that the state of a $G_2$ solution at a fixed time is represented 
by an arc of points in the SH state space.
We have seen in Chapter~\ref{chap:explicit} that if this arc straddles
 the stable manifold of a saddle point then the solution
will develop steep spatial gradients. 
In the language of triggers,
if a trigger variable has both signs, then
when the trigger activates,
 steep spatial gradients will develop in the neighbourhood of
 the point where the trigger variable is zero.

In this section we discuss the steep spatial gradients in more detail,
with a view to confirming the validity of BKL conjecture II.
We shall see that a mild initial inhomogeneity plus a zero in 
a trigger variable can lead to a strong asymptotic inhomogeneity, via 
local SH dynamics.

\subsection*{Step-like structures}

\begin{figure}
\begin{center}
    \epsfig{file=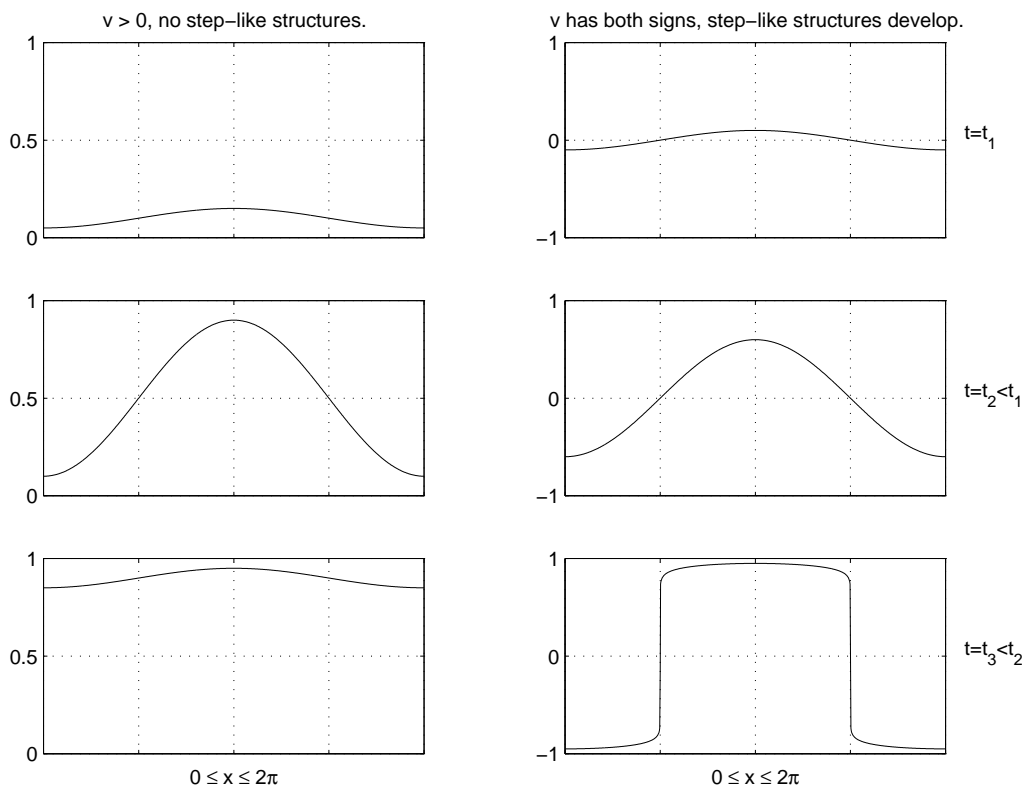,width=\textwidth}
\end{center}
\caption{The tilt variable $v$ can develop step-like structures
        into the past.}\label{fig:v_step}
\end{figure}

The tilt variable $v$ assumes a value close to 0 or $\pm1$ when it is 
inactive, and
 undergoes transitions between 0 and $\pm1$ when it activates.
If $v$ has 
both signs
on an arc of the Kasner circle $\mathcal{K}$
 when  it activates, it develops step-like structures, 
as illustrated schematically in Figure~\ref{fig:v_step}.
This is not unlike the step-like structures in the LRS dust solution
in Section~\ref{sec:lrs_dust},
since the tilt transition will change $v=\eps>0$ to $v=1-\eps$,
and $v=-\eps<0$ to $v=-1+\eps$, where $0 < \eps \ll 1$.

\subsection*{Spiky structures}

\begin{figure}
\begin{center}
    \epsfig{file=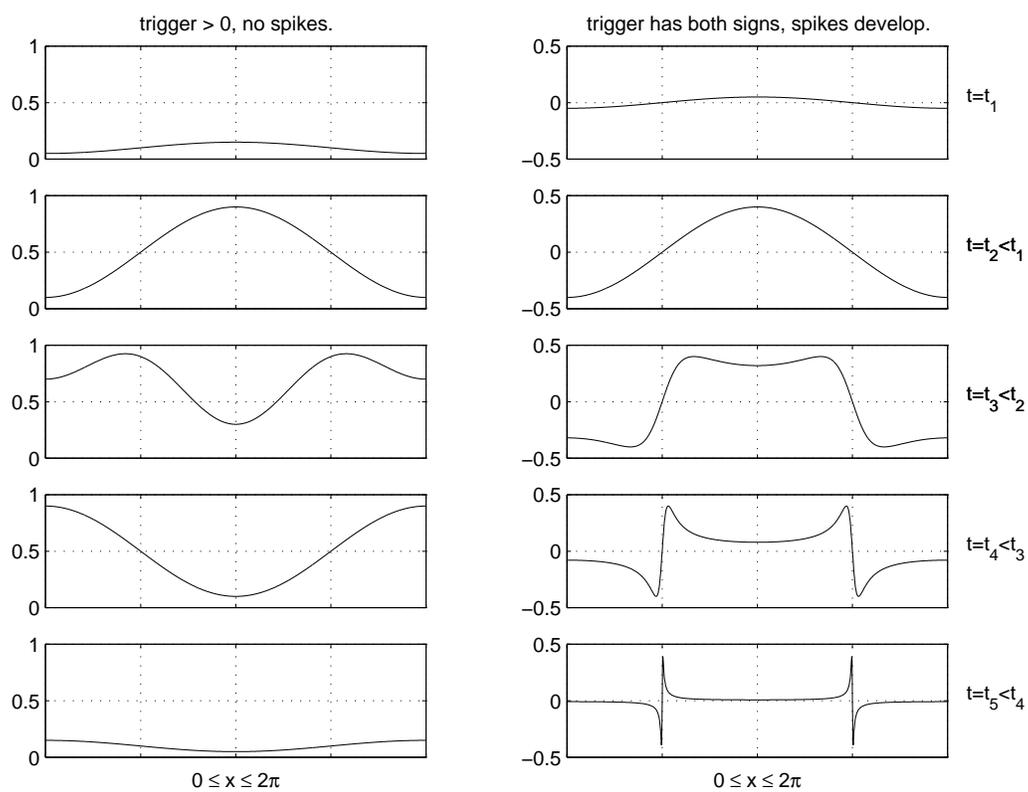,width=\textwidth}
\end{center}
\caption[The triggers $\Nm$ and $\Sc$ can develop spikes into the past.]
	{The triggers $\Nm$ and $\Sc$ can develop spikes
        into the past, as illustrated schematically 
	in the above snapshots.}\label{fig:Nm_spike}
\end{figure}

The curvature trigger $\Nm$ and the frame triggers $\Sc$ and $\St$ assume 
a value close to zero when they are inactive, and become of order unity 
when they activate.
If such a trigger has both signs on a Kasner arc
when it activates, it develops spikes,
as illustrated schematically in Figure~\ref{fig:Nm_spike}.
The spikes shown in  
Figure~\ref{fig:WM_spike} for the transformed WM solution
are a simple prototype.

In the case of the transformed WM solution 
(\ref{transformed_WM_1})--(\ref{WM_cs}), the spikes are 
\emph{permanent} -- their amplitude does not tend to zero  
 as $t \rightarrow -\infty$.
The spikes in the transformed WM solution form in a neighbourhood of 
isolated points
where $f(e^{2t}-2x)=0$, called ``spike points", defined by
\be
	f(e^{2t}-2x_{\rm spike}(t))=0.
\ee
Since $t$ appears in the argument of $f$,
a spike point depends on $t$ and moves toward a limiting spike 
point $x_{\rm spike}(-\infty)$ as $t \rightarrow -\infty$:
\be
	x_{\rm spike}(t) = x_{\rm spike}(-\infty) + \tfrac{1}{2}e^{2t}.
\ee
Despite the fact that $\EEE \rightarrow 0$ exponentially fast, the steep 
slopes of the spikes ensure that
$M(\Y) \EEE \partial_x \Y$ in (\ref{symbolic_eq1})
is comparable to $g(\Y)$ in size,
as shown by the asymptotic expansion of the 
transformed WM solution below.

The decay rates as $t \rightarrow -\infty$ along
the timeline 
$x=x_{\rm spike}(-\infty)$ are as follows (dropping tildes in 
(\ref{transformed_WM_1})--(\ref{transformed_WM_4})):
\begin{align}
\label{WM_spike__1}
\Sm &= -\tfrac{\rty}{2} + \bigO(e^{2t})
\\
\Nc &= -\tfrac{2\rty}{3} \fp^2\ e^{2t} + \bigO(e^{4t})
\\
\Sc &= \tfrac{\rty}{3} \fp  \ e^t + \bigO(e^{3t})
\\
\Nm &=  \tfrac{\rty}{3} (\fp^3-\tfrac{1}{2}\fpp)\ e^{3t} +\bigO(e^{5t})
\end{align}
\begin{xalignat}{3}
\dt \Sm &= -\tfrac{2\rty}{3} \fp^2 \ e^{2t} + \bigO(e^{4t})
	&
\di \Nc &= \tfrac{4\rty}{3} \fp^2\ e^{2t} + \bigO(e^{4t})
\label{spike_1}
\\
\dt \Nc &= -\tfrac{4\rty}{3} \fp^2 \ e^{2t} + \bigO(e^{4t})
	&
\di \Sm &= \tfrac{4\rty}{3} (-\fp^4+\tfrac{3}{2}\fpp\fp) \ e^{4t} +
                \bigO(e^{6t})
\\
\dt \Sc &= \tfrac{\rty}{3} \fp \ e^{t} + \bigO(e^{3t})
	&
\di \Nm &= -\tfrac{2\rty}{3} \fp \ e^t + \bigO(e^{3t})
\label{spike_3}
\\
\dt \Nm &= \rty (\fp^3-\tfrac{1}{2}\fpp) \ e^{3t} +\bigO(e^{5t})
	&
\di \Sc &= \tfrac{2\rty}{3} ( 4 \fp^3 - \fpp)\
                e^{3t} +\bigO(e^{5t})\ ,
\label{spike_4}
\end{xalignat}
where $\fp=f'(-2x_{\rm spike}(-\infty))$ and
$\fpp=f''(-2x_{\rm spike}(-\infty))$.

For comparison, the following are the decay rates as $t \rightarrow 
-\infty$ along a typical timeline $x=a$, where $f(-2a)\neq0$:
\begin{align}
\label{WM_typical__1}
\Sm &= -\tfrac{\rty}{6} + \bigO(e^{2t})
\\
\Nc &= -\tfrac{2\rty}{3} \tfrac{\fp}{\fz} \ e^{2t} + \bigO(e^{4t})
\\
\Sc &= -\tfrac{\rty}{3} \fp \ e^{t} + \bigO(e^{3t})
\\
\Nm &= \tfrac{\rty}{3} \left[ \fp +\tfrac{1}{\fz} \right] e^t + 
\bigO(e^{3t})
\end{align}
\begin{xalignat}{3}
	\dt \Sm &= -\tfrac{2\rty}{3} \left[ 2\tfrac{\fp}{\fz} 
		+ \tfrac{1}{\fz^2} \right]
                e^{2t} + \bigO(e^{4t})
	&
\di \Nc &= \tfrac{4\rty}{3} \left[ -\tfrac{\fp^2}{\fz^2} 
				+\tfrac{\fpp}{\fz}\right]
	e^{4t} + \bigO(e^{6t})
\label{typical_1}
\\
\dt \Nc &= -\tfrac{4\rty}{3} \tfrac{\fp}{\fz}\ e^{2t} + \bigO(e^{4t})
	&
\di \Sm &= \tfrac{4\rty}{3} \left[ -\tfrac{\fp}{\fz^3}-\tfrac{\fp^2}{\fz^2} 
	+\tfrac{\fpp}{\fz} \right]
	e^{4t} + \bigO(e^{6t})
\\
\dt \Sc &= -\tfrac{\rty}{3} \fp \ e^t + \bigO(e^{3t})
	&
\di \Nm &= -\tfrac{2\rty}{3} \left[ -\tfrac{\fp}{\fz^2}+\fpp \right]
                 e^{3t} + \bigO(e^{5t})
\\
\dt \Nm &= \tfrac{\rty}{3} \left[ \fp + \tfrac{1}{\fz} \right]
                e^t + \bigO(e^{3t})
	&
\di \Sc &= \tfrac{2\rty}{3} \fpp \  e^{3t} + \bigO(e^{5t})\ ,
\label{typical_4}
\end{xalignat}
where $\fp=f'(-2a)$, $\fpp=f''(-2a)$ and $\fz=f(-2a)$.
%
%

For typical points, from (\ref{typical_1})--(\ref{typical_4}), we see that 
the spatial derivative terms are negligible compared with the time 
derivative and algebraic terms in the evolution equations 
(\ref{WM2})--(\ref{WM5}).
This supports the BKL conjecture II in Section~\ref{sec:goals} that the 
spatial derivatives in the EFEs
for a typical cosmological model are not dynamically significant in the 
past asymptotic regime.

For spike points, however, (\ref{spike_1}), (\ref{spike_3}) and 
(\ref{spike_4}) show that
despite asymptotic silence,
 the spatial derivative terms are of the same order as
 the algebraic terms in the evolution equations
(\ref{WM2}), (\ref{WM4}) and (\ref{WM5}) respectively.
This limits the validity of the BKL conjecture II within this class of 
models to typical timelines.

For a general class of vacuum OT $G_2$ solutions,
Kichenassamy \& Rendall
1998~\cite[equations (5)--(6)]{art:KichenassamyRendall1998}
gave an asymptotic expansion for the metric functions $P$ and $Q$ 
(see (\ref{PQ_metric})--(\ref{PQ_rel}))
as $t \rightarrow -\infty$.
Rendall \& Weaver 2001~\cite{art:RendallWeaver2001}
transformed this expansion
to produce expansions for 
$\tilde{P}$, $\tilde{Q}_t$ and $\tilde{Q}_x$ for
a spiky solution.
In order to obtain asymptotic expansions for the derivative terms in the 
evolution equations (\ref{WM2})--(\ref{WM5}), we found it necessary to 
increase the accuracy of their asymptotic expansions, as is done in 
Appendix~\ref{expansion}.

For typical points, the leading orders for the time and spatial 
derivatives are
\footnote{The big $\bigt$ below is defined as follows: $f(t) = 
\bigt(g(t))$ means there exist constants $c_1$ and $c_2$ such that
$f(t)$ satisfies $c_1 g(t) < f(t) < c_2 g(t)$ for all $t$ greater (or 
less) than $t_0$.}
\begin{xalignat}{2}
\label{KR_typical_1}
	\ptl_t \Smt &= \bigt(e^{4kt}+e^{4(1-k)t})
	&
	\parb_1 \Nct &= \bigt(-t e^{4t})
\\
	\ptl_t \Nct &= \bigt(-t e^{2t})
	&
	\parb_1 \Smt &= \bigt(e^{2t})
\\
	\ptl_t \Sct &= \bigt(e^{2(1-k)t})
	&
	\parb_1 \Nmt &= \bigt(-t e^{2(1+k)t})
\\
	\ptl_t \Nmt &= \bigt(e^{2kt})
	&
	\parb_1 \Sct &= \bigt(-t e^{2(2-k)t})\ ,
\end{xalignat}
where $k(x)$ satisfies $0 < k < 1$.
We see that the spatial derivatives terms 
tend to zero as $t \rightarrow -\infty$ faster than the corresponding time 
derivatives
 in the evolution 
equations (\ref{WM2})--(\ref{WM5}).
\footnote{$\parb_1 \Smt$ has a marginal impact on $\ptl_t \Nct$.}

For spike points,
the leading orders for the time and spatial
derivatives are   
\begin{xalignat}{2}
        \ptl_t \Smt &= \bigt(e^{4(1-k)t})
        &
        \parb_1 \Nct &= \bigt(e^{4(1-k)t})
\label{KR_spike_1}
\\
        \ptl_t \Nct &= \bigt(-t e^{2t})
        &
        \parb_1 \Smt &= \bigt(e^{2t})
\\
        \ptl_t \Sct &= \bigt(e^{2(1-k)t})
        &
        \parb_1 \Nmt &= \bigt(e^{2(1-k)t})
\label{KR_spike_3}
\\
        \ptl_t \Nmt &= \bigt(e^{6kt}+e^{2(2-k)t})
        &
        \parb_1 \Sct &= \bigt(-t e^{2(2-k)t})\ ,
\label{KR_spike_4}
\end{xalignat}
where $k(x)$ satisfies $0 < k < 1$.
Equations
(\ref{KR_spike_1}) and (\ref{KR_spike_3}) show 
that,
despite asymptotic silence,
 the spatial derivative terms are of the same order as
 the algebraic terms in the evolution equations
(\ref{WM2}) and (\ref{WM4}) respectively.
On the other hand,
(\ref{KR_spike_4}) 
\emph{shows that $\parb_1 \Sct$ dominates $\ptl_t \Nmt$ in 
the case $k \geq \tfrac{1}{2}$.}
This casts more doubts on the BKL conjecture II along 
the timelines corresponding to the spike points.

It should be noted that the results 
(\ref{KR_typical_1})--(\ref{KR_spike_4})
do not reduce exactly to the WM results because the WM solutions are 
special in a number of respects.
For proper reduction, set $k(x)=\frac{1}{2}$ in the more detailed 
expansions in Appendix~\ref{expansion}.

The main investigators of spikes (so-called Gowdy spikes in the context 
of vacuum $G_2$ cosmologies)
are Berger, Moncrief, Isenberg and their collaborators.
Berger \& Moncrief 1993~\cite{art:BergerMoncrief1993}  first discovered 
spikes in numerical experiments.
A more detailed numerical analysis is performed by
Berger \& Garfinkle 1998~\cite{art:BergerGarfinkle1998}.

\section{Effects of spatial inhomogeneity II: role of spatial derivative 
	terms}\label{sec:past_spike_trans}

In this section, we discuss inherently non-SH dynamics.

\subsection*{Spike transitions}

Not all spikes are permanent.
In Garfinkle \& Weaver 2003~\cite{art:GarfinkleWeaver2003},
numerical simulations show that for vacuum OT $G_2$ cosmologies,
 spikes which develop during certain Kasner 
epochs are transient in the sense that they recede to a smooth 
profile via \emph{spike transitions}.
\footnote{More precisely, spikes (as identified by their peak in the 
profile of $\Smh$) 
that form near the Kasner arc spanning from $T_2$ counter-clockwise to the 
point $(\Sph,\Smh)= (-\tfrac{1}{7},-\tfrac{4\sqrt{3}}{7})$ are transient.}
Eventually, however, new spikes may appear at exactly the same spike 
point $x_{\rm spike}(t)$, and may be permanent.
Unfortunately we have been unable to provide analytical heuristics to 
predict spike transitions.

In Chapter~\ref{chap:sim}, we shall illustrate transient spikes
in $G_2$ cosmologies with generic $G_2$ action, and describe spike 
transitions in more detail. 
We shall see that $\parb_1 \Nm$ is larger than at permanent 
spikes, enough to change the sign of $\ptl_t \Sc$ and initiate a spike 
transition.

\subsection*{Shock waves}

Shock waves can form if there are short-wavelength fluctuations
with large enough amplitude
 in the perfect fluid variables.
Heuristically, this can be explained as follows.
For $G_2$ cosmologies with one tilt degree of freedom,
the evolution equation (\ref{G2_v_1}) for the tilt $v$ has the following 
form:
\be
	\dt v - \frac{[(3\gamma-4)-(\gamma-1)(4-\gamma)v^2]}{G_+G_-}
		\,v\,\EEE \dx v = \cdots\ .
\label{burger_v}
\ee
While other terms are not negligible, the left hand side 
of (\ref{burger_v}) behaves like an inviscid Burgers' equation:
\be
	\dt u + u \dx u =0.
\ee
It is well-known that solutions of Burgers' equation can develop shock 
waves, depending on the initial conditions
(see, for example, Whitham 1974~\cite{book:Whitham1974}, in which 
Chapter 4 is dedicated to Burgers' equation).
We predict that $G_2$ cosmologies can develop shock waves in the 
perfect fluid variables if the conditions are right, namely
the occurrence of short-wavelength fluctuations with large enough 
amplitude in the perfect fluid variables $(\Omega,v)$. 
In other words, the larger $\EEE \ptl_x v$ is, the sooner $v$ develops 
shock waves.

%% file: eqs.tex
\chapter{Equations for numerical simulations}\label{chap:G2}

In the remaining two chapters,
we show
a variety of numerical simulations to support our analysis in 
Chapter~\ref{chap:past}, and to explore new phenomena.

For simplicity we restrict our considerations to perfect fluid models with 
one tilt degree of freedom:
\be
	v_2=v_3=0,
\label{one_tilt}
\ee
although for some experiments we shall use vacuum models.
In doing so, the shear spatial gauge (\ref{G2_spatial_gauge}) simplifies 
to
\be
\label{common_gauge}
        \Sigma_{3}=0,\quad 
	\Np = \sqrt{3}\Nm,\quad
	R = -\sqrt{3} \Sc\ .
\ee

As discussed in Section~\ref{sec:SH_var} in the context of SH dynamics, 
the stability of the constraints is of great concern when doing numerical 
simulations.
The concern is even greater for simulations of PDEs, which are less 
accurate than simulations of ODEs.
Numerical experiments of $G_2$ cosmologies show that the Gauss constraint 
(\ref{C_G}) is the most unstable one during the transient regime, and the 
resulting error is unacceptably large.
We remedy this problem by using (\ref{C_G}) to determine $\Sp$.
The decision to use the $(\mathcal{C}_{\rm C})_1$ constraint to determine 
$r$ (see (\ref{r})) also turns out to be equally important.
The remaining $(\mathcal{C}_{\rm C})_3$ and
$(\mathcal{C}_\beta)$ constraints yield small, acceptable errors during 
the transient regime, and are stable in the asymptotic regimes.
Ideally, the errors should be minimized, and the difficulties in doing so 
are discussed in Section~\ref{sec:unsatisfactory_aspects}.

With the restrictions (\ref{one_tilt})--(\ref{common_gauge}) and solving 
$(\mathcal{C}_G)=0$ for $\Sp$,
the state vector (\ref{state_vector}) reduces to
the following for numerical purposes:
\be
        \X =(\EEE,A,\Sigma_-,\Nc,\Sigma_\times,\Nm,\Sigma_2,
        \Omega,v_1,\Omega_\Lambda),
\ee
together with the parameter $\gamma$, and we drop the index on $v_1$.
The full system of 10 evolution equations and 2 constraint equations is
given in Appendix~\ref{appG2}.
We write the system of evolution equations
(\ref{be_numeq})--(\ref{ee_numeq})
 symbolically as
\begin{gather}
        \dt \X + M(\X) \partial_x \X = g(\X).
\label{symbolic_eqs}
\end{gather}

Boundary conditions are required for simulations of PDEs. 
We shall prescribe \emph{periodic boundary condition}, i.e. the numerical 
solutions are required to satisfy
\be
	\X(t,0)=\X(t,2\pi)\quad\text{for all $t$.}
\label{BC}
\ee
It is customary to do so for numerical simulations of $G_2$ cosmologies.

By identifying $x=2\pi$ with $x=0$, we have used up some but not all of 
the freedom in the choice of $x$-coordinate.
A change of $x$-coordinate that satisfies (\ref{f_condition}) preserves 
this identification.
We want to use all the freedom in the choice of $x$ when specifying 
the initial conditions.
Making $\EEE=const.$ in the initial conditions achieves this, and also 
makes the simulations more efficient.

\section{Specifying the initial conditions}\label{sec:IC}

To perform a numerical simulation, we must first specify initial 
conditions
that satisfy the boundary condition
(\ref{BC}) and the remaining constraints $(\mathcal{C}_{\rm C})_3$ and
$(\mathcal{C}_\beta)$:
\begin{align}
\label{eqs_CC3}
        0 &= (\mathcal{C}_{\rm C})_3 = \EEE \ptl_x \St 
		-(r+3A-\sqrt{3}\Nc)\St
\\
\label{eqs_C_beta}
        0 &= (\mathcal{C}_\beta) =
        (\EEE \ptl_x - 2r)\Oml\ .
\end{align}

A good way to specify the initial conditions is the following.
We first specify the following eight variables:
\be
\label{init}
	(\EEE,\ r,\ A,\ \Sm,\ \Nc,\ \Sc,\ \Nm,\ \Omega)\
\ee
on the interval $x \in [ 0,2\pi]$, and the parameter $\gamma$.
\footnote{Although during the simulation we use (\ref{r}) to calculate
$r$ in terms of the state variables $\X$, when
specifying the initial conditions it is more convenient to specify $r$ and
use (\ref{r}) to calculate $Q$ and hence $v$.}
Because we are using the periodic boundary condition (\ref{BC}), the 
initial data must be $2\pi$-periodic.
Then we calculate successively the following four variables
\be
	(\Oml,\ \St,\ \Sp,\ Q),
\ee
where
\be
\label{Q_v}
	Q=\frac{\gamma v}{G_+} \Omega\ ,\quad
        G_+ = 1 + (\gamma-1) v^2\ ,
\ee
from the constraints (\ref{eqs_C_beta}), (\ref{eqs_CC3}), (\ref{C_G}) and 
(\ref{r}) respectively:
\begin{align}
\label{define_Oml}
        \Oml &= (\Oml)_0 \exp \left( \int_0^x \frac{2r}{\EEE}\, dx \right)
\\
\label{St_init}
       \St &=  (\St)_0
        \exp \left( \int_0^x \frac{r+3A-\sqrt{3}\Nc}{\EEE} \,dx \right)
\\
	\Sp &= \tfrac{1}{2}
	\big[ 1-\Sm^2-\Sc^2-\St^2 -\Om_k-\Omega-\Oml \big]
\label{define_Sp_init}
\\
	Q &= \tfrac{2}{3}\big[ -3A\Sp-3\Nc\Sm+3\Nm\Sc-r \big]\ ,
\label{define_Q}
\end{align}
where $(\Oml)_0$ and
$(\St)_0$ are freely specifiable positive constants, and
\be
	\Om_k = 
         \Nm^2  + \Nc^2 - \tfrac{2}{3} (\EEE \ptl_x -r)A + A^2\ .
\ee
We often specify initial conditions which are so simple that $\Oml$ and  
$\St$ can be computed by hand.
We must make sure that (\ref{define_Oml}) and (\ref{St_init}) yield
$2\pi$-periodic $\Oml$ and
$\St$, i.e. we must ensure that (\ref{init}) satisfies
\footnote{If $\Lambda=0$, then $r$ does not have to satisfy 
(\ref{periodic1}). Likewise if $\St=0$, then $r+3A-\sqrt{3}\Nc$ does not 
have to satisfy (\ref{periodic2}).} 
\begin{gather}
	\int_0^{2\pi} \frac{2r}{\EEE}\, dx =0
\label{periodic1}
\\
	\int_0^{2\pi} \frac{r+3A-\sqrt{3}\Nc}{\EEE}\, dx =0\ .
\label{periodic2}
\end{gather}
The tilt $v$ is obtained by solving (\ref{Q_v}), which is a quadratic 
equation for $v$, giving
\be
\label{v_init}
        v = 
	\frac{2 (Q/\Omega)}{
		\gamma + \sqrt{\gamma^2 -4(\gamma-1)(Q/\Omega)^2}}\ .\
\footnote{The second solution for $v$ does not satisfy $v^2 \leq 1$.}
\ee
We must make sure that (\ref{init}) and $(\St)_0$ yield a $v$ that 
satisfies
\be
\label{v_range}
	-1 < v < 1,\quad \text{or equivalently,}\quad
	-1 < \tfrac{Q}{\Omega} < 1\ .\
\footnote{One can prove this by plotting (\ref{v_init}) on the box
$ -1 \leq \tfrac{Q}{\Omega} \leq 1$, $1 \leq \gamma \leq 2$, and
by plotting (\ref{Q_v}) on the box $-1 \leq v \leq 1$, $1 \leq \gamma \leq 
2$ using {\tt MAPLE}.}
\ee
One way to ensure this is to specify a large $\Omega$, as we now explain.

Consider the following format for the initial data, which we shall use in 
most numerical simulations in the next chapter:
\begin{gather}
        \EEE=(\EEE)_0\ ,\quad 
        r=\eps \sin x,\quad
        A = A_0 + \epsilon \sin x,\quad
        \Sm = (\Sm)_0 + \epsilon \sin x,
\notag\\
        \Nc = \sqrt{3}A_0 + \tfrac{\epsilon}{\sqrt{3}} \sin x,\quad
        \Nm = (\Nm)_0 + \epsilon \sin x,\quad
        \Sc = (\Sc)_0 + \epsilon \sin x,
\notag\\
        \Omega = \Om_0 + \epsilon \sin x,
\label{IC}
\end{gather}
where the subscript 0 indicates a constant, and $\epsilon$ is a small 
positive constant.
Note that we have made the choice $r_0 =0$ and
$(\Nc)_0 = \sqrt{3}A_0$ in order to 
satisfy (\ref{periodic1})--(\ref{periodic2}).
The remaining variables are determined by 
(\ref{define_Oml})--(\ref{define_Q}) and (\ref{v_init}).
Equations (\ref{define_Oml})
and (\ref{St_init}) simplify to
\be
	\Oml = (\Oml)_0 \exp\left[-2\tfrac{\epsilon}{(\EEE)_0}\cos x \right],
	\quad
        \St = (\St)_0 \exp\left[-3\tfrac{\epsilon}{(\EEE)_0}\cos x \right]\ ,
\label{Oml_St}
\ee
while $\Sp$ and $v$ have complicated expressions.
Expanding $\tfrac{Q}{\Omega}$ as a power series in $\epsilon$, we obtain
\begin{multline}
	\tfrac{Q}{\Omega} =
	A_0 + 2(\Nm)_0(\Sc)_0-2\sqrt{3}A_0(\Sm)_0
\\
	-A_0\big[1-4A_0^2-(\Nm)_0^2-(\Sm)_0^2-(\Sc)_0^2-(\St)_0^2
	\big]/\Omega_0
	+ \bigO(\epsilon).
\end{multline}
One can ensure $v^2 <1$ by  choosing a sufficiently small $\epsilon$ and a 
sufficiently large $\Omega_0$.
It also helps to specify a small $A_0$ (assuming $A_0 > 0$).

\subsubsection*{Timelike area gauge $A=0$}

If we use the timelike area gauge ($A=0$), there is a simpler scheme. 
We first specify
\be
        (\EEE,\ \Sm,\ \Nc,\ \Sc,\ \Nm,\ \Omega,\ v)
\ee
and $\gamma$.
Then we calculate successively
\be
	(r,\ \Omega_\Lambda,\ \St,\ \Sp),
\ee
from (\ref{r}), (\ref{eqs_C_beta}),
(\ref{eqs_CC3}) and (\ref{C_G}) respectively:
\begin{align}
        r &= -3(\Nc\Sm -\Nm\Sc)
                -\frac{3}{2} \frac{\gamma \Omega}{G_+} v
\\
	\Oml &= (\Oml)_0 \exp \left( \int_0^x \frac{2r}{\EEE}\, dx \right)
\\
       \St &=  (\St)_0
        \exp \left( \int_0^x \frac{r-\sqrt{3}\Nc}{\EEE} dx \right)
\\
        \Sp &= \tfrac{1}{2}
	\big[ 1-\Sm^2-\Sc^2-\St^2 -\Nm^2-\Nc^2-\Omega-\Oml\big]
	\ ,
\end{align}
where $(\Oml)_0$ and $(\St)_0$ are freely specifiable positive constants.
Note that in order to ensure that $\Omega_\Lambda$ and $\St$ are periodic,
equations (\ref{periodic1})--(\ref{periodic2}) with $A=0$ must be 
satisfied.

\subsubsection*{Vacuum subcase}

The scheme for the generic case does not apply to vacuum initial 
conditions, while the scheme for the $A=0$ subcase above does.
We now provide a way to specify vacuum initial conditions with $A > 0$.
We choose $\Nc=0$ and specify
\be
	(\EEE,\ A,\ \Sc,\ \Nm,\ \St).
\ee
Then the $(\mathcal{C}_{\rm C})_3$ and $(\mathcal{C}_{\rm G})$
constraints (\ref{CC_3}) and (\ref{C_G})
are solved for $\Sp$ and $\Sm$ in sequence, yielding
\begin{align}
\label{vac_Sp}
        \Sp &= 1-(\tfrac{1}{3}\EEE \ptl_x \ln\St-\Nm\Sc)/A
\\
        \Sm &= \left[ 1-\Om_k - 2\Sp - \Sc^2 - \St^2 \right]^{1/2},
\end{align}
where
\be
        \Om_k =\Nm^2  - \tfrac{2}{3} (\EEE \ptl_x -r)A + A^2,\quad
	r = - 3A\Sp + 3\Nm\Sc \ .
\label{vac_r}
\ee

\section{Is the system hyperbolic?}\label{sec:hyperbolic}

We ask the following question: does the system of equations 
(\ref{symbolic_eqs})
lead to a well-posed initial value problem?
To answer this we rely on a theorem that states that
if (\ref{symbolic_eqs}) is symmetric hyperbolic,
then it leads to a well-posed IVP
(see Friedrich \& Rendall 2000~\cite[pages 
147--157]{inbook:FriedrichRendall2000}).
By symmetric hyperbolic we mean $M(\X)$ is symmetric, or more 
generally, 
that 
(\ref{symbolic_eqs}) can be multiplied by a symmetric positive definite 
matrix $A(\X)$ 
so that $A(\X)M(\X)$ is symmetric
(see Friedrich \& Rendall 2000~\cite[pages 
148, 157]{inbook:FriedrichRendall2000}).
One important notion of hyperbolicity is strong hyperbolicity, which 
means $M(\X)$ is diagonalizable.
Strongly hyperbolic systems are also symmetric hyperbolic in the general 
sense (see Friedrich \& Rendall 2000~\cite[page
161]{inbook:FriedrichRendall2000}).
To our knowledge there are no other proofs of well-posedness for 
quasi-linear systems of PDEs that satisfy other notions of hyperbolicity.
See Gustafsson \etal 1995~\cite[page 119]{book:Gustafssonetal1995} for
these notions in the context of linear PDEs.

The presence of $\ptl_x A$ (contained in $q$ and $\Sp$)
 in $M(\X)$ complicates the matrix.
As a result we have been unable to determine whether
the system
(\ref{symbolic_eqs})
 is symmetric hyperbolic.%
\footnote{The system is not strongly hyperbolic, because although 
$M(\X)$ 
has real eigenvalues, it is not diagonalizable on a set of points with 
measure zero.}
We comment that if we use the timelike area gauge, where
$A$ is identically zero, then the system is symmetric hyperbolic,%
\footnote{This is proved by van Elst \etal 2002~\cite{art:vEUW2002} for OT 
$G_2$ cosmologies.
This result generalizes to the class of generic $G_2$ cosmologies with one 
tilt degree of freedom, because the evolution equations for OT $G_2$ are
aumented by the $\ptl_t \St$ equation (\ref{app_St}), which has a zero 
characteristic speed of propagation.}
and thus has a well-posed IVP.
Since
we find that the numerical results obtained in this gauge and in the 
separable area gauge are qualitatively the same, we are confident that the 
separable area gauge is suitable for simulations.

\section{Accuracy of numerical schemes}\label{sec:acc}

In our numerical experiments, we make use of
the following three numerical schemes:
\begin{enumerate}[i)]
\item	the fourth order Runge-Kutta method ({\tt RK4}),
\item	the 3-step iterative Crank-Nicholson method ({\tt ICN}), and
\item	the {\tt CLAWPACK} code
\end{enumerate}
(see Gustafsson \etal 1995~\cite[page 241]{book:Gustafssonetal1995},
Teukolsky 2000~\cite{art:Teukolsky2000}, and LeVeque
2002~\cite{book:LeVeque2002} respectively). 
{\tt RK4} is a classic numerical solver for ODEs, but is suggested for 
application to PDEs by Gustafsson \etal.
{\tt ICN} is a new method developed by Choptuik, but is not 
well-documented.
{\tt CLAWPACK} is developed by LeVeque, and is well-documented in 
\cite{book:LeVeque2002}.

We now explain the algorithm for each method briefly,
and describe some tests of their accuracy when applied to the system 
(\ref{symbolic_eqs}).
To conform with the notation in the literature on numerical algorithms for 
PDEs,
we denote the state vector $\X$ by $y$
and rewrite the symbolic equation (\ref{symbolic_eqs}) as
\be
\label{symbolic_f}
	\dt y = f(y,\ \partial_x y).
\ee
We denote the grid size by $\Delta x$ and the step size by $\Delta t$.
First,
we have to decide how to discretize the spatial partial
derivative $\partial_x y$.
The central differencing scheme for the derivative gives
\be
\partial_x y(t,x) = \frac{y(t,x+\Delta x) - y(t,x-\Delta x)}{2\Delta x}
		+ \bigO(\ (\Delta x)^2),
\ee
which is second order accurate,
i.e. the error is of order $(\Delta x)^2$.
The central differencing on five grid points is fourth order 
accurate:
\begin{multline}
\partial_x y(t,x) = 
	\frac{4}{3}\frac{y(t,x+\Delta x) - y(t,x-\Delta x)}{2\Delta x}
\\
	- \frac{1}{3}\frac{y(t,x+2\Delta x) - y(t,x-2\Delta x)}{4\Delta x}
                + \bigO(\ (\Delta x)^4)
\end{multline}
(see, e.g., Gustafsson \etal 1995~\cite[pages 19, 
247]{book:Gustafssonetal1995}).

Secondly, we have to give a scheme for time-stepping, i.e.
to compute $y(t+\Delta t,x)$, given $y(t,x)$.
Let $y_n$ and $y_{n+1}$ denote the computed values for
$y(t)$ and $y(t+\Delta t)$ respectively.
The fourth order Runge-Kutta method ({\tt RK4}) is described as follows
(see, e.g., Gustafsson \etal 1995~\cite[page 
241]{book:Gustafssonetal1995}).
For brevity we write $f(y_n, \partial_x y_n)$ as $f(y_n)$ (or think of $f$ 
as an operator).
\begin{align*}
	k_1 &= \Delta t f(y_n)
\\
	k_2 &= \Delta t f(y_n + \tfrac{1}{2}k_1)
\\
        k_3 &= \Delta t f(y_n + \tfrac{1}{2}k_2)  
\\
        k_4 &= \Delta t f(y_n + k_3)  
\\
	y_{n+1} &= y_n + \tfrac{1}{6}(k_1 + 2 k_2 + 2 k_3 + k_4).
\end{align*}
Using Taylor polynomials, it is straightforward to show that
{\tt RK4} is fourth order accurate 
(the error in the spatial derivatives may for the moment be ignored):
\be
	y_{n+1} = y(t+\Delta t,x) + \bigO(\ (\Delta t)^5),
\ee
i.e. the power of $\Delta t$ in the error term is four orders higher than 
that in the formula for $y_{n+1}$.
Now combining {\tt RK4} with the 5-grid central difference scheme, one 
obtains 
a fourth order accurate finite difference scheme.

The 3-step iterative Crank-Nicholson method ({\tt ICN}) is described as 
follows:
\begin{align*}
        k_1 &= \Delta t f(y_n)
\\
        k_2 &= \Delta t f(y_n + \tfrac{1}{2}k_1)
\\
        k_3 &= \Delta t f(y_n + \tfrac{1}{2}k_2)
\\
        y_{n+1} &= y_n + k_3.
\end{align*}
{\tt ICN} is second order accurate.%
\footnote{See Teukolsky 2000~\cite{art:Teukolsky2000}.
While an $n$-th order Runge-Kutta method is $n$-th order 
accurate,
Teukolsky's analysis shows that
 all {\tt ICN} methods are only second order accurate, and the 
3-step {\tt ICN} method is the cheapest stable method among them (see 
Teukolsky 2000~\cite{art:Teukolsky2000}). 
We note that there are more sophisticated
implementations of the iterative Crank-Nicholson method (Choptuik 
2004~\cite{com:Choptuik2004}).
Teukolsky claims that there is no benefit in applying more iterations. 
Choptuik, however, has reservations about this claim, on the basis that 
Teukolsky's idealized analysis is incomplete.}
Now combining {\tt ICN} with the 3-grid central difference scheme, one 
obtains a second order accurate finite difference scheme.
Since {\tt RK4} evaluates the equations four times at each 
time step while {\tt ICN} only three,
 it takes {\tt RK4} $4/3$ times as long to 
run as it takes {\tt ICN}.

The {\tt CLAWPACK} code (see
LeVeque 2002~\cite{book:LeVeque2002})
 handles the evolution equations
(\ref{symbolic_eqs}) in two stages for
each time step, using the so-called Godunov splitting
(see LeVeque 2002~\cite[pages 380-388]{book:LeVeque2002}).
In the first stage, {\tt CLAWPACK} solves the following system
(referring to (\ref{symbolic_eqs})):
\be
\label{symb_1}
        \dt y + M(y) \partial_x y =0,
\ee
where $M(y)$ must be diagonalizable. i.e. it can be written as
\be
\label{QDQ}
        M = Q D Q^{-1},
\ee
where $D$ is a diagonal matrix of real eigenvalues, and $Q$ is a matrix
constructed from the corresponding eigenvectors.
The upwind finite volume method
(see
LeVeque 2002~\cite[page 73]{book:LeVeque2002})
 is then used to
discretize (\ref{symb_1}).
To explain the discretization, consider
rewriting (\ref{symb_1}) as
\be
        \dt w + D \partial_x w =0,\quad w = Q^{-1} y,
\ee     
pretending that $Q$ is constant.
Since $D$ is diagonal, this system decouples into independent advection
equations, and the eigenvalues determine the velocities (and hence the 
upwind directions) of
the waves in the characteristic eigenfields $w$. 
So we are able to 
determine the upwind direction for each wave.
In finite volume methods, an $x$-interval is divided into cells, and
fluxes are computed at the interface of two adjacent cells.
To obtain the data at the interface, some averaging scheme must be used 
(see LeVeque 2002~\cite[Section 15.3]{book:LeVeque2002}),
and here we simply take the arithmetic average.
The discretization of (\ref{symb_1}) yields
\be
        y_i^{n+1} = y_i^n - \frac{\Delta t}{\Delta x}\left[
        \mathcal{A}^{+n}_{i-\frac{1}{2}} (y_i^n-y_{i-1}^n)
        + \mathcal{A}^{-n}_{i+\frac{1}{2}} (y_{i+1}^n-y_i^n) \right],
\ee
where
\be
        \mathcal{A}^{\pm n}_{i-\frac{1}{2}}
        = Q_{i-\frac{1}{2}}^n D^{\pm n}_{i-\frac{1}{2}}
          (Q_{i-\frac{1}{2}}^n)^{-1},
\ee
and $D=D^+ + D^-$, where
$D^+$ contains only positive eigenvalues while
$D^-$ contains only negative eigenvalues.
The vector
$\mathcal{A}^+_{i-\frac{1}{2}} (y_i-y_{i-1})/\Delta x$ describes fluxes to
the
right from the $(i-1)$-th cell to the $i$-th cell, etc.
For details see
LeVeque 2002~\cite[pages 78--82]{book:LeVeque2002}.
The first stage is first order accurate.   
   
In the second stage, {\tt CLAWPACK} solves the system
\be
        \dt y = g(y).
\ee
This part can be discretized by any ODE solver.
We solve this part by using the explicit Euler method,
\[
	y_{n+1} = y_n + \Delta t\, g(y_n),
\]
which is first order accurate.
        
Overall, {\tt CLAWPACK} is first order accurate.
While {\tt RK4} and {\tt ICN} do poorly in simulating shock waves
(Taylor polynomials do not apply at points of discontinuity),
{\tt CLAWPACK} sacrifices high accuracy for smooth solutions for the
ability to simulate shock waves reasonably well. The error analysis of
{\tt CLAWPACK} near shock waves is based on conservative systems of 
the form:
\[
        \dt y + \partial_x( G(y) ) = g(y).
\]
However, the system (\ref{be_numeq})--(\ref{ee_numeq}) is not
conservative, due to the presence of inhomogeneity in
$E_1{}^1$ and the non-conservative pair $(\ln\Omega, \arctanh\ v)$.
 This will likely reduce the accuracy of
{\tt CLAWPACK} when simulating shock waves.
Another more serious abuse of {\tt CLAWPACK} is that we have put $\ptl_x 
A$
(contained in $q$ and $\Sp$)
in the right hand side of the equations ($g(y)$ becomes $g(y,\partial_x
y)$),%
\footnote{{\tt CLAWPACK} requires a coefficient matrix $M(y)$ that is
diagonalizable. By putting $\ptl_x A$   
 in the left hand side, {\tt CLAWPACK}
is liable to crash due to unbounded determinant of the $Q$
matrix in (\ref{QDQ}).}
 which makes it unsuitable for shock waves
simulations. 
The best we can do is to simulate shock waves using the timelike area 
gauge $A=0$.
The abuse becomes relatively harmless if $\EEE$ is small.
For example, {\tt CLAWPACK} does a better job than the 
other two schemes at simulating permanent spikes under low resolution
(see Section~\ref{sec:unsatisfactory_aspects}),
so there is value in keeping {\tt CLAWPACK}.

To measure the rate of convergence for each method,
we make three runs, using 512, 1024 and 4096 spatial grid points. 
The data using 4096 spatial grid points serves as the reference, and the 
errors of the two runs with cruder grids are compared. 
We use both $L_2$ and $L_\infty$ norms to measure the errors.%
\footnote{Let $y_{i}^m$ be the $m$-th component of the vector $y$ 
evaluated 
at the $i$-th grid point. Then the $L_2$ norm of the error in $y$ is given 
by 
\[
	\| y - y_{\rm ref} \|_2 =  \sqrt{\frac{1}{N}
		\sum_{i,m} (y_{i}^m - (y_{\rm ref})_{i}^m)^2 }\ ,
\]
where $N$ is the number of grid points.
The $L_\infty$ norm of the error in $y$ is given by
\[
	\| y - y_{\rm ref} \|_\infty =
	\max_{i,m} | y_{i}^m - (y_{\rm ref})_{i}^m |\ .
\]
}

\begin{table}[h]
\begin{spacing}{1.1}
\caption{The rate of convergence of numerical methods:
errors relative to the solution on 4096 grid points as measured
 in the $L_\infty$ norm.}
                \label{tab:error}
\begin{center}
\begin{tabular*}{\textwidth}%
        {@{\extracolsep{\fill}}cccrc}
\hline
Method & error (512) & error (1024) & ratio & convergence \\
\hline
{\tt RK4} & $2.9746 \times 10^{-7}$ & $1.9852 \times 10^{-8}$ 
	& 14.9839 & fourth order
\\
{\tt ICN} & $2.2908 \times 10^{-3}$ & $5.3947 \times 10^{-4}$
	& 4.2464 & second order
\\
{\tt CLAWPACK} & $1.7974 \times 10^{-1}$ & $7.8783 \times 10^{-2}$
        & 2.2814 & first order
\\
\hline
\end{tabular*}
\end{center}
\end{spacing}
\end{table}

In Table~\ref{tab:error} we present the errors in $L_\infty$ 
norm.
A method is said to be $n$-th order 
accurate in time if its	error (approximately) shrinks by a factor of 
$2^{-n}$ when the number of grid points is doubled. 
 Errors in $L_2$ norm are of similar size and are not presented.
The initial data is chosen to be (\ref{IC}) with $\gamma=\tfrac{4}{3}$ and
\begin{gather}
        \epsilon= 0.01,\quad
        (\EEE)_0= 1  ,\quad  
        A_0     = 0.2,\quad
        (\Sm) _0= 0.2,\quad
        (\Sc) _0  = 0.4,
\notag\\
        (\Nm) _0  =-0.4,\quad
        \Om _0  = 0.5,\quad
        (\Oml)_0 = 0.7,\quad
        (\St)_0  = 0.1.
\end{gather}
The simulations run from $t=0$ to $t=-1$. Longer simulations will see the 
formation of spikes. The order of accuracy is derived from Taylor 
polynomial approximation. 
With decreasing resolution around the spikes, the approximation becomes 
poorer and poorer, and eventually becomes totally useless. Higher 
resolution will be needed to prolong simulations.
The CFL stability condition
(see Gustafsson \etal 1995~\cite[page 54]{book:Gustafssonetal1995})
 is set at $\Delta t \leq \Delta x/\max\EEE$. 
This serves as a cap for the step size $\Delta t$ to maintain the 
stability of the solution.%
\footnote{We also impose an absolute cap to the time step: $\Delta t \leq 
0.1$ for stability.}
The preferred value for $\Delta t$ is set at
$\Delta t = 0.9 \Delta x/\max\EEE$,
i.e. the step size is computed as follows:
\be
	\Delta t = 0.9 \Delta x / \max\EEE(t,x).
\ee
After the data at time $t+\Delta t$ is obtained, we check that
\be
	\Delta t \leq \Delta x / \max\EEE(t+\Delta t,x).
\ee
If this is satisfied, we continue; 
otherwise we recalculate the data with a smaller step size
\be
        \Delta t_{\rm smaller} = 0.9 \Delta x / \max\EEE(t+\Delta t,x).
\ee
This smaller step size will satisfy the CFL condition:
\be
        \Delta t_{\rm smaller} \leq \Delta x / \max\EEE(t+\Delta 
		t_{\rm smaller},x).
\ee



Now since the reference solution is computed by the same method, we can 
only conclude that each method converges at the respective rate. Do they 
all converge to the real solution?
To answer this we need an exact solution against which the 
error of the methods can be measured.
Unfortunately there are no exact generic $G_2$ solutions available thus 
far.
The most suitable possible test solution is the WM 
solution (see Section~\ref{sec:WMspike}),
which is an exact vacuum OT $G_2$ solution,
containing one arbitrary function $f:\mathbb{R} \rightarrow \mathbb{R}$.
It is given by (\ref{WM_sol}):
\be
	\Sm = -\tfrac{1}{2\sqrt{3}},\quad
	\Nc = 0 = A,\quad
	\Sc = -\Nm = \tfrac{1}{\sqrt{3}}e^t f(e^{2t} - 2x),\quad
	\EEE= e^{2t}.
\ee
For the test, we choose
\be
	f(e^{2t} - 2x) = \sin(e^{2t} - 2x) \quad\text{with}\quad
	0 \leq x \leq 2\pi.
\ee
The solution then satisfies periodic boundary conditions.
The simulations run from $t=0$ to $t=-1$. The CFL condition is set at
$\Delta t \leq \Delta x/\max\EEE$,
and the preferred value for $\Delta t$ at
$\Delta t = 0.9 \Delta x/\max\EEE$.
In Table~\ref{tab:error_WM} we present the errors in $\Sc$ in the 
$L_\infty$ norm. Errors in the  $L_2$ norm are similar and are not 
presented.

\begin{table}[h]
\begin{spacing}{1.1}
\caption{The accuracy of numerical methods: errors in the simulated $\Sc$ 
	for the WM solution as measured in the  $L_\infty$ norm.}
                \label{tab:error_WM}
\begin{center}
\begin{tabular*}{\textwidth}%
        {@{\extracolsep{\fill}}cccrc}
\hline
Method & error (512) & error (1024) & ratio & accuracy \\
\hline
{\tt RK4} & $9.5834 \times 10^{-9}$ & $6.2809 \times 10^{-10}$
        & 15.2581 & fourth order
\\
{\tt ICN} & $9.8275 \times 10^{-6}$ & $2.7601 \times 10^{-6}$
        & 3.5605 & second order
\\
{\tt CLAWPACK} & $4.3212 \times 10^{-3}$ & $2.2423 \times 10^{-3}$
        & 1.9272 & first order
\\
\hline
\end{tabular*}
\end{center}
\end{spacing}
\end{table}

\begin{table}[h]
\begin{spacing}{1.1}
\caption{The accuracy of numerical methods: errors in the simulated $\Sm$ 
	for the transformed WM solution as measured in the $L_\infty$ norm.}  
                \label{tab:error_WMspike}
\begin{center}
\begin{tabular*}{\textwidth}%
        {@{\extracolsep{\fill}}cccrc}
\hline
Method & error (512) & error (1024) & ratio & accuracy \\
\hline
{\tt RK4} & $1.5157 \times 10^{-5}$ & $9.7292 \times 10^{-7}$
        & 15.5789 & fourth order
\\
{\tt ICN} & $2.6242 \times 10^{-3}$ & $6.6930 \times 10^{-4}$
        & 3.9208 & second order
\\
{\tt CLAWPACK} & $2.4204 \times 10^{-2}$ & $1.2511 \times 10^{-2}$   
        & 1.9347 & first order
\\
\hline
\end{tabular*}
\end{center}
\end{spacing}
\end{table}

The WM solution can be transformed into a new solution
 (see Section~\ref{sec:WMspike}). 
We test the codes against this 
solution, using the initial condition for the WM solution above.
In Table~\ref{tab:error_WMspike} we present the errors in $\Sm$ in 
$L_\infty$ norm.

The tests confirm that the methods produce solutions that converge to 
the exact solution at the expected rates, which
gives us some confidence that the methods are reliable.

\section{Unsatisfactory aspects of the numerical 
simulations}\label{sec:unsatisfactory_aspects}

\subsubsection*{Instability of $(\mathcal{C}_{\rm C})_3$}

The evolution equation for $(\mathcal{C}_{\rm C})_3$ is given by
\be
        \dt (\mathcal{C}_{\rm C})_3 = (2q-2+3\Sp+\sqrt{3}\Sm) 
(\mathcal{C}_{\rm C})_3.
\ee     
This constraint is stable into the past near Kasner equilibrium points.
However, when $\St$ is large, $(\mathcal{C}_{\rm C})_3$ grows as
$t\rightarrow -\infty$.
This is a cause for concern especially when $(\mathcal{C}_{\rm C})_3$ 
grows at
the beginning of the simulation (i.e. before entering the asymptotic
regime). Using finer grids or taking smaller time steps reduces the size
of $(\mathcal{C}_{\rm C})_3$, but is very inefficient.
Unfortunately, we find that
adding the constraint in the evolution equations 
\footnote{For example,
\[
	\ptl_t \St = ( q -2 +\sqrt{3}\Sm ) \St + (\mathcal{C}_{\rm C})_3\ 
.
\]
}
does not change its instability.
Constraint instability is a recurring problem in numerical relativity.
For other methods in controlling constraint instability, see Lindblom 
\etal 2004~\cite{art:Lindblometal2004}.

\subsubsection*{Resolution}

As we have not implemented an adaptive mesh refinement scheme, 
high-resolution simulations are expensive for the numerical schemes 
we have. 
This imposes a severe limitation when spikes occur.
Without mesh refinement, we can only adequately simulate spikes for a 
short time before the spikes become too narrow to resolve numerically.
If $\EEE$ is sufficiently large near a spike,
{\tt RK4} can generate spurious oscillations, while {\tt ICN} also 
generate some overshoot. Only {\tt CLAWPACK} handles spikes reasonably
well in all situations.

%

\subsubsection*{The dust case}

For dust ($\gamma=1$) the equations are not strongly hyperbolic (the 
matrix $M(\X)$ is not diagonalizable at any point in spacetime), 
regardless of whether $A=0$ or not.
We shall not simulate the dust case in this thesis.

\subsubsection*{The Taub Kasner point $T_1$}

The Taub Kasner point $T_1$ given by $\beta=0$
 causes difficulties for numerical 
simulations. It causes blow-up in the $\beta$-normalized variables because 
$\beta$ is zero here. This problem cannot be avoided by using 
Hubble-normalized variables or other gauges, because the Taub Kasner point 
is qualitatively different from other Kasner points, namely it is 
non-silent in the $x$-direction.
We can only simulate solutions that do 
not get too close to the Taub Kasner point during the period of interest.
As a general rule of thumb,
the more inhomogeneous the initial condition is, the more likely we will 
run into the Taub Kasner point at an early stage in numerical simulations.

%% file: simulations.tex
        \chapter{Numerical simulations}\label{chap:sim}

In this chapter, we present a variety of numerical simulations of $G_2$ 
cosmologies that
confirm the analytical predictions made in Chapter~\ref{chap:past}. 
The numerical results also illustrate the spike transitions, which we were
unable to predict analytically in Chapter~\ref{chap:past}.
In addition we describe simulations of the future asymptotic approach to 
de Sitter, and the close-to-FL regime.

For most of the simulations,
\footnote{The exceptions are the simulation of vacuum models in 
Section~\ref{sec:multiple}, 
the shock wave simulation in Section~\ref{sec:shock},
and the close-to-FL simulation in 
Section~\ref{sec:intermediate}.}
we shall specify initial conditions by following the format (\ref{IC}) 
introduced in Chapter~\ref{chap:G2}:
\begin{gather}
        \EEE=(\EEE)_0\ ,\quad
        r=\eps \sin x,\quad
        A = A_0 + \epsilon \sin x,\quad
        \Sm = (\Sm)_0 + \epsilon \sin x,
\notag\\
        \Nc = \sqrt{3}A_0 + \tfrac{\epsilon}{\sqrt{3}} \sin x,\quad
        \Nm = (\Nm)_0 + \epsilon \sin x,\quad
        \Sc = (\Sc)_0 + \epsilon \sin x,
\notag\\
        \Omega = \Om_0 + \epsilon \sin x,
\label{IC_format}
\end{gather}
where the subscript 0 indicates a constant.
$\Oml$ and $\St$ are given by
(\ref{Oml_St}):
\be
        \Oml = (\Oml)_0 \exp\left[-2\tfrac{\epsilon}{(\EEE)_0}\cos x 
\right],
        \quad
        \St = (\St)_0 \exp\left[-3\tfrac{\epsilon}{(\EEE)_0}\cos x 
\right]\ ,
\label{Oml_St_sim}
\ee
while $\Sp$ and $v$ are computed numerically using 
(\ref{Sp}), (\ref{define_Q}) and (\ref{v_init}), and have
complicated expressions.
For more special $G_2$ actions, we set the appropriate variables to 
zero.

While $\beta$-normalized variables are used in the simulations, 
Hubble-normalized variables are displayed in most of the plots.

\section{Asymptotic silence}\label{sec:sim_silent}

In this section we shall provide evidence to support the occurrence of
 asymptotic silence not only as $t \rightarrow -\infty$, 
namely,
\be
	\lim_{t \rightarrow -\infty} \EEE =0\ ,\quad
	\lim_{t \rightarrow -\infty} r =0\ ,
\ee
but also as $t \rightarrow \infty$ in the presence of a cosmological 
constant.

\begin{figure}
\begin{center}
    \epsfig{file=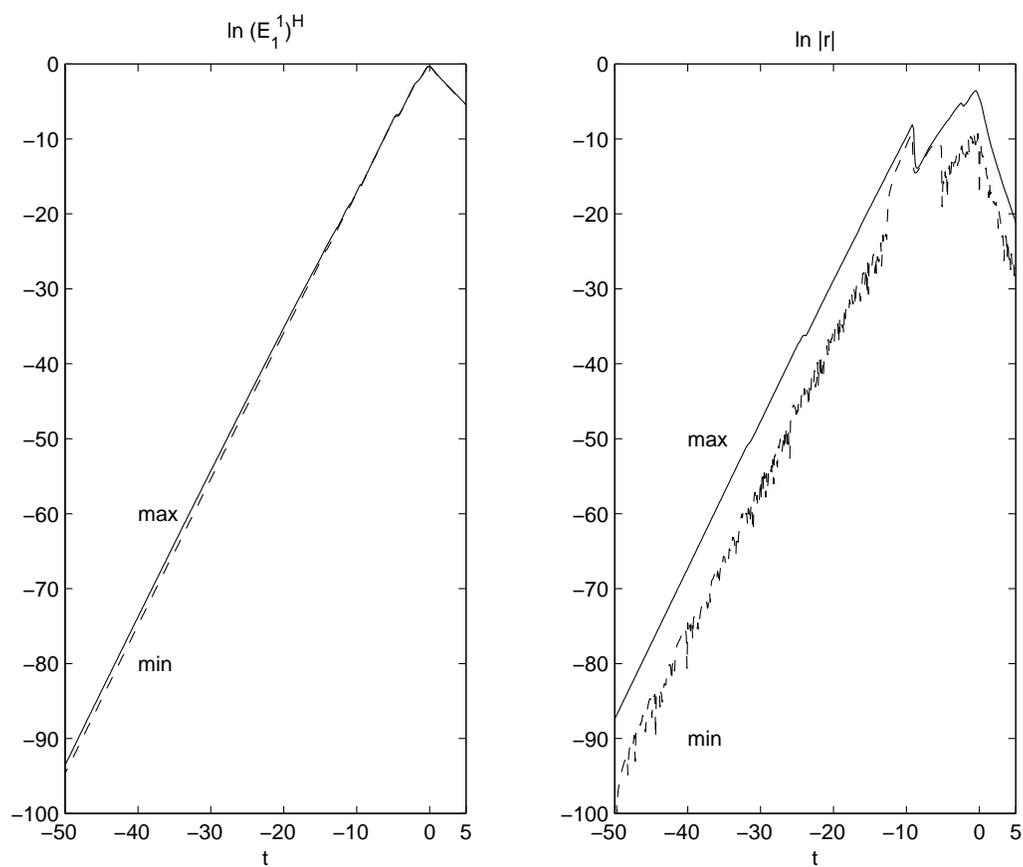,width=\textwidth}
\end{center}
\caption[Exponential decay of $(\EEE)^H$ and $r$ as $t \rightarrow -
\infty$ and as $t \rightarrow  \infty$.]
	{Exponential decay of $(\EEE)^H$ and $r$ as $t \rightarrow - 
\infty$ and as $t \rightarrow  \infty$,
showing the maximum and minimum values of $\ln(\EEE)^H$ and $\ln |r|$.
The graph for $\min\limits_{0 \leq x \leq 2\pi} \ln |r|$ is broken because 
$r$ has both signs.}\label{fig:lnEEEh_lnr}
\end{figure}

Using the format (\ref{IC_format}) for initial conditions,
consider the following initial data at $t=0$, with $\gamma  = 
\tfrac{4}{3}$:
\begin{gather}
	\epsilon= 0.01,\quad
	(\EEE)_0= 1  ,\quad
	A_0	= 0.2,\quad
	(\Sm) _0= 0.2,\quad
        (\Sc) _0  = 0.4,
\notag\\
        (\Nm) _0  = 0.4,\quad
        \Om _0  = 0.5,\quad
        (\Oml)_0 = 0.7,\quad
        (\St)_0  = 0.1.
\label{IC_1}
\end{gather}

We ran {\tt ICN} with 512 grid points, from $t=0$ to $t=-50$, and from 
$t=0$ to $t=5$ storing data at intervals of $t=0.1$.
First, we plot the maximum and minimum of $\ln\, (\EEE)^H$ and
$\ln\, |r|$ 
against $t$ in Figure~\ref{fig:lnEEEh_lnr}
We see that $(\EEE)^H$ and $r$ decay exponentially 
both into the past and into the future, supporting the conjecture that 
the past and future dynamics are asymptotically silent.%
\footnote{Actually, this is only true along almost all timelines.
In Section~\ref{subsec:spike_transition}, we shall see that along some 
timelines $r$ can become large, but the spatial resolution in this run is 
too low to simulate it correctly.}
Thus it is reasonable to say that the local background 
dynamics of $G_2$ cosmologies is the SH dynamics.

\section{Background dynamics}

In this section we illustrate the role of SH dynamics in $G_2$ dynamics.

\subsection{Local Mixmaster dynamics}

\begin{figure}
\begin{center}
    \epsfig{file=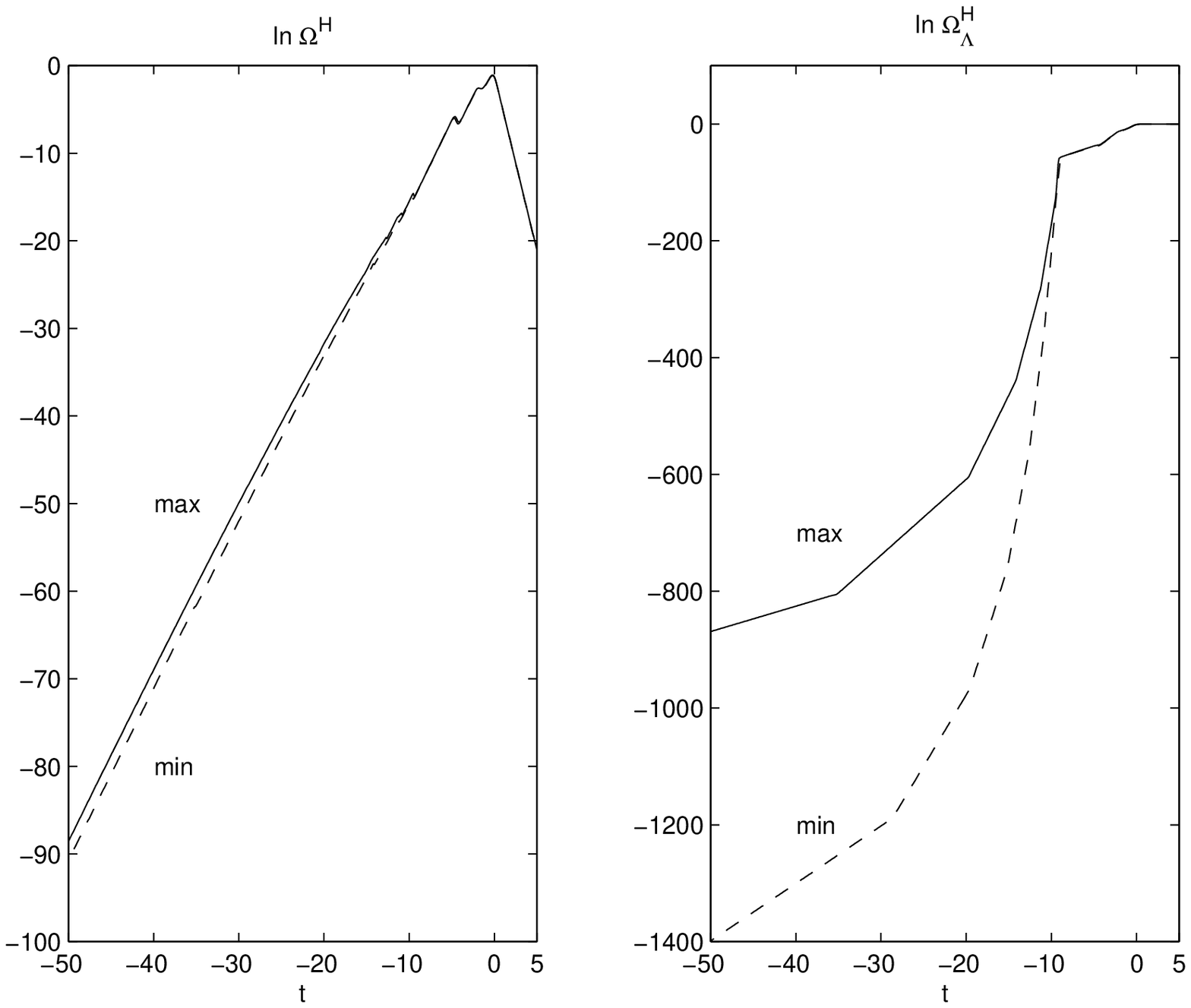,width=\textwidth}
\end{center}
\caption[Exponential decay of $\Omega^H$ and $\Oml^H$ as $t \rightarrow
        -\infty$.]
	{Exponential decay of $\Omega^H$ and $\Oml^H$ as $t \rightarrow 
	-\infty$,
showing the maximum and minimum values of $\ln\Omega^H$ and $\ln \Oml^H$.}
\label{fig:BKL_I}
\end{figure}

Still using the initial data (\ref{IC_1}),
we plot the maximum and minimum of $\ln(\Omega^H)$ and $\ln(\Oml^H)$ 
against $t$ in Figure~\ref{fig:BKL_I}. We see that $\Omega^H$ and $\Oml^H$ 
decay exponentially into the past, supporting the BKL conjecture I.

\begin{figure}
\begin{center}
    \epsfig{file=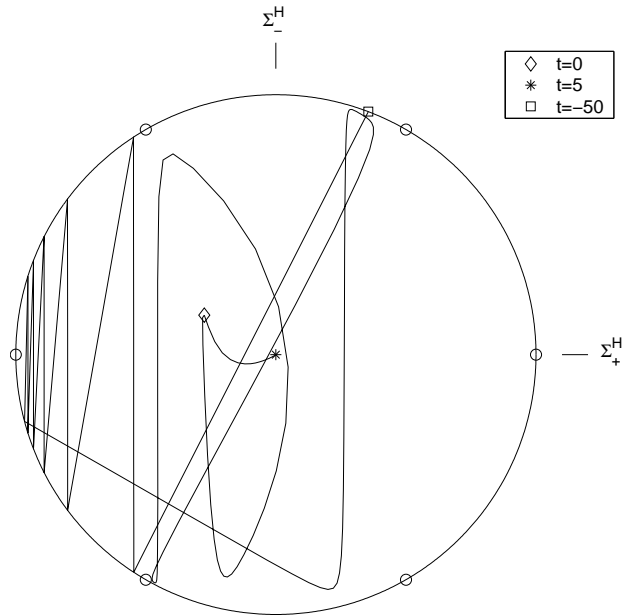,width=0.6\textwidth}
\end{center}
\caption{Typical evolution along a fixed timeline, showing the transition 
orbits.}\label{fig:May26_state}
\end{figure}

\begin{figure}
\begin{center}
    \epsfig{file=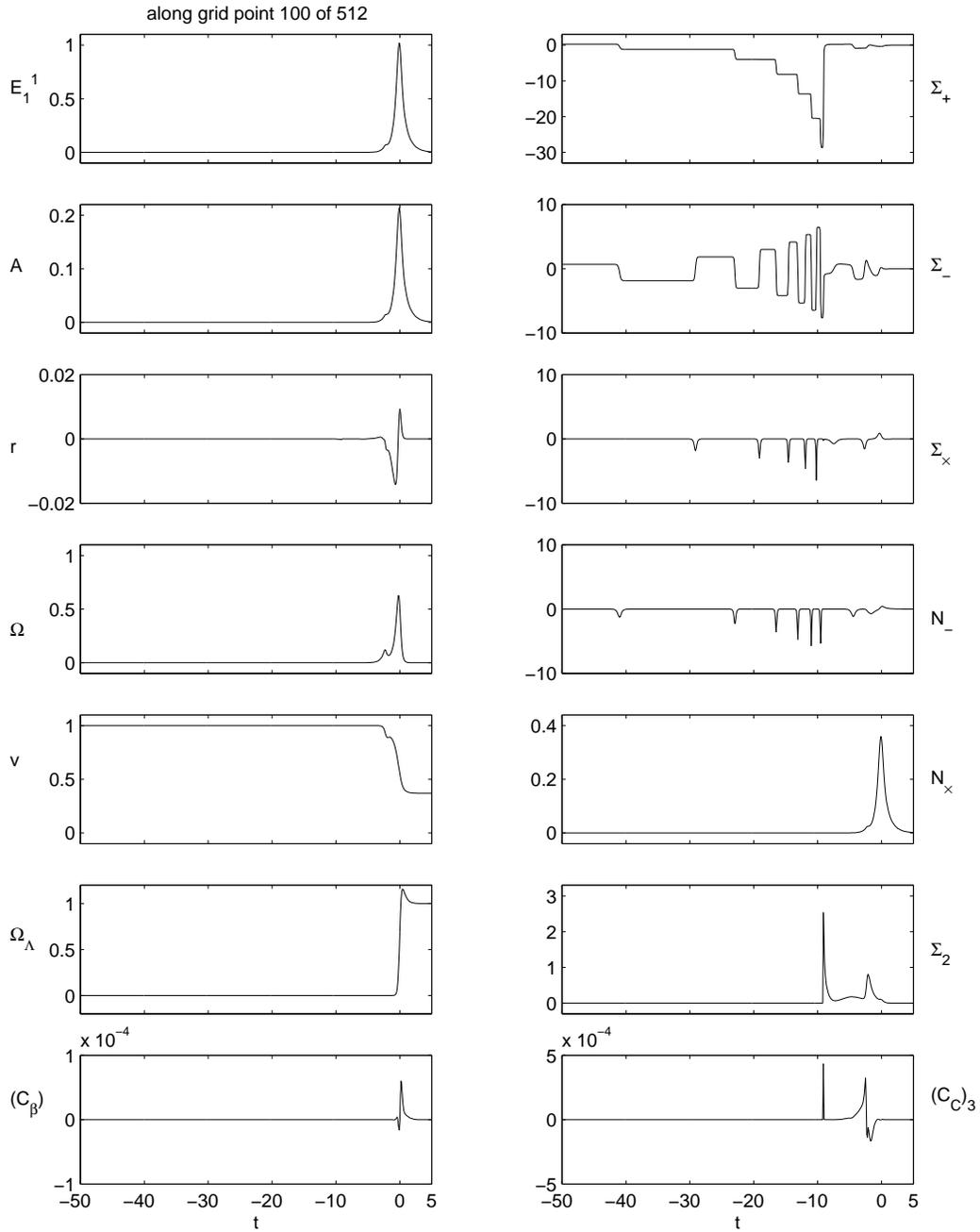,width=\textwidth}
\end{center}
\caption{Typical evolution along a fixed timeline.}\label{fig:May26}
\end{figure}

Figures~\ref{fig:May26_state}--\ref{fig:May26} show the evolution along a 
fixed timeline $x=const.$
 (the grid point 100 of 512). 
They show the generalized Mixmaster dynamics into the past and 
the approach to de Sitter into the future.

Evolving into the past from $t=0$, the solution enters the Kasner regime 
at about $t=-3$. One prominent $\St$ transition is observed at about 
$t=-9$, followed by an alternating sequence of $\Nm$ and $\Sc$ 
transitions (see also Figure~\ref{fig:May26} below).

Figure~\ref{fig:May26} shows that $v$ is stuck at $v=1$ as $t \rightarrow 
-\infty$, and tends to a non-zero limit as $t \rightarrow \infty$.
For more details about the future asymptotics, see 
Section~\ref{sec:future}.

Table~\ref{tab:rates} shows the expected and actual growth rates into the 
past (positive value indicates growth into the past) along grid point 100, 
at 
two selected times. 
The actual rates are in good agreement with the expected rates, which can 
be inferred from the $\beta$-normalized 
version of (\ref{lin1})--(\ref{lin3}).

\begin{table}[h]
\begin{spacing}{1.1}
\caption[Predicted and actual growth rates of the triggers]
	{Predicted and actual growth rates of the triggers into the past 
	(positive value indicates growth into the past) along grid point 
	100 of 512.}
                \label{tab:rates}
\begin{center}
\begin{tabular}{|c|cccc|}
\hline
Triggers & $\Nm$ & $\Sc$ & $\St$ & $\sqrt{1-v^2}$
\\
\hline
\multicolumn{3}{l}{At $t=-50$ ($\phi = 1.2076$ or $69.2^\circ$)}
\\
\hline
Expected rate &     $-4.3894$ &  2.3894 &   $-0.4083$ &  0.7864
\\
Actual rate   &	    $-4.4174$ &  2.3920 &   $-0.4083$ &  0.7864
\\
\hline
\multicolumn{3}{l}{At $t=-35$ ($\phi = 4.1341$ or $236.9^\circ$)}
\\
\hline
Expected rate &    4.3976 &  $-6.3976$ &  $-0.4173$ &  $-3.6161$
\\
Actual rate   &    4.4040 &  $-6.4323$ &  $-0.4173$ &  $-3.6161$
\\
\hline
\end{tabular}
\end{center}
\end{spacing}
\end{table}

\newpage
\subsection{Spatial variation in the background dynamics}
	\label{subsec:Mexico}

\begin{figure}
\begin{center}
    \epsfig{file=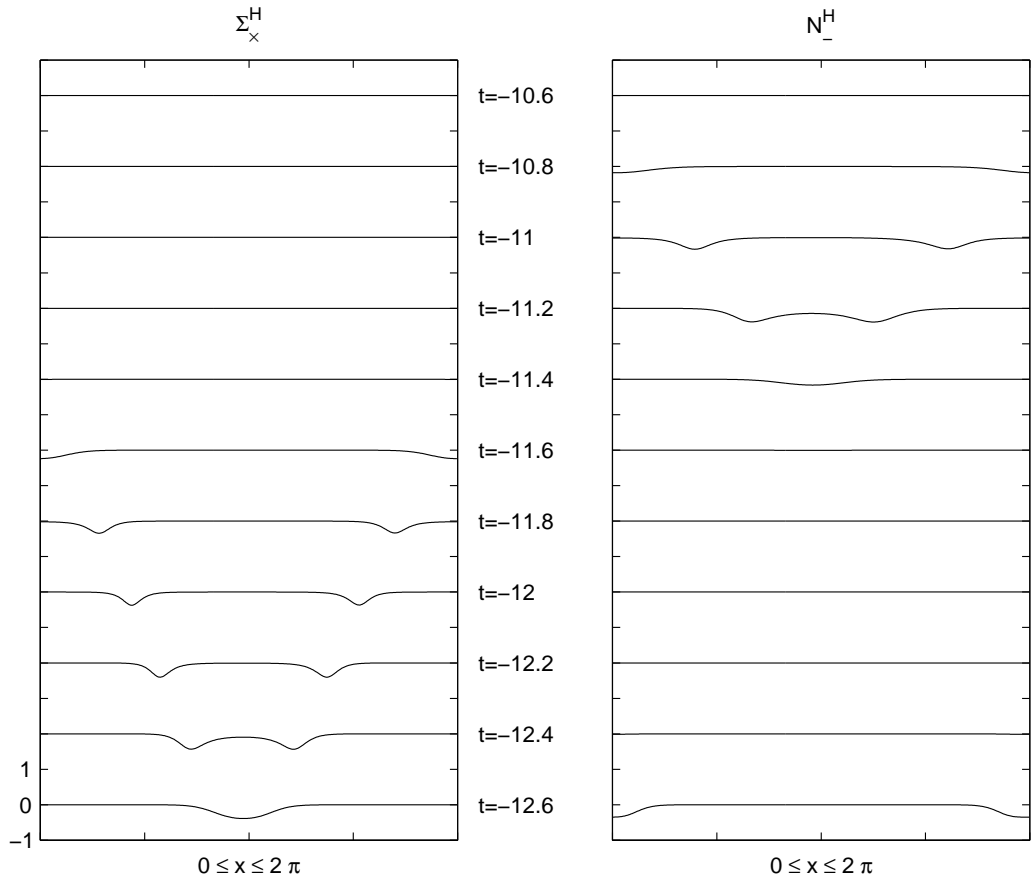,width=\textwidth}
\end{center}
\caption{Snapshots of triggers $\Sigma_\times^H$ and 
	$N_-^H$, showing pulses 
apparently travelling at superluminal speeds.}\label{fig:May26_triggers}
\end{figure}

The timing of the transitions in Mixmaster dynamics varies 
from one timeline to another.
Continuing with the initial data (\ref{IC_1}),
Figure~\ref{fig:May26_triggers} shows some snapshots of the triggers 
$\Sc^H$ and $\Nm^H$, illustrating this variation in the timing.
The variation in the timing leads to a pulse-like profile of the triggers.

As discussed in Section~\ref{sec:past_as},
the particle horizon is given by $\int_{-\infty}^0 \EEE dt$.
Since $\EEE=1$ at $t=0$ here and decays like $e^{2t}$,
$\int_{-\infty}^0 \EEE dt$ is approximately equal to $\tfrac{1}{2}$,
and so information can travel only a finite distance of $\Delta 
x\approx\tfrac{1}{2}$ 
as $t\rightarrow \infty$.
But the pulses in Figure~\ref{fig:May26_triggers} travel more than 
$\Delta x=\tfrac{1}{2}$, which is only possible if they propagate faster 
than light.
How is this possible?
In fact, the apparent superluminal speed is an illusion created by the 
variation in the background SH dynamics from one timeline to another (e.g. 
variation in the timing of the transitions in Mixmaster dynamics). 
An analogy to this phenomenon is a Mexican wave in a stadium.

\newpage
\subsection{Multiple transitions}\label{sec:multiple}

It is simplest to analyze multiple transitions in the absence of 
spikes or step-like structures.
Since it is
 difficult to produce a spike-free solution over a very long run 
in a perfect fluid solution,%
\footnote{That the initial condition for $\Omega$ is large (see 
the comment following (\ref{v_range})) makes 
it difficult to predict whether a solution will be spike-free and will 
stay away from the Taub Kasner point for a long period of time.}
we resort to simulating vacuum solutions.

Using the scheme for the vacuum case in Section~\ref{sec:IC},
we choose $\Nc=0$ and specify the following vacuum initial condition:
\begin{gather}
	\EEE = A = 0.1 + 0.01 \sin x,\quad
\notag\\
	\Sc = -\Nm = 0.8 + 0.1 \sin x,\quad
	\St = 0.1\ .
\label{IC_multi}
\end{gather}
Then $\Sp$, $\Sm$ and $r$ are given by (\ref{vac_Sp})--(\ref{vac_r}).

\begin{figure}
\begin{center}
    \epsfig{file=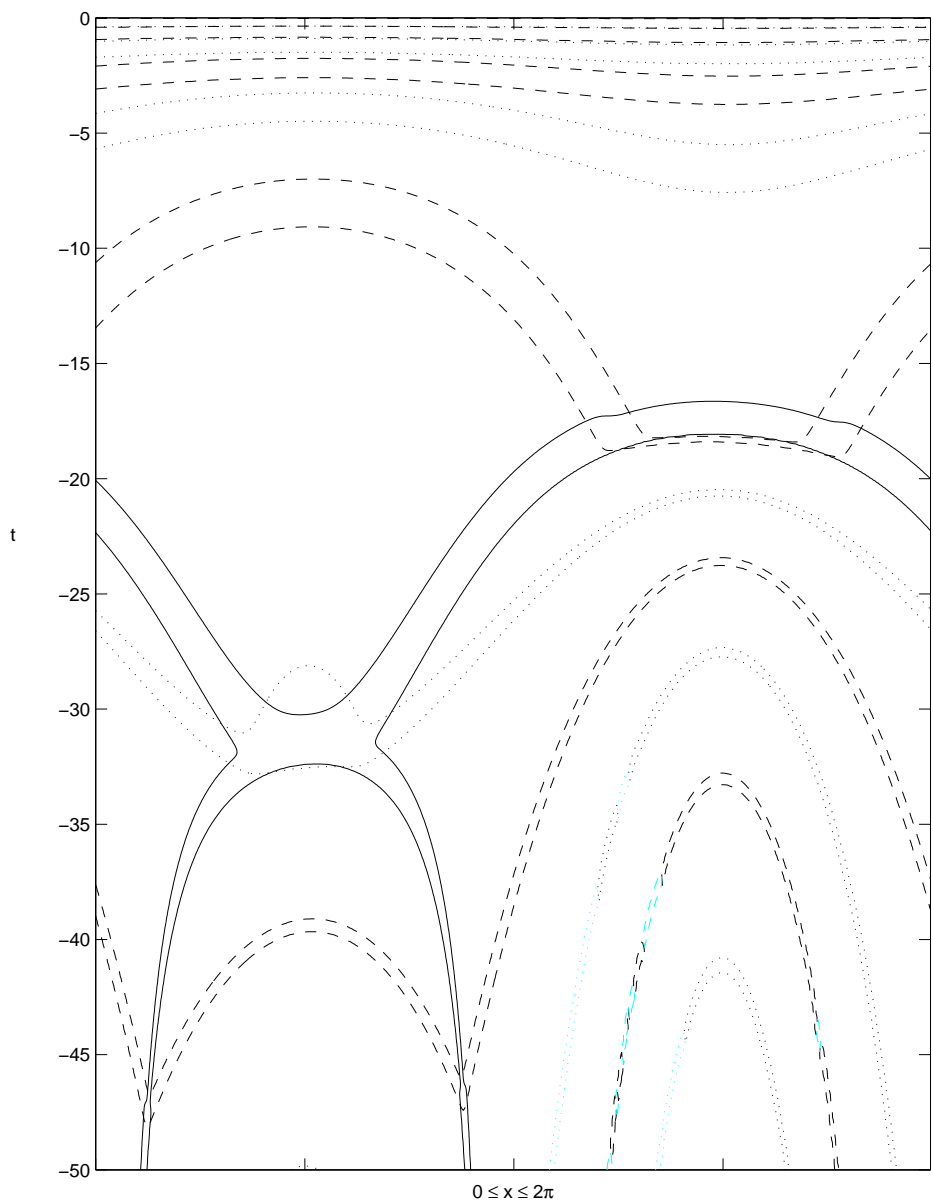,width=0.9\textwidth}
\end{center}
\caption[Multiple transitions.]
	{The level curves of $\Sigma_2^H=0.1$ (solid), $\Nm^H=-0.1$ 
(dashed) and $\Sc^H=0.1$ (dotted), indicating the activation of the 
triggers. Multiple transitions occur when two bands overlap.
For example, in a neighbourhood of $t=-32$ and $x=\pi/4$, the solid band 
and the dotted band overlap, indicating a multiple transition involving 
$\Sigma_2^H$ and $\Sc^H$.}
\label{fig:multi_band}
\end{figure}

We ran {\tt CLAWPACK} with 512 grid points, from $t=0$ to $t=-50$, storing 
data at intervals of $t=0.1$.
Figure~\ref{fig:multi_band} shows bands of level curves indicating the 
activation of the triggers $\St$, $\Sc$ and $\Nm$. The result is 
spike-free up to $t=-33$.
Spikes occur in the lower right quarter of 
Figure~\ref{fig:multi_band}.

In Figure~\ref{fig:May26_triggers} of the previous section,
we saw that there is variation in the timing of the transitions in
Mixmaster dynamics, due to spatial inhomogeneity.
Figure~\ref{fig:multi_band} shows that the
sequence of the transitions can also vary from one timeline to another.
For example, proceeding along the timeline $x=\pi$ as $t \rightarrow 
-\infty$, the solution undergoes an $\Nm$ transition first at about 
$t=-10$ before undergoing a $\St$ transition at about $t=-20$.
In comparison, proceeding along the timeline $x=\tfrac{3\pi}{4}$ as $t 
\rightarrow 
-\infty$, the solution undergoes a $\St$ transition first at about
$t=-17$ before undergoing an $\Nm$ transition at about $t=-18$.
Thus,
by continuity, there must be timelines between $x=\pi$ and 
$x=\tfrac{3\pi}{4}$ that undergo both $\St$ and $\Nm$ transitions at the 
same time, i.e. they undergo a multiple transition.

The intersection of the bands indicates multiple transitions.
We see that such intersections are inevitable in the global picture.
Thus multiple transitions will occur from time to time globally, as $t
\rightarrow -\infty$.
Locally along a fixed timeline, it is possible that
the probability of multiple transitions
occurring will tend to zero as $t \rightarrow -\infty$, in agreement with
the multiple transitions conjecture in SH dynamics, but we cannot make a
strong claim in this regard.

\newpage

\section{Effects of spatial inhomogeneity I: role of saddles}

In this section, we illustrate the role of saddle points in creating 
step-like or spiky structures. We also take this opportunity to illustrate 
the past attractors of diagonal and OT $G_2$ cosmologies.

\subsection{Step-like structures}

We now illustrate step-like structures in the tilt variable $v$ in
diagonal $G_2$ cosmologies, 
and provide evidence to support the claim that the
 past attractor is given by 
(\ref{G2_diag_at}):
\be
        \mathcal{A}^- = \mathcal{K}\
        {\rm arc}(\Sph >-\tfrac{1}{2}(3\gamma-4) )\
           \cup\ \mathcal{K}_{\pm1}\
        {\rm arc}(\Sph <-\tfrac{1}{2}( 3\gamma-4) ).
\ee
Recall that diagonal $G_2$ cosmologies satisfy (\ref{diag_G2}):
\be
	\St=0=\Sc=\Nm\ .
\label{diag_sim}
\ee

Consider the following initial data at $t=0$ 
with $\gamma  = \tfrac{4}{3}$
(in conjunction with (\ref{IC_format}) and (\ref{diag_sim})):
\begin{gather}
        \epsilon= 0.3,\quad
        (\EEE)_0= 1  ,\quad
        A_0     = 0,\quad
	(\Sm)_0 = 0.2,\quad
        \Sc \equiv 0,
\notag\\
        \Nm \equiv 0,\quad
        \Om _0  = 0.6,\quad
        (\Oml)_0 = 0.1,\quad
        \St \equiv 0.
\label{IC_diag}
\end{gather}

\begin{figure}
\begin{center}
    \epsfig{file=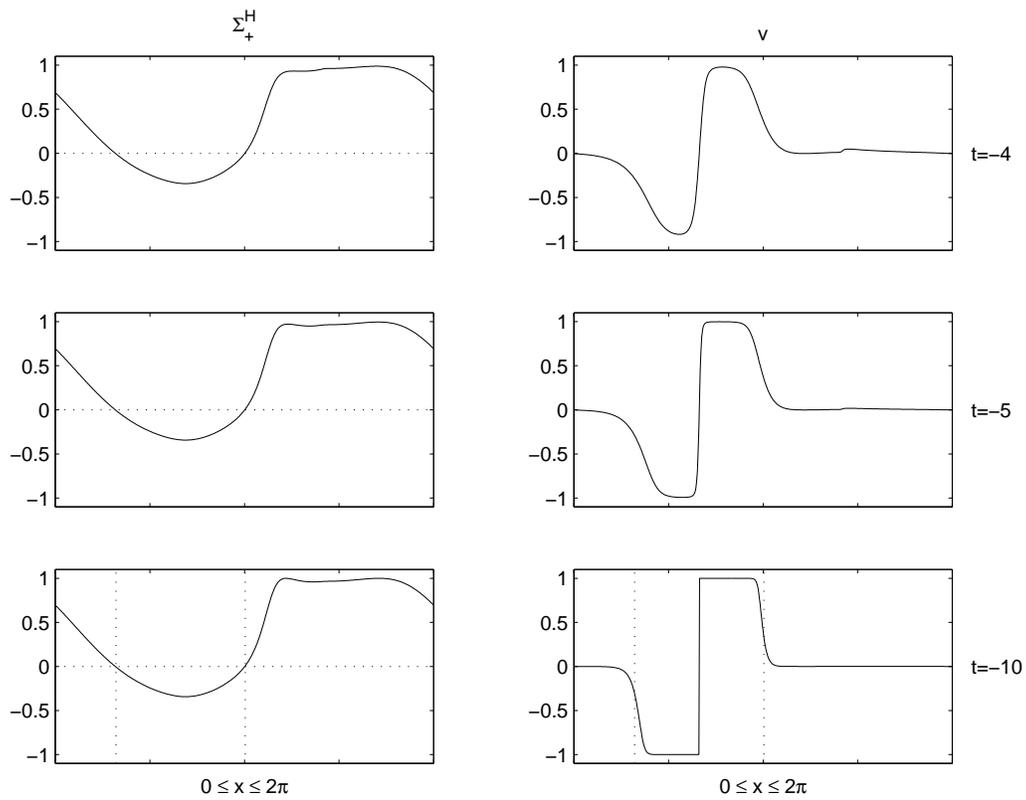,width=\textwidth}
\end{center}
\caption{Snapshots of $\Sph$ and $v$ in a 
	diagonal $G_2$ cosmology, showing step-like structures in $v$.}
\label{fig:diag_snap}
\end{figure}

We ran {\tt CLAWPACK} with 512 grid points, from $t=0$ to $t=-10$, storing 
data at intervals of $t=0.1$.
Figure~\ref{fig:diag_snap} shows the snapshots of the tilt variable $v$ at 
$t=-4$, $t=-5$ and $t=-10$. We see the strong suggestion that three 
step-like structures 
develop as $t \rightarrow -\infty$:
one between $v=\pm1$, and two between $v=0$ and $v=1$.
The one between $v=\pm1$ involves an unstable state $v=0$, while the two 
between $v=0$ and $v=1$ involve a string of Kasner equilibrium points with 
$\Sph=0$ and $-1 \leq v \leq 1$ that connect the stable arcs
$\mathcal{K}\ {\rm arc}(\Sph>0)$ and 
$\mathcal{K}_{\pm1}\ {\rm arc}(\Sph <0)$.

\begin{figure}
\begin{center}
    \epsfig{file=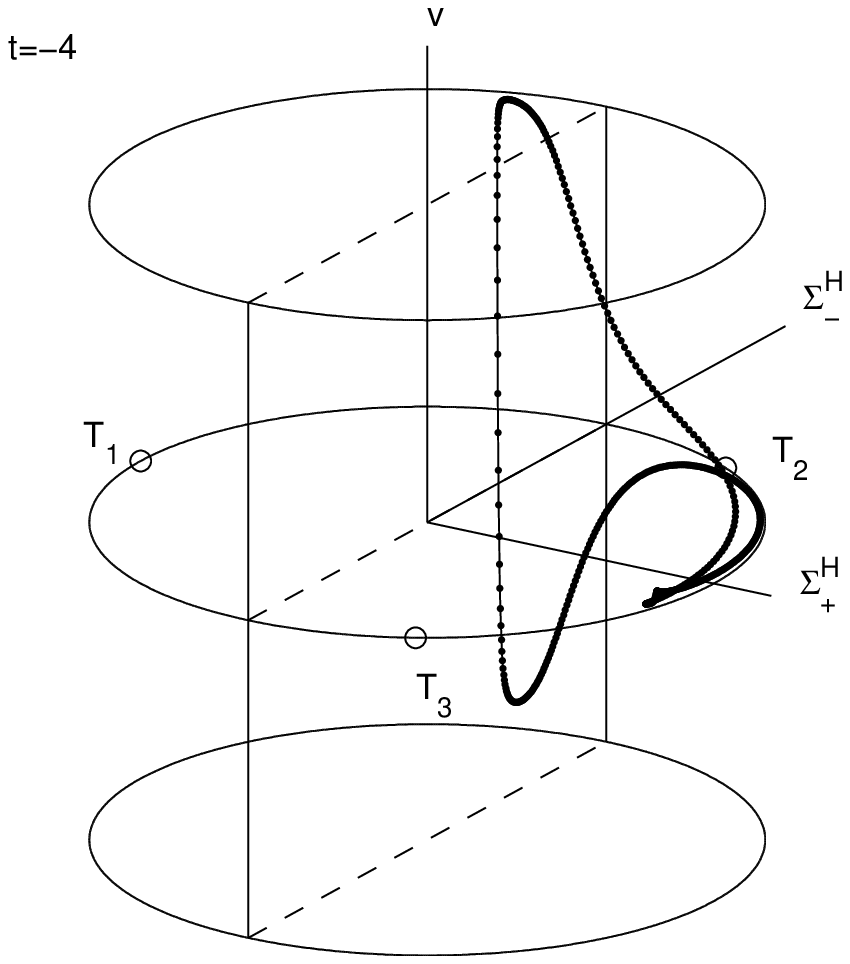,width=0.7\textwidth}  
\end{center}
\caption{Snapshot of the variables $(\Sph,\Smh,v)$ 
	in a diagonal $G_2$ cosmology at $t=-4$.}\label{fig:diag_step_4}
\end{figure}

\begin{figure}
\begin{center}
    \epsfig{file=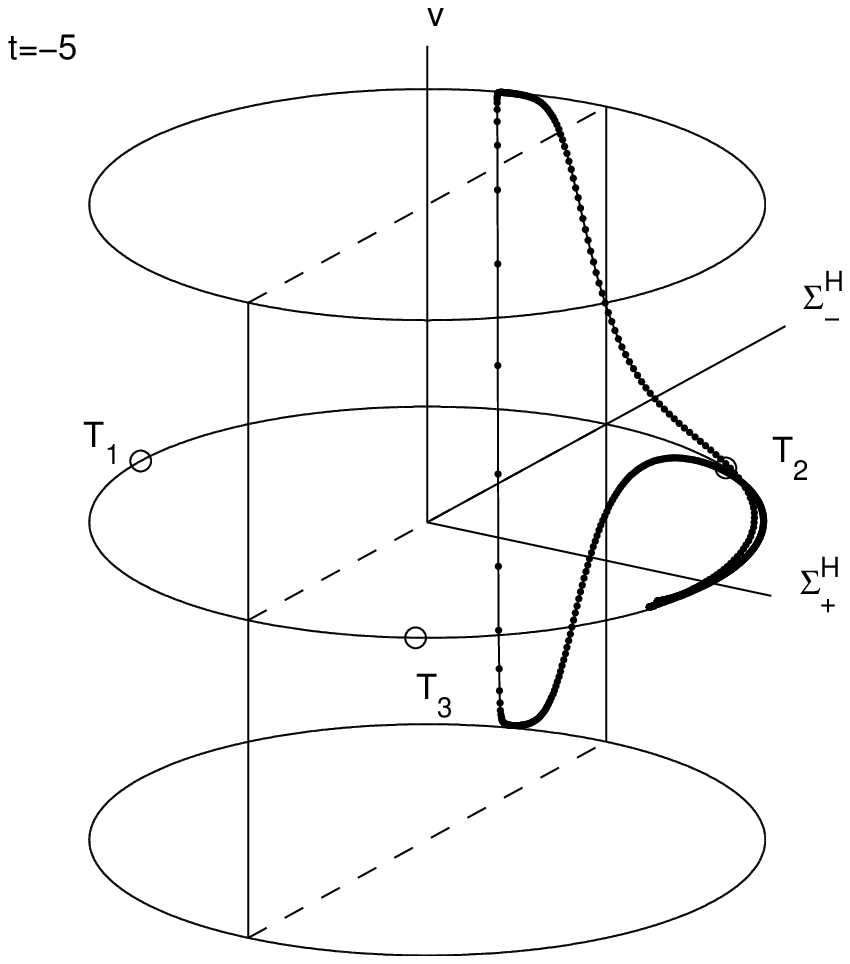,width=0.7\textwidth}
\end{center}
\caption{Snapshot of the variables $(\Sph,\Smh,v)$
        in a diagonal $G_2$ cosmology at $t=-5$.}\label{fig:diag_step_5}
\end{figure}

\begin{figure}
\begin{center}
    \epsfig{file=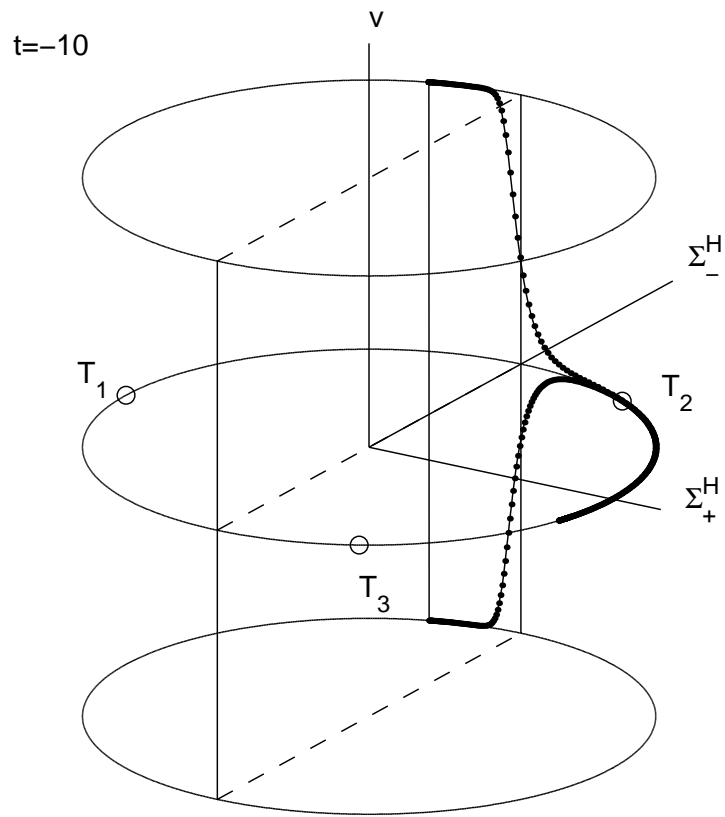,width=0.7\textwidth}
\end{center}
\caption{Snapshot of the variables $(\Sph,\Smh,v)$
        in a diagonal $G_2$ cosmology at $t=-10$.}\label{fig:diag_step_10}
\end{figure}

Figures~\ref{fig:diag_step_4}--\ref{fig:diag_step_10} show the snapshots 
of the variables $(\Sph,\Smh,v)$ in the $(\Sph,\Smh,v)$ space. Each dot
represents the state at a grid point.
We see that the state vector approaches the past attractor 
(\ref{SH_diag_at}).

\newpage
\subsection{Spiky structures}

We now illustrate spiky structures in the trigger $\Nm$ in
OT $G_2$ cosmologies,
and provide evidence to support the claim that the
 past attractor is given by
(\ref{G2_OT_at}):
\be
\label{sim_G2_OT_at}
        \mathcal{A}^- = \mathcal{K}\ \text{arc}(T_3 Q_1).
\ee

Recall that OT $G_2$ cosmologies satisfy (\ref{OT_G2}):
\be
        \St=0\ .
\label{OT_sim}
\ee

Consider the following initial data at $t=0$ with
$\gamma  = \tfrac{4}{3}$
(in conjunction with (\ref{IC_format}) and (\ref{OT_sim})):
\begin{gather}
        \epsilon= 0.1,\quad
        (\EEE)_0= 1  ,\quad
        A_0     = 0,\quad
        (\Sm)_0 = 0.2,\quad
        (\Sc)_0 = 0.2,
\notag\\
        (\Nm)_0 = 0,\quad
        \Om _0  = 0.6,\quad
        (\Oml)_0 = 0.1,\quad
        \St \equiv 0.
\label{IC_OT}
\end{gather}

\begin{figure}
\begin{center}
    \epsfig{file=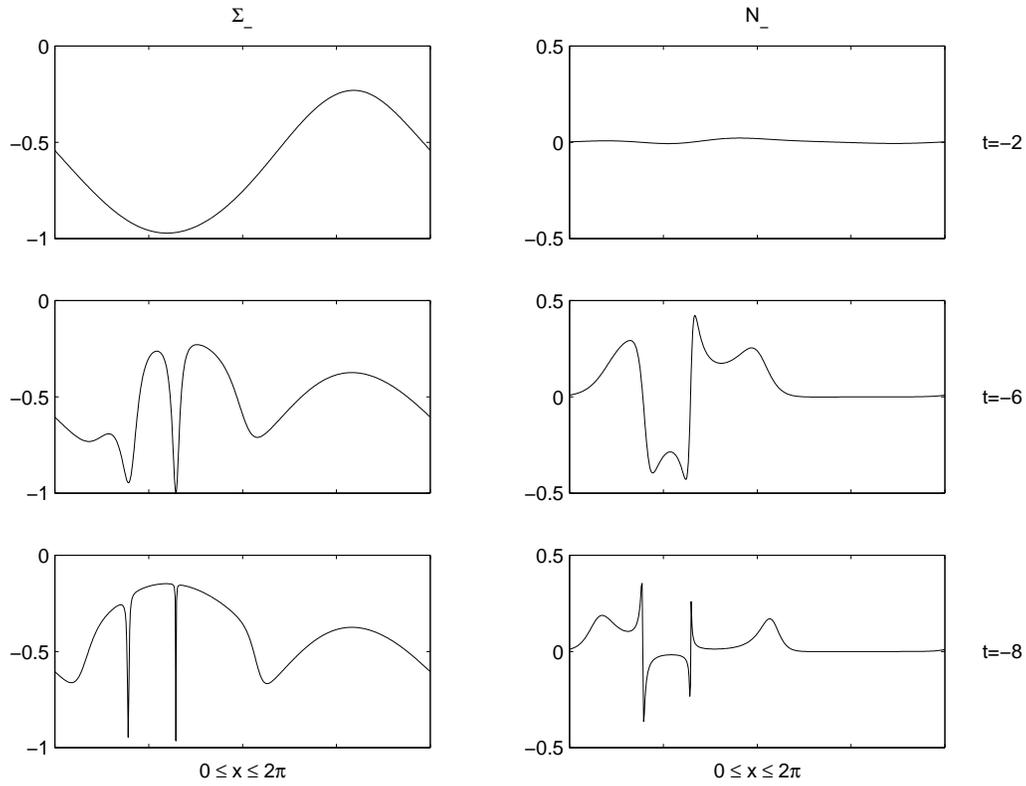,width=\textwidth}
\end{center}
\caption[Snapshots of $\Sm$ and the trigger $\Nm$.]
	{Snapshots of $\Sm$ and the trigger
$\Nm$ at $t=-2$, $t=-6$ and $t=-8$,
showing the asymptotic signature $(\Sm)_{\rm sig}(x)$ and the formation of 
permanent spikes in $\Nm$.}\label{fig:OT_snap}
\end{figure}

\begin{figure}
\begin{center}
    \epsfig{file=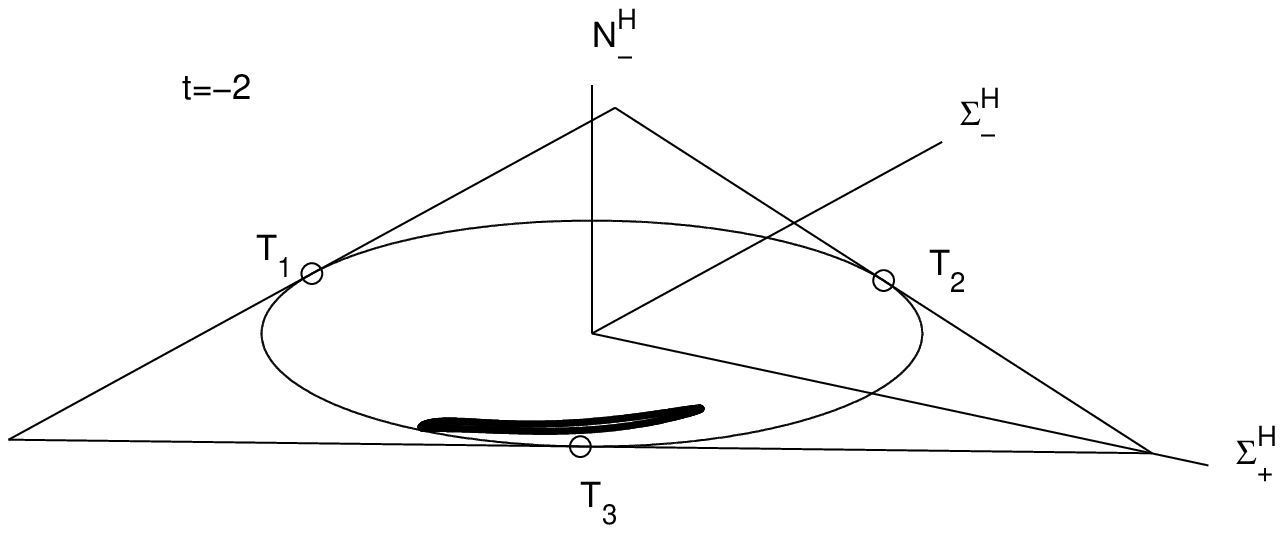,width=0.7\textwidth}
    \epsfig{file=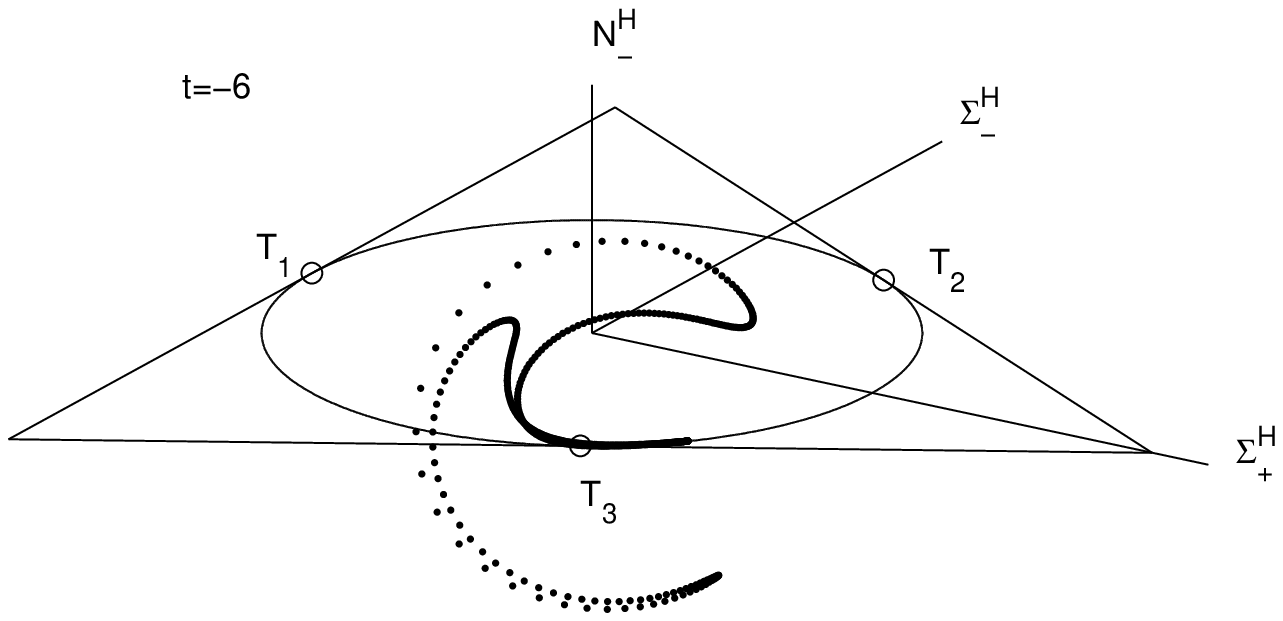,width=0.7\textwidth}
    \epsfig{file=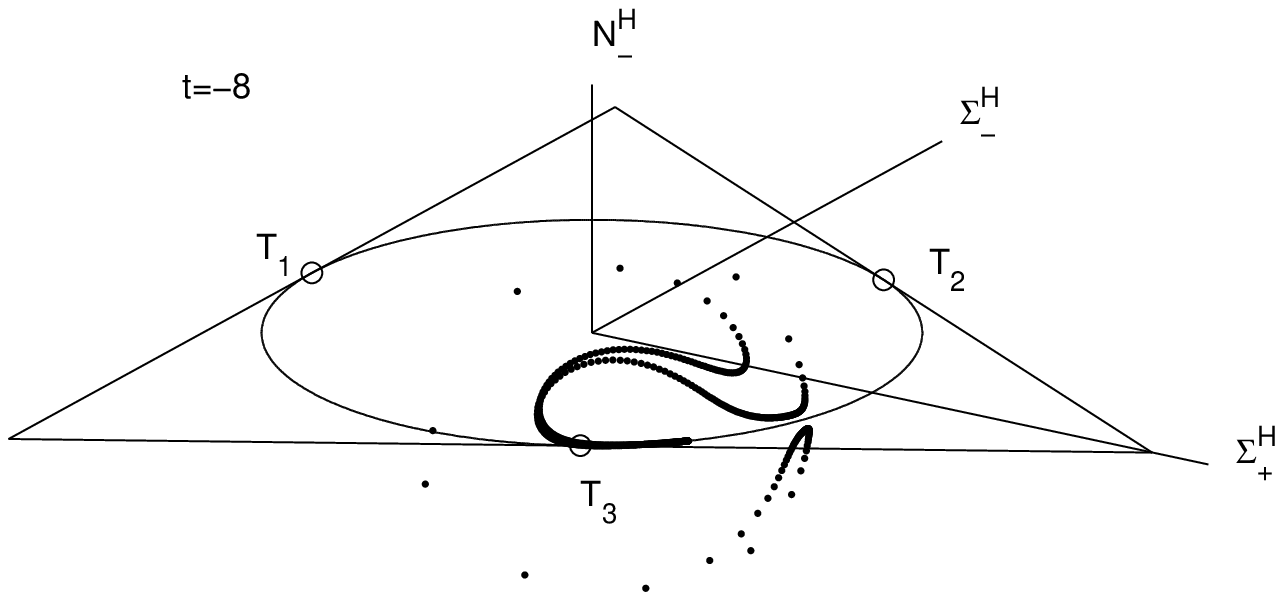,width=0.7\textwidth}
\end{center}
\caption{Snapshots of the variables $(\Sph,\Smh,\Nm^H)$ in the 
	$(\Sph,\Smh,\Nm^H)$ space. Each dot represents the solution at a 
	grid point.}
\label{fig:OT_spike_268} 
\end{figure}

We ran {\tt CLAWPACK} with 512 grid points, from $t=0$ to $t=-10$, storing 
data at intervals of $t=0.1$.
Figure~\ref{fig:OT_snap} shows the snapshots of $\Sm$ and the trigger 
$\Nm$ at $t=-2$, $t=-6$ and $t=-8$.
We see the strong suggestion that spikes develop as
$t \rightarrow -\infty$.
Figures~\ref{fig:OT_spike_268} shows the snapshots
of the variables $(\Sph,\Smh,\Nm^H)$ in the $(\Sph,\Smh,\Nm^H)$ space. 
Each dot represents the state at a grid point.
We see that the solution shadows the Taub vacuum Bianchi II orbits on the 
sphere (\ref{Nm_sphere}):
\[
        1 = (\Nm^H)^2 + (\Sp^H)^2 + (\Sm^H)^2
\]
along some grid points, as it
 approaches the stable Kasner arc between $T_3$ and $Q_1$. 
This may be compared
 with the Taub vacuum Bianchi II orbits shadowed by the transformed 
WM solution in Figure~\ref{fig:WM_3d}.
There are two points stuck on the arc to the left of the $T_3$ Taub Kasner 
point, creating two spikes. Along all other grid points, the orbits 
approach the past attractor (\ref{sim_G2_OT_at}).
The points closest to $T_3$ take the longest to traverse the Taub vacuum 
Bianchi II orbits.

The simulations also confirm that
\be
	\lim_{t \rightarrow -\infty} \parb_1 \X =0,
\ee
along spike points, despite large spatial gradients.
Thus asymptotic silence holds along spike points.
Our result has been mentioned in
Andersson \etal 2004~\cite[page S50]{art:Anderssonetal2004}.

\newpage

\section{Effects of spatial inhomogeneity II: role of spatial derivative 
terms}

In this section, we illustrate the role of spatial derivative terms in 
creating spike transitions and shock waves.

\subsection{Spike transitions}\label{subsec:spike_transition}

Consider the following initial data at $t=0$ with $\gamma  = \tfrac{4}{3}$
(in conjunction with (\ref{IC_format})):
\begin{gather}
        \epsilon= 0.1,\quad
        (\EEE)_0= 1  ,\quad
        A_0     = 0,\quad
        (\Sm) _0  = 0.2,\quad
        (\Sc) _0  = 0.2,
\notag\\
        (\Nm) _0  = 0,\quad
        \Om _0  = 0.6,\quad
        (\Oml)_0 = 0.1,\quad
        (\St)_0  = 0.1.
\label{IC_spike_trans}
\end{gather}

High resolution is essential to simulate a spike transition correctly.
We ran {\tt CLAWPACK} with 16384 grid points, from $t=0$ to $t=-4$,
storing data at intervals of $t=0.1$.
\footnote{The simulation in Section~\ref{sec:sim_silent} does not have 
high enough resolution to correctly simulate spike transitions, but was 
acceptable for the typical background SH dynamics.}

\begin{figure}
\begin{center}
    \epsfig{file=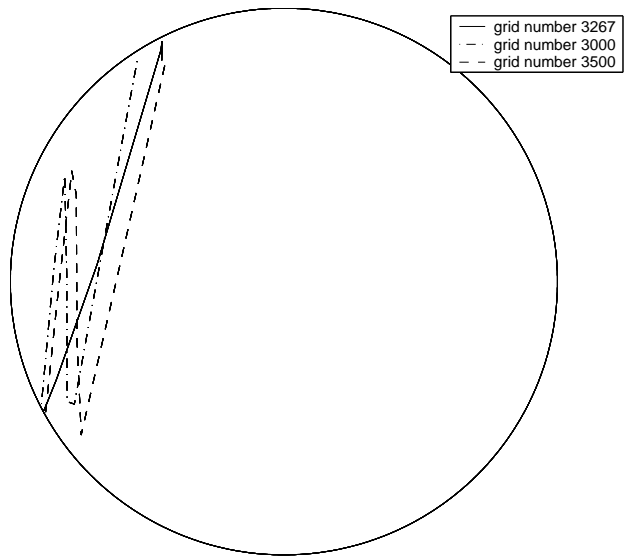,width=0.6\textwidth}
\end{center}
\caption[The orbit of an $N_-$ spike transition along $x_{\rm spike}$,
        from $t=-2.2$ to $t=-3.3$, projected onto
        the $(\Sph,\Smh)$ plane.]
	{The orbit of an $N_-$ spike transition along $x_{\rm spike}$, 
	from $t=-2.2$ to $t=-3.3$, projected onto 
	the $(\Sph,\Smh)$ plane, shown as a solid line. The orbits of
	a sequence of $\Nm$--$\Sc$--$\Nm$ transitions along two nearby 
	grid points are shown as a dashed line and a dash-dot line.}
\label{fig:spike_trans_path}
\end{figure}

Figure~\ref{fig:spike_trans_path} shows the orbits along three grid 
points: two typical points at grid number 3000 and 3500, and the spike 
point at grid number 3267. The typical points undergo a sequence of three 
transitions, while the spike point undergoes a spike transition.
The spike transition orbit is a straight line, taking the solution to the 
Kasner point that is reached by the sequence of $\Nm$--$\Sc$--$\Nm$ 
transitions, in about the same time.

\begin{figure}
\begin{center}
    \epsfig{file=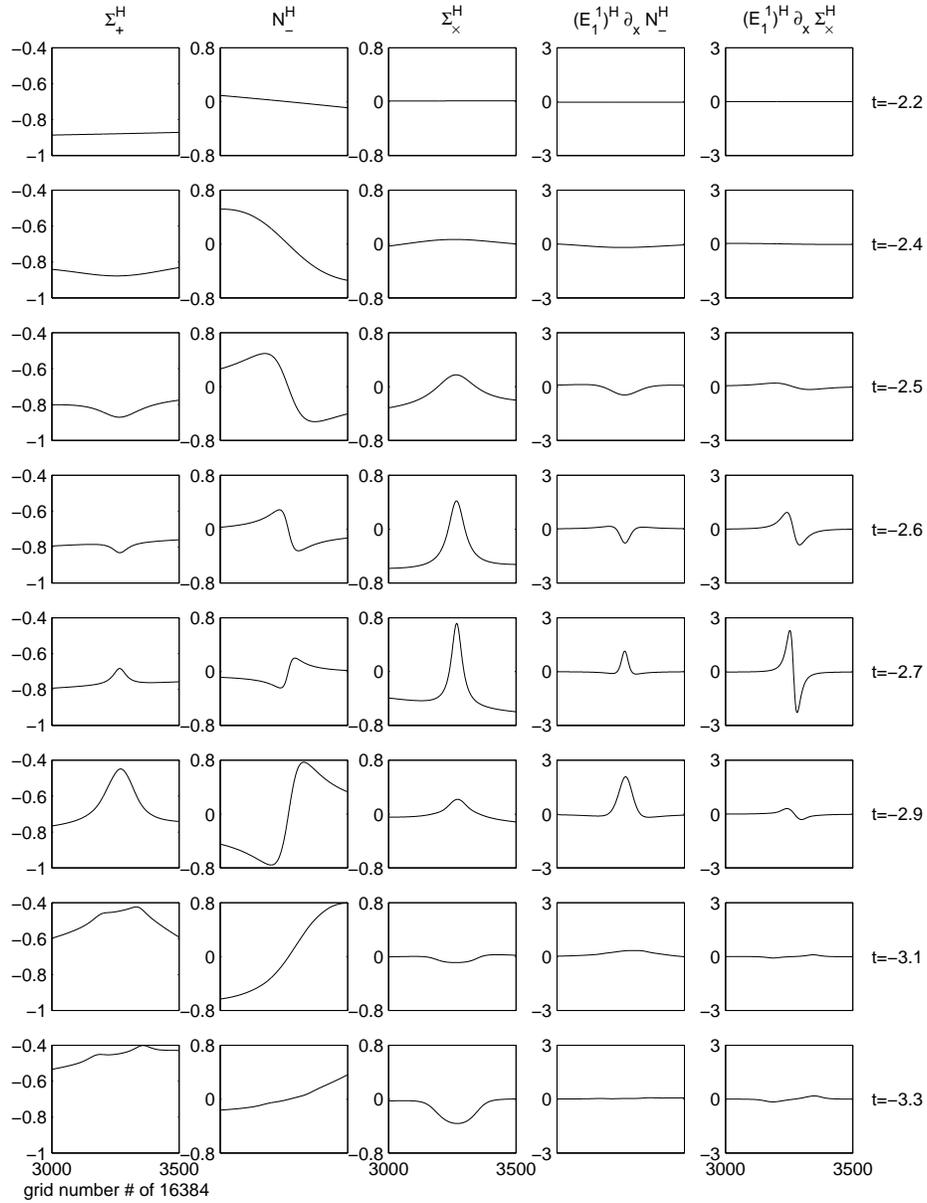,width=0.9\textwidth}
\end{center}
\caption[An $N_-$ spike transition.]
	{An $N_-$ spike transition. 
	Hubble-normalized variables are plotted.
	A reversal of the $N_-$ spike occurs between $t=-2.6$ and 
	$t=-2.7$.}
\label{fig:trans_snap}
\end{figure}

Figure~\ref{fig:trans_snap} shows the snapshots of $\Sph$, the 
triggers $\Nm^H$, $\Sc^H$ and 
the spatial derivative terms
$(\EEE)^H\, \ptl_x \Nm^H$,
$(\EEE)^H\, \ptl_x \Sc^H$ on the interval
between grid point 3000 and grid point 3500. 
A spike forms at $t=-2.6$
in all of the variables, 
but $\Sc^H$ also activates at the spike point 
(grid point 3267), and develops a very steep gradient.
In addition, 
$(\EEE)^H\, \ptl_x \Nm^H$
and
$(\EEE)^H\, \ptl_x \Sc^H$ develop spikes that have bigger amplitudes than 
the spikes in both $\Nm^H$ and $\Sc^H$.
A reversal of the spike in $\Nm^H$ occurs some time between $t=-2.6$ and 
$t=-2.7$.
The dynamics appears to be inherently inhomogeneous.

\begin{figure}
\begin{center}
    \epsfig{file=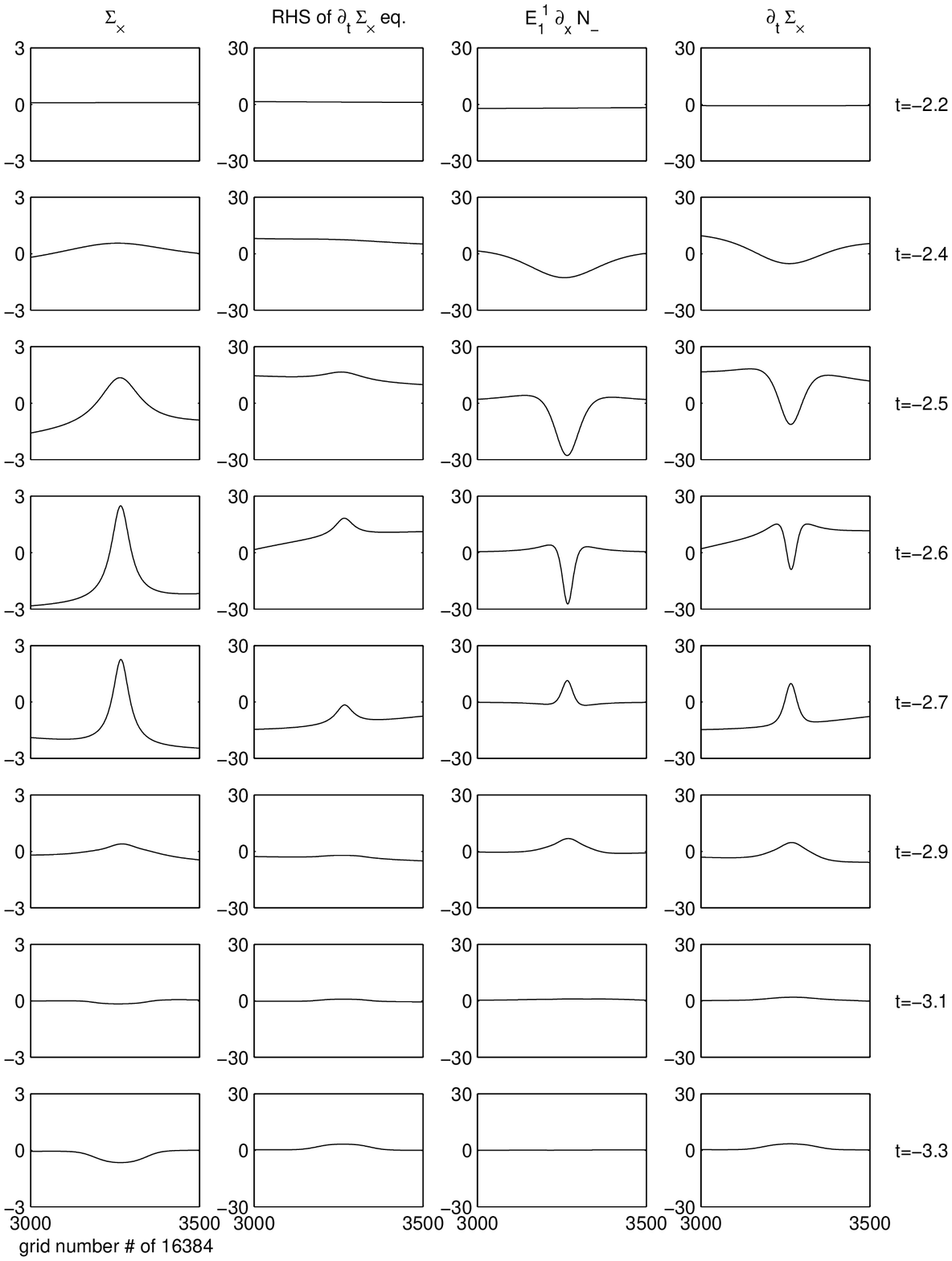,width=0.9\textwidth}
\end{center}
\caption[Influence of $\parb_1 \Nm$ on $\ptl_t \Sc$.]
	{Influence of $\EEE \ptl_x \Nm$ on $\ptl_t \Sc$. At $t=-2.4$, 
$\EEE \ptl_x
\Nm$ is large enough along the spike point that $\ptl_t \Sc$ changes sign, 
causing $\Sc$ to develop a spike.}
	\label{fig:trans_snap3}
\end{figure}

\begin{figure}
\begin{center}
    \epsfig{file=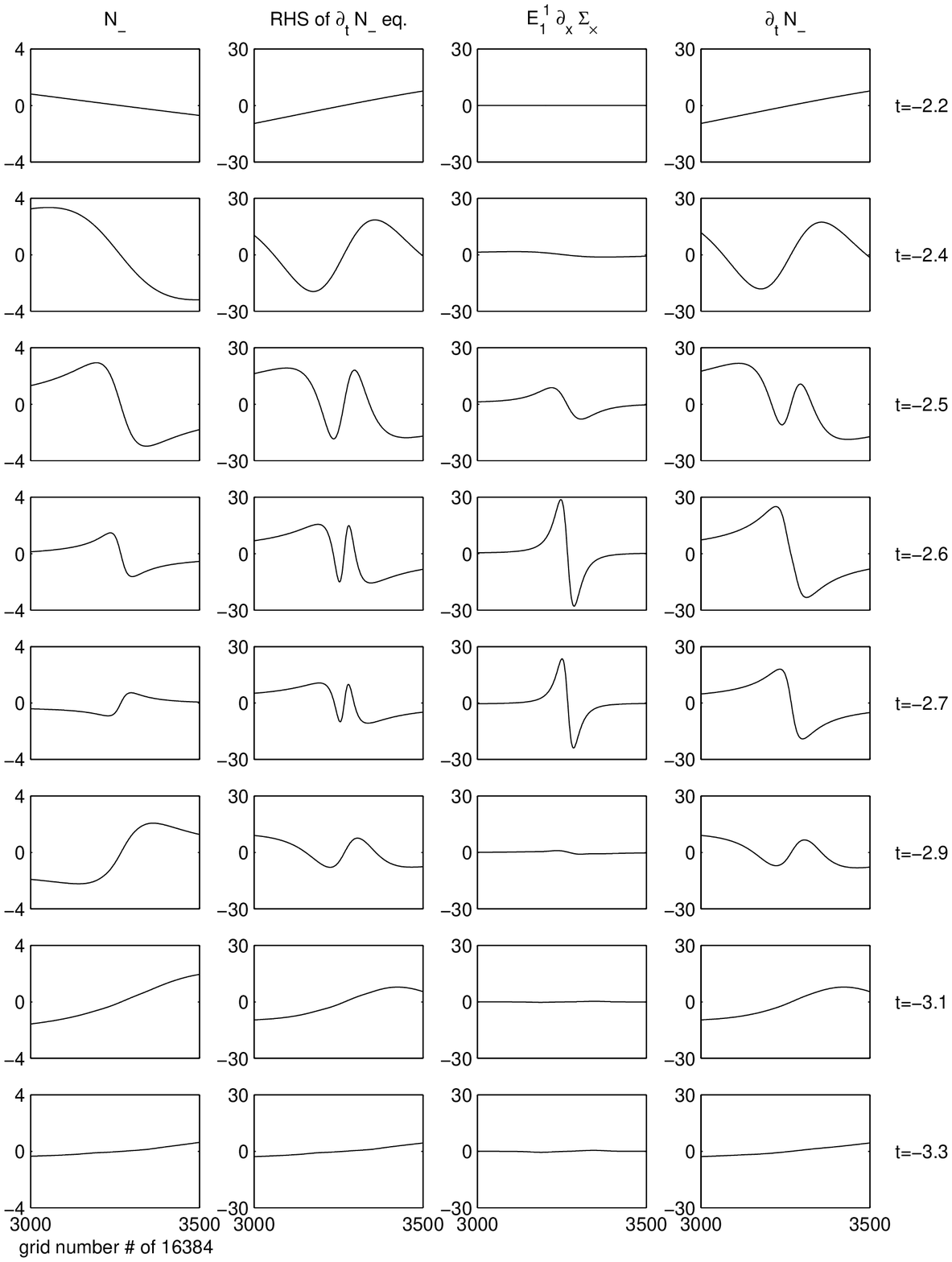,width=0.9\textwidth}
\end{center}
\caption[Influence of $\parb_1 \Sc$ on $\ptl_t \Nm$.]
	{Influence of $\EEE \ptl_x \Sc$ on $\ptl_t \Nm$.
	The $\Sc$ spike that was started by a large $\EEE \ptl_x \Nm$ at 
	$t=-2.4$, now becomes large enough at $t=-2.6$ that the slope of 
	$\ptl_t \Nm$ changes sign, causing the reversal of the $N_-$ 
	spike at $t=-2.7$.}
	\label{fig:trans_snap4}
\end{figure}

We now take a closer look at the dynamics.
Figures~\ref{fig:trans_snap3}--\ref{fig:trans_snap4} show the 
$\beta$-normalized terms in the evolution equations 
(\ref{app_Sc})--(\ref{app_Nm})
 for the pair $(\Sc,\Nm)$.
First, a spike forms in $\Nm$ with the same mechanism as permanent $\Nm$ 
spikes.
From Figure~\ref{fig:trans_snap3}, we see that $\parb_1 \Nm$ is 
significant enough that, when added to the algebraic terms on the right 
hand side of the $\dt \Sc$ equation, it leads to a sign change in $\dt 
\Sc$ at the spike point, which then activates $\Sc$ at the spike point to 
start the spike transition.
From Figure~\ref{fig:trans_snap4}, we see that
an active $\Sc$ at the spike point in turn gives a $\parb_1 \Sc$ large 
enough that, when added to the algebraic terms on the right   
hand side of the $\dt \Nm$ equation, it leads to a sign change in $\dt
\Nm$ at the spike point, which causes the reversal of the spike in $\Nm$.

What separates transient spikes like these from the permanent spikes that 
we observed in OT $G_2$ cosmologies?
Here, $\parb_1 \Nm$ is large enough to activate $\Sc$ at the spike point. 
In permanent spikes, $\parb_1 \Nm$ is too small, so $\Sc$ still decays to 
zero at the spike point. Whether $\parb_1 \Nm$ is large enough depends on 
the Kasner point where the spike occurs (see 
Section~\ref{sec:past_spike_trans}).

The terminology \emph{spike transition}
is introduced by the author and was first referred to by
Andersson \etal 2004~\cite[page S50]{art:Anderssonetal2004},
who gave a survey of regular transitions in vacuum OT $G_2$ cosmologies, 
using Hubble-normalized variables.
By examining the interplay between $\Nm$ and $\Sc$, we 
shed new light on the mechanism of spike 
transitions, building on the work of
Garfinkle \& Weaver 2003~\cite{art:GarfinkleWeaver2003}.
What we have seen here is the transition of a so-called high-velocity 
spike.
\footnote{The so-called velocity is equal to $\sqrt{3}\sqrt{\Sm^2+\Sc^2}$, 
and a 
spike transition is called a ``slide-down" in the velocity in 
\cite{art:GarfinkleWeaver2003} because 
$\Sm^2+\Sc^2$ decreases by a predicted amount during a spike 
transition. Garfinkle \& Weaver also recognized that a sign change in 
the evolution equations is crucial for a spike transition.}
 In OT $G_2$ cosmologies, a high-velocity spike is by definition a 
transient spike, which either reappears as a permanent, low-velocity 
spike eventually or disappears \cite{art:GarfinkleWeaver2003}.

Similarly, a transient spike can occur in $\Sc$, and will undergo a spike 
transition that is equivalent to a sequence of $\Sc$--$\Nm$--$\Sc$ 
transitions.

The simulation also suggests that spike transitions may violate asymptotic 
silence: 
as $t \rightarrow -\infty$, $\EEE$ tends to zero, 
but $r$ becomes large during a spike transition, due to the term $\Nm 
\Sc$ (see equation (\ref{r_app})).
\footnote{When doing linearization on the Kasner circles, we dropped the
spatial derivative terms. The linearized evolution equations for $\Nm$ and 
$\Sc$ indicate that $\Nm\Sc$ tends to zero into the past.
This linearization does not hold when the spatial derivative terms are 
large enough, as in the case of spike transitions.}
In other words, during a spike transition, the orbit does not shadow the 
silent boundary.
If spike transitions occur indefinitely as $t \rightarrow -\infty$, then 
asymptotic silence is violated.
Inadequate numerical spatial resolution confines the simulations to very 
short times, so we have not been able to accurately simulate two 
consecutive spike transitions. 
\footnote{For a fixed grid size $\Delta x$, the largest gradient of a jump 
in a variable $Y$ that can be represented is
 $(Y_{\max} - Y_{\min})/\Delta x$. 
A finer grid size is needed if the numerical gradient of a spike is 
observed to approach this upper bound
and $\EEE (Y_{\max} - Y_{\min})/\Delta x$ is observed to be growing, 
otherwise the numerical solution can underestimate $\parb_1 Y$ 
and thus can delay or even prevent a spike transition from occurring.
This can mislead one to predict that spike transitions will eventually 
stop as $t \rightarrow -\infty$, or that $\parb_1 Y$ tends to zero
as $t \rightarrow -\infty$ (see Andersson \etal 
2004b~\cite{art:Anderssonetal2004b}, Figure 2).}
We shall leave this for future research.

There are still questions about what happens to low-velocity spikes in 
generic $G_2$ cosmologies. Will they be permanent or will they undergo 
transitions? We have been unable to specify initial conditions that 
produce such transitions before numerical resolution becomes inadequate.
We shall also leave this for future research.

\newpage
\subsection{Shock waves}\label{sec:shock}

Numerical experiments with shock waves show that the constraints
$(\mathcal{C}_{\rm C})_3$
and
$(\mathcal{C}_\beta)$
(see (\ref{C_St})--(\ref{C_beta_app}))
 are
unstable numerically near shock waves.
To avoid this problem, we shall set $\Oml$ and $\St$ to zero.
As mentioned in Section~\ref{sec:hyperbolic}, the presence of $\parb_1 A$ 
on the right hand side may prevent the system from being symmetric 
hyperbolic, which may 
be bad for simulations of shock waves. To avoid this problem we shall set 
$A=0$.

Consider the following initial data at $t=0$ with $\gamma  = \tfrac{4}{3}$
(using the scheme for the case $A=0$ in Section~\ref{sec:IC}.
\begin{gather}
	A=0=\St=\Oml
\notag\\
	\EEE = 100,\quad
        \Sm = 0.1,\quad
        \Sc = 0.01,
\notag\\
        \Nm  = 0.01,\quad
	\Nc  = 0.01,\quad
        \Om  = 0.01,\quad
	v = 0.5 + 0.3 \sin x\ .
\label{IC_shock}
\end{gather}
Here, a large $\EEE$ is chosen to create inhomogeneities with a 
short-wavelength (relative to the horizon). The amplitude in the 
fluctuation of $v$ is also chosen to be large. These conditions are 
favourable for the formation of shock waves.

\begin{figure}
\begin{center}
    \epsfig{file=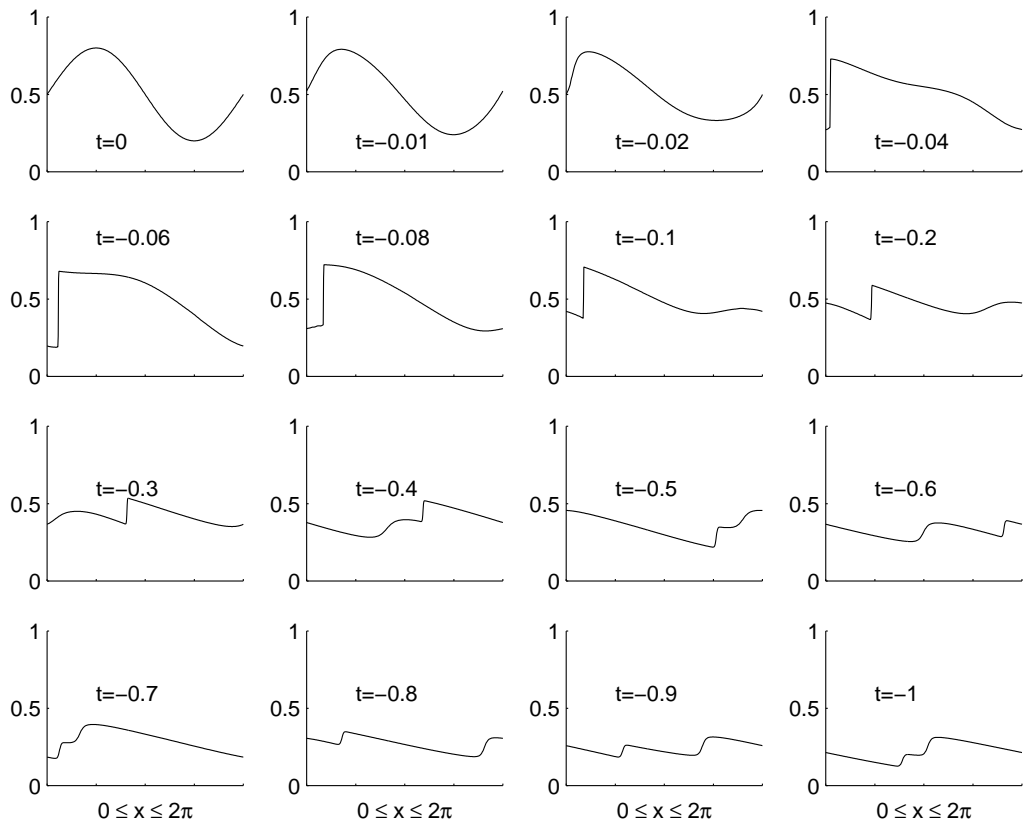,width=\textwidth}
\end{center}
\caption{Shock wave in $v$ into the past.}\label{fig:shock_snap}
\end{figure}

\begin{figure}
\begin{center}
    \epsfig{file=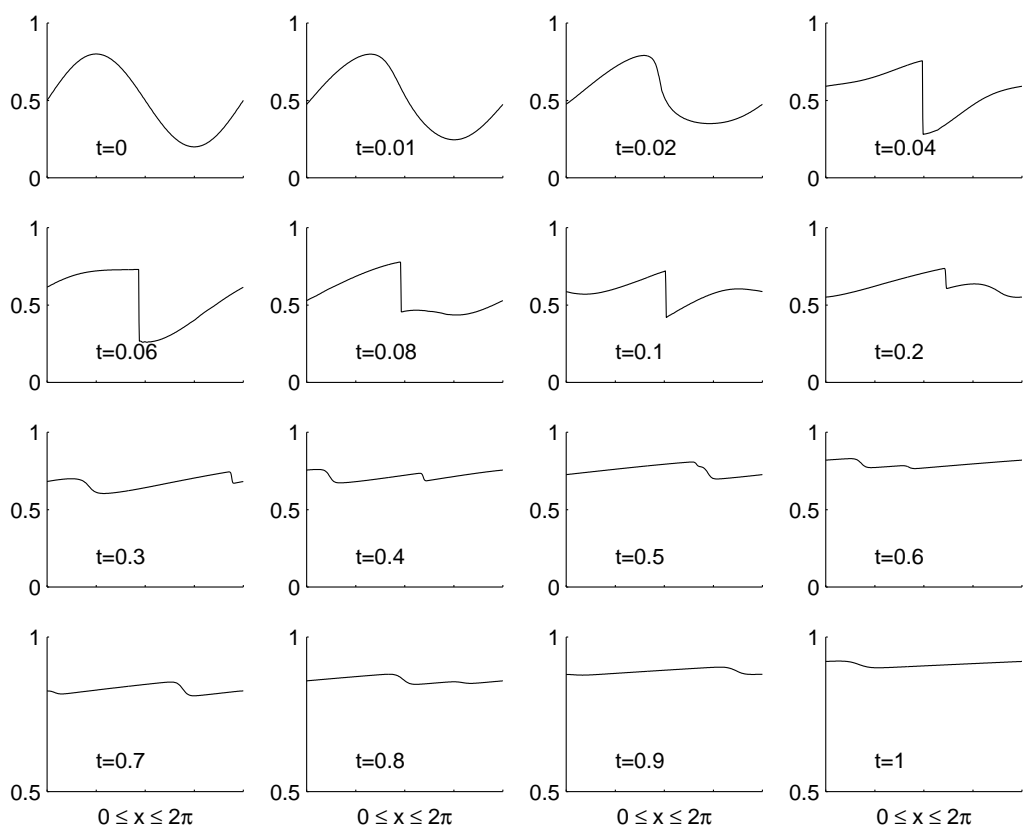,width=\textwidth}
\end{center}
\caption{Shock wave in $v$ into the future.}\label{fig:shock_snap_f}
\end{figure}

We ran {\tt CLAWPACK} with 512 grid points, from $t=0$ to $t=-1$,
and from $t=0$ to $t=1$
storing data at intervals of $t=0.01$.
Figures~\ref{fig:shock_snap}--\ref{fig:shock_snap_f}
show the formation of a shock wave in the tilt variable $v$ into the past 
and into the future respectively. 

The shock wave subsequently smoothes out or dissipates asymptotically.
The mechanism of this dissipation is not clear. SH dynamics cannot 
dissipate a discontinuity, but would a small $\parb_1 \X$ term dissipate 
it?

\begin{figure}
\begin{center}
    \epsfig{file=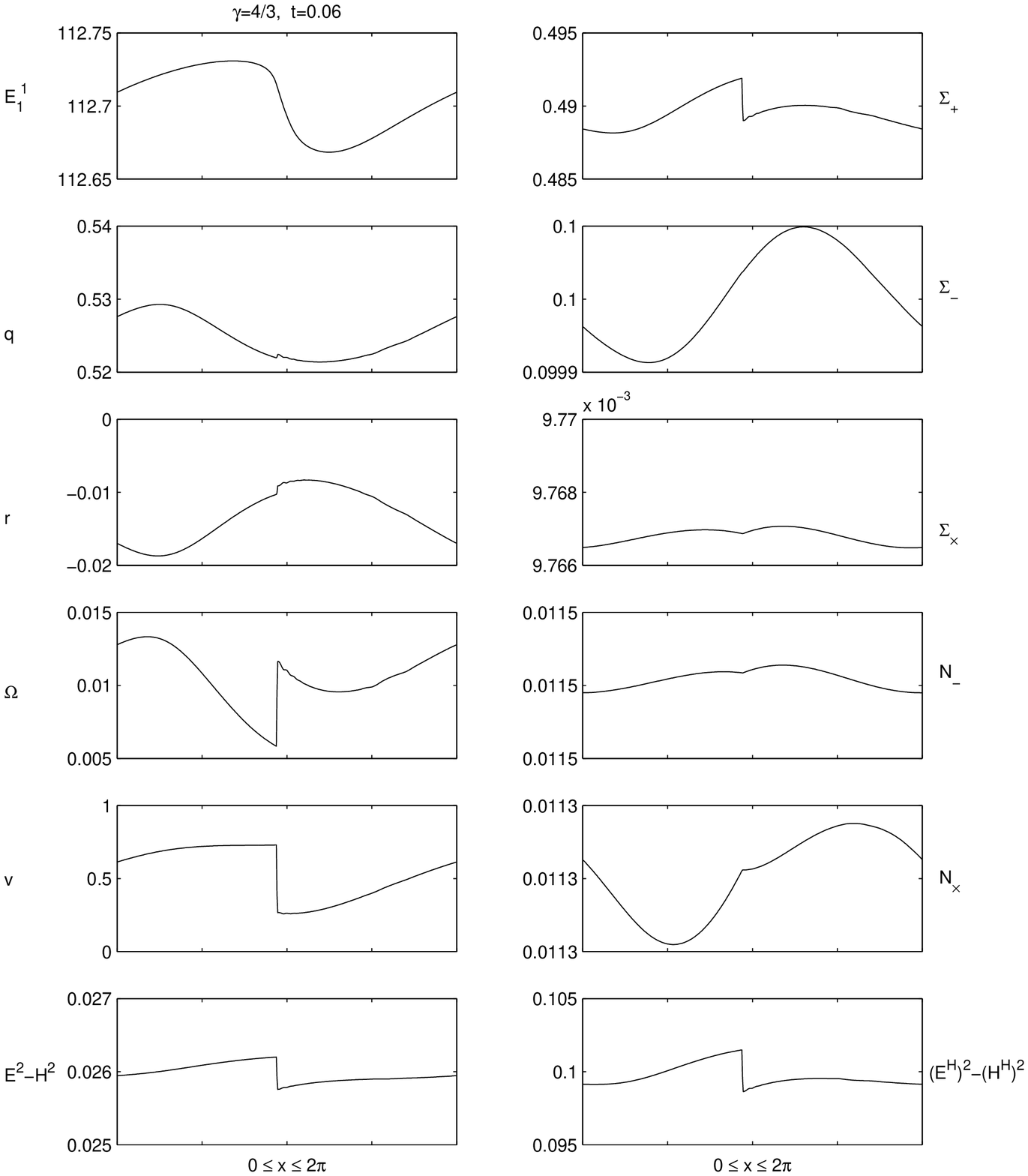,width=\textwidth}
\end{center}
\caption[Snapshot showing shock waves or non-differentiability.]
	{Snapshot of all variables, showing 
	shock waves or non-differentiability 
	into the future.}\label{fig:snap_all_shock}
\end{figure}

Figure~\ref{fig:snap_all_shock} shows a snapshot of all the variables into 
the future (at $t=0.06$).
We see that
shock waves also appear in the profiles of $\Omega$, $\Sp$, $q$ and $r$, 
but 
not in $\Sc$, $\Nm$ and $\Nc$, which seem to
become non-differentiable at the location of the shock wave, 
but still continuous. The profiles of $\EEE$ and $\Sm$ seem smooth, but it 
could be that their non-differentiability is barely visible.
Also plotted in Figure~\ref{fig:snap_all_shock} are the $\beta$-normalized 
Weyl scalar $\mathcal{E}^2-\mathcal{H}^2$ and the Hubble-normalized Weyl 
scalar $(\mathcal{E}^H)^2-(\mathcal{H}^H)^2$.

\newpage
\section{Future asymptotics}\label{sec:future}

In this section, we illustrate the future 
asymptotic behaviour of the tilt variable $v$, as predicted by 
(\ref{de_Sitter_2}).

Consider three sets of the following initial data at $t=0$, with $\gamma = 
1.2$, $\tfrac{4}{3}$ and $1.5$
(in conjunction with (\ref{IC_format})):
\begin{gather}
        \epsilon= 0.09,\quad
        (\EEE)_0= 1  ,\quad
        A_0     = 0.1,\quad
        (\Sm) _0  = 0.2,\quad
        (\Sc) _0  = 0.2,
\notag\\
        (\Nm) _0  = 0.2,\quad
        \Om _0  = 0.4,\quad
        (\Oml)_0 = 0.01,\quad
        (\St)_0  = 0.1.
\label{IC_d_S}
\end{gather}
Changing $\gamma$ affects only the initial condition for $v$.

We ran {\tt ICN} with 512 grid points, from $t=0$ to $t=10$, storing data 
at intervals of $t=0.1$.

\begin{figure}
\begin{center}
    \epsfig{file=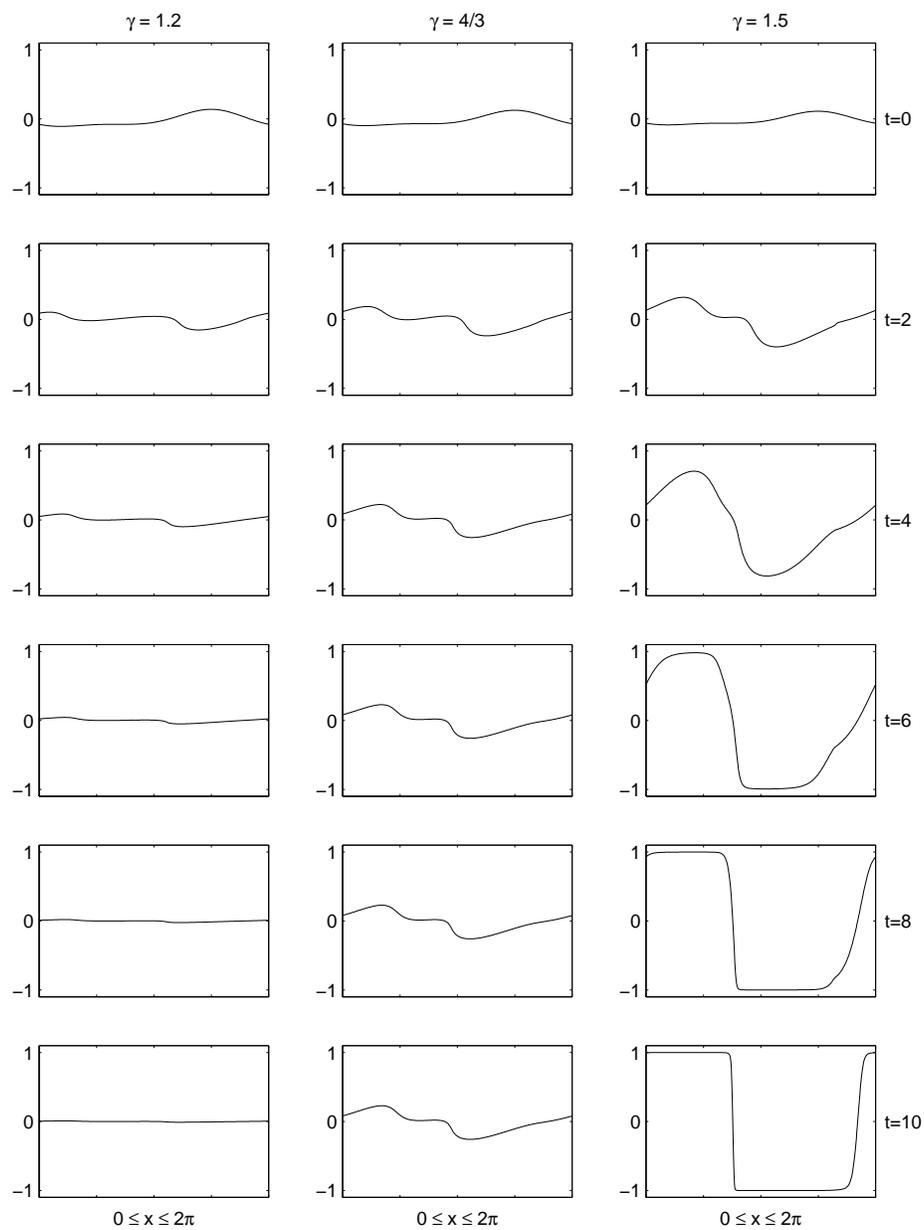,width=0.9\textwidth}
\end{center}
\caption{Snapshots of $v$ for three solutions with 
	$\gamma =1.2$, $\tfrac{4}{3}$ and $1.5$,
	showing the asymptotic signature $v_{\rm sig}(x)$ in the case 
	$\gamma \geq \tfrac{4}{3}$.}\label{fig:v_deSitter}
\end{figure}

As predicted by (\ref{de_Sitter_2}),
the limit of the tilt variable $v$ depends on the parameter $\gamma$.
Figure~\ref{fig:v_deSitter} plots the snapshots of $v$ for the three 
solutions.
This confirms that if $\gamma < \tfrac{4}{3}$, $v$ tends to zero;
if $\gamma = \tfrac{4}{3}$, $v$ tends to a function of $x$ that depends 
on the initial condition;
for $\gamma > \tfrac{4}{3}$, $v$ tends to $\pm1$, depending on the 
initial condition.
\footnote{That step-like structures can develop in the case 
$\gamma 
>\tfrac{4}{3}$ was first predicted by Rendall 
2003~\cite{art:Rendall2003}.}

As mentioned in Section~\ref{sec:sim_silent}, the evolution is 
asymptotically silent as $t \rightarrow \infty$:
\be
        \EEE \rightarrow 0,\quad
        r \rightarrow 0\ ,
\ee
and the solutions tend to
\be
        \Oml = 1,\quad
        (\Omega,A,\Sp,\Sm,\Sc,\St,\Nm,\Nc) = \mathbf{0}.
\ee
The numerical decay rates agree with the analytical results of Lim \etal 
2004~\cite{art:Limetal2004}.
Table~\ref{tab:deSitter_rates} gives the predicted growth rates from 
\cite{art:Limetal2004}, and the typical absolute errors of the numerical 
decay rates. The errors are small, indicating excellent agreement between 
numerical and analytical results.
Note that for the case $\tfrac{4}{3}$, since $v$ tends to an arbitrary
function, we take $v(10,x)$ as the signature function, and
determine the decay rate of $v(t,x)-v(10,x)$ at $t=5$.

\begin{table}[h]
\begin{spacing}{1.1}
\caption[Predicted and actual growth rates at $t=10$.]
	{Predicted and actual growth rates at $t=10$. Errors are obtained 
along typical timelines, ignoring irregular growth rates along timelines 
where a variable crosses zero.}
                \label{tab:deSitter_rates}
\begin{center}
\begin{tabular}{cclll}
\hline
Variable & predicted growth rate & \multicolumn{3}{c}{Absolute error (order)}
\\
	& & $\gamma=1.2$ & $\gamma = \tfrac{4}{3}$ & $\gamma=1.5$
\\
\hline
$\EEE$	& $-1$ & $10^{-8}$ & $10^{-8}$ & $10^{-8}$
\\
$A$	& $-1$ & $10^{-8}$ & $10^{-8}$ & $10^{-8}$
\\
$r$	& $-1$ & $10^{-5}$ & $10^{-3}$ & $10^{-3}$
\\
$\Sp$	& $-2$ & $10^{-3}$ & $10^{-3}$ & $10^{-3}$
\\
$\Sm$	& $-2$ & $10^{-5}$ & $10^{-5}$ & $10^{-3}$
\\
$\Sc$	& $-2$ & $10^{-3}$ & $10^{-3}$ & $10^{-3}$
\\
$\St$	& $-3$ & $10^{-9}$ & $10^{-9}$ & $10^{-9}$
\\
$\Nm$	& $-1$ & $10^{-5}$ & $10^{-5}$ & $10^{-5}$
\\
$\Nc$	& $-1$ & $10^{-5}$ & $10^{-5}$ & $10^{-5}$
\\
$\Oml-1$ & $-2$ & $10^{-3}$ & $10^{-5}$ & $10^{-3}$
\\
\hline
$\Om$	& $-3\gamma$ & $10^{-5}$
\\
	& $-4$ & & $10^{-5}$
\\
	& $-4$ & & & $10^{-3}$
\\
\hline
$v$	& $3\gamma-4$ & $10^{-3}$
\\
$v-v_{\rm sig}(x)$ & $-1$ & & $10^{-2}$
\\
$1-v^2$ & $\tfrac{2(3\gamma-4)}{2-\gamma}$ & &  & $10^{-4}$
\\
\hline
\end{tabular}
\end{center}
\end{spacing} 
\end{table}

\newpage
\section{Close-to-FL epoch}\label{sec:intermediate}

In this section, we describe the evolution of  a cosmological model that 
is close to the flat FL solution at $t=0$.
By ``close to flat FL" we mean the 
Hubble-normalized
shear and the Weyl scalars are 
sufficiently small, and $\Omega$ is sufficiently close to 1
(see WE, page 62).

Consider the following initial data at $t=0$ 
for an OT $G_2$ cosmology ($\St=0$)
with $\gamma  = \tfrac{4}{3}$
(using the scheme for the case $A=0$ in Section~\ref{sec:IC}).
\begin{gather}
        \EEE = 1,\quad
        \Sm = 0.01 \sin 3x,\quad
        \Sc = 0.01 \sin 5x,\quad
        \Nm  = 0,\quad
        \Nc  = 0,
\notag\\
	v = 0.01 \sin x,\quad
	\Om = 0.98 G_+,\quad G_+ = 1 + (\gamma-1)v^2,
\notag\\
	r = -2(0.98)(0.01)\sin x,
\notag\\
	\Oml = 10^{-6} \exp[
	- \tfrac{8}{\EEE}(0.98)(0.01) \sin^2 \tfrac{x}{2} ].
\label{IC_wave}
\end{gather}  
The initial value for $\Om$ is chosen
so that the initial value for $\Sp$ is close to zero (see 
Figure~\ref{fig:t_0}).

We ran {\tt CLAWPACK} with 512 grid points, from $t=0$ to $t=-15$,
storing data at intervals of $t=0.1$,
and from $t=0$ to $t=15$,
storing data at intervals of $t=0.005$.

\begin{figure}
\begin{center}
    \epsfig{file=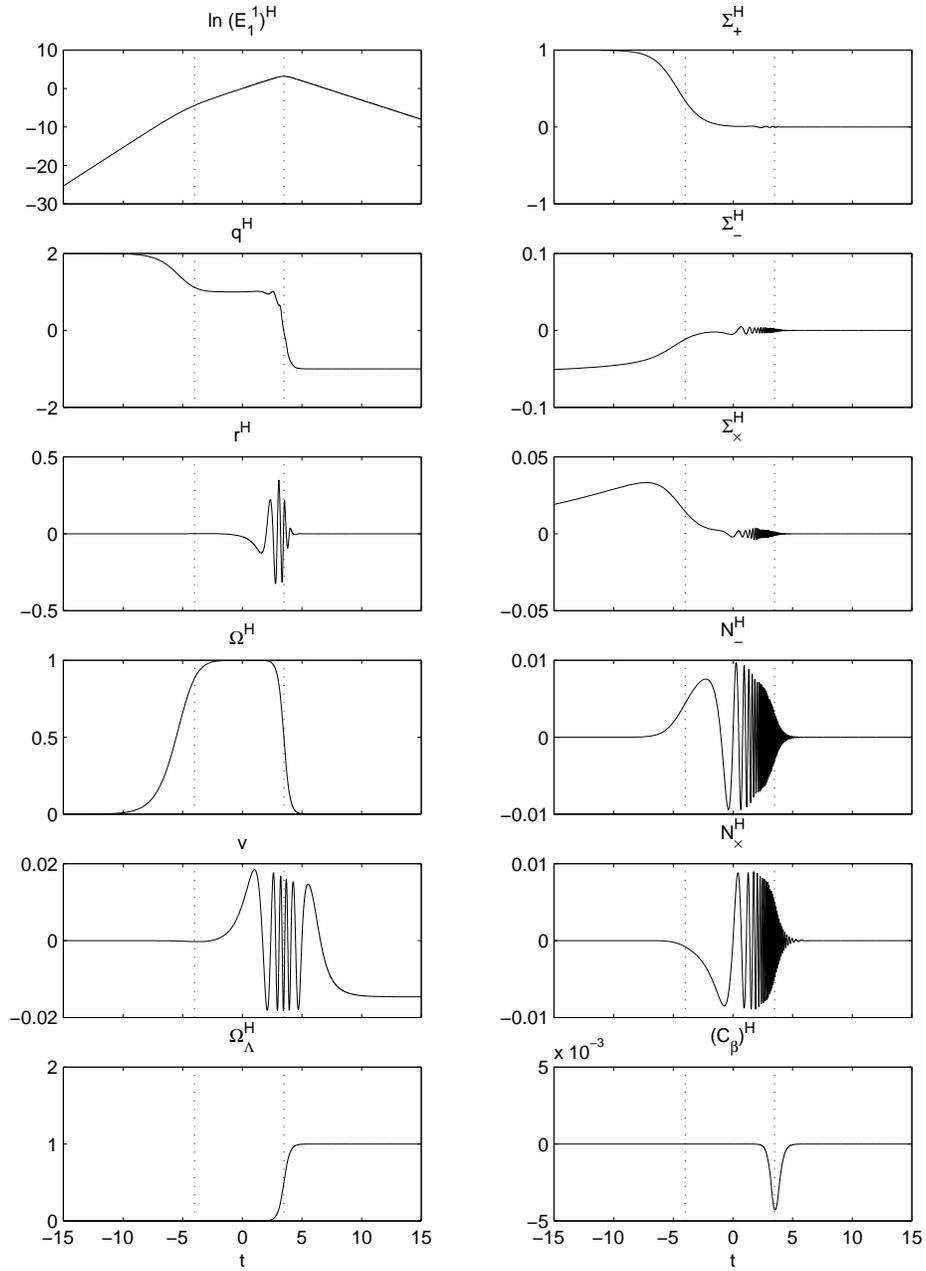,width=0.9\textwidth}
\end{center}
\caption{Evolution along a fixed timeline -- from the
past asymptote to the future asymptote.}\label{fig:wave_snap_2}
\end{figure}

Figure~\ref{fig:wave_snap_2} shows the evolution along the grid point 100
of 512.
The graphs of $\ln(\EEE)^H$ and $q^H$ show three distinct epochs -- two
asymptotic and one intermediate.
They are the Kasner asymptotic epoch from $t=-15$ to $t=-4$,
the flat FL intermediate epoch from $t=-4$ to $t=3.48$,
and the de Sitter asymptotic epoch from $t=3.48$ to $t=15$.
The fact that the flat FL epoch is of finite duration is a reflection of 
the fact that the flat FL model is unstable
\footnote{In the Hubble-normalized SH state space, the flat FL solution is 
described by an equilibrium point that is a saddle point, reflecting its 
instability within the SH context. The simulations also reflect unstable 
inhomogeneous modes.}
into the future and into the past.
The transition from the flat FL epoch to the de Sitter asymptotic epoch is 
rapid, with steep slopes in the graphs of $\Omega^H$ and $\Oml^H$.
The transition from the Kasner epoch to the flat FL epoch is less rapid, 
with a milder slope in the graph of $\Omega^H$.
The growth rate of $\EEE$ is as expected, since
$\EEE = (\EEE)_0 e^{2t} $ for the Kasner solutions,
$\EEE = (\EEE)_0 e^t $ for the flat FL solution with
$\gamma=\tfrac{4}{3}$, and
$\EEE = (\EEE)_0 e^{-t} $ for the de Sitter solution.
As $t$ increases,
$\EEE$ grows exponentially until $\Omega\approx
\tfrac{1}{2}$ at
$t=3.48$, when $\EEE$ reaches a maximum of $23.7902$ and starts to 
decrease exponentially.
\footnote{Notice that the $(\mathcal{C}_\beta)^H$ constraint has
an unacceptably large
 error at about $t=3.48$. As a result, 
the simulation loses accuracy temporarily.
As mentioned in Section~\ref{sec:unsatisfactory_aspects},
stability of constraints is still an open issue in numerical simulations.}

Evolving into the past from $t=0$, the solution enters the Kasner
asymptotic regime and approaches the Kasner arc $(T_3 Q_1)$.
A snapshot of the spatial profile at $t=-15$ would show two spikes forming 
in $\Sc$, and the asymptotic spatial signature, i.e. the
dependence of $\Sm$ on $x$.
This asymptotic behaviour is similar to the one shown in 
Figure~\ref{fig:OT_snap}.

Evolving far into the future, the solution approaches the de Sitter
solution.
A snapshot of the spatial profile at $t=15$ would show the asymptotic 
spatial signature, i.e. the
dependence of $v$ on $x$.
This asymptotic behaviour is qualitatively the same as the one shown in 
Figure~\ref{fig:v_deSitter}.

\begin{figure}
\begin{center}
    \epsfig{file=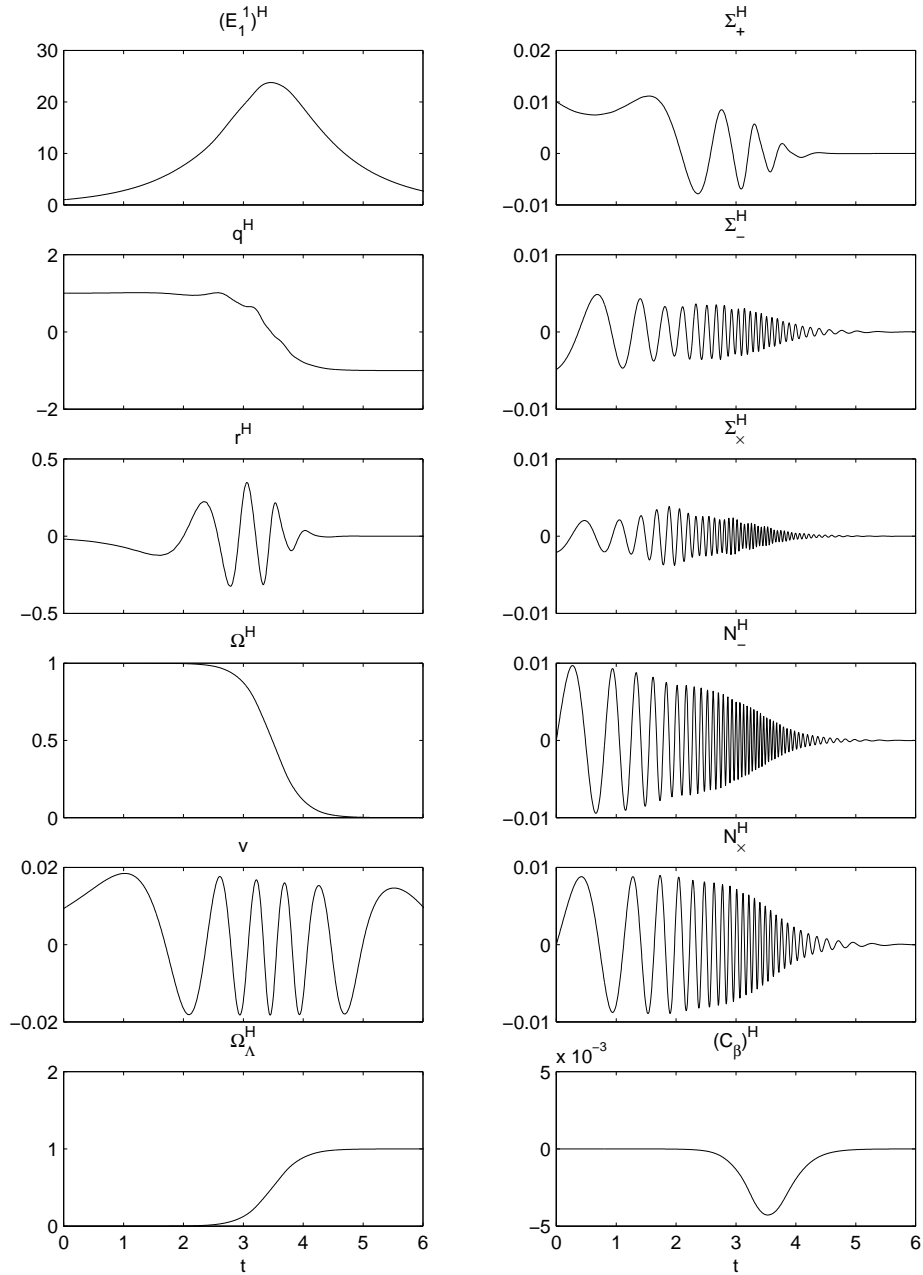,width=0.9\textwidth}
\end{center}  
\caption{Evolution along a fixed timeline -- 
details of the close-to-FL epoch.}\label{fig:narrow_snap}
\end{figure}

We now take a detailed look at the close-to-FL intermediate epoch.
Evolving into the future from $t=0$, we observe oscillations in the 
variables.
Figure~\ref{fig:narrow_snap} shows 
in detail the evolution during the close-to-FL epoch and 
the 
earlier part of the de Sitter epoch,
during which these oscillations occur.
We see 45 oscillations in the pair $(\Sc,\Nm)$, 27 in $(\Sm,\Nc)$, and 5 
in 
$v$.
\footnote{There are also oscillations in $\Omega$, approximately of 
amplitude $0.02$, but these are not visible in the graph because of the 
scale.}
How do these numbers relate to the wavelengths of spatial 
inhomogeneities?
Figure~\ref{fig:t_p348} shows the snapshot at $t=3.48$, 
revealing that the wavelength of spatial inhomogeneities is
$2\pi/5$ in the pair $(\Sc,\Nm)$, $2\pi/3$ in $(\Sm,\Nc)$, and $2\pi$ in
$(\Omega,v)$.
$(\Sm,\Nc,\Sc,\Nm)$ are gravitational field variables, whose 
characteristic 
speed of propagation is the speed of light, while $v$ is a matter field 
variable, whose characteristic
speed of propagation is the speed of sound, given by
$\sqrt{\gamma-1}$ times the speed of light.
\footnote{For characteristic eigenfields and their corresponding 
characteristic velocities of propagation, see 
van Elst \etal 2002~\cite{art:vEUW2002}.}
Comparing the numbers of oscillations with the wavelengths and taking 
into account the speeds of propagation, we have excellent agreement --
we see $\frac{3}{5}$ as many oscillations in $(\Sm,\Nc)$ as in $(\Sc,\Nm)$,
because the wavelength of spatial inhomogeneities in $(\Sm,\Nc)$ is 
$\frac{5}{3}$ times as long.
Since $\gamma=\tfrac{4}{3}$, we can predict that there should be
$\sqrt{\frac{4}{3}-1} \times \tfrac{1}{3} \approx 0.5774 \times \tfrac{1}{3}$ as
many oscillations in
$v$ as in $(\Sm,\Nc)$, because waves in $v$ travel $0.5774$
as fast as $(\Sm,\Nc)$ do, and the wavelength of spatial inhomogeneities 
in $v$ is $3$ times as long.
The simulations agree with the prediction.

We also observe that the oscillations are most rapid at $t=3.48$, when 
$\EEE$ reaches its maximum. This is expected, because the frequency is 
equal to the speed of propagation divided by the wavelength. Recall from 
(\ref{slope}) that the coordinate speed of light is $\EEE\mathcal{N}$.
Since we have chosen a $t$-coordinate such that $\mathcal{N}=1$,
for a fixed wavelength the frequency is proportional to $\EEE$.
We note that the frequency depends on the choice of $t$-coordinate, and
the physical frequency should be computed with respect to the clock time, 
for which $\mathcal{N}=\beta$. Since $\beta \rightarrow \infty$ into the 
past and $\beta \rightarrow \sqrt{\frac{\Lambda}{3}}$ into the future, the 
physical frequency actually decreases with time.

\begin{figure}
\begin{center}
    \epsfig{file=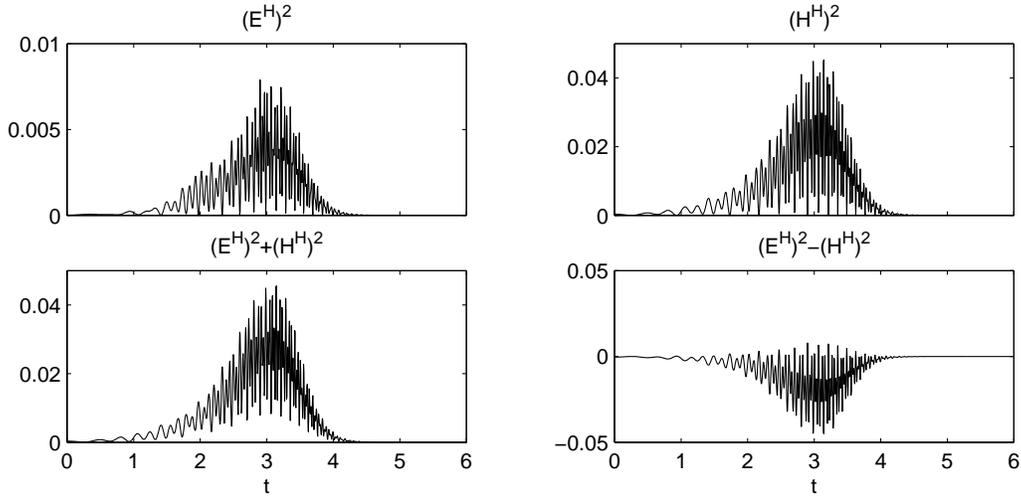,width=\textwidth}
\end{center}
\caption[Evolution of the Weyl scalars along a fixed timeline.]
	{Evolution of the Hubble-normalized Weyl scalars along a fixed 
	timeline, showing growth as $\EEE$ increases.}
	\label{fig:narrow_snap_Weyl}
\end{figure}

\begin{figure}
\begin{center}
    \epsfig{file=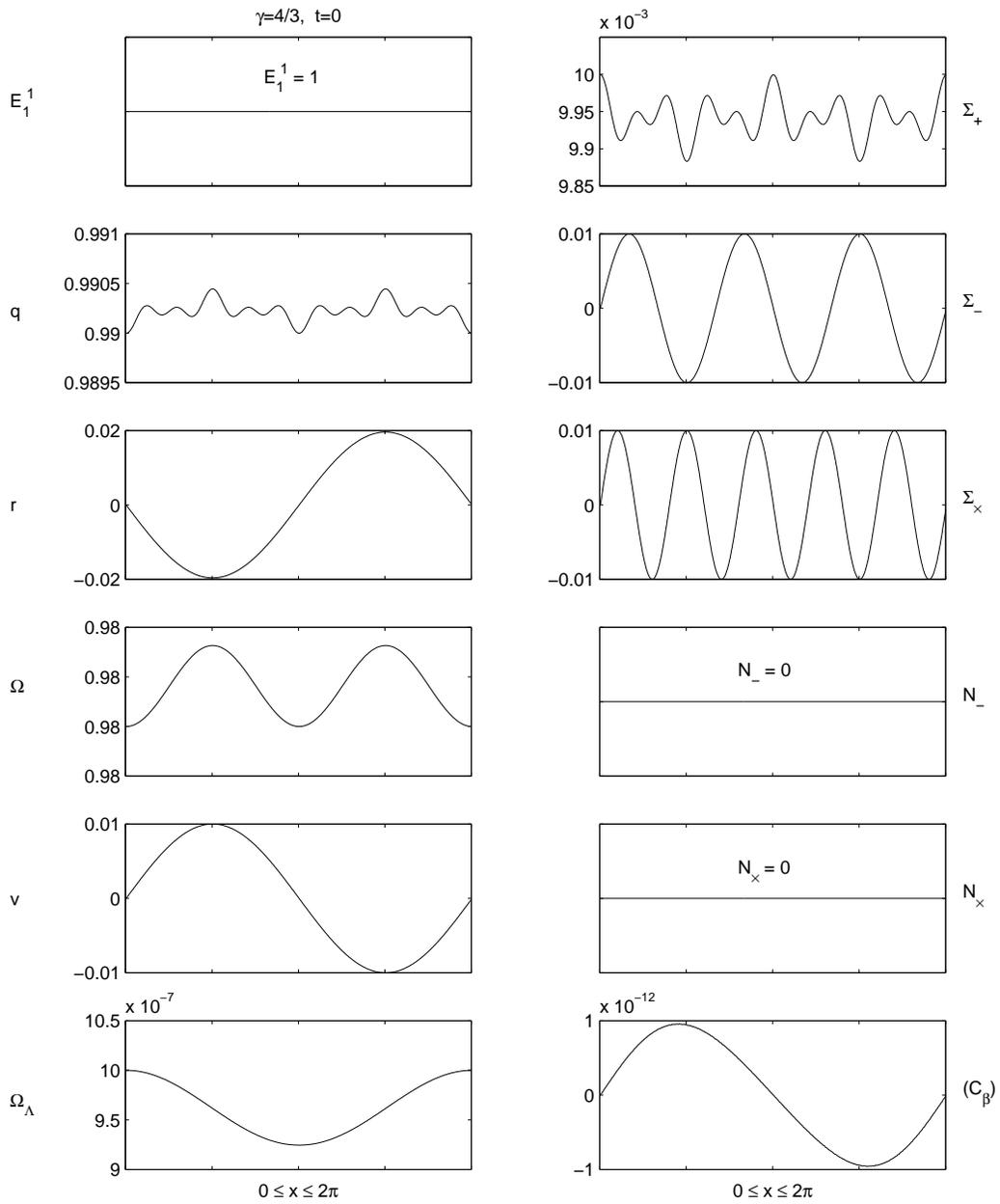,width=\textwidth}
\end{center}
\caption{Snapshot of the initial condition at $t=0$, showing a small 
	$\Sp$.}\label{fig:t_0}
\end{figure}

\begin{figure}
\begin{center}
    \epsfig{file=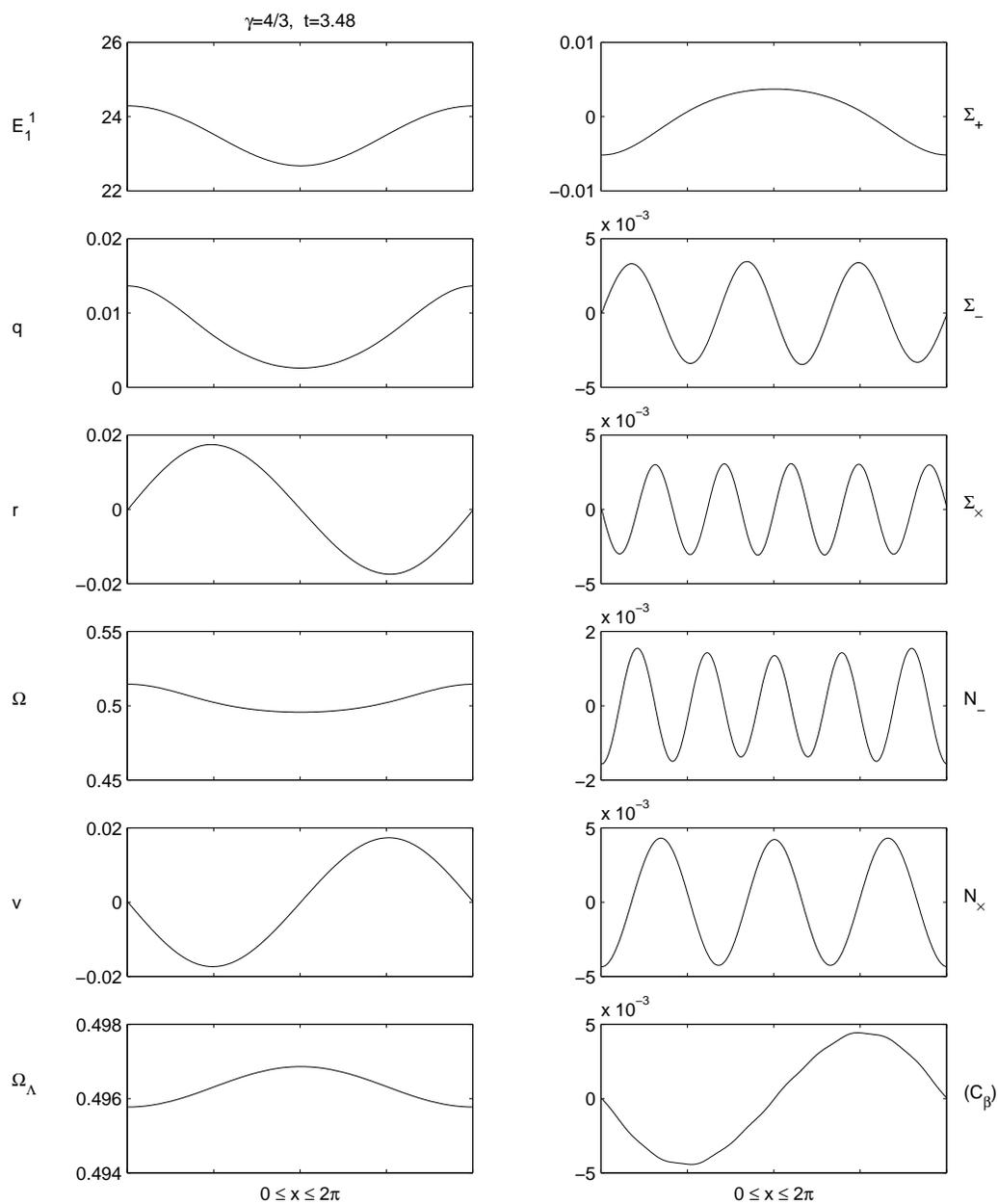,width=\textwidth}
\end{center}  
\caption{Snapshot at $t=3.48$, showing the wavelengths 
	in the different modes.}\label{fig:t_p348}
\end{figure}

With the shear and spatial curvature variables staying close to zero 
during the flat FL epoch
and subsequently during the de Sitter epoch, the instability of the 
flat FL model appears to be primarily due to the growth of $\EEE$ into the 
future, leading to growth in the Weyl curvature, which describes 
anisotropy in the free gravitational field.
\footnote{Such behaviour can also occur in SH cosmologies of Bianchi type 
VII$_0$
(see Nilsson 
\etal 1999~\cite{art:Nilssonetal1999}),
i.e. the Hubble-normalized Weyl curvature can be large even though the 
Hubble-normalized shear is close to zero.}
The spatial derivative terms are the largest at $t=3.48$, when $\EEE$ 
attains 
a maximum.
The largest spatial derivative terms are $\EEE \ptl_x \Sc$ and 
$\EEE \ptl_x \Nm$, both of which have magnitudes approximately equal to 
\begin{align*}
	\text{magnitude}\ &\approx \ 
\max \Big(\EEE(\Sc,\Nm)\Big) \times\ \frac{2\pi}{\rm wavelength}
\\
	&\approx (24 \times 0.005) \times 5 = 0.6\ .
\end{align*} 
Thus the Weyl scalars $\mathcal{E}$ and $\mathcal{H}$
(see (\ref{scalars}) and (\ref{Ep})--(\ref{Ht})), which contain spatial 
derivative terms, are approximately $\tfrac{1}{3} \times 0.6 = 0.2$ at 
$t=3.48$.
This growth in the Weyl curvature scalars mirrors the growth of $\EEE$ 
and the decrease of the wavelength of inhomogeneities relative to the 
particle horizon.
From $t=3.48$ onwards, the Weyl curvature scalars decrease, reflecting 
the decrease of $\EEE$   
and the increase of the wavelength of inhomogeneities relative to the 
event horizon.
Figure~\ref{fig:narrow_snap_Weyl} shows the evolution of the Weyl scalars 
along the grid point 100 of 512, confirming the above argument.

We have performed a variety of simulations of $G_2$ cosmologies with a 
close-to-FL epoch.
Decreasing the initial size of $\Oml$ will increase the duration of the 
flat FL epoch, defined by $\Omega \approx 1$, $\Sigma \approx0$, into the 
future.
On the other hand, this change will
 allow more time for $\EEE$ to grow, 
which will in turn increase the Weyl scalars significantly.
Increasing the initial value of $\EEE$ will also increase the Weyl 
scalars.
We comment that there are three modes (with two fields in each mode) 
that develop oscillations, namely
$(\Omega,v)$ describing acoustic waves, and 
$(\Sm,\Nc)$ and $(\Sc,\Nm)$ describing gravitational waves.
In the initial condition (\ref{IC_wave}) we introduced three different 
wavelengths into the three modes ($\sin x$, $\sin 3x$ and $\sin 5x$).
Coupling between these three modes does occur, but is rather weak.
For example if $v=0$ initially the gravitational wave modes will 
subsequently turn $v$ on, with $\sin 3x$ and $\sin 5x$ modes.
We note that the two fields in each mode can have different wavelengths.

In conclusion we note that
the simulations during the close-to-FL epoch $-4 \leq t \leq 3.48$ can 
be viewed as non-linear perturbations of the flat FL model.
It would be interesting to use simulations of this nature to test the 
validity of the linear perturbations of the flat FL model that are used 
extensively in cosmology.

%% file: conclusion.tex
	\chapter{Conclusion}\label{chap:conclusion}

In this thesis, by
constructing a unified framework,
analyzing explicit solutions,
providing heuristic arguments, and
performing numerical simulations,
we have made considerable 
progress in understanding the dynamics of inhomogeneous cosmologies with 
one spatial degree of freedom, the so-called $G_2$ cosmologies.
We have highlighted the role of spatially homogeneous cosmologies, in 
particular 
the tilted models, in providing the background dynamics.

The dynamics of tilted SH cosmologies is still not well-understood.
In Chapter~\ref{chap:G2hie}, we have given a unified
classification scheme for $G_2$ cosmologies and for 
$G_2$-compatible SH cosmologies, based on the $G_2$ action and on the tilt 
degrees of freedom.
In Chapter~\ref{chap:SH}, this classification leads to
 a unified system of equations and 
spatial gauge choice that is
valid for $G_2$-compatible SH cosmologies of all 
Bianchi types.
By analyzing the stability of the Kasner equilibrium points, we
have shown that the occurrence of Mixmaster dynamics is 
determined by the $G_2$ action, and that the tilt determines whether the 
fluid vorticity is dynamically significant near the initial singularity.

The notion of asymptotic silence provides the link between $G_2$ and SH 
dynamics.
This link enabled us to establish that 
Mixmaster dynamics occurs in a 
$G_2$ cosmology if and only if the $G_2$ action is generic, 
by making use of the analysis of Mixmaster dynamics in SH cosmologies.
We have also given the past attractors for subclasses
of $G_2$ cosmologies
 that do not permit 
Mixmaster dynamics.

We have made significant progress in understanding the development of 
steep spatial gradients.
In Chapter~\ref{chap:explicit}, we have given a detailed analysis of two 
classes of explicit solutions, the first one with step-like structures, 
and the second one with spikes.
The second one is new, and is the only known explicit $G_2$ solution with 
spikes.
We have made a detailed comparison of the asymptotic size of different 
terms in the evolution equations, which revealed that spatial derivative 
terms are not always negligible.
Furthermore, the growth of spatial derivative terms triggers 
the so-called spike transitions.
As a result, the BKL conjecture II does not hold in general for $G_2$ 
cosmologies.

All numerical simulations of $G_2$ cosmologies to date have considered 
vacuum models, and have used a metric formulation.
Chapter~\ref{chap:sim} is one of two pieces of work, the other being
Andersson \etal
2004~\cite{art:Anderssonetal2004b}~\cite{art:Anderssonetal2004c},
in which numerical simulations of $G_2$ cosmologies have been
performed in a dynamical system framework.
Our work complements that of Andersson \etal in that they consider the
past asymptotic dynamics of vacuum $G_2$ cosmologies, whereas we have
considered the full range of the dynamics for non-vacuum models in which
the matter content is a radiation perfect fluid and a positive
cosmological constant.

Our simulations show that,
for models that are close to FL at some time, the evolution consists of 
three distinct epochs, namely
the past and future asymptotic epochs and the close-to-FL intermediate
epoch.
The numerical results support the following conclusions, within these 
epochs.

\begin{itemize}
\item	Past asymptotic silence,
	future asymptotic silence (if $\Lambda>0$), the BKL 
	conjecture I, and the Kasner 
	Attractivity Conjecture are valid along almost all orbits of a 
	typical $G_2$ cosmology.

\item	Mixmaster dynamics occurs along a typical timeline of a typical
	$G_2$ cosmology in the past asymptotic regime, provided that the
	group action is generic.

\item	The dynamics in asymptotically silent regimes consists of the 
	background SH dynamics and a spectrum of inhomogeneous phenomena, 
	in which the trigger variables and the wavelengths of 
	inhomogeneities play crucial roles.

\item	A trigger variable that has both signs causes step-like or spiky 
	structures to form. 

\item	A spike transition occurs if the spatial 
	derivative of a trigger causes the time derivative in the 
	evolution equation to change sign.

\item	Sufficiently short wavelength spatial inhomogeneities in the 
	matter variables can lead to the formation of shock waves.

\item	A long enough close-to-FL epoch allows the wavelength of spatial 
	inhomogeneities to decrease, which leads to wave-like phenomena
	and potentially a large Weyl curvature.

\item	Wave-like phenomena disappear as the solution enters the past 
	asymptotic regime or the de Sitter future asymptotic regime, which 
	is asymptotically silent.
\end{itemize}

We conclude by commenting on future research that this thesis leads 
to.
Firstly, the unified formulation that we have given provides a framework 
for proving the conjectures, in particular the asymptotic silence 
conjecture, the BKL conjecture I, and the Kasner Attractivity Conjecture,
that appear to govern the asymptotic dynamics of tilted SH and $G_2$ 
cosmologies.
Secondly, the analysis of the tilted SH cosmologies can now be completed 
using the unified formulation.
Thirdly,
it is possible that further analysis of the
close-to-FL simulations, which represent non-linear perturbations of the
flat FL model,
may shed light on an important question: how reliable are
the standard linear perturbation analyses of the flat FL model, which
play a fundamental role in cosmology?

%% file: app_action.tex
	\chapter{$G_2$ action}\label{app:HO}

In this appendix, we prove Proposition~\ref{prop3.1} in 
Section~\ref{sec:G2tree}
about $G_2$ models with a hypersurface-orthogonal KVF.
We split the proposition into two parts: Proposition~\ref{propA.1} and its 
corollary.

\begin{prop}\lb{propA.1}
Consider a $G_2$ model presented in a group-invariant orbit-aligned frame
with $\mathbf{e}_2$ and $\mathbf{e}_3$ tangent to the $G_2$ orbits.
Then the $G_2$ admits a hypersurface-orthogonal KVF if and only if
\be
        \tilde{\sigma}_{AB} \tilde{n}^{AB}=0,\quad
        \tilde{n}_{AB} \sigma^{1A} \sigma^{1B} =0,\quad
        {}^*\tilde{\sigma}_{AB} \sigma^{1A} \sigma^{1B} =0.
\label{HO_grav_app}
\ee
\end{prop}

\begin{proof}
``$\Rightarrow$": Let $\pmb{\xi}$ be a HO KVF, i.e.
\be
	        \xi_{[a;b} \xi_{c]} =0.
\ee
Expanding this yields
\begin{align}
	\xi_C \epsilon^{CB} \me_0(\xi_B) +
	\epsilon^{AB} \xi_B \gamma^C{}_{0A} \xi_C &=0
\label{HO_proof_1}
\\
	\xi_C \epsilon^{CB} \me_1(\xi_B) +
        \epsilon^{AB} \xi_B \gamma^C{}_{1A} \xi_C &=0
\label{HO_proof_2}
\\
	\gamma^C{}_{01} \xi_C &=0.
\label{HO_proof_3}
\end{align}
Using (\ref{eq:giab}), equations (\ref{HO_proof_3}) simplifies to
\be
	\sigma_{1A} \xi^A =0.
\label{xi_3}
\ee
The group-invariant frame yields
\begin{align}
\label{Killing_0}
	\me_0(\xi^B) + \gamma^B{}_{0A} \xi^A &=0
\\
        \me_1(\xi^B) + \gamma^B{}_{1A} \xi^A &=0,
\label{Killing_1}
\end{align}
which are then contracted with $\xi_C \epsilon^C{}_B$ and substituted into
(\ref{HO_proof_1}) and (\ref{HO_proof_2}) to obtain
\be
	{}^*\tilde{\sigma}_{AB} \xi^A \xi^B=0
	\quad\text{and}\quad
	\tilde{n}_{AB} \xi^A \xi^B=0
\label{xi_12}
\ee
respectively.
Equations
(\ref{xi_3}) and (\ref{xi_12}) combine to yield (\ref{HO_grav_app}).
\qed

\

\noindent
``$\Leftarrow$":
Assuming (\ref{HO_grav_app}), use the rotation freedom to set
\footnote{For convenience we introduce the following notation:
\be
\label{action_notation}
        n_C{}^C = 2 n_+,\quad
        \tilde{n}_{AB} = \sqrt{3}
                \left(
                \begin{matrix}
                        n_- & n_\times
                \\
                        n_\times & -n_-
                \end{matrix}
                \right),
\ee
and
\be
        \sigma_{12} = \sqrt{3}\sigma_3,\quad
        \sigma_{13} = \sqrt{3}\sigma_2.
\ee
}
\be
	n_-=0.
\label{HO_proof_6}
\ee
Then (\ref{HO_grav_app}) yields
\be
	\sigma_{12}=0,\quad
	\sigma_\times=0.
\ee
The evolution equations for $n_-$, $\sigma_{12}$ and $\sigma_\times$, and 
the $(C_{\rm C})_2$ constraint then imply that
\be
	n_+=0,\quad \Omega_1=0.
\label{HO_proof_8}
\ee
Equations
(\ref{HO_proof_6})--(\ref{HO_proof_8}) imply that the frame vector $\me_2$ 
is hypersurface-orthogonal.
We also have, trivially,
\be
	\Omega_1 = - \sqrt{3} \sigma_\times,\quad
	n_+ = \sqrt{3}n_-.
\ee
Using Lemma 1 below, this means $\me_2$ is parallel to a KVF.
Thus this KVF is also hypersurface-orthogonal, and is independent of the 
choice (\ref{HO_proof_6}).
\end{proof}

\newpage

\noindent
{\bf Lemma 1}

\noindent
{\it
Consider a $G_2$ model presented in a group-invariant orbit-aligned frame
with $\mathbf{e}_2$ and $\mathbf{e}_3$ tangent to the $G_2$ orbits.
Then
\be
        \Omega_1 = - \sqrt{3} \sigma_\times,\quad
        n_+ = \sqrt{3}n_-
\label{lemma}
\ee
if and only if $\me_2$ is parallel to a KVF.
}

\begin{proof}
Let $\pmb{\xi}$ and $\pmb{\eta}$ be KVFs.
As seen in the proof of the Proposition, the group-invariant frame yields
\begin{align}
\label{lemm1a}
        \me_0(\xi^B) + \gamma^B{}_{0A} \xi^A &=0
\\
        \me_1(\xi^B) + \gamma^B{}_{1A} \xi^A &=0,
\label{lemm1b}
\end{align}
and likewise for $\pmb{\eta}$.

\noindent
``$\Leftarrow$": Assuming $\pmb{\xi}$ is parallel to $\me_2$, we have 
$\xi_3=0$. Then (\ref{lemm1a})--(\ref{lemm1b}) imply that
\be
        \gamma^3{}_{02} =0=\gamma^3{}_{12},
\ee
or equivalently (\ref{lemma}), as required to prove.
\qed

\

\noindent
``$\Rightarrow$": Assuming (\ref{lemma}), or equivalently,
\be
	\gamma^3{}_{02} =0=\gamma^3{}_{12},
\ee
we obtain the following:
\be
	\me_0(\xi_3) + \gamma^3{}_{03} \xi_3 =0,\quad
        \me_1(\xi_3) + \gamma^3{}_{13} \xi_3 =0,
\label{lemm2}
\ee
and likewise for $\pmb{\eta}$.
Equation (\ref{lemm2}) has a very special form.
If $\xi_3=0$ at one point in spacetime, then (\ref{lemm2}) implies that
$\xi_3=0$ at every point in spacetime.

If $\xi_3=0$ or $\eta_3=0$, then $\pmb{\xi}$ or $\pmb{\eta}$ is parallel 
to $\me_2$ and we are done.
If not, pick any point P and a constant K so that
\be
	\xi_3 + K \eta_3=0 \quad \text{at P.}
\ee
But $\xi_3 + K \eta_3$ satisfies the equations
\be          
        \me_0(\xi_3 + K \eta_3) + \gamma^3{}_{03} (\xi_3 + K \eta_3) =0,\quad
        \me_1(\xi_3 + K \eta_3) + \gamma^3{}_{13} (\xi_3 + K \eta_3) =0,
\ee
which now implies
\be
        \xi_3 + K \eta_3=0
\ee
at every point in spacetime.
Thus $\pmb{\xi} + K \pmb{\eta}$ is parallel to $\me_2$ and we are done.  
\end{proof}

\

\noindent
{\bf Corollary of Proposition~\ref{propA.1}}

\noindent
{\it
Consider a $G_2$ model presented in a group-invariant orbit-aligned frame
with $\mathbf{e}_2$ and $\mathbf{e}_3$ tangent to the $G_2$ orbits.
If the $G_2$ admits a hypersurface-orthogonal KVF, then
\be
        \epsilon^{AB} q_A \sigma_{1B} =0,\quad
        \epsilon^{AB} \pi_{1A} \sigma_{1B} =0,\quad
        {}^* \tilde{\pi}_{AB} \sigma^{1A} \sigma^{1B} =0.\
\label{HOKVF_matter_app}
\ee
}

\begin{proof}
Using (\ref{xi_3}) and (\ref{xi_12}), we simplify 
(\ref{Killing_0}) and (\ref{Killing_1}) to the following:
\begin{align}
\label{Killing_0_c}
        \me_0(\xi^A) &= \Omega_1 \epsilon_{AB} \xi^B
\\
        \me_1(\xi^A) &= \tfrac{1}{2} n_C{}^C \epsilon_{AB} \xi^B.
\label{Killing_1_c}
\end{align}
We shall then differentiate (\ref{xi_3}) and (\ref{xi_12}).
First, using (\ref{evo_sigma_1A}) and (\ref{Killing_0_c}),
$\me_0(\sigma_{1A}\xi^A)=0$ yields
\be
	\pi_{1A} \xi^A =0,
\ee
which implies
\be
        \epsilon^{AB} \pi_{1A} \sigma_{1B} =0.
\ee
Next, using (\ref{c_c_A}) and (\ref{Killing_1_c}),
$\me_1(\sigma_{1A}\xi^A)=0$ yields
\be
	q_A \xi^A =0,
\ee
which implies
\be
        \epsilon^{AB} q_A \sigma_{1B} =0.
\ee
Similarly, using (\ref{evo_sigma_AB}) and (\ref{Killing_0_c}),
$\me_0({}^*\tilde{\sigma}_{AB}\xi^A\xi^B)=0$ and
$\me_1(\tilde{n}_{AB}\xi^A\xi^B)=0$ combine to yield
\be
	{}^* \tilde{\pi}_{AB}\xi^a\xi^B=0,
\ee
which implies
\be
        {}^* \tilde{\pi}_{AB} \sigma^{1A} \sigma^{1B} =0,
\ee
as required.
It also follows from (\ref{HO_grav_app}) and (\ref{HOKVF_matter_app})
that
\be
        {}^* \tilde{\pi}_{AB} \tilde{\sigma}^{AB}=0,\quad
        \tilde{\pi}_{AB} \tilde{n}^{AB}=0.\
\footnote{To show this, use the notation (\ref{action_notation}) and
introduce the notation
\[
	(n_-,n_\times) = (n\cos\phi,n\sin\phi),\quad
	(\sigma_{12},\sigma_{13})=(\sigma \cos\theta,\sigma\sin\theta),
\]
and similarly for $\tilde{\pi}_{AB}$.}
\ee
\end{proof}

%% file: rotation.tex
\chapter{Rotation formulae}\label{app_rotation}

For $G_2$ cosmologies,
the spatial gauge freedom is represented by a position-dependent 
rotation of the spatial frame vectors $\{\me_2,\me_3\}$:
\begin{equation}
\left(\begin{array}{c}
      \hat{\mathbf{e}}_{2} \\
      \hat{\mathbf{e}}_{3}
      \end{array}\right)
= \left(\begin{array}{cc}
        \cos\varphi & \sin\varphi \\
       -\sin\varphi & \cos\varphi
        \end{array}\right)
\left(\begin{array}{c}  
       \mathbf{e}_{2} \\
       \mathbf{e}_{3}
       \end{array}\right) \ , \quad
\varphi = \varphi(t,x) \ .
                \label{B2}
\end{equation}
For $G_2$-compatible SH cosmologies, the rotation is restricted to 
$\varphi = \varphi(t)$.

Writing the commutators using (\ref{B2}) gives the formulae for the 
transformation of the variables (regardless of $\beta$- or 
Hubble-normalization) under a rotation:
\begin{align}
\label{R_rot}
        \hat{R}     &= R - \ptl_t \varphi\ ,
\\
        \hat{N}_+    &= \Np + \parb_1 \varphi\ ,
\\
\left(\begin{array}{c}   
      \hat{\Sigma}_3 \\  
      \hat{\Sigma}_2 
      \end{array}\right)  
&= \left(\begin{array}{cc}
        \cos\varphi & \sin\varphi \\
       -\sin\varphi & \cos\varphi
        \end{array}\right)
\left(\begin{array}{c}
       \Stt \\
       \St
       \end{array}\right) \ ,
\\
\left(\begin{array}{c}
      \hat{v}_3 \\
      \hat{v}_2     
      \end{array}\right)
&= \left(\begin{array}{cc}
        \cos\varphi & \sin\varphi \\
       -\sin\varphi & \cos\varphi
        \end{array}\right)
\left(\begin{array}{c}
       v_2 \\
       v_3
       \end{array}\right) \ ,
\\
\left(\begin{array}{c}
      \hat{\Sigma}_- \\   
      \hat{\Sigma}_\times 
      \end{array}\right)
&= \left(\begin{array}{cc}
        \cos2\varphi & \sin2\varphi \\
       -\sin2\varphi & \cos2\varphi
        \end{array}\right)
\left(\begin{array}{c}
       \Sm \\
       \Sc
       \end{array}\right) \ ,
\label{Sigma_rot}
\\
\left(\begin{array}{c}
      \hat{N}_- \\
      \hat{N}_\times
      \end{array}\right)
&= \left(\begin{array}{cc}
        \cos2\varphi & \sin2\varphi \\
       -\sin2\varphi & \cos2\varphi
        \end{array}\right)
\left(\begin{array}{c}    
       \Nm \\
       \Nc
       \end{array}\right) \ .
\label{N_rot}
\end{align}
Other variables, namely $\EEE$, $A$, $\Sp$, $r$, $q$, $\Omega$, $v_1$
and $\Oml$, are not affected by the rotation.

Note that the determinant of $N_{AB}$, given by
\be
	\det N_{AB} = \Np^2 -3(\Nm^2+\Nc^2),
\ee
is invariant if $\varphi=\varphi(t)$, but will change if $\varphi$ also 
depends on $x$. The sign of $\det N_{AB}$ is thus significant for 
$G_2$-compatible SH cosmologies, but not for $G_2$ cosmologies.

%% file: expansion.tex

\chapter{Asymptotic expansions at spike points}\label{expansion}

For a general vacuum OT $G_2$ solution,
Kichenassamy \& Rendall
1998~\cite[equations (5)--(6)]{art:KichenassamyRendall1998}
gave a generic expansion for the metric functions $P$ and $Q$ (see
(\ref{PQ_metric})--(\ref{PQ_rel})):
\begin{align}
        P &= -2k(x)t-\phi(x) + \cdots
\label{K_R_P}
\\
        Q &= X_0(x) + e^{4k(x)t}\psi(x) + \cdots\ ,
\label{K_R_Q}
\end{align}
where $k(x)$ satisfies $ 0 < k(x) < 1$. 
The above expansion is inadequate in satisfying the evolution equations
(\ref{WM2})--(\ref{WM5}). 
To obtain an adequate expansion, we start with the vacuum Bianchi II 
solution
\be
	\Sm = \frac{-\frac{2-k}{\rty} - \frac{k}{\rty} F^2}{1+F^2}\ ,\quad
	\Nm = \pm \frac{\frac{2}{\rty}(1-k) F}{1+F^2}\ ,\quad
	\Sc=\Nc=0\ ,
\ee
where $F = F_0 e^{-2(1-k)t}$.
We replace the constants $k$ and $F_0$ by arbitrary functions 
of $x$ and integrate the evolution equations (\ref{WM2})--(\ref{WM5}) to 
obtain updated
$\Nc$, $\Sc$, $\Nm$ and $\Sm$ in succession. Finally, we 
integrate the asymptotic expansion for $\Sm$ and $\Sc$ to obtain the 
asymptotic expansion for $P$ and $Q$, using (\ref{PQ_rel}).

The necessary expansion for $P$ and $Q$ is as follows:
\begin{align}
        P &= -2k(x)t-\phi(x) +A(x) e^{4k(x)t} + B(x) e^{4(1-k)t} 
	+C(x) e^{8k(x)t}
	+ \cdots
\label{our_P}
\\
        Q &= X_0(x) + \psi(x) e^{4k(x)t}
                + \poi(x) e^{8k(x)t}
                + \chi(x) e^{12k(x)t}
\notag\\
        &\quad
		+ \xi(x) t e^{4t} + \zeta(x) e^{4t}
		+ \cdots\ ,
\label{our_Q}
\end{align}
where
\begin{gather}
        A = \psi^2 e^{-2\phi},\quad
        B = -\frac{(X_0')^2 e^{-2\phi}}{16(1-k)^2}\ ,\quad
	C = -\tfrac{1}{2}A^2,
\label{coefficient_A_B}
\\
	\poi = -A \psi\ ,\quad 
	\chi = A^2 \psi\ ,\quad
	\xi = - \frac{k' X_0'}{4(1-k)}\ ,
\\
	\zeta = -2 k B \psi + \frac{X_0''}{16(1-k)} 
		- \frac{\phi' X_0'}{8(1-k)}
	- \frac{(2-k)}{4(1-k)} \xi\ .
\end{gather}

\enlargethispage{\baselineskip}

Rendall \& Weaver 2001~\cite{art:RendallWeaver2001}
transform the Kichenassamy-Rendall expansion (\ref{K_R_P})--(\ref{K_R_Q}) 
to produce expansions for $\tilde{P}$, $\tilde{Q}_t$ and $\tilde{Q}_x$ for
the transformed solution.%
\footnote{The temporal and spatial coordinates used by
Rendall \& Weaver 2001~\cite{art:RendallWeaver2001} are $\tau$ and
$x_{RW}$, which are related to ours according to
$
        \tau = -2t,\ x_{RW} = 2x.
$
For example, $X_0'(x)$ here would
correspond to $2 X_0'(x_{RW})$ in \cite{art:RendallWeaver2001}.
}
We shall transform (\ref{our_P})--(\ref{our_Q}) to obtain a consistent
expansion for the transformed solutions $(\Smt,\Nct,\Sct,\Nmt)$.
The resulting expansions for the transformed solution are as follows.
Along $x=x_{\rm spike}(-\infty)$, where $X_0(x_{\rm spike}(-\infty))=0$:
\footnote{It turns out that the asymptotic expansions for $\ptl_t \Smt$, 
$\ptl_t \Nct$, $\ptl_t \Sct$ and $\ptl_t \Nmt$ are the same as the time 
derivative of the asymptotic expansions 
(\ref{Smt_spike})--(\ref{Nmt_spike}), and so we do not list them.}
\begin{align}
\lb{Smt_spike}
        \Smt &= -\tfrac{1}{\rty}(1+k)
        + \tfrac{2}{\rty} (1-k)B e^{4(1-k)t}
        + \bigO(e^{12kt} + e^{8(1-k)t} - t e^{4t})
\\ 
        \Nct &= - \tfrac{1}{\rty}k'\ (-t e^{2t})
        + \tfrac{1}{2\sqrt{3}} (2X_0'\psi e^{-2\phi} + \phi') e^{2t}
        + \bigO(-t e^{2(1+2k)t})
\\
        \Sct &= - \tfrac{1}{2\sqrt{3}} X_0' e^{-\phi} e^{2(1-k)t}
                + \bigO(e^{2(1+k)t}+ e^{6(1-k)t})
\\
        \Nmt &= \tfrac{2}{\rty} (1-k) \xi e^{-\phi}  (-t e^{2(2-k)t})
\notag\\
        &\quad
		- \tfrac{1}{2\rty} e^{-\phi} \left[
		\xi +4(1-k)\zeta + 8(1-k) B \psi \right] e^{2(2-k)t}
\notag\\
        &\quad
		+ \bigO(e^{14kt} - t e^{2(4-3k)t} - t e^{2(2+k)t})
\lb{Nmt_spike}
\end{align}
\begin{align}
        \di \Nct &= \tfrac{1}{\rty} (X_0')^2 e^{-2\phi} e^{4(1-k)t}
                    + \bigO(e^{8(1-k)t} - t e^{4t})
\\
        \di \Smt &= -\tfrac{1}{\rty} k' e^{2t}
		+ \tfrac{4}{\rty}(1-k)\left[X_0' \xi e^{-2\phi} + 2 k' B 
			\right] (-t e^{2(3-2k)t})
\notag\\
        &\quad
		+ \Big[ - \tfrac{1}{\rty} X_0' \xi e^{-2\phi} 
	+ \tfrac{2}{\rty} (1-k) B' - \tfrac{2}{\rty} [ 4(1-k)X_0' 
		\psi e^{-2\phi} +k' ] B
\notag\\
        &\quad\qquad
	- \tfrac{4}{\rty}(1-k) X_0' \zeta e^{-2\phi} \Big] e^{2(3-2k)t}
\notag\\
        &\quad
	+ \bigO(-t e^{2(5-4k)t} - t e^{2(1+6k)t})
\\
        \di \Nmt &= \tfrac{2}{\rty} kX_0'e^{-\phi} e^{2(1-k)t}
                                        + \bigO( e^{6(1-k)t})
\\
        \di \Sct &= -\sqrt{3} k' X_0' e^{-\phi} (-t e^{2(2-k)t})
\notag\\   
        &\quad
        + \tfrac{1}{2\sqrt{3}}
        \left[
        4(X_0')^2\psi e^{-3\phi} -X_0'' e^{-\phi} + 3 X_0' \phi' e^{-\phi}
        \right]
        e^{2(2-k)t}
\notag\\
        &\quad
        + \bigO(- t e^{2(4-3k)t} - t e^{2(2+k)t})\ .
\end{align}
Along typical points:
\footnote{It turns out that the asymptotic expansions for the temporal and 
spatial derivatives of $\Smt$, $\Nct$, $\Sct$ and $\Nmt$
are the same as the temporal and spatial
derivatives of the asymptotic expansions
(\ref{Smt_typ})--(\ref{Nmt_typ}), and so we do not list them.}
\begin{align}
        \Smt &= -\tfrac{1}{\rty}(1-k)
        - \tfrac{2}{\rty}
                k(\tfrac{e^{2\phi}}{X_0^2}+2\tfrac{\psi}{X_0}+A) e^{4kt} 
        - \tfrac{2}{\rty} (1-k) B e^{4(1-k)t}
\notag\\
        &\quad
        + \bigO(e^{8kt}+ e^{8(1-k)t} - t e^{4t})
\lb{Smt_typ}
\\
        \Nct &= \tfrac{1}{\rty}k'\ (-t e^{2t}) 
        + \tfrac{1}{2\sqrt{3}}\left[ 2\tfrac{X_0'}{X_0}-\phi' \right] 
e^{2t}
        + \bigO( -t e^{2(1+2k)t} - t e^{2(3-2k)t})
\\
        \Sct &=  \tfrac{1}{2\sqrt{3}} X_0' e^{-\phi} e^{2(1-k)t}  
        + \bigO(-t e^{2(1+k)t} + e^{6(1-k)t})
\\
        \Nmt &= \tfrac{2}{\rty}
                k ( \tfrac{e^{\phi}}{X_0}+\psi e^{-\phi}) e^{2kt}
	+ \bigO(e^{6kt} -t e^{2(2-k)t})\ .
\lb{Nmt_typ}
%
\end{align}
The above expansion is consistent with the evolution equations
(\ref{WM2})--(\ref{WM5}).
To simplify to the expansion for the transformed WM solution in 
Section~\ref{sec:spike}, set
\be
        k = \tfrac{1}{2},\quad
        \phi =0,\quad
        \psi = f_1,\quad  
        X_0' = -2f_1,\quad
	X_0'' = 4 f_2\ .
\label{RW_WM}
\ee
In this case, the $A(x)$ and $B(x)$ terms have the same power and they
cancel each other in the expansion for $P$, since $A(x)=-B(x)$ from 
(\ref{coefficient_A_B}) and (\ref{RW_WM}).

The above expansion is
a new result, and provides the asymptotic expansion in the framework of   
the orthonormal frame formalism.

%% file: appG2.tex
\chapter[Equations for numerical simulations
	]{Equations for numerical simulations of $G_2$ models}\label{appG2}

Following the derivation in Section~\ref{sec:G2}, we arrive at a
system of equations (\ref{be_sys})--(\ref{ee_sys})
 with $\beta$-normalized variables for $G_2$ cosmologies.
As mentioned at the beginning of Chapter~\ref{chap:G2}, we shall restrict 
our considerations
to $G_2$ cosmologies with one tilt degree of freedom.
In doing so, the shear spatial gauge (\ref{G2_spatial_gauge}) simplifies 
to (\ref{common_gauge}):
\be
        \Sigma_{3}=0,\quad
        \Np = \sqrt{3}\Nm,\quad
        R = -\sqrt{3} \Sc\ .
\ee
We drop the index on the remaining tilt variable $v_1$, and simplify 
the evolution equation (\ref{G2_vsq}) for $v^2$ by collecting
the $\di (v^2)$, $\di \ln G_+$ and $\di v_1$ terms into an 
$\EEE \ptl_x v$ 
term.
As explained in Chapter~\ref{chap:G2}, we solve the constraint (\ref{C_G}) 
to determine $\Sp$.

Since $\EEE$ and $A$ have the same evolution equation,
we can save numerical resources by writing
\be
	A = B(x) \EEE
\ee
and evolve $\EEE$ only, and obtain $B(x)$ from the initial 
conditions. See Section~\ref{sec:IC} for more discussion of the initial 
conditions.

Numerical experiments show that $\EEE$, $\St$, $\Omega$ and $v$ are prone 
to getting out of range, e.g. $\Omega$ becoming negative.
To prevent this we evolve $\ln\EEE$, $\ln\St$, $\ln\Omega$ and 
$\arctanh\ v$ instead.

The resulting system for numerical simulations, 
given below, consists of 9 evolution equations
(\ref{be_numeq})--(\ref{ee_numeq}) and two constraint equation
(\ref{C_St})--(\ref{C_beta_app}).

\noindent
{\it Evolution equations:}
\begin{align}
\label{be_numeq}
        \dt \ln \EEE &= q+3\Sp
\\
        \dt \Sm + \EEE \ptl_x \Nc &= (q+3\Sp-2)\Sm 
                                +2A\Nc
				+ 2\sqrt{3}(\Sc^2-\Nm^2)
				- \sqrt{3}\St^2
\\
        \dt \Nc + \EEE \ptl_x \Sm &= (q+3\Sp  )\Nc
\\
\label{app_Sc}
        \dt \Sc - \EEE \ptl_x \Nm &= (q+3\Sp-2-2\sqrt{3}\Sm)\Sc
                                -2A\Nm
				-2\sqrt{3}\Nc\Nm
\\
\label{app_Nm}
        \dt \Nm - \EEE \ptl_x \Sc &= 
(q+3\Sp+2\sqrt{3}\Sm)\Nm+2\sqrt{3}\Sc\Nc
\\
\label{app_St}
        \dt \ln\St &= q -2 +\sqrt{3}\Sm
\\
        \dt \ln\Omega &+ \frac{\gamma v}{G_+} \EEE \ptl_x \ln\Omega
                        + \frac{\gamma G_-(1-v^2)}{{G_+}^2} \EEE \ptl_x 
(\arctanh\ 
v)
\notag\\
                &=\frac{2\gamma}{G_+}\left[
                        \frac{G_+}{\gamma}(q+1)-\tfrac12(1-3\Sp)(1+v^2)-1
                                + A v \right]
\label{dtlnOmega}
\\
        \dt (\arctanh\ v) &+ \frac{(\gamma-1)(1-v^2)}{\gamma G_-} 
\EEE \ptl_x\ln\Omega  
\notag\\ 
	&
    - \frac{[(3\gamma-4)-(\gamma-1)(4-\gamma)v^2] v}{G_+G_-} \EEE \ptl_x 
(\arctanh\ v)
\notag\\
                &= -(\gamma G_-)^{-1} \Big[
                        (2-\gamma)G_+ r
                        - \gamma v [3\gamma-4+3(2-\gamma)\Sp]
\notag\\
                &\hspace{2.5cm}   +2\gamma(\gamma-1)A v^2 \Big]
\label{dtatanhv}
\\
	\dt \ln \Oml &= 2(q+1)\ , 
\label{ee_numeq}
\end{align}
where
\begin{align}
        q+3\Sp&= 2(1-A^2) + \EEE \ptl_x A -3\St^2
        -\tfrac{3}{2}(2-\gamma)\frac{(1-v^2)}{G_+}\Omega -3\Oml
\\
        \Sp &= \Big[ 1-A^2+\tfrac{2}{3} \EEE \ptl_x A 
	+ A \frac{\gamma v}{G_+} \Omega
	+ 2A(\Nc\Sm-\Nm\Sc)
\notag\\
        &\qquad -\Sm^2-\Sc^2-\St^2-\Nc^2-\Nm^2-\Omega -\Oml
		\Big]/ [2(1-A^2)]
\label{Sp}
\\
        r &= -3A\Sp-3\Nc\Sm+3\Nm\Sc-\tfrac{3}{2} \frac{\gamma v}{G_+} \Omega
\label{r_app}
\\
        G_\pm &= 1 \pm (\gamma-1)v^2\ .
\end{align}

\noindent
{\it Constraint equations:}
\begin{align}
\label{C_St}
        0 &= (\mathcal{C}_{\rm C})_3 =  \EEE \ptl_x \St 
-(r+3A-\sqrt{3}\Nc)\St
\\
\label{C_beta_app}
        0 &= (\mathcal{C}_\beta) =
        (\EEE \ptl_x - 2r)\Oml\ .
\end{align}

For simulations in the de Sitter asymptotic regime (where $\Oml-1$ and 
$q+1$ tend to zero), we use $\Oml-1$ and $q+1$ as variables to maintain 
numerical precision.

